\documentclass[a4paper, traditabstract,longauth]{aa} 
\usepackage{graphicx,amsmath}
\usepackage{epsf}
\usepackage{ifthen}
\usepackage[table,usenames,dvipsnames]{xcolor}
\usepackage{fixltx2e}

\usepackage{color}
\usepackage{epsfig}

\definecolor{linkcolor}{rgb}{0.6,0,0}
\definecolor{citecolor}{rgb}{0,0,0.75}
\definecolor{urlcolor}{rgb}{0.12,0.46,0.7}
\usepackage[breaklinks, colorlinks, urlcolor=urlcolor,citecolor=citecolor,linkcolor=linkcolor]{hyperref} 
\hypersetup{linktocpage}

\usepackage[nonamebreak]{natbib}

\usepackage[varg]{txfonts}

\usepackage{etex}
\reserveinserts{28}

\pdfoutput=1
\pdfmapfile{+txfonts.map}

\bibpunct{(}{)}{;}{a}{}{,} 

\defcitealias{planck2013-p09a}{PCNG13}
\defcitealias{planck2014-a19}{PCNG15}

\def\setsymbol#1#2{\expandafter\def\csname #1\endcsname{#2}}
\def\getsymbol#1{\csname #1\endcsname}

\def\Planck{\textit{Planck}}

\def\alltwentythirteenresultspapers{\nocite{planck2013-p01, planck2013-p02, planck2013-p02a, planck2013-p02d, planck2013-p02b, planck2013-p03, planck2013-p03c, planck2013-p03f, planck2013-p03d, planck2013-p03e, planck2013-p01a, planck2013-p06, planck2013-p03a, planck2013-pip88, planck2013-p08, planck2013-p11, planck2013-p12, planck2013-p13, planck2013-p14, planck2013-p15, planck2013-p05b, planck2013-p17, planck2013-p09, planck2013-p09a, planck2013-p20, planck2013-p19, planck2013-pipaberration, planck2013-p05, planck2013-p05a, planck2013-pip56, planck2013-p06b, planck2013-p01a}}

\newbox\tablebox    \newdimen\tablewidth
\def\leaderfil{\leaders\hbox to 5pt{\hss.\hss}\hfil}
\def\endPlancktable{\tablewidth=\columnwidth 
    $$\hss\copy\tablebox\hss$$
    \vskip-\lastskip\vskip -2pt}
\def\endPlancktablewide{\tablewidth=\textwidth 
    $$\hss\copy\tablebox\hss$$
    \vskip-\lastskip\vskip -2pt}
\def\tablenote#1 #2\par{\begingroup \parindent=0.8em
    \abovedisplayshortskip=0pt\belowdisplayshortskip=0pt
    \noindent
    $$\hss\vbox{\hsize\tablewidth \hangindent=\parindent \hangafter=1 \noindent
    \hbox to \parindent{$^#1$\hss}\strut#2\strut\par}\hss$$
    \endgroup}
\def\doubleline{\vskip 3pt\hrule \vskip 1.5pt \hrule \vskip 5pt}

\def\L2{\ifmmode L_2\else $L_2$\fi}

\def\DeltaT{\ifmmode \Delta T\else $\Delta T$\fi}
\def\deltat{\ifmmode \Delta t\else $\Delta t$\fi}
\def\fknee{\ifmmode f_{\rm knee}\else $f_{\rm knee}$\fi}
\def\Fmax{\ifmmode F_{\rm max}\else $F_{\rm max}$\fi}
\def\solar{\ifmmode{\rm M}_{\mathord\odot}\else${\rm M}_{\mathord\odot}$\fi}
\def\Msolar{\ifmmode{\rm M}_{\mathord\odot}\else${\rm M}_{\mathord\odot}$\fi}
\def\Lsolar{\ifmmode{\rm L}_{\mathord\odot}\else${\rm L}_{\mathord\odot}$\fi}
\def\inv{\ifmmode^{-1}\else$^{-1}$\fi}
\def\mo{\ifmmode^{-1}\else$^{-1}$\fi}
\def\sup#1{\ifmmode ^{\rm #1}\else $^{\rm #1}$\fi}
\def\expo#1{\ifmmode \times 10^{#1}\else $\times 10^{#1}$\fi}
\def\,{\thinspace}
\def\lsim{\mathrel{\raise .4ex\hbox{\rlap{$<$}\lower 1.2ex\hbox{$\sim$}}}}
\def\gsim{\mathrel{\raise .4ex\hbox{\rlap{$>$}\lower 1.2ex\hbox{$\sim$}}}}

\def\simprop{\mathrel{\raise .4ex\hbox{\rlap{$\propto$}\lower 1.2ex\hbox{$\sim$}}}}
\def\deg{\ifmmode^\circ\else$^\circ$\fi}
\def\pdeg{\ifmmode $\setbox0=\hbox{$^{\circ}$}\rlap{\hskip.11\wd0 .}$^{\circ}
          \else \setbox0=\hbox{$^{\circ}$}\rlap{\hskip.11\wd0 .}$^{\circ}$\fi}
\def\arcs{\ifmmode {^{\scriptstyle\prime\prime}}
          \else $^{\scriptstyle\prime\prime}$\fi}
\def\arcm{\ifmmode {^{\scriptstyle\prime}}
          \else $^{\scriptstyle\prime}$\fi}
\newdimen\sa  \newdimen\sb
\def\parcs{\sa=.07em \sb=.03em
     \ifmmode \hbox{\rlap{.}}^{\scriptstyle\prime\kern -\sb\prime}\hbox{\kern -\sa}
     \else \rlap{.}$^{\scriptstyle\prime\kern -\sb\prime}$\kern -\sa\fi}
\def\parcm{\sa=.08em \sb=.03em
     \ifmmode \hbox{\rlap{.}\kern\sa}^{\scriptstyle\prime}\hbox{\kern-\sb}
     \else \rlap{.}\kern\sa$^{\scriptstyle\prime}$\kern-\sb\fi}
\def\ra[#1 #2 #3.#4]{#1\sup{h}#2\sup{m}#3\sup{s}\llap.#4}
\def\dec[#1 #2 #3.#4]{#1\deg#2\arcm#3\arcs\llap.#4}
\def\deco[#1 #2 #3]{#1\deg#2\arcm#3\arcs}
\def\rra[#1 #2]{#1\sup{h}#2\sup{m}}

\def\dots{\relax\ifmmode \ldots\else $\ldots$\fi}
\def\WHzsr{\ifmmode $W\,Hz\mo\,sr\mo$\else W\,Hz\mo\,sr\mo\fi}
\def\mHz{\ifmmode $\,mHz$\else \,mHz\fi}
\def\GHz{\ifmmode $\,GHz$\else \,GHz\fi}
\def\mKs{\ifmmode $\,mK\,s$^{1/2}\else \,mK\,s$^{1/2}$\fi}
\def\muKs{\ifmmode \,\mu$K\,s$^{1/2}\else \,$\mu$K\,s$^{1/2}$\fi}
\def\muKRJs{\ifmmode \,\mu$K$_{\rm RJ}$\,s$^{1/2}\else \,$\mu$K$_{\rm RJ}$\,s$^{1/2}$\fi}
\def\muKHz{\ifmmode \,\mu$K\,Hz$^{-1/2}\else \,$\mu$K\,Hz$^{-1/2}$\fi}
\def\MJysr{\ifmmode \,$MJy\,sr\mo$\else \,MJy\,sr\mo\fi}
\def\MJysrmK{\ifmmode \,$MJy\,sr\mo$\,mK$_{\rm CMB}\mo\else \,MJy\,sr\mo\,mK$_{\rm CMB}\mo$\fi}
\def\microns{\ifmmode \,\mu$m$\else \,$\mu$m\fi}

\def\muK{\ifmmode \,\mu$K$\else \,$\mu$\hbox{K}\fi}
\def\microK{\ifmmode \,\mu$K$\else \,$\mu$\hbox{K}\fi}
\def\muW{\ifmmode \,\mu$W$\else \,$\mu$\hbox{W}\fi}
\def\kms{\ifmmode $\,km\,s$^{-1}\else \,km\,s$^{-1}$\fi}
\def\kmsMpc{\ifmmode $\,\kms\,Mpc\mo$\else \,\kms\,Mpc\mo\fi}
\providecommand{\sorthelp}[1]{}
 
\newcommand{\fnl}{f_{\rm{NL}}}

\newcommand{\gnl}{g_{\rm{NL}}}

\def\WMAP{{WMAP}}

\newcommand\ba{\begin{eqnarray}}
\newcommand\ea{\end{eqnarray}}
\newcommand\bea{\begin{eqnarray}}
\newcommand\eea{\end{eqnarray}}

\newcommand\be{\begin{equation}}
\newcommand\ee{\end{equation}}

\newcommand{\boldvec}[1]{{{\vec{#1}}}}

\newcommand{\vk}{\boldvec{k}}

\newcommand{\vx}{\boldvec{x}}

\newcommand{\itT}{\textit{T}}

\newcommand{\itE}{\textit{E}}

\newcommand{\itTpE}{\textit{T+E}}

\def\Commander{\texttt{Commander}}

\def\NILC{\texttt{NILC}}
\def\SMICA{\texttt{SMICA}}
\def\SEVEM{\texttt{SEVEM}}

\def\={\nonumber &=}
\def\nn{\nonumber}

\def\({\left(}
\def\){\right)}
\def\[{\left[}
\def\]{\right]}
\def\<{\left\langle}
\def\>{\right\rangle}

\def\curl{\mathcal}

\def\eq{\begin{eqnarray}}
\def\qe{\end{eqnarray}}

\def\and{\quad \mbox{and} \quad}

\def\fnl{f_\textrm{NL}}

\def\bfnl{\kern2pt\overline{\kern-2ptf}_\textrm{NL}}

\def\exp{\textrm{exp}}

\def\kall{k_1,k_2,k_3}

\def\barQ{\kern2pt\overline{\kern-2pt\curl{Q}}}

\def\bargamma{\kern2pt\overline{\kern-2pt\gamma}}
\def\barzeta{\kern2pt\overline{\kern-2pt\zeta}}

\def\barR{\kern2pt\overline{\kern-2pt\curl{R}}}

\def\klist{k_1,k_2,k_3}
\def\eqref#1{(\ref{#1})}

\def\leaderfi1{\leaders\hbox to 5pt{\hss.\hss}\hfil}

\def\setsize{\csname @setfontsize\endcsname \setsize}

\def\gnlloc{g_{\rm NL}^{\rm local}}
\def\gnldotpi4{g_{\rm NL}^{\dot\sigma^4}}
\def\gnldpi4{g_{\rm NL}^{(\partial\sigma)^4}}
\def\gnlB{g_{\rm NL}^{\dot\sigma^2 (\partial\sigma)^2}}

\begin{document}

\title{\vglue -10mm\textit{Planck} 2018 results. IX. Constraints on primordial non-Gaussianity}

\titlerunning{Constraints on primordial non-Gaussianity}
\authorrunning{Planck Collaboration}

\author{\small
Planck Collaboration: Y.~Akrami\inst{12, 51, 53}
\and
F.~Arroja\inst{55}
\and
M.~Ashdown\inst{60, 4}
\and
J.~Aumont\inst{89}
\and
C.~Baccigalupi\inst{73}
\and
M.~Ballardini\inst{19, 37}
\and
A.~J.~Banday\inst{89, 7}
\and
R.~B.~Barreiro\inst{56}
\and
N.~Bartolo\inst{25, 57}\thanks{Corresponding author: Nicola~Bartolo \url{nicola.bartolo@pd.infn.it}}
\and
S.~Basak\inst{79}
% ...
}
\institute{\small
AIM, CEA, CNRS, Universit\'{e} Paris-Saclay, Universit\'{e} Paris-Diderot, Sorbonne Paris Cit\'{e}, F-91191 Gif-sur-Yvette, France\goodbreak
\and
APC, AstroParticule et Cosmologie, Universit\'{e} Paris Diderot, CNRS/IN2P3, CEA/lrfu, Observatoire de Paris, Sorbonne Paris Cit\'{e}, 10, rue Alice Domon et L\'{e}onie Duquet, 75205 Paris Cedex 13, France\goodbreak
\and
African Institute for Mathematical Sciences, 6-8 Melrose Road, Muizenberg, Cape Town, South Africa\goodbreak
\and
Astrophysics Group, Cavendish Laboratory, University of Cambridge, J J Thomson Avenue, Cambridge CB3 0HE, U.K.\goodbreak
\and
Astrophysics \& Cosmology Research Unit, School of Mathematics, Statistics \& Computer Science, University of KwaZulu-Natal, Westville Campus, Private Bag X54001, Durban 4000, South Africa\goodbreak
\and
CITA, University of Toronto, 60 St. George St., Toronto, ON M5S 3H8, Canada\goodbreak
\and
CNRS, IRAP, 9 Av. colonel Roche, BP 44346, F-31028 Toulouse cedex 4, France\goodbreak
\and
California Institute of Technology, Pasadena, California, U.S.A.\goodbreak
\and
Centre for Theoretical Cosmology, DAMTP, University of Cambridge, Wilberforce Road, Cambridge CB3 0WA, U.K.\goodbreak
\and
Computational Cosmology Center, Lawrence Berkeley National Laboratory, Berkeley, California, U.S.A.\goodbreak
\and
D\'{e}partement de Physique Th\'{e}orique, Universit\'{e} de Gen\`{e}ve, 24, Quai E. Ansermet,1211 Gen\`{e}ve 4, Switzerland\goodbreak
\and
D\'{e}partement de Physique, \'{E}cole normale sup\'{e}rieure, PSL Research University, CNRS, 24 rue Lhomond, 75005 Paris, France\goodbreak
\and
Departamento de Astrof\'{i}sica, Universidad de La Laguna (ULL), E-38206 La Laguna, Tenerife, Spain\goodbreak
\and
Departamento de F\'{\i}sica, Universidad de Oviedo, C/ Federico Garc\'{\i}a Lorca, 18 , Oviedo, Spain\goodbreak
\and
Department of Astrophysics/IMAPP, Radboud University, P.O. Box 9010, 6500 GL Nijmegen, The Netherlands\goodbreak
\and
Department of General Education, National Institute of Technology, Kagawa College, 355 Chokushi-cho, Takamatsu, Kagawa 761-8058, Japan\goodbreak
\and
Department of Mathematics, University of Stellenbosch, Stellenbosch 7602, South Africa\goodbreak
\and
Department of Physics \& Astronomy, University of British Columbia, 6224 Agricultural Road, Vancouver, British Columbia, Canada\goodbreak
\and
Department of Physics \& Astronomy, University of the Western Cape, Cape Town 7535, South Africa\goodbreak
\and
Department of Physics and Astronomy, University of Sussex, Brighton BN1 9QH, U.K.\goodbreak
\and
Department of Physics, Gustaf H\"{a}llstr\"{o}min katu 2a, University of Helsinki, Helsinki, Finland\goodbreak
\and
Department of Physics, Princeton University, Princeton, New Jersey, U.S.A.\goodbreak
\and
Department of Physics, University of California, Santa Barbara, California, U.S.A.\goodbreak
\and
Department of Physics, University of Illinois at Urbana-Champaign, 1110 West Green Street, Urbana, Illinois, U.S.A.\goodbreak
\and
Dipartimento di Fisica e Astronomia G. Galilei, Universit\`{a} degli Studi di Padova, via Marzolo 8, 35131 Padova, Italy\goodbreak
\and
Dipartimento di Fisica e Scienze della Terra, Universit\`{a} di Ferrara, Via Saragat 1, 44122 Ferrara, Italy\goodbreak
\and
Dipartimento di Fisica, Universit\`{a} La Sapienza, P. le A. Moro 2, Roma, Italy\goodbreak
\and
Dipartimento di Fisica, Universit\`{a} degli Studi di Milano, Via Celoria, 16, Milano, Italy\goodbreak
\and
Dipartimento di Fisica, Universit\`{a} degli Studi di Trieste, via A. Valerio 2, Trieste, Italy\goodbreak
\and
Dipartimento di Fisica, Universit\`{a} di Roma Tor Vergata, Via della Ricerca Scientifica, 1, Roma, Italy\goodbreak
\and
European Space Agency, ESAC, Planck Science Office, Camino bajo del Castillo, s/n, Urbanizaci\'{o}n Villafranca del Castillo, Villanueva de la Ca\~{n}ada, Madrid, Spain\goodbreak
\and
European Space Agency, ESTEC, Keplerlaan 1, 2201 AZ Noordwijk, The Netherlands\goodbreak
\and
Gran Sasso Science Institute, INFN, viale F. Crispi 7, 67100 L'Aquila, Italy\goodbreak
\and
HEP Division, Argonne National Laboratory, Lemont, IL 60439, USA\goodbreak
\and
Haverford College Astronomy Department, 370 Lancaster Avenue, Haverford, Pennsylvania, U.S.A.\goodbreak
\and
Helsinki Institute of Physics, Gustaf H\"{a}llstr\"{o}min katu 2, University of Helsinki, Helsinki, Finland\goodbreak
\and
INAF - OAS Bologna, Istituto Nazionale di Astrofisica - Osservatorio di Astrofisica e Scienza dello Spazio di Bologna, Area della Ricerca del CNR, Via Gobetti 101, 40129, Bologna, Italy\goodbreak
\and
INAF - Osservatorio Astronomico di Padova, Vicolo dell'Osservatorio 5, Padova, Italy\goodbreak
\and
INAF - Osservatorio Astronomico di Trieste, Via G.B. Tiepolo 11, Trieste, Italy\goodbreak
\and
INAF, Istituto di Radioastronomia, Via Piero Gobetti 101, I-40129 Bologna, Italy\goodbreak
\and
INAF/IASF Milano, Via E. Bassini 15, Milano, Italy\goodbreak
\and
INFN - CNAF, viale Berti Pichat 6/2, 40127 Bologna, Italy\goodbreak
\and
INFN, Sezione di Bologna, viale Berti Pichat 6/2, 40127 Bologna, Italy\goodbreak
\and
INFN, Sezione di Ferrara, Via Saragat 1, 44122 Ferrara, Italy\goodbreak
\and
INFN, Sezione di Milano, Via Celoria 16, Milano, Italy\goodbreak
\and
INFN, Sezione di Roma 1, Universit\`{a} di Roma Sapienza, Piazzale Aldo Moro 2, 00185, Roma, Italy\goodbreak
\and
INFN, Sezione di Roma 2, Universit\`{a} di Roma Tor Vergata, Via della Ricerca Scientifica, 1, Roma, Italy\goodbreak
\and
Imperial College London, Astrophysics group, Blackett Laboratory, Prince Consort Road, London, SW7 2AZ, U.K.\goodbreak
\and
Institut d'Astrophysique Spatiale, CNRS, Univ. Paris-Sud, Universit\'{e} Paris-Saclay, B\^{a}t. 121, 91405 Orsay cedex, France\goodbreak
\and
Institut d'Astrophysique de Paris, CNRS (UMR7095), 98 bis Boulevard Arago, F-75014, Paris, France\goodbreak
\and
Institute Lorentz, Leiden University, PO Box 9506, Leiden 2300 RA, The Netherlands\goodbreak
\and
Institute of Astronomy, University of Cambridge, Madingley Road, Cambridge CB3 0HA, U.K.\goodbreak
\and
Institute of Theoretical Astrophysics, University of Oslo, Blindern, Oslo, Norway\goodbreak
\and
Instituto de Astrof\'{\i}sica de Canarias, C/V\'{\i}a L\'{a}ctea s/n, La Laguna, Tenerife, Spain\goodbreak
\and
Instituto de Astrof\'{\i}sica e Ci\^{e}ncias do Espa\c{c}o, Faculdade de Ci\^{e}ncias da Universidade de Lisboa, Campo Grande, PT1749-016 Lisboa, Portugal\goodbreak
\and
Instituto de F\'{\i}sica de Cantabria (CSIC-Universidad de Cantabria), Avda. de los Castros s/n, Santander, Spain\goodbreak
\and
Istituto Nazionale di Fisica Nucleare, Sezione di Padova, via Marzolo 8, I-35131 Padova, Italy\goodbreak
\and
Jet Propulsion Laboratory, California Institute of Technology, 4800 Oak Grove Drive, Pasadena, California, U.S.A.\goodbreak
\and
Jodrell Bank Centre for Astrophysics, Alan Turing Building, School of Physics and Astronomy, The University of Manchester, Oxford Road, Manchester, M13 9PL, U.K.\goodbreak
\and
Kavli Institute for Cosmology Cambridge, Madingley Road, Cambridge, CB3 0HA, U.K.\goodbreak
\and
Kavli Institute for the Physics and Mathematics of the Universe (Kavli IPMU, WPI), UTIAS, The University of Tokyo, Chiba, 277- 8583, Japan\goodbreak
\and
Laboratoire d'Oc{\'e}anographie Physique et Spatiale (LOPS), Univ. Brest, CNRS, Ifremer, IRD, Brest, France\goodbreak
\and
Laboratoire de Physique Subatomique et Cosmologie, Universit\'{e} Grenoble-Alpes, CNRS/IN2P3, 53, rue des Martyrs, 38026 Grenoble Cedex, France\goodbreak
\and
Laboratoire de Physique Th\'{e}orique, Universit\'{e} Paris-Sud 11 \& CNRS, B\^{a}timent 210, 91405 Orsay, France\goodbreak
\and
Low Temperature Laboratory, Department of Applied Physics, Aalto University, Espoo, FI-00076 AALTO, Finland\goodbreak
\and
Max-Planck-Institut f\"{u}r Astrophysik, Karl-Schwarzschild-Str. 1, 85741 Garching, Germany\goodbreak
\and
Mullard Space Science Laboratory, University College London, Surrey RH5 6NT, U.K.\goodbreak
\and
NAOC-UKZN Computational Astrophysics Centre (NUCAC), University of KwaZulu-Natal, Durban 4000, South Africa\goodbreak
\and
National Centre for Nuclear Research, ul. A. Soltana 7, 05-400 Otwock, Poland\goodbreak
\and
Nicolaus Copernicus Astronomical Center, Polish Academy of Sciences, Bartycka 18, 00-716 Warsaw, Poland\goodbreak
\and
Perimeter Institute for Theoretical Physics, Waterloo ON N2L 2Y5, Canada\goodbreak
\and
Purple Mountain Observatory, No. 8 Yuan Hua Road, 210034 Nanjing, China\goodbreak
\and
SISSA, Astrophysics Sector, via Bonomea 265, 34136, Trieste, Italy\goodbreak
\and
San Diego Supercomputer Center, University of California, San Diego, 9500 Gilman Drive, La Jolla, CA 92093, USA\goodbreak
\and
School of Chemistry and Physics, University of KwaZulu-Natal, Westville Campus, Private Bag X54001, Durban, 4000, South Africa\goodbreak
\and
School of Physics and Astronomy, Cardiff University, Queens Buildings, The Parade, Cardiff, CF24 3AA, U.K.\goodbreak
\and
School of Physics and Astronomy, Sun Yat-sen University, 2 Daxue Rd, Tangjia, Zhuhai, China\goodbreak
\and
School of Physics and Astronomy, University of Nottingham, Nottingham NG7 2RD, U.K.\goodbreak
\and
School of Physics, Indian Institute of Science Education and Research Thiruvananthapuram, Maruthamala PO, Vithura, Thiruvananthapuram 695551, Kerala, India\goodbreak
\and
School of Physics, The University of New South Wales, Sydney NSW 2052, Australia\goodbreak
\and
Simon Fraser University, Department of Physics, 8888 University Drive, Burnaby BC, Canada\goodbreak
\and
Sorbonne Universit\'{e}, Observatoire de Paris, Universit\'{e} PSL, \'{E}cole normale sup\'{e}rieure, CNRS, LERMA, F-75005, Paris, France\goodbreak
\and
Sorbonne Universit\'{e}-UPMC, UMR7095, Institut d'Astrophysique de Paris, 98 bis Boulevard Arago, F-75014, Paris, France\goodbreak
\and
Space Research Institute (IKI), Russian Academy of Sciences, Profsoyuznaya Str, 84/32, Moscow, 117997, Russia\goodbreak
\and
Space Science Data Center - Agenzia Spaziale Italiana, Via del Politecnico snc, 00133, Roma, Italy\goodbreak
\and
Space Sciences Laboratory, University of California, Berkeley, California, U.S.A.\goodbreak
\and
The Oskar Klein Centre for Cosmoparticle Physics, Department of Physics, Stockholm University, AlbaNova, SE-106 91 Stockholm, Sweden\goodbreak
\and
UPMC Univ Paris 06, UMR7095, 98 bis Boulevard Arago, F-75014, Paris, France\goodbreak
\and
Universit\'{e} de Toulouse, UPS-OMP, IRAP, F-31028 Toulouse cedex 4, France\goodbreak
% ...
}

\date{Received xxxx, Accepted xxxxx}


\abstract{\vglue -3mm
We analyse the \Planck\ full-mission cosmic microwave background (CMB)
temperature and $E$-mode polarization maps to obtain constraints
on primordial non-Gaussianity (NG). We compare estimates obtained
from separable template-fitting, binned, and optimal modal
bispectrum estimators, finding consistent values for the
local, equilateral, and orthogonal bispectrum amplitudes. Our combined
temperature and polarization analysis produces the following final
results: $f_{\rm NL}^{\rm local} = -0.9 \pm 5.1$;
$f_{\rm NL}^{\rm equil} = -26 \pm 47$;
and $f_{\rm NL}^{\rm ortho} = - 38 \pm 24$
(68\,\%~CL, statistical).  These results include the low-multipole
($4 \leq \ell < 40$) polarization data, not included in our
previous analysis, pass an extensive battery of tests (with
additional tests regarding foreground residuals compared to 2015), and are
stable with respect to our 2015 measurements (with small fluctuations, at
the level of a fraction of a standard deviation, consistent with
changes in data processing).  Polarization-only bispectra display a
significant improvement in robustness; they can now be used
independently to set primordial NG constraints with a sensitivity
comparable to \WMAP\ temperature-based results, and giving excellent
agreement.  In addition to the analysis of the standard local, equilateral, and
orthogonal bispectrum shapes, we consider a large number of additional
cases, such as scale-dependent feature and resonance bispectra,
isocurvature primordial NG, and parity-breaking models, where we also place
tight constraints
but do not detect any signal. The non-primordial lensing bispectrum
is, however, detected with an improved significance compared to 2015,
excluding the null hypothesis at $3.5\,\sigma$. Beyond
estimates of individual shape amplitudes, we also present
model-independent reconstructions and analyses of the \Planck\ CMB
bispectrum. Our final constraint on the local primordial trispectrum shape is
$g_{\rm NL}^{\rm local} = (-5.8 \pm 6.5) \times 10^4$ (68\,\%~CL, statistical),
while constraints for other trispectrum shapes are also determined.
Exploiting the tight limits on various bispectrum and trispectrum shapes, we 
constrain the parameter space of different early-Universe scenarios that
generate primordial NG, including general single-field models of inflation,
multi-field models (e.g., curvaton models), models of inflation with axion
fields producing parity-violation bispectra in the tensor sector, and
inflationary models involving vector-like fields with directionally-dependent
bispectra. 
Our results provide a high-precision test for structure-formation scenarios,
showing complete agreement with the basic picture of the $\Lambda$CDM
cosmology regarding the statistics of the initial conditions,
with cosmic structures arising from 
adiabatic, passive, Gaussian, and primordial seed perturbations.}


\maketitle

\tableofcontents

\section {Introduction}
\label{sec:intro}

This paper, one of a set associated with the 2018 release (also known as ``PR3'') of data from the 
\Planck\footnote{\Planck\ (\url{http://www.esa.int/Planck}) is a project of the European Space Agency  (ESA) with instruments provided by two scientific consortia funded by ESA member states and led by Principal Investigators from France and Italy, telescope reflectors provided through a collaboration between ESA and a scientific consortium led and funded by Denmark, and additional contributions from NASA (USA).} 
mission \citep{planck2016-l01}, 
\alltwentythirteenresultspapers\ presents the data analysis and constraints on primordial 
non-Gaussianity (NG) obtained using the Legacy {\it Planck} cosmic microwave background (CMB) 
maps. It also includes some implications for inflationary models driven by the 2018 NG constraints. 
This paper updates the earlier study based on the temperature data from the
nominal \Planck\ operations period, including the first 14 months of
observations \citep[hereafter \citetalias{planck2013-p09a}]{planck2013-p09a},
and a later study that used temperature data and a first set of polarization
maps from the full \Planck\ mission---29 and 52 months of observations for the
HFI (High Frequency Instrument) and LFI (Low Frequency Instrument),
respectively  \citep[hereafter \citetalias{planck2014-a19}]{planck2014-a19}.
The analysis described in this paper sets the most stringent constraints on primordial NG to date, which are near what is ultimately possible from using only CMB temperature data. The results of this paper are mainly based on the measurements of the CMB angular bispectrum, complemented with the next higher-order NG correlation function, i.e., the trispectrum. For notations and conventions relating to (primordial) bispectra and trispectra we refer the reader to the two previous {\it Planck} papers on primordial NG \citepalias{planck2013-p09a,planck2014-a19}.
This paper also complements the precise characterization of inflationary models \citep{planck2016-l10} and cosmological parameters \citep{planck2016-l06}, with specific statistical estimators that go beyond the constraints on primordial power spectra.  
It also complements the statistical and isotropy tests on CMB anisotropies of \cite{planck2016-l07}, focusing on the interpretation of specific, well motivated, non-Gaussian models of inflation.  
These models span from the irreducible minimal amount of primordial NG predicted by standard single-field models of slow-roll inflation, to various classes of inflationary models that constitute the prototypes of 
extensions of the standard inflationary picture and of physically motivated mechanisms able to generate a higher level of primordial NG measurable in the CMB anisotropies. 
This work establishes the most robust constraints on some of the most well-known and studied types of primordial NG, namely the local, equilateral, and orthogonal shapes. Moreover, this 2018 
analysis includes a better characterization of the constraints coming from CMB polarization data.  Besides focusing on these major goals, we re-analyse a variety of other NG signals, investigating also some new aspects of primordial NG. For example, we perform for the first time an analysis of the running of NG using \Planck\ data in the context of some well defined inflationary models.  Additionally, we constrain primordial NG predicted by theoretical scenarios on which much attention has been focused recently, such as,
bispectrum NG generated in the tensor (gravitational wave) sector. For a detailed analysis of oscillatory features that combines power spectrum and bispectrum constraints see \cite{planck2016-l10}. As in the last data release (``PR2''), as well as extracting the constraints on NG amplitudes for specific shapes, we also provide a model-independent reconstruction of the CMB angular bispectrum by using various methods. Such a reconstruction can help in pinning down interesting features in the CMB bispectrum signal beyond those captured by existing parameterizations.

The paper is organized as follows. In Sect.~\ref{sec:models} we recall the main primordial NG models tested in this paper. Section~\ref{sec:SEP} briefly describes the bispectrum estimators that we use, as well as details of
the data set and our analysis procedures.
In Sect.~\ref{sec:npNG} we discuss detectable non-primordial contributions to the CMB bispectrum, namely those arising from lensing and point sources.
In Sect.~\ref{sec:Results} we constrain $f_{\rm NL}$ for the local, equilateral, and orthogonal bispectra. We also report the results for scale-dependent NG models and other selected bispectrum shapes, including NG in the tensor (primordial gravitational wave) sector; in this section reconstructions and model-independent analyses of the CMB bispectrum are also provided. In Sect.~\ref{sec:Sec_valid_data} these results are validated through a series of null tests on the data, with the goal of assessing the robustness of the results. This includes in particular a first analysis of Galactic dust and thermal SZ residuals. \Planck\  CMB trispectrum limits are obtained and discussed in Sect.~\ref{sec:tau_gnl}.  In Sect.~\ref{sec:Implications} we derive the main implications of \Planck's constraints on primordial NG for some specific early Universe models. We conclude in Sect.~\ref{sec:Conc}.

\section {Models}
\label{sec:models}


Primordial NG comes in with a variety of {\it shapes}, corresponding to well motivated {\it classes\/} of inflationary model. For each class, a common physical mechanism is responsible for the generation of the corresponding type of primordial NG. Below we briefly summarize the main types of primordial NG that are constrained in this paper, providing the precise shapes that are used for data analysis. For more details about specific realizations of inflationary models within each class, see the previous two \Planck\ papers on primordial NG \citepalias{planck2013-p09a,planck2014-a19} and reviews \citep[e.g.,][]{Bartolo:2004if,2010AdAst2010E..73L,2010AdAst2010E..72C,2010CQGra..27l4010K,2010AdAst2010E..71Y}. We give a more expanded description only of those shapes of primordial NG analysed here for the first time with \Planck\ data (e.g., running of primordial NG).

\subsection{General single-field models of inflation}
The parameter space of single-field models is well described by the so called equilateral and orthogonal templates \citep{2006JCAP...05..004C,2007JCAP...01..002C,2010JCAP...01..028S}. 
The equilateral shape is
\begin{align}
\label{equilateralBis}
\nonumber
& B_{\Phi}^{\rm equil}(k_1,k_2,k_3)= 6A^2 f_{\rm NL}^{\rm equil}\\
\nonumber
& \times \left\{
-\frac1{k^{4-n_{\rm s}}_1k^{4-n_{\rm s}}_2}-\frac1{k^{4-n_{\rm s}}_2k^{4-n_{\rm s}}_3}
-\frac1{k^{4-n_{\rm s}}_3k^{4-n_{\rm s}}_1} 
-\frac2{(k_1k_2k_3)^{2(4-n_{\rm s})/3}}
\right. \\
& \qquad\qquad\qquad
\left.+\left[\frac1{k^{(4-n_{\rm s})/3}_1k^{2(4-n_{\rm s})/3}_2k^{4-n_{\rm s}}_3}
+\mbox{5 perms.}\right]\right\}\, , 
\end{align}
while the orthogonal NG is described by
\begin{align}\label{orthogonalBis}
\nonumber
& B^{\rm ortho}_{\Phi}(k_1,k_2,k_3)= 6A^2 f_{\rm NL}^{\rm ortho}\\
\nonumber
& \times \left\{
-\frac3{k^{4-n_{\rm s}}_1k^{4-n_{\rm s}}_2}-\frac3{k^{4-n_{\rm s}}_2k^{4-n_{\rm s}}_3}
-\frac3{k^{4-n_{\rm s}}_3k^{4-n_{\rm s}}_1}\right.
-\frac8{(k_1k_2k_3)^{2(4-n_{\rm s})/3}}
\\
& \qquad\qquad\qquad
\left.+\left[\frac3{k^{(4-n_{\rm s})/3}_1k^{2(4-n_{\rm s})/3}_2k^{4-n_{\rm s}}_3}
+\mbox{5 perms.}\right]\right\}\, .
\end{align}
Here the potential $\Phi$ is defined in relation to the comoving curvature perturbation $\zeta$ by $\Phi \equiv (3/5) \zeta$ on superhorizon scales (thus corresponding to Bardeen's gauge-invariant gravitational 
potential \citep{1980PhRvD..22.1882B} during matter domination on superhorizon scales). $P_{\Phi}(k)=A/k^{4-n_{\rm s}}$ is the Bardeen gravitational potential power spectrum, with normalization $A$ and scalar spectral index $n_{\rm s}$.
A typical example of this class is provided by models of inflation where there is a single scalar field driving inflation and generating the primordial perturbations, characterized by a non-standard kinetic term or 
more general higher-derivative interactions. In the first case the inflaton Lagrangian is ${\mathcal L}=P(X, \phi)$, where $X=g^{\mu \nu} \partial_\mu \phi \, \partial_\nu \phi$, with at most one derivative on $\phi$ \citep{2007JCAP...01..002C}. Different higher-derivative interactions of the inflaton field characterize, ghost inflation \citep{2004JCAP...04..001A} or models of inflation based on Galileon symmetry 
\citep[e.g.,][]{2011JCAP...01..014B}. The two amplitudes $f_{\rm NL}^{\rm equil}$ and $f_{\rm NL}^{\rm ortho}$ usually depend on the sound speed $c_{\rm s}$ at which the inflaton field fluctuations propagate and on a second independent amplitude measuring the inflaton self-interactions. The Dirac-Born-Infeld (DBI) models of inflation \citep{2004PhRvD..70j3505S,2004PhRvD..70l3505A} are a string-theory-motivated example of the $P(X, \phi)$ models, predicting an almost equilateral type NG with $f^{\rm equil}_{\rm NL} \propto c_{\rm s}^{-2}$ for $c_{\rm s} \ll 1$. 
More generally, the effective field theory (EFT) approach to inflationary perturbations \citep{2008JHEP...03..014C,2010JCAP...01..028S,2010JCAP...08..008B} yields NG shapes that can be mapped into the equilateral and orthogonal template basis. The EFT approach allows us to draw generic conclusions about single-field inflation. We will discuss them using one example in Sect.~\ref{sec:Implications}. Nevertheless, we shall also explicitly search for such EFT shapes, analysing their exact non-separable predicted shapes, $B^{\rm EFT1}$ and $B^{\rm EFT2}$, along with those of DBI, $B^{\rm DBI}$, and ghost inflation, $B^{\rm ghost}$ \citep{2004JCAP...04..001A}.


\subsection{Multi-field models}
\label{Sec:Multifield_models}
The bispectrum for multi-field models is typically of the local type\footnote{See, e.g., \citet{2010AdAst2010E..76B} for a review on this type of model in the context of primordial NG. Early papers discussing primordial local bispectra given by Eq.~(\ref{localBis}) include \citet{1993ApJ...403L...1F}, \citet{1994ApJ...430..447G}, \citet{2000MNRAS.313..323G}, \citet{2000MNRAS.313..141V}, \citet{2000PhRvD..61f3504W}, and \citet{2001PhRvD..63f3002K}.}
\begin{align}
\label{localBis}
B^{\rm local}_{\Phi}(k_1,k_2,k_3)&=2 f^{\rm local}_{\rm NL}
 \Big[ P_{\Phi}(k_1) P_{\Phi}(k_2) \nonumber\\
& \qquad\qquad +P_{\Phi}(k_1) P_{\Phi}(k_3) +P_{\Phi}(k_2) P_{\Phi}(k_3) \Big]
 \nonumber \\
&= 2 A^2 f^{\rm local}_{\rm NL} \left[ \frac{1}{k_1^{4-n_{\rm s}}k_2^{4-n_{\rm s}}} +{\rm cycl.} \right]\, .
\end{align}
This usually arises when more scalar fields drive inflation and give rise to the primordial curvature perturbation (``multiple-field inflation''), or when extra light scalar fields, different from the inflaton field driving inflation, 
determine (or contribute to) the final curvature perturbation. In these models initial isocurvature perturbations are transferred on super-horizon scales to the curvature perturbations. Non-Gaussianities if present are transferred too. This, along with nonlinearities in the transfer mechanism itself, is a potential source of significant NG (\citealt{2002PhRvD..65j3505B,2002PhRvD..66j3506B,2006JCAP...05..019V,2006PhRvD..73h3522R,2007PhRvD..76h3512R,2005PhRvL..95l1302L,Tzavara:2010ge,Jung:2016kfd}). The bispectrum of Eq.~(\ref{localBis}) mainly correlates large- with small-scale modes, peaking in the ``squeezed'' configurations 
$k_1 \ll k_2\approx k_3$. This is a consequence of the transfer mechanism taking place on superhorizon scales and thus generating a localized point-by-point primordial NG in real space. 
The curvaton model \citep{1990PhRvD..42..313M,1997PhRvD..56..535L,2002NuPhB.626..395E,2002PhLB..524....5L,2001PhLB..522..215M} is a clear example where local NG is generated in this way \citep[e.g.,][]{2002PhLB..524....5L,2003PhRvD..67b3503L,2004PhRvD..69d3503B}.
In the minimal adiabatic curvaton scenario $f_{\rm NL}^{\rm local}=(5/4r_\mathrm{D})-5r_\mathrm{D}/6-5/3$ \citep{2004PhRvD..69d3503B,2004PhRvL..93w1301B}, in the case when the curvaton field potential is purely quadratic \citep{2002PhLB..524....5L,2003PhRvD..67b3503L,2005PhRvL..95l1302L,2006JCAP...09..008M,2006PhRvD..74j3003S}. 
Here $r_{\rm D}=[3\rho_{\rm curv}/(3 \rho_{\rm curv}+4\rho_{\rm rad})]_{\rm D}$ represents the ``curvaton decay fraction'' at the epoch of the curvaton decay, employing the sudden decay approximation. 
Significant NG can be produced \citep{2004PhRvD..69d3503B,2004PhRvL..93w1301B} for low values of $r_{\rm D}$; a different modelling of the curvaton scenario has been discussed by \citet{2006JCAP...04..009L} and \citet{2006PhRvD..74j3003S}. We update the limits on both models in Sect.~\ref{sec:Implications}, using the local NG constraints. More general models with a curvaton-like spectator field have also been intensively investigated recently \citep[see, e.g.,][]{2018PhRvD..98f3525T}. Notice that through a similar mechanism to the curvaton mechanism, local bispectra can be generated from nonlinear dynamics during the preheating and reheating phases \citep{2005PhRvL..94p1301E,2008PhRvL.100d1302C,2006PhRvD..73j6012B,2009PhRvL.103g1301B} or due to fluctuations in the decay rate or interactions of the inflaton field, as realized in modulated (p)reheating and modulated hybrid inflationary models \citep{2003astro.ph..3614K,2004PhRvD..69h3505D,2004PhRvD..69b3505D,2004PhRvD..70h3004B,Zaldarriaga:2003my,2005JCAP...11..006L,2005PhRvD..72l3516S,2006PhRvL..97l1301L,2006PhRvD..73b3522K,2012JCAP...05..039C}.  We will also explore whether there is any evidence for dissipative effects during warm inflation, with a signal which changes sign in the squeezed limit \citep[see e.g.,][]{2014JCAP...12..008B}.


\subsection{Isocurvature non-Gaussianity}
\label{Sec:isocurv_NG_intro}

In most of the models mentioned in this section the focus is on
primordial NG in the adiabatic curvature perturbation~$\zeta$. However,
in inflationary scenarios with multiple scalar fields, isocurvature
perturbation modes can be produced as well.
If they survive until recombination, these
will then contribute not only to the power
spectrum, but also to the bispectrum, producing in general both a pure
isocurvature bispectrum and mixed bispectra because of the cross-correlation
between isocurvature and adiabatic perturbations \citep{2002astro.ph..6039K,
2002PhRvD..65j3505B,Komatsu:2003iq,2008JCAP...11..019K,Langlois:2008vk,
Kawasaki:2008pa,Hikage:2008sk,2011JCAP...01..008L,Langlois:2011hn,
2012JCAP...07..037K,Langlois:2012tm,2013JCAP...07..007H,2013JCAP...03..020H}.

In the context of the $\Lambda$CDM cosmology, there are at the time of
recombination four possible
distinct isocurvature modes (in addition to the adiabatic mode), namely the
cold-dark-matter (CDM) density, baryon-density, neutrino-density, and 
neutrino-velocity isocurvature modes \citep{2000PhRvD..62h3508B}.
However, the baryon isocurvature mode behaves identically to the CDM
isocurvature mode, once rescaled by factors of $\Omega_{\rm b}/\Omega_{\rm c}$,
so we will only consider the other three isocurvature modes in this paper.
Moreover, we will only investigate isocurvature NG of the {\em local} type,
since this is the most relevant case in multi-field inflation models, which
we require in order to produce isocurvature modes. We will also
limit ourselves to studying just one type of isocurvature mode
(considering each of the three types separately) together with the adiabatic
mode, to avoid the number of free parameters becoming so large
that no meaningful limits can be derived. Finally, for simplicity we assume
the same spectral index for the primordial isocurvature power spectrum and the
adiabatic-isocurvature cross-power spectrum as for
the adiabatic power spectrum, again to reduce the number of free parameters.
As shown by \citet{Langlois:2011hn}, under these assumptions we have in
general six independent $f_\mathrm{NL}$ parameters: the usual purely adiabatic
one; a purely isocurvature one; and four correlated ones. 

The primordial isocurvature bispectrum templates are a generalization of the
local shape in Eq.~(\ref{localBis}):
\begin{align}
B^{IJK}(k_1, k_2, k_3) = &
\,2 f_{\rm NL}^{I, JK} P_\Phi(k_2) P_\Phi(k_3) 
+ 2 f_{\rm NL}^{J, KI} P_\Phi(k_1) P_\Phi(k_3) \nonumber\\
& \qquad\qquad + 2 f_{\rm NL}^{K, IJ} P_\Phi(k_1)P_\Phi(k_2), 
\label{localBis_isocurv}
\end{align}
where $I,J,K$ label the different adiabatic and isocurvature modes.
The invariance under the simultaneous exchange of two of these indices
and the corresponding momenta means that $f_{\rm NL}^{I, JK} = f_{\rm NL}^{I, KJ}$,
which reduces the number of independent parameters from eight to six in the
case of two modes (and explains the notation with the comma).
The different CMB bispectrum templates derived from these primordial shapes
vary most importantly in the different types of radiation transfer functions
that they contain. For more details, see in particular \citet{Langlois:2012tm}.

An important final remark is that, unlike
the case of the purely adiabatic mode, polarization improves 
the constraints on the isocurvature NG significantly, up to a factor of about
6 as predicted by \citet{Langlois:2011hn,Langlois:2012tm} and confirmed by
the 2015 \Planck\ analysis \citepalias{planck2014-a19}. The reason for this
is that while
the isocurvature temperature power spectrum (to which the local bispectrum
is proportional) becomes very quickly negligible compared to the adiabatic
one as $\ell$ increases (already around $\ell\,{\approx}\,50$ for CDM), the
isocurvature polarization power spectrum remains comparable to the adiabatic
one to much smaller scales (up to $\ell\,{\approx}\,200$ for CDM). Hence there
are many more polarization modes than temperature modes that are relevant for
determining these isocurvature $\fnl$ parameters. For more details, again
see \citet{Langlois:2012tm}.


\subsection{Running non-Gaussianity}
\label{Sec:Running}

We briefly describe inflationary models that predict a mildly scale-dependent bispectrum, which is also known in the literature as the running of the bispectrum \citep[see e.g.,][]{2005PhRvD..72l3518C,2006PhRvD..73d3505L,2010JCAP...04..024K,2010JCAP...02..034B,2010JCAP...10..004B,2011JCAP...03..017S}. In inflationary models this running is as natural as the running of the power spectrum, i.e., the spectral index $n_{\rm s}$. Other models with strong scale dependence, e.g., oscillatory models, will be discussed in Sect.~\ref{oscillatory}. Further possibilities for strong scale dependence exist \citep[see e.g.,][]{2009JCAP...07..026K,2011PhRvD..83d1301R}, but we will not consider these in our study. The simplest model (single-field slow-roll, with canonical action and initial conditions) predicts that both the amplitude and scale dependence are of the order of the slow-roll parameters (this is true except in some very particular models, see, e.g.,~\cite{2013EL....10259001C}), i.e., they are small and currently not observable. However, other elaborate but theoretically well-motivated models make different predictions and these can be used to confirm or exclude such models. Measuring the running of the non-Gaussianity parameters with scale is important because this running carries information about, for instance, the number of inflationary fields and their interactions. This information may not be accessible with the power spectrum alone.
The first constraints on the running of a local model were obtained with WMAP7 data in \cite{2012PhRvL.109l1302B}. Forecasts of what would be feasible with future data were performed by for instance \citet{2008JCAP...04..014L}, \citet{2009JCAP...12..022S}, \citet{2011JCAP...01..006B}, \citet{2012MNRAS.422.2854G}, and \citet{2012JCAP...12..034B}. 

\subsubsection{Local-type scale-dependent bispectrum \label{subsubsec:Local}}

We start by describing models with a local-type mildly scale-dependent bispectrum.
Assuming that there are multiple scalar fields during inflation with canonical kinetic terms, that their correlators are Gaussian at horizon crossing, and using the slow-roll approximation and the $\delta N$ formalism, \citet{2010JCAP...10..004B} found a quite general expression for the power spectrum of the primordial potential perturbation:
\begin{equation}
P_{\Phi}(k)=\frac{2\pi^2}{k^3}\mathcal{P}_{\Phi}(k)=\frac{2\pi^2}{k^3}\sum_{ab}\mathcal{P}_{ab}(k),
\end{equation}
where the indexes $a,b$ run over the different scalar fields. 

The nonlinearity parameter then reads \citep{2010JCAP...10..004B}
\begin{align}
f_{\rm NL}(k_1,k_2,k_3)&=\frac{B_{\Phi}(k_1,k_2,k_3)}{2 \left[P_{\Phi}(k_1)P_{\Phi}(k_2)+2\,\mathrm{perms.}\right]}\nonumber\\
&=\frac{\sum_{abcd}(k_1k_2)^{-3}\mathcal{P}_{ac}(k_1)\mathcal{P}_{bd}(k_2)f_{cd}(k_3)+2\,\mathrm{perms.}}{(k_1k_2)^{-3}\mathcal{P}(k_1)\mathcal{P}(k_2)+2\,\mathrm{perms.}}\,,\nonumber\\
\end{align}
where the last line is the general result valid for any number of slow-roll fields. The functions $f_{cd}$ (as well as the functions $\mathcal{P}_{ab}$) can be parameterized as power laws.
In the general case, $f_\mathrm{NL}$ can also be written as
\begin{equation}
f_{\rm NL}(k_1,k_2,k_3)=\sum_{ab}f_{\rm NL}^{ab}\frac{(k_1k_2)^{n_{{\rm multi},a}}k_3^{3+n_{f,ab}}+2\,\mathrm{perms.}}{k_1^3+k_2^3+k_3^3}\,,
\end{equation}
where $n_{{\rm multi},a}$ and $n_{f,ab}$ are parameters of the models that are proportional to the slow-roll parameters.
It is clear that in the general case there are too many parameters to be constrained.  Instead we will consider two simpler cases, which will be among the three models of running non-Gaussianity that will be analysed in 
Sect.~\ref{running_analysis}. 

Firstly, when the curvature perturbation originates from only one of the scalar fields (e.g., as in the simplest curvaton scenario) the bispectrum simplifies to \citep{2010JCAP...10..004B}
\begin{equation}
\label{running1}
B_\Phi(k_1,k_2,k_3)\propto (k_1k_2)^{n_{\rm s}-4} k_3^{n_{\rm NG}}+2\,\mathrm{perms.}
\end{equation}
In this case 
\begin{equation}
f_{\rm NL}(k_1,k_2,k_3)=f^p_{\rm NL}\frac{k_1^{3+n_{\rm NG}}+k_2^{3+n_{\rm NG}}+k_3^{3+n_{\rm NG}}}{k_1^3+k_2^3+k_3^3},
\label{singlesource}
\end{equation}
where $n_{\rm NG}$ is the \textit{running parameter} which is sensitive to the third derivative of the potential. If the field producing the perturbations is not the inflaton field but an isocurvature field subdominant during inflation, then neither the spectral index measurement nor the running of the spectral index are sensitive to the third derivative. Therefore, those self-interactions can uniquely be probed by the running of $f_{\rm NL}$. 

The second class of models are two-field models where both fields contribute to the generation of the perturbations but the running of the bispectrum is still given by one parameter only (by choosing some other parameters appropriately) as \citep{2010JCAP...10..004B}
\begin{equation}
\label{running2}
B_\Phi(k_1,k_2,k_3)\propto (k_1k_2)^{n_{\rm s}-4+(n_{\rm NG}/2)} +2\,\mathrm{perms.}
\end{equation}
Comparing the two templates (\ref{running1}) and (\ref{running2}) one sees that there are multiple ways to generalize (with one extra parameter) the constant local $f_{\rm NL}$ model, even with the same values for $f_{\rm NL}$ and $n_{\rm NG}$. If one is able to distinguish observationally between these two shapes then one could find out whether the running originated from single or multiple field effects for example.

\citet{2010JCAP...10..004B} further assumed that $|n_{f_{\rm NL}}\ln \left(k_{\rm max}/k_{\rm min}\right)|\ll1$. 
In our case, $\ln \left(k_{\rm max}/k_{\rm min}\right)\la8$ and $n_{\rm NG}$ can be at most of order $0.1$. If the observational constraints on $n_{\rm NG}$ using the previous theoretical templates turn out to be weaker, then one cannot use those constraints to limit the fundamental parameters of the models because the templates are being used in a region where they are not applicable. However, from a phenomenological point of view, we wish to argue that the previous templates are still interesting cases of scale-dependent bispectra, even in that parameter region.
\cite{2010JCAP...10..004B} also computed the running of the trispectra amplitudes $\tau_{\rm NL}$ and $g_{\rm NL}$. For general single-source models they showed that $n_{\tau_{\rm NL}}=2n_{\rm NG}$, analogous to the well-known consistency relation $\tau_{\rm NL}(k)=\left(\frac{6}{5}f_{\rm NL}(k)\right)^2$, providing a useful consistency check. 

\subsubsection{Equilateral type scale-dependent bispectrum \label{subsubsec:Equil}}

General single-field models that can produce large bispectra having a significant correlation with the equilateral template also predict a mild running non-Gaussianity. 
A typical example is DBI-inflation, as studied, e.g., by \citet{2005PhRvD..72l3518C} and \citet{2007JCAP...01..002C}, with a generalization within the effective field theory of inflation in \citet{2010JCAP...12..026B}. Typically in these models 
a running NG arises of the form
\begin{equation}
\label{rNGequil}
f_\mathrm{NL}\rightarrow f_\mathrm{NL}^*\left(\frac{k_1+k_2+k_3}{3k_{piv}}\right)^{n_{\rm NG}}\, ,
\end{equation}
where $n_{\rm NG}$ is the running parameter and $k_{\rm piv}$ is a pivot scale
needed to constrain the amplitude.
For example, in the case where the main contribution comes from a small sound speed of the inflaton field, $n_{\rm NG}=-2 s$, where $s=\dot{c_{\rm s}}/(H c_{\rm s})$, and therefore running NG allows us to constrain the time dependence of the sound speed.\footnote{This would help in further breaking (via primordial NG) some degeneracies among the parameters determining the curvature power spectrum in these modes. For a discussion and an analysis of this type, see \citet{planck2013-p17} and \citet{planck2014-a24}.} The equilateral NG with a running of the type given in Eq.~\eqref{rNGequil} is a third type of running NG analysed in Sect.~\ref{running_analysis} (together with the local single-source model of Eq.~\ref{running1} and the local two-field model of Eq.~\ref{running2}). We refer the reader to Sect.~\ref{runningmethod} for the details on the methodology adopted to analyse these models.


\subsection{Oscillatory bispectrum models}
\label{oscillatory}

Oscillatory power spectrum and bispectrum signals are possible in a variety of well-motivated inflationary models, including those with an imposed shift symmetry or if there are sharp features in the inflationary potential.  Our first \Planck\ temperature-only non-Gaussian analysis \citepalias{planck2013-p09a} included a search for the simplest resonance and feature models, while the second \Planck\ temperature with polarization analysis \citepalias{planck2014-a19} substantially expanded the frequency range investigated, while also encompassing a much wider class of oscillatory models. These phenomenological bispectrum shapes had free parameters designed to capture the main properties of the key extant oscillatory models, thus surveying for any oscillatory signals present in the data at high significance. Our primary purpose here is to use the revised \Planck\ 2018 data set to determine the robustness of our second analysis, so we only briefly introduce the models studied, referring the reader to our previous work \citepalias{planck2014-a19} for more detailed information.

\subsubsection{Resonance and axion monodromy}
\label{sec:resonance}
Motivated by the UV completion problem facing large-field inflation, effective shift symmetries can be used to preserve the flat potentials required by inflation, with a prime example being the periodically modulated potential of axion monodromy models.  This periodic symmetry can cause resonances during inflation, imprinting logarithmically-spaced oscillations in the power spectrum, bispectrum and beyond \citep{2010JCAP...06..009F,Hannestad:2009yx,2011JCAP...01..017F}.  For the bispectrum, to a good approximation, these models yield the simple oscillatory shape \citep[see e.g.,][]{2010AdAst2010E..72C}
\eq\label{eq:resBprim}
B_\Phi^{\rm res}(\klist) = \frac{6A ^2\fnl^{\rm res} }{(k_1 k_2 k_3)^2}\sin\Big[ \omega \ln({k_1+k_2+k_3}) + \phi\Big]\,, 
\qe
where the constant $\omega$ is an effective frequency associated with the underlying periodicity of the model and $\phi$ is a phase.  The units for the wavenumbers, $k_i$, are arbitrary as any specific choice can be absorbed into the phase which is marginalised over in our results.  There are more general resonance models that naturally combine properties of inflation inspired by fundamental theory, notably a varying sound speed $c_{\rm s}$ or an excited initial state.  These tend to modulate the oscillatory signal on $K = k_1+k_2+k_3$ constant slices, with either equilateral or flattened shapes, respectively \citep[see e.g.,][]{Chen:2010bka}, which we take to have the form
\eq\label{eq:equilBprim}
S^{\rm eq} (\klist) = \frac{\tilde k_1\tilde k_2\tilde k_3}{k_1k_2k_3}\,, \qquad S^{\rm flat} = 1 - S^{\rm eq}\,,
\qe
where $\tilde k_1 \equiv k_2+k_3 - k_1$.  Note that $S^{\rm eq}$ correlates closely with the equilateral shape in \eqref{equilateralBis} and $S^{\rm flat}$ with the orthogonal shape in \eqref{orthogonalBis}, since the correction from the spectral index $n_{\rm s}$ is small.  The resulting generalized resonance shapes for which we search are then
\begin{align}
B^{\rm res-eq}(\klist) &\equiv S^{\rm eq} (\klist) \times B^{\rm res} (\klist)\,,\nn\\
\label{eq:resflatBprim}
B^{\rm res-flat}(\klist) &\equiv S^{\rm flat} (\klist) \times B^{\rm res} (\klist)\,.
\end{align}
This analysis does not exhaust resonant models associated with a non-Bunch-Davies initial state, which can have a more sharply flattened shape (``enfolded'' models), but it should help identify this tendency if present in the data.  In addition, the discussion in \citet{Flauger:2014ana} showed that the resonant frequency can ``drift'' slowly over time with a correction term to the frequency being proposed, but again we leave that for future analysis. Finally, we note that there are multifield models in which sharp corner-turning can result in residual oscillations with logarithmic spacing, thus mimicking resonance models \citep{2011JCAP...01..030A,2013JCAP...05..006B}.  However, these oscillations are more strongly damped and can be searched for by modulating the resonant shape (Eq.~\ref{eq:resBprim}) with a suitable envelope, as discussed for feature models below. 


\subsubsection{Scale-dependent oscillatory features}
\label{sec:feature}
Sharp features in the inflationary potential can generate oscillatory signatures \citep{2007JCAP...06..023C}, as can rapid variations in the sound speed $c_{\rm s}$ or fast turns in a multifield potential.  Narrow features in the potential induce a corresponding signal in the power spectrum, bispectrum, and trispectrum; to a first approximation, the oscillatory bispectrum has a simple sinusoidal behaviour given by \citep{2007JCAP...06..023C}
\begin{align}
\label{eq:featureBprim}
\bar B^{\rm feat}(\kall) = \frac{6A ^2\fnl^{\rm feat} }{(k_1 k_2 k_3)^2}\sin\Big[{\omega(k_1+k_2+k_3)} + \phi\Big]\,,
\end{align}
where $\omega$ is a frequency determined by the specific feature properties and $\phi$ is a phase. The wavenumbers $k_i$ are in units of $Mpc^{-1}$.
A more accurate analytic bispectrum solution has been found that includes a damping envelope taking the form
\citep{Adshead:2011jq}
 \eq\label{eq:Adsetal1}
B^{K^2\cos}(\kall) =\frac{6A ^2 \fnl^{\rm K^2\cos} }{(k_1 k_2 k_3)^2} K^2 D(\alpha \omega K) \cos (\omega K)\,,
\qe
where $K= k_1+ k_2 +k_3$ and the envelope function is given by $D(\alpha \omega K) =\alpha \omega/ \big(K \sinh (\alpha\omega K)\big)$.  Here, the model-dependent parameter $\alpha$ determines the large wavenumber cutoff, with $\alpha=0$ for no envelope in the limit of an extremely narrow feature.  Oscillatory signals generated instead by a rapidly varying sound speed $c_{\rm s}$ take the form
\eq\label{eq:Adsetal2}
B^{K\sin}(\kall) =\frac{6A ^2 \fnl^{\rm K\sin} }{(k_1 k_2 k_3)^2} K \,D(\alpha \omega K) \sin (\omega K)\,.
\qe

In order to encompass the widest range of physically-motivated feature models, we will also modulate the predicted signal (Eq.~\ref{eq:featureBprim}) with equilateral and flattened shapes, as defined in Eq.~\eqref{eq:equilBprim}, i.e.,
\begin{align}
\label{eq:featequilBprim}
B^{\rm feat-eq}(\klist) &\equiv S^{\rm eq} (\klist) \times B^{\rm feat} (\klist)\,,\\
\label{eq:featflatBprim}
B^{\rm feat-flat}(\klist) &\equiv S^{\rm flat} (\klist) \times B^{\rm feat} (\klist)\,.
\end{align}
Like our survey of resonance models, this allows the feature signal to have arisen in inflationary models with (slowly) varying sound speeds or with excited initial states.  In the latter case, it is known that very narrow features can mimic non-Bunch Davies bispectra with a flattened or enfolded shape \citep{2007JCAP...06..023C}.


\subsection{Non-Gaussianity from excited initial states}
\label{Sec:NBD_models}
Inflationary perturbations generated by ``excited'' initial states (non-Bunch-Davies vacuum states) generically create non-Gaussianities with a distinct enfolded shape \citep[see e.g.,][]{2007JCAP...01..002C, Holman:2007na, Meerburg:2009ys}, that is, where the bispectrum signal is dominated by flattened configurations with $k_1+k_2\approx k_3$ (and cyclic permutations).  In the present analysis, we investigate all the non-Bunch-Davies (NBD) models discussed in the previous \Planck\ non-Gaussian papers, where explicit equations can be found for the bispectrum shape functions.  In the original analysis \citepalias{planck2013-p09a}, we described: the vanilla flattened shape model $B^{\rm flat}$$\,\propto\,$$ S^{\rm flat}$ already given in Eq.~\eqref{eq:equilBprim}; a more realistic flattened model $B^{\rm NBD}$ \citep{2007JCAP...01..002C} from power-law $k$-inflation, with excitations generated at time $\tau_{\rm c}$, yielding an oscillation period (and cut-off) $k_{\rm c} \approx (\tau_{\rm c} c_{\rm s})^{-1}$; the two leading-order shapes for excited canonical single-field inflation labelled $B^{\rm NBD1\hbox{-}cos}$ and $B^{\rm NBD2\hbox{-}cos}$ \citep{Agullo:2011xv}; and a non-oscillatory sharply flattened model $B^{\rm NBD3}$, with large enhancements from a small sound speed $c_{\rm s}$ \citep{2010AdAst2010E..72C}.  In the second (temperature plus polarization) analysis \citepalias{planck2014-a19}, we also studied additional NBD shapes including a sinusoidal version of the original NBD bispectrum $B^{\rm NBD\hbox{-}sin}$ \citep{Chen:2010bka}, and similar extensions for the single-field excited models \citep{Agullo:2011xv}, labelled $B^{\rm NBD1-sin}$ and $B^{\rm NBD2-sin}$, with the former dominated by oscillatory squeezed configurations.


\subsection{Directional-dependent NG} 
\label{sec:vector_models}
The standard local bispectrum in the squeezed limit ($k_1 \ll k_2\approx k_3$) has an amplitude that is the same for all different angles between the large-scale mode with wavevector $k_1$ and the small-scale modes parameterized by wavevector $\vec k_s =(\vec k_2-\vec k_3)/2$. More generally, we can consider ``anisotropic'' bispectra, in the sense of an angular dependence on the orientation of the large-scale and small-scale modes,  where in the squeezed limit the bispectrum depends on all even powers of $\mu =\hat {\vec k}_1\cdot\hat{\vec k}_s$ (where $\hat{\vec k} = {\vec k}/k$, and all odd powers vanish by symmetry even out of the squeezed limit). Expanding the squeezed bispectrum into Legendre polynomials with even multipoles $L$, the $L>0$ shapes then be used to cleanly isolate new physical effects. 
In the literature it is more common to expand in the angle $\mu_{12} =\hat {\vec k}_1\cdot\hat{\vec k}_2$, which makes some aspects of the analysis simpler while introducing non-zero odd $L$ moments 
(since they no longer vanish when defined using the non-symmetrized small-scale wavevector). We then parameterize variations of local NG using \citep{Shiraishi:2013vja}: 
\eq\label{vectorBis}
B_\Phi(\kall) = \sum_L c_L [P_L(\mu_{12})P_{\Phi}(k_1)P_{\Phi}(k_2)+\hbox{2\,perms.}]\, ,
\qe
where $P_L(\mu)$ is the Legendre polynomial with $P_0=1, \;P_1=\mu$, and $P_2=\textstyle {\frac{1}{2} (3 \mu^2-1)}$. For instance, in the $L=1$ case the shape is given by 
\eq\label{eq:dbiS}
 B_\Phi^{L=1}(\kall) = \frac{2A ^2\fnl^{L=1}}{(k_1 k_2 k_3)^2}\,\left[\frac{k_3^2}{k_1^2k_2^2}(k_1^2+k_2^2-k_3^2) + \hbox{2\,perms.}\right].
 \qe
Bispectra of the directionally dependent class in general peak in the squeezed limit ($k_1 \ll k_2\approx k_3$), but they feature a non-trivial dependence on the parameter $\mu_{12} =\hat {\vec k}_1\cdot\hat{\vec k}_2$. 
The local NG template corresponds to $c_i= 2 f_{\rm NL} \delta_{i0}$. The nonlinearity parameters $f_{\rm NL}^L$ are related to the $c_L$ coefficients by $c_0=2 f_{\rm NL}^{L=0}$, $c_1=-4 f_{\rm NL}^{L=1}$, and $c_2=-16 f_{\rm NL}^{L=2}$. The $L=1$ and 2 shapes are characterized by sharp variations in the flattened limit, e.g., for $k_1+k_2\approx k_3$, while in the squeezed limit, $L=1$ is suppressed, unlike $L=2$, which grows like the local bispectrum shape (i.e., the $L=0$ case). 

Bispectra of the type in Eq.~\eqref{vectorBis} can arise in different inflationary models, e.g., models where anisotropic sources contribute to the curvature perturbation. Bispectra of this type are indeed a general and unavoidable 
outcome of models that sustain long-lived superhorizon gauge vector fields during inflation \citep{Bartolo:2012sd}. A typical example is the case of the inflaton field $\varphi$ coupled to the kinetic term $F^2$ of a $U(1)$ gauge field $A^\mu$, via the interaction term $I^2(\varphi) F^2$, where $F_{\mu\nu}=\partial_\mu A_\nu-\partial_\nu A_\mu$ and the coupling $I^2(\varphi) F^2$ can allow for scale invariant vector fluctuations to be generated on superhorizon scales \citep{Barnaby:2012tk,Bartolo:2012sd}.\footnote{Notice that indeed these models generate bispectra (and power spectra) that break statistical isotropy and, after an angle average, the bispectrum takes the above expression (Eq.~\ref{vectorBis}).}
Primordial magnetic fields sourcing curvature perturbations can also generate a dependence on both $\mu$ and $\mu^2$ \citep{2012JCAP...06..015S}. The $I^2(\varphi) F^2$ models predict $c_2=c_0/2$, while models where the primordial curvature perturbations are sourced by large-scale magnetic fields produce $c_0$, $c_1$, and $c_2$.  The so-called ``solid inflation'' models (\citealt{2012arXiv1210.0569E}; see also \citealt{2013JCAP...08..022B,2014PhRvD..90f3506E,2014PhRvD..89l3509S,2014JCAP...11..009B}) also predict bispectra of the form Eq.~\eqref{vectorBis}. In this case $c_2 \gg c_0$ \citep{2012arXiv1210.0569E,2014PhRvD..90f3506E}. Inflationary models that break rotational invariance and parity also generate this kind of NG with the specific prediction $c_0 : c_1 : c_2 = 2 : -3 : 1$ \citep{2015JCAP...07..039B}.
Therefore, measurements of the $c_i$ coefficients can test for the existence of primordial vector fields during inflation, fundamental symmetries, or non-trivial structure underlying the inflationary model (as in solid inflation). 
\noindent

Recently much attention has been focused on the possibility of testing the presence of higher-spin particles via their imprints on higher-order inflationary correlators. Measuring primordial NG can allow us to pin down masses and spins of the particle content present during inflation, making inflation a powerful cosmological collider \citep{2010AdAst2010E..72C,2010JCAP...04..027C,2013JHEP...06..051N,2015arXiv150308043A,2018JHEP...04..140B,Arkani-Hamed:2018kmz}. In the case of long-lived superhorizon higher-spin (effectively massless or partially massless higher spin fields) bispectra like in Eq.~(\ref{vectorBis}) are generated, where even coefficients up to $c_{n=2s}$ are excited, $s$ being the spin of the field \citep{2018PhRvD..98d3533F}.  A structure similar to Eq.~(\ref{vectorBis}) arises in the case of massive spin particles, where the coefficients $c_i$ has a specific non-trivial dependence on the mass and spin of the particles \citep{2015arXiv150308043A,2018JHEP...04..140B,2018JCAP...05..013M}.

\subsection{Parity-violating tensor non-Gaussianity motivated by pseudo-scalars} 
 
 In some inflationary scenarios involving the axion field, there are chances to realize the characteristic NG signal in the tensor-mode sector. In these cases a non-vanishing bispectrum of primordial gravitational waves, $B_{\rm h}^{s_1 s_2 s_3}$, arises via the nonlinear interaction between the axion and the gauge field. Its magnitude varies depending on the shape of the axion-gauge coupling, and, in the best-case scenario, the tensor mode can be comparable in size to or dominate the scalar mode \citep{2013JCAP...11..047C,2016JCAP...01..041N,2018PhRvD..97j3526A}.

The induced tensor bispectrum is polarized as $B_{\rm h}^{+++} \gg B_{\rm h}^{++-}, B_{\rm h}^{+--}, B_{\rm h}^{---}$ (because the source gauge field is maximally chiral), and peaked at around the equilateral limit (because the tensor-mode production is a subhorizon event). Its size is therefore quantified by the so-called tensor nonlinearity parameter,
\begin{equation}
f_{\rm NL}^{\rm tens} \equiv \lim_{k_i \to k} \frac{B_{\rm h}^{+++}(\vec{k}_1, \vec{k}_2, \vec{k}_3)}{ F_{\zeta}^{\rm equil}(k_1, k_2, k_3)} ~, \label{eq:fnltens}
\end{equation}
with $F_\zeta^{\rm equil} \equiv (5/3)^3 B_\Phi^{\rm equil} / f_{\rm NL}^{\rm equil}$.

In this paper we constrain $f_{\rm NL}^{\rm tens}$ by measuring the CMB temperature and $E$-mode bispectra computed from $B_{\rm h}^{+++}$ (for the exact shape of $B_{\rm h}^{+++}$ see \citetalias{planck2014-a19}). By virtue of their parity-violating nature, the induced CMB bispectra have non-vanishing signal for not only the even but also the odd $\ell_1 + \ell_2 + \ell_3$ triplets \citep{2013JCAP...11..051S}. Both are investigated in our analysis, yielding more unbiased and accurate results. The \Planck\ 2015 paper \citepalias{planck2014-a19} found a best limit of $f_{\rm NL}^{\rm tens} = (0 \pm 13) \times 10^2$ (68\%\,CL), from the foreground-cleaned temperature and high-pass filtered $E$-mode data, where the $E$-mode information for $\ell < 40$ was entirely discarded in order to avoid foreground contamination. This paper updates those limits with additional data, including large-scale $E$-mode information.

\section{Estimators and data analysis procedures}
\label{sec:SEP}

\subsection{Bispectrum estimators}\label{sec:bispec_estim}

We give here a short description of the data-analysis procedures used in this paper. For 
additional details, we refer the reader to the primordial NG analysis associated with previous 
\Planck\ releases \citepalias{planck2013-p09a,planck2014-a19} and to references provided below.

For a rotationally invariant CMB sky and even parity bispectra (as is the case for combinations of $T$ and $E$),  
the angular bispectrum can be written as
\begin{equation}\label{eq:Bred}
\langle a^{X_1}_{\ell_1 m_1}a^{X_2}_{\ell_2 m_2}a^{X_3}_{\ell_3 m_3}\rangle = 
\curl{G}^{\ell_1 \ell_2 \ell_3}_{m_1 m_2 m_3} \,b^{X_1 X_2 X_3}_{\ell_1 \ell_2
\ell_3}\, ,
\end{equation} 
where $b^{X_1 X_2 X_3}_{\ell_1 \ell_2\ell_3}$ defines the ``reduced bispectrum,'' and $\curl{G}^{\ell_1 \ell_2 \ell_3}_{m_1 m_2 m_3}$ 
 is the Gaunt integral, i.e., the integral over solid angle of the product of three spherical harmonics,
\begin{equation}\label{eq:Gaunt}
\curl{G}^{\ell_1 \ell_2 \ell_3}_{m_1 m_2 m_3} \equiv\int Y_{\ell_1 m_1}(\vec{\hat{n}}) \, Y_{\ell_2 m_2}(\vec{\hat{n}}) \, Y_{\ell_3 m_3}(\vec{\hat{n}}) \, d^2\vec{\hat{n}} \; .
\end{equation}
The Gaunt integral (which can be expressed as a product of Wigner 3$j$-symbols) enforces rotational symmetry. It satisfies
both a triangle inequality and a limit
given by some maximum experimental resolution $\ell_\mathrm{max}$. This defines a tetrahedral domain of allowed bispectrum triplets, $\{\ell_1,\ell_2,\ell_3\}$.

In order to estimate the $\fnl$ value for a given primordial shape, we need to compute a theoretical prediction of the corresponding CMB bispectrum ansatz $b^{\rm th}_{\ell_1 \ell_2 \ell_3}$ and fit it to the observed 3-point function \citep[see e.g.,][]{2001PhRvD..63f3002K}.

Optimal cubic bispectrum estimators were first discussed in \cite{1998MNRAS.299..805H}. It was then shown that, in the limit of small NG, the optimal polarized $\fnl$ estimator is described by \citep{2006JCAP...05..004C}
\begin{align}\label{eq:optimalestimator}
\hat{f}_{\textrm{NL}}  = &
 \frac{1}{N} \sum_{X_i, X'_i} \sum_{\ell_i,m_i}
 \sum_{\ell'_i, m'_i} \curl{G}^{\,\,\ell_1\; \ell_2\; \ell_3}_{m_1 m_2 m_3 }
 b^{X_1 X_2 X_3, \, \rm th}_{\ell_1 \ell_2 \ell_3}
 \Biggl\{ \biggl[ \Big(\tens{C}^{-1}_{\ell_1 m_1, \ell_1' m_1'}\Bigr)^{X_1 X'_1}
 a^{X'_1}_{\ell_1'm_1'}
 \nonumber \\
& \qquad\qquad \times
 \Bigl(\tens{C}^{-1}_{\ell_2 m_2, \ell_2' m_2'} \Bigr)^{X_2 X'_2}
 a^{X'_2}_{\ell_2'm_2'}
 \Bigl(\tens{C}^{-1}_{\ell_3 m_3, \ell_3' m_3'}\Bigr)^{X_3 X'_3}
 a^{X'_3}_{\ell_3'm_3'} \biggr] \nonumber \\
& - \biggl[ \,\Bigl(\tens{C}^{-1}_{\ell_1 m_1, \ell_2 m_2}\Bigr)^{X_1 X_2}
 \Bigl(\tens{C}^{-1}_{\ell_3 m_3,\ell_3' m_3'}\Bigr)^{X_3 X'_3} a^{X'_3}_{\ell_3'm_3'}
  + \mathrm{cyclic} \biggr] \Biggr\} \; , 
\end{align}
where the normalization $N$ is fixed by requiring unit response to $ b_{\ell_1 \ell_2 \ell_3}^{\rm th}$ when $\fnl=1$.
 $\tens{C}^{-1}$ is the inverse of the block matrix:
\begin{equation}
\tens{C}=\left(
\begin{array}{c|c}
\tens{C}^{TT} & \tens{C}^{TE} \\
\hline
\tens{C}^{ET} & \tens{C}^{EE}
\vphantom{C^{C^{C}}}
\end{array}\right) \, .
\end{equation}
The blocks represent the full \textit{TT}, \textit{TE}, and \textit{EE} covariance matrices, with $\tens{C}^{ET}$ being the transpose of $\tens{C}^{TE}$.
CMB $a_{\ell m}$ coefficients, bispectrum templates, and covariance matrices in the previous relation are assumed to include instrumental beam and noise.

As shown in the formula above, these estimators are always characterized by the presence of two distinct contributions. One is cubic in the observed multipoles, and computes the correlation between the observed bispectrum and the theoretical template $b_{\ell_1 \ell_2 \ell_3}^{\rm th}$.
This is generally called the ``cubic term'' of the estimator. The other is instead linear in the observed multipoles. Its role is that of correcting for mean-field contributions to the uncertainties, generated by the breaking of rotational invariance, due to the presence of a mask or to anisotropic/correlated instrumental noise \citep{2006JCAP...05..004C,Yadav:2007ny}. 
 
Performing the inverse-covariance filtering operation implied by Eq.~\eqref{eq:optimalestimator} is numerically very demanding \citep{2009JCAP...09..006S,2012arXiv1211.0585E}.
An alternative, simplified approach, is that of working in the ``diagonal covariance approximation,''  yielding \citep{Yadav:2007rk}
\begin{flalign}\label{eq:diagcovestimator}
\hat{f}_{\textrm{NL}}  =  & \frac{1}{N} \sum_{X_i, X'_i} \sum_{\ell_i,m_i} \curl{G}^{\,\,\ell_1\; \ell_2\; \ell_3}_{m_1 m_2 m_3 } (\tens{C}^{-1})_{\ell_1}^{X_1 X'_1} (\tens{C}^{-1})_{\ell_2} ^{X_2 X'_2}  (\tens{C}^{-1})_{\ell_3}^{X_3 X'_3} b^{X_1 X_2 X_3, \, \rm th}_{\ell_1 \ell_2 \ell_3}
\;\;\;\; \nonumber
\\ & \times \left[ a^{X'_1}_{\ell_1 m_1}\, a^{X'_2}_{\ell_2 m_2}\,a^{X'_3}_{\ell_3 m_3} - \tens{C}_{\ell_1 m_1,\ell_2 m_2}^{X'_1 X'_2} a^{X'_3}_{\ell_3 m_3}  - \tens{C}_{\ell_1 m_1,\ell_3 m_3}^{X'_1 X'_3} a^{X'_2}_{\ell_2 m_2} \right. \nonumber
\\ & \qquad\qquad\left. - \tens{C}_{\ell_2 m_2,\ell_3 m_3}^{X'_2 X'_3} a^{X'_1}_{\ell_1 m_1} \right] \, .
\end{flalign}
Here, $\tens{C}^{-1}_\ell$ represents the inverse of the following $2\times2$ matrix:
\begin{equation}
\tens{C}_\ell=\left(
\begin{array}{cc}
\tens{C}^{TT}_\ell & \tens{C}_\ell^{TE} \\
\tens{C}_\ell^{ET} & \tens{C}_\ell^{EE}
\vphantom{C^{C^{C}}}
\end{array}\right) \, .
\end{equation}

As already described in \citetalias{planck2013-p09a}, we find that this simplification, while avoiding the covariance-inversion operation, still leads to uncertainties that are very close to optimal, provided that the multipoles are pre-filtered using a simple diffusive inpainting method. As in previous analyses, we stick to this approach here.

A brute-force implementation of Eq.~\eqref{eq:diagcovestimator} would require the evaluation of all the possible bispectrum configurations in our data set. This is completely unfeasible, as it would scale as $\ell_{\rm max}^{\,5}$. The three different bispectrum estimation pipelines employed in this analysis are characterized by the different approaches used to address this issue. 

Before describing these methods in more detail in the following sections, we would like to stress here, the importance of having these multiple approaches. The obvious advantage is that this redundancy enables a stringent cross-validation of our results.  There is, however, much more than that, as different methods allow a broad range of applications, beyond $\fnl$ estimation, such as, for example, model-independent reconstruction of the bispectrum in different decomposition domains, precise characterization of spurious bispectrum components, monitoring direction-dependent NG signals, and so on. 

\subsubsection{KSW and skew-$C_\ell$ estimators} 
\label{sec:KSW_est}

Komatsu-Spergel-Wandelt (KSW) and skew-$C_\ell$ estimators \citep{Komatsu:2003iq,2010MNRAS.401.2406M} can be 
applied to bispectrum templates that can be {\em factorized}, i.e., they can be written or well approximated as a linear combination of separate products of 
functions. This is the case for the standard local, equilateral, and orthogonal shapes, which cover a large range of theoretically motivated scenarios.
The idea is that factorization leads to a massive reduction in computational time, via reduction of the three-dimensional summation over $\ell_1$, $\ell_2$, $\ell_3$  into a product of three separate one-dimensional sums over each multipole.

The skew-$C_\ell$ pipeline differs from KSW essentially in that, before collapsing the estimate into the $\fnl$ parameter, 
 it initially determines the so called ``bispectrum-related power spectrum'' (in short, ``skew-$C_\ell$'') function \citep[see][]{2010MNRAS.401.2406M} for details). The slope of this function is 
shape-dependent, which makes the skew-$C_\ell$ extension very useful to separate and monitor multiple and spurious NG components in the map.

\subsubsection{Running of primordial non-Gaussianity}
\label{runningmethod}
In the previous $2015$ analysis, the KSW pipeline was used only to constrain the separable local, equilateral, orthogonal, and lensing templates. In the current analysis we extend its scope by adding the capability to constrain running of non-Gaussianity, encoded in the spectral index of the nonlinear amplitude $f_{\rm NL}$, denoted $n_{\rm NG}$.

In our analysis we consider both the two local running templates, described by Eqs.~\eqref{running1} and \eqref{running2} in Sect.~\ref{subsubsec:Local}, and the general parametrizazion for equilateral running of Sect.~\ref{subsubsec:Equil}, which reads:
\begin{equation}
f_{\rm NL} \to  f^*_{\rm NL}\left(\frac{k_1+k_2+k_3}{3k_{\rm piv}}\right)^{n_{\rm NG}},
\label{eq:fnlam}
\end{equation}
where $n_{\rm NG}$ is the running parameter and $k_{\rm piv}$ is a pivot scale needed to constrain the amplitude. Contrary to the two local running shapes this expression is not explicitly separable. To make it suitable for the KSW estimator (e.g., to preserve the factorizability over $k_i$), we can use a Schwinger parametrization and rearrange it as
\begin{equation}
f_{\rm NL} \to \frac{f^*_{\rm NL}}{3k_{\rm piv}^{n_{\rm NG}}}\, \frac{ k_{\rm sum}}{\Gamma(1-n_{\rm NG})}\int_0^\infty\mathrm{d}t\,t^{-n_{\rm NG}}e^{-tk_{\rm sum}},
\label{eq:fnlsdsw}
\end{equation}
where $k_{\rm sum}=k_1+k_2+k_3$.

Alternatively, but not equivalently, factorizability can be preserved by replacing the arithmetic mean of the three wavenumbers with the geometric mean \citep{2009JCAP...12..022S}:
\begin{equation}
f_{\rm NL} \to f^*_{\rm NL}\left(\frac{k_1k_2k_3}{k_{\rm piv}^3}\right)^{\frac{n_{\rm NG}}{3}}.
\label{eq:fnlgm}
\end{equation}
Making one of these substitutions immediately yields the scale-dependent version of any bispectrum shape. 
 Analysis in \cite{2018JCAP...05..045O} has shown strong correlation between the two templates, where the former behaves better numerically and is the template of choice for the running in this analysis.

A generalization of the local model, taking into account the scale dependence of $f_{\rm NL}$, can be found in \cite{2010JCAP...10..004B}, as summarized in Sect.~\ref{subsubsec:Local}.

Unlike $f_{\rm NL}$, the running parameter $n_{\rm NG}$ cannot be estimated via direct template fitting.
The optimal estimation procedure, developed in \citet{2012PhRvL.109l1302B} and extended to all the scale-dependent shapes treated here in \citet{2018JCAP...05..045O}, is based instead on the reconstruction of the likelihood function, with respect $n_{\rm NG}$.
The method exploits the KSW estimator to obtain estimates of $f^*_{\rm NL}$ for different values of the running, using explicitly separable bispectrum templates. 
With these values in hand, the running parameter probability density function (PDF) is computed from its analytical expression.

The computation of the marginalized likelihood depends on the choice of the prior distributions; in \citet{2012PhRvL.109l1302B} and \citet{2018JCAP...05..045O} a flat prior on $f^*_{\rm NL}$ was assumed.
This prior depends on the choice of the arbitrary pivotal scale $\vec{k}_{\rm piv}$, since a flat prior on $f^*_{\rm NL}$ defined at a certain scale, corresponds to a non-flat prior for another scale. 
The common solution is to select the pivot scale that minimizes the correlation between the parameters.
This is in general a good choice, and would work properly in the case of a significant detection of a bispectrum signal. 
In the absence of a clear detection, however, it is worth noting some caveats. 
Since the range of scales available is obviously finite, a fit performed at a certain pivot scale will tend to favour particular values of $n_{\rm NG}$. Therefore, there is not a perfectly ``fair'' scale for the fit.
As a consequence, statistical artefacts can affect the estimated constraints in the case of low significance of the measured $f^*_{\rm NL}$ central value.
To prevent this issue, we resort to two additional approaches that make the final $n_{\rm NG}$ PDF pivot independent: the implementation of a parametrization invariant Jeffreys prior; and frequentist likelihood profiling.
Assuming that the bispectrum configurations follow a Gaussian distribution, the likelihood can be written as \citep[see][for a derivation]{2012PhRvL.109l1302B} 
\begin{equation}
    \mathcal{L}(n_{\rm NG},f^*_{\rm NL})\propto \exp\,{\left[ -\frac{N(f^*_{\rm NL}-\hat f_{\rm NL})^2}{2} \right]}\,\,\exp\,{\left( \frac{\hat f_{\rm NL}^2N}{2}\right)},
    \label{eq:likefn}
\end{equation}
where $\hat{f}_{\rm NL}$ is the value of the NG amplitude recovered from the KSW estimator for a fixed $n_{\rm NG}$ value of the running, and $N$ is the KSW normalization factor.
Integrating this expression with respect to $f^*_{\rm NL}$ we obtain the marginalized likelihood.  Assuming a constant prior we obtain
\begin{equation}
\mathcal{L}(n_{\rm NG})\propto\frac{1}{\sqrt{N}}\,\exp\,\left(\frac{\hat{f}^2_{\rm NL}N}{2} \right).
\label{eq:likeconstp}
\end{equation}
The Jeffreys prior is defined as the square root of the determinant of the Fisher information matrix $\mathcal{I}(f_{\rm NL},n_{\rm NG})$. In the case of separable scale-dependent bispectra, the Fisher matrix is
 \begin{align}
 \mathcal{I}_{\alpha,\beta}\equiv&\sum_{\ell_1\leq\ell_2\leq\ell_3}\frac{(2\ell_1+1)(2\ell_2+1)(2\ell_3+1)}{4\pi}\begin{pmatrix} \ell_1 & \ell_2 & \ell_3 \\ 0 & 0 & 0 \end{pmatrix}^2 \nonumber \\
 &\qquad\qquad\times\cfrac{1}{\sigma^2_{\ell_1\ell_2\ell_3}}\, \frac{\partial b_{\ell_1\ell_2\ell_3}}{\partial \theta_\alpha}\, \frac{\partial b_{\ell_1\ell_2\ell_3}}{\partial \theta_\beta},
 \label{eq:fishbisred}
\end{align}
where $\theta_\alpha$ and $\theta_\beta$ correspond to $f^*_{\rm NL}$ or $n_{\rm NG}$ (depending on the value of the index), $b_{\ell_1\ell_2\ell_3}$ is the reduced bispectrum, and the matrix is a Wigner-3j symbol.

We search an expression for the posterior distribution marginalized over $f^*_{\rm NL}$. 
Assuming the Jeffreys prior for both parameters and integrating over $f_{\rm NL}$, we obtain the marginalized posterior
\begin{align}
     \mathcal{P}(n_{\rm NG})\propto& \left[ \hat f_{\rm NL} \sqrt{\frac{2\pi}{N}}\,\,\exp\,{\left( \frac{\hat f_{\rm NL}^2N}{2}\right)}\,\,{\rm erf}\left( \hat f_{\rm NL} \sqrt{\frac{N}{2}}\right) +\frac{2}{N}\right] \nonumber \\
     &\qquad\qquad\times \sqrt{\det(\mathcal{I}(f^*_{\rm NL}=1,n_{\rm NG}))}.
     \label{eq:pdfjef}
\end{align}
The implementation of this expression in the estimator is straightforward; the only additional step is the numerical computation of the Fisher matrix determinant for each value of $n_{\rm NG}$ considered. 
The derived expression is independent of the pivot scale.

Alternatively, in the frequentist approach, instead of marginalizing over $f^*_{\rm NG}$, the likelihood is sampled along its maximum for every $n_{\rm NG}$ value. For fixed $n_{\rm NG}$, the maximum likelihood $f^*_{\rm NG}$ is given exactly by the KSW estimator $\hat f_{\rm NL}$. From Eq.~\eqref{eq:likefn}, we see that for this condition the first exponential is set to 1 (since $f^*_{\rm NG}=\hat f_{\rm NL}$ at the maximum), and the profile likelihood reduces to
\begin{equation}
    \mathcal{L}(n_{\rm NG})\propto \exp\,{\left( \frac{\hat f_{\rm NL}^2 N}{2}\right)}.
    \label{eq:marglik}
\end{equation}
Notice that this expression also does not depend on the pivot scale. We will additionally use this expression to perform a likelihood ratio test between our scale-dependent models and the standard local and equilateral shapes.

\subsubsection{Modal estimators}  
\label{sec:modal_est}

Modal estimators \citep{2010PhRvD..82b3502F,2012JCAP...12..032F} are based on constructing complete, orthogonal 
bases of separable bispectrum templates (``bispectrum modes'') and finding their amplitudes by fitting them to the data. 
This procedure can be made fast, due to the separability of the modes, via a KSW type of approach. The vector of estimated mode 
amplitudes is referred to as the ``mode spectrum.'' This mode spectrum is theory independent and it contains all the information that
needs to be extracted from the data. It is also possible to obtain theoretical mode spectra, by expanding primordial shapes in the same 
modal basis used to analyse the data. This allows us to measure $\fnl$ for any given primordial bispectrum template, by correlating 
the theoretical mode vectors, which can be quickly computed for any shape, with the data mode spectrum. This feature makes modal techniques 
ideal for analyses of a large number of competing models. Also important is that non-separable bispectra are expanded with 
arbitrary precision into separable basis modes. Therefore the treatment of non-separable shapes is always numerically efficient in the modal approach. 
Finally, the data mode spectrum can be used, in combination with measured mode amplitudes, to build linear combinations of basis templates, which 
provide a model-independent reconstruction of the full data bispectrum. This reconstruction is of course smoothed in practice, since we use a finite number 
of modes. The modal bispectrum presented here follows the same approach as in $2015$. In particular we use two modal pipelines, ``Modal~1''
and ``Modal~2,'' characterized both by a different approach to the decomposition of polarized bispectra and by a different choice of basis, as detailed 
in \citetalias{planck2013-p09a}, \citetalias{planck2014-a19}, and at the end of Sect.~\ref{sec:settings}.

\subsubsection{Binned bispectrum estimator}
\label{sec:binned_est} 

The ``Binned'' bispectrum estimator \citep{Bucher:2009nm, Bucher:2015ura}
is based on the exact optimal $\fnl$ estimator, in combination with the
observation that many bispectra of interest are relatively smooth functions in
$\ell$ space. This means that data and templates can be binned in $\ell$ space
with minimal loss of information, but with large computational gains.
As a consequence, no KSW-like approach is required, and the theoretical
templates and observational bispectra are computed and stored completely
independently, and only combined at the very last stage in
a sum over the bins to obtain $\fnl$. This has several advantages: the
method is fast; it is easy to test additional shapes without having to rerun
the maps; the bispectrum of a map can be studied on its own in a non-parametric
approach (a binned reconstruction of the full data bispectrum is provided, which
can additionally be smoothed); and the dependence of $\fnl$ on $\ell$ can be
investigated for free, simply by leaving out bins from the final sum. All of
these advantages are used to good effect in this paper.

The Binned bispectrum estimator was described in more detail in the papers
associated with the 2013 and
2015 \Planck\ releases, and full details can be found in \cite{Bucher:2015ura}.
The one major change made to the Binned estimator code compared
to the 2015 release concerns the computation of the linear correction term,
required to make the estimator optimal in the case that rotational invariance is
broken, as it is in the \Planck\ analysis because of the mask and anisotropic
noise. The version of the code used in 2015, while fast to compute the linear
correction for a single map, scaled poorly with the number of maps, as the
product of the data map with all the Gaussian maps squared had to be
recomputed for each data map. Hence computing real errors, which requires
analysing a large set of realistic simulations, was slow. The new code
can precompute the average of the Gaussian maps squared, and then quickly
apply it to all the data maps. For the full \Planck\ analysis, with errors
based on 300 simulations, one gains an order of magnitude in computing time
\citep[see][for more details]{Bucher:2015ura}.

\subsection{Data set and analysis procedures}

\subsubsection{Data set and simulations}
\label{sec:dataset}

For our temperature and polarization data analyses we use the \Planck\ 2018 CMB
maps, as constructed with the four component-separation methods, \SMICA, \SEVEM,
\NILC, and \Commander\ \citep{planck2016-l04}. We also make much use of
simulated maps, for several different purposes, 
from computing errors to evaluating the linear mean-field 
correction terms for our estimators, as well as for performing data-validation
checks. 
Where not otherwise specified we will use the FFP10 simulation data set 
described in \citet{planck2016-l02}, \citet{planck2016-l03}, and \citet{planck2016-l04}, which are
the most realistic \Planck\ simulations
currently available. The maps we consider have been 
processed through the same four component-separation pipelines.
The same weights used by the different pipelines on actual data have been
adopted to combine different simulated frequency channels.
   
Simulations and data are masked using the common masks of the \Planck\ 
2018 release in temperature and polarization; see \citet{planck2016-l04} for a
description of how these masks have been produced. The sky coverage fractions
are, $f_{\rm sky} = 0.779$ in temperature and 
$f_{\rm sky} = 0.781$ in polarization.

\subsubsection{Data analysis details}\label{sec:settings}

Now we describe the setup adopted for the analysis 
of \Planck\ 2018 data by the four different $\fnl$ estimators described 
earlier in this section.

In order to smooth mask edges and retain optimality, as explained earlier, we 
inpaint the mask via a simple diffusive inpainting
method \citep{Bucher:2015ura}. First, we fill the masked regions with the
average value of the non-masked part of the map. Then we replace each masked
pixel with the average value of its neighbours and iterate this 2000 times. 
This is exactly the same procedure as adopted in $2013$ and $2015$.
 
Linear correction terms and $\fnl$ errors are obtained from
the FFP10 simulations, processed through the four component-separation pipelines. To this end, 
all pipelines use all the available $300$ FFP10 noise realizations, except both modal estimators, which use 
only $160$ maps (in order to speed up the computation). The good 
convergence of the modal pipelines with $160$ maps was thoroughly tested in previous releases. 
There, we showed with a large number of tests on realistic simulations
that the level of agreement between all our bispectrum estimators was perfectly consistent with 
theoretical expectations.
Accurate tests have found some mismatch between the noise levels in the data and that of the FFP10 simulations 
\citep{planck2016-l03,planck2016-l04}. This is roughly at the $3\,\%$ level in the noise power spectrum at $\ell \approx 2000$, in temperature, and at the percent level, 
in polarization. We find (see Sect.~\ref{sec:noisetest} for details) that this mismatch does not play a significant role in our analysis, and can safely be ignored.

All theoretical quantities (e.g., bispectrum templates and
lensing bias) are computed assuming the \Planck\ 2018 best-fit cosmology and
making use of the CAMB computer code\footnote{\url{http://camb.info/}}
\citep{Lewis:1999bs} to compute radiation transfer functions and theoretical
power spectra. The HEALPix
computer code\footnote{\url{http://healpix.sourceforge.net/}}
\citep{gorski2005} is used to perform spherical harmonic transforms.

As far as temperature is concerned, we maintain the same multipole ranges as in the $2013$ and $2015$ analyses, which is 
$2 \leq \ell \leq 2000$ for the KSW and modal estimators and
$2 \leq \ell \leq 2500$ for the Binned estimator. The different choice of $\ell_{\rm max}$ does not produce 
any significant effect on the results, since the $2000 < \ell \leq 2500$ range is noise dominated and the measured 
value of $\fnl$ remains very stable in that range, as confirmed by validation tests discussed in Sect.~\ref{sec:Sec_valid_data}. The angular resolution (beam FWHM) of both the cleaned temperature and polarization maps is 5 arcminutes.

The main novelty of the current analysis is the use of the low-$\ell$ polarization 
multipoles that were not exploited in $2015$ ($\ell <40$ polarization multipoles were removed by means of a 
high-pass filter). More precisely, KSW and modal estimators work in the polarization multipole 
range $4 \leq \ell \leq 1500$, while the Binned estimator considers 
$4 \leq \ell \leq 2000$. For the same reasons as above, this different choice
of $\ell_\mathrm{max}$ does not have any impact on the results.

The choice of using $\ell_{\rm min} = 4$ for polarization, thus removing the first two polarization multipoles, is instead dictated 
by the presence of some anomalous results in tests on simulations. When $\ell = 2$ and $\ell = 3$ are included, 
we observe some small bias arising in the local $\fnl$ measurement extracted from FFP10 maps, together with a spurious increase of the uncertainties.
We also notice larger discrepancies between the 
different estimation pipelines than expected from either theoretical arguments or previous validation tests on simulations. This can 
be ascribed to the presence of some small level of non-Gaussianity in the polarization noise at very low $\ell$.  We stress 
that the choice of cutting the first two polarization multipoles does not present any particular issue, since it is performed a priori, before looking 
at the data (as opposed to the simulations), and generates an essentially negligible loss of information.

In addition, the Binned bispectrum estimator removes from the analysis all
bispectrum $TEE$ configurations (i.e., those involving one temperature mode and
two polarization modes) with the temperature mode in the bin $[2,3]$.
This is again motivated by optimality considerations: with these modes included
the computed errors are much larger than the optimal Fisher errors,
while after removing them the errors are effectively optimal. As errors
are computed from simulations, this is again an a priori choice, made
before looking at the data. It is not clear why only the Binned bispectrum
estimator requires this additional removal, but of course the estimators are
all quite different, with different sensitivities, which is exactly one of the
strengths of having multiple estimators for our analyses.

The Binned bispectrum estimator uses a binning that is identical to the one
in 2015, with 57 bins. The boundary values 
of the bins are 2, 4, 10, 18, 30, 40, 53, 71, 99, 126, 154, 211, 243, 281, 
309, 343, 378, 420, 445, 476, 518, 549, 591, 619, 659, 700, 742, 771, 800, 
849, 899, 931, 966, 1001, 1035, 1092, 1150, 1184, 1230, 1257, 1291, 1346, 
1400, 1460, 1501, 1520, 1540, 1575, 1610, 1665, 1725, 1795, 1846, 1897, 2001, 
2091, 2240, and 2500 (i.e., the first bin is [2, 3], the second [4, 9], etc., 
while the last one is [2240, 2500]). This binning was determined in 2015 by
minimizing the increase in the theoretical variance for the primordial shapes
due to the binning.

As in our $2015$ analysis, we use two different polarized modal estimators. The ``Modal~1'' pipeline expands separately the 
 \textit{TTT}, \textit{EEE}, \textit{TTE}, and \textit{EET} bispectra \citep{2019arXiv190402599S}. It then writes the estimator normalization in separable expanded form and estimates 
$\fnl$ via a direct implementation of Eq.~\eqref{eq:diagcovestimator}.
The ``Modal~2'' pipeline uses a different approach \citep[see][for details]{2014arXiv1403.7949F}. It first orthogonalizes $T$ and $E$ multipoles to produce new, uncorrelated, $\hat{a}^T_{\ell m}$ and $\hat{a}^E_{\ell m}$ coefficients. It then builds uncorrelated bispectra out of these coefficients, which are constrained independently, simplifying the form and reducing the number of terms in the estimator. However, the rotation procedure does not allow a direct estimation of the \textit{EEE} bispectrum. Direct \textit{EEE} reconstruction is generally useful for validation purposes and can be performed with the Modal~1 estimator.

As in $2015$, Modal~1 is used to study in detail the local, equilateral, and orthogonal shapes, as well as to perform a large number of validation and robustness tests. Modal~2 is mostly dedicated to a thorough study of non-standard shapes having a large parameter space (like oscillatory bispectra). The two pipelines are equipped with modal bases optimized for their respective purposes. Modal~1 uses 600 polynomial modes, augmented with radial modes extracted from the KSW expansion of the local, equilateral, and orthogonal templates, in order to speed up convergence for these shapes. The Modal~2 expansion uses a 
higher-resolution basis, including 2000 polynomial modes and a Sachs-Wolfe local template, to improve efficiency in the squeezed limit. 

For oscillating non-Gaussianities we also use two specialized estimators \citep{Munchmeyer:2014nqa,Munchmeyer:2014cca} that specifically target the high-frequency range of shapes, which cannot be covered by the modal pipelines or the Binned estimator. Both of these estimators are equivalent to those used in~\citetalias{planck2014-a19}.

\section{Non-primordial contributions to the CMB bispectrum}
\label{sec:npNG}

In this section we investigate those non-primordial contributions to the CMB
bispectrum that we can detect in the cleaned maps, namely lensing and
extragalactic point sources. These then potentially have to be taken into
account when determining the constraints on the various primordial NG shapes in
Sect.~\ref{sec:Results}. On the other hand, the study of other non-primordial
contaminants (that we do not detect in the cleaned maps) is part of the
validation work in Sect.~\ref{sec:Sec_valid_data}.

\subsection{Non-Gaussianity from the lensing bispectrum}\label{subsec:lensingISW} 
CMB lensing generates a significant CMB bispectrum \citep{Hanson:2009gu,2009PhRvD..80l3007M,2011JCAP...03..018L,2013arXiv1303.1722M}.
In temperature, this is due to correlations between the lensing potential and the integrated Sachs-Wolfe \citep[ISW;][]{Sachs:1967er} contribution to the CMB anisotropies. In polarization, the dominant contribution instead comes from correlations between the lensing potential and \itE\ modes generated by scattering at reionization. Both the ISW and reionization contributions affect large scales, while lensing is a small-scale effect, so that the resulting bispectra peak on squeezed configurations. Therefore, they can significantly contaminate primordial NG measurements, especially for the local shape.
It has been known for a while
that for a high-precision experiment like \Planck\ the effect is large enough
that it must be taken into account. The temperature-only 2013 \Planck\ results \citepalias[][\citealt{planck2013-p14}, \citealt{planck2013-p12}]{planck2013-p09a} showed the first detection of the lensing CMB temperature bispectrum and the associated bias. This was later confirmed in the 2015
\Planck\ results \citepalias[\citealt{planck2014-a26}]{planck2014-a19}
both for $T$-only and for the full \itTpE\ results.
The template of the lensing bispectrum\footnote{As a reminder, let us stress that the expression ``lensing bispectrum'' in this paper always refers to the 3-point function generated by correlations between the lensing potential and ISW or reionization contributions, as explained in the main text. We are therefore not referring here to NG lensing signatures arising from the deflection potential alone, such as those considered in the context of CMB lensing reconstruction (and producing a leading trispectrum contribution).}
is given by \citep{2000PhRvD..62d3007H,2011JCAP...03..018L},
\begin{align}
b_{\ell_1 \ell_2 \ell_3}^{X_1 X_2 X_3,\,\mathrm{lens}} = &\
f_\mathrm{NL}^\mathrm{lens} \Bigl(
C_{\ell_2}^{X_2\phi} \tilde{C}_{\ell_3}^{X_1X_3} f_{\ell_1 \ell_2 \ell_3}^{X_1}
+ C_{\ell_3}^{X_3\phi} \tilde{C}_{\ell_2}^{X_1X_2} f_{\ell_1 \ell_3 \ell_2}^{X_1} \nonumber\\
& + C_{\ell_1}^{X_1\phi} \tilde{C}_{\ell_3}^{X_2X_3} f_{\ell_2 \ell_1 \ell_3}^{X_2}
+ C_{\ell_3}^{X_3\phi} \tilde{C}_{\ell_1}^{X_1X_2} f_{\ell_2 \ell_3 \ell_1}^{X_2} \nonumber\\
& + C_{\ell_1}^{X_1\phi} \tilde{C}_{\ell_2}^{X_2X_3} f_{\ell_3 \ell_1 \ell_2}^{X_3}
+ C_{\ell_2}^{X_2\phi} \tilde{C}_{\ell_1}^{X_1X_3} f_{\ell_3 \ell_2 \ell_1}^{X_3} \Bigr)\,
\label{LISW_redbisp_pol}
\end{align}
where the $X_i$ are either \itT\ or \itE. The tilde on $\tilde{C}_\ell^{X_iX_j}$
indicates that it is the lensed power spectrum, while
$C_\ell^{T\phi}$ and $C_\ell^{E\phi}$ are the temperature/polarization-lensing 
potential cross-power spectra.
The functions $f_{\ell_1 \ell_2 \ell_3}^{T,E}$ are defined by
\begin{align}
f_{\ell_1 \ell_2 \ell_3}^T = &\
\frac{1}{2} \bigl[ \ell_2 (\ell_2 + 1) + \ell_3 (\ell_3 + 1) - \ell_1 (\ell_1+1)
\bigr], \nonumber\\
f_{\ell_1 \ell_2 \ell_3}^E = &\ 
\frac{1}{2} \left[ \ell_2 (\ell_2 + 1) + \ell_3 (\ell_3 + 1) - \ell_1 (\ell_1+1)
\right ] \nonumber\\
&\ \times 
\left(\begin{array}{ccc} \ell_1 & \ell_2 & \ell_3 \\ 2 & 0 & -2 \end{array}\right)
\left(\begin{array}{ccc} \ell_1 & \ell_2 & \ell_3 \\ 0 & 0 & 0 \end{array}\right)^{-1},
\end{align}
if the sum $\ell_1+\ell_2+\ell_3$ is even and $\ell_1,\ell_2,\ell_3$ satisfies
the triangle inequality, and zero otherwise.
Unlike for all other templates, the amplitude parameter
$f_\mathrm{NL}^\mathrm{lens}$ is not unknown, but should be exactly equal
to 1 in the context of the assumed $\Lambda$CDM cosmology.

\begin{table}[htbp!]               
\begingroup
\newdimen\tblskip \tblskip=5pt
\caption{Results for the amplitude of the lensing bispectrum 
$f_{\rm NL}^{\rm lens}$ from the
\SMICA, \SEVEM, \NILC, and \Commander\ foreground-cleaned CMB maps,
for different bispectrum estimators. Uncertainties are 68\,\%~CL.}
\label{tab:fNL_lisw}                            
\nointerlineskip
\vskip -6mm
\footnotesize
\setbox\tablebox=\vbox{
   \newdimen\digitwidth
   \setbox0=\hbox{\rm 0}
   \digitwidth=\wd0
   \catcode`*=\active
   \def*{\kern\digitwidth}
   \newdimen\signwidth
   \setbox0=\hbox{+}
   \signwidth=\wd0
   \catcode`!=\active
   \def!{\kern\signwidth}
\halign{\hbox to 0.65in{#\leaderfil}\tabskip 1em&
\hfil#\hfil\tabskip 0.7em&
\hfil#\hfil\tabskip 0.7em&
\hfil#\hfil\tabskip 0.7em&
\hfil#\hfil\tabskip 0pt\cr
\noalign{\doubleline}
\omit&\multispan4\hfil Lensing amplitude\hfil\cr
\noalign{\vskip -4pt}
\omit&\multispan4\hrulefill\cr
\omit\hfil Estimator\hfil&\SMICA&\SEVEM&\NILC&\Commander\cr
\noalign{\vskip 4pt\hrule\vskip 4pt}
\omit\hfil \itT\hfil&&\cr
Binned&  $0.64\pm0.33$& $0.42\pm0.33$& $0.65\pm0.33$& $0.45\pm0.33$\cr
Modal~1& $0.74\pm0.33$& $0.59\pm0.32$& $0.72\pm0.32$& $0.54\pm0.33$\cr
Modal~2& $0.73\pm0.27$& $0.61\pm0.27$& $0.73\pm0.27$& $0.62\pm0.27$\cr
\noalign{\vskip 4pt\hrule\vskip 4pt}
\omit\hfil \textit{T+E}\hfil&&\cr
Binned&  $0.81\pm0.27$& $0.62\pm0.27$& $0.77\pm0.27$& $0.67\pm0.27$\cr
Modal~1& $0.90\pm0.26$& $0.82\pm0.25$& $0.83\pm0.25$& $0.73\pm0.26$\cr
\noalign{\vskip 3pt\hrule\vskip 4pt}}}
\endPlancktable                    
\endgroup
\end{table}                        

\begin{table}[htbp!]               
\begingroup
\newdimen\tblskip \tblskip=5pt
\caption{Results for the amplitude $f_{\rm NL}^{\rm lens}$
  of the lensing bispectrum from the
  \SMICA\ ``no-SZ'' temperature map, combined with the standard \SMICA\
  polarization map, for the Binned and Modal~1 bispectrum estimators.
  Uncertainties are 68\,\%~CL.}
\label{tab:lisw_noSZ}
\nointerlineskip
\vskip -3mm
\footnotesize
\setbox\tablebox=\vbox{
   \newdimen\digitwidth
   \setbox0=\hbox{\rm 0}
   \digitwidth=\wd0
   \catcode`*=\active
   \def*{\kern\digitwidth}
   \newdimen\signwidth
   \setbox0=\hbox{+}
   \signwidth=\wd0
   \catcode`!=\active
   \def!{\kern\signwidth}
\halign{\hbox to 0.85in{#\leaderfil}\tabskip 1em&
\hfil#\hfil\tabskip 0pt\cr
\noalign{\doubleline}
\omit\hfil Estimator\hfil& Lensing amplitude\cr
\noalign{\vskip 4pt\hrule\vskip 4pt}
\omit\hfil \itT\hfil&\cr
Binned&  $0.83\pm0.35$\cr
Modal~1& $0.90\pm0.34$\cr
\noalign{\vskip 4pt\hrule\vskip 4pt}
\omit\hfil \textit{T+E}\hfil&\cr
Binned&  $0.90\pm0.28$\cr
Modal~1& $1.03\pm0.27$\cr
\noalign{\vskip 3pt\hrule\vskip 4pt}}}
\endPlancktable                    
\endgroup
\end{table}                        

The results for $f_\mathrm{NL}^\mathrm{lens}$ can be found in
Table~\ref{tab:fNL_lisw}.
Error bars have been determined based on FFP10 simulations.
As we have seen in the previous releases, the results for $T$-only are on the
low side. \SMICA\ and \NILC\ have remained stable compared to 2015 and are
marginally consistent with the expected value at the $1\,\sigma$ level. \SEVEM\
and \Commander, on the other hand, have both decreased compared to 2015 and
are now further than $1\,\sigma$ away from unity. However, when polarization is
added all results increase and become mostly consistent with unity at the
$1\,\sigma$ level. Using the \SMICA\ map (which we often focus on
in the rest of the paper, see Sect.~\ref{sec:Sec_valid_data} for discussion)
and the Modal~1 estimator (because it is one of the two
estimators for which the lensing template has been implemented in both
\itT\ and \itE\ and the Binned estimator is slightly less well-suited for this
particular shape, since it is a difficult template to bin), we conclude that
we have a significant detection of the lensing bispectrum; the hypothesis
of having no lensing bispectrum is excluded at $3.5\,\sigma$ using the full
temperature and polarization data.

It was pointed out in \cite{Hill:2018ypf} that the coupling between the ISW effect and the 
thermal Sunyaev-Zeldovich (tSZ) effect can produce a significant ISW-tSZ-tSZ
temperature bispectrum, which peaks for squeezed modes, and can therefore contaminate 
especially our local and lensing results. The semi-analytic approach in \cite{Hill:2018ypf} 
makes predictions for single frequency channels,  and cannot be directly applied to the multi-frequency component-separated 
data we are using. However, it is interesting that such an approach shows an anti-correlation at all frequencies 
between the ISW-tSZ-tSZ contamination and the lensing bispectrum shape. This could be 
a possible explanation for the slightly low \itT-only value of $f_\mathrm{NL}^\mathrm{lens}$
observed in the final data, across all component-separated maps.
We therefore investigate this issue further, by measuring the amplitude of the lensing bispectrum 
using \SMICA\ maps in which the tSZ signal has been subtracted in addition to the usual components.
The results for this
``no-SZ'' map are given in Table~\ref{tab:lisw_noSZ}. It is clear that the
results have increased and are now closer to unity. Using these values,
the hypothesis of having no lensing bispectrum is excluded at
$3.8\,\sigma$ using the full temperature and polarization data, and the Modal~1 estimator.

The hypothesis that ISW-tSZ-tSZ residuals are contributing to the temperature-only lensing bispectrum 
amplitude result is reinforced by the fact that our  $f_\mathrm{NL}^\mathrm{lens}$ measurements from \itTpE\
are systematically closer to $1$ (polarization does not correlate with ISW and helps in debiasing the result).
The SZ-removed (hereafter ``no-SZ'') \SMICA\ measurements of local $f_{\rm NL}$, however,
do not support this hypothesis, since they do not display any large shift, whereas a residual ISW-tSZ-tSZ bispectrum should correlate with all shapes that peak in the squeezed limit. Still, one cannot exclude the possibility that multi-frequency component-separation affects the two cases differently.
Both our local no-SZ results and a further discussion of this effect are presented in Sect.~\ref{sec:noSZ}, where the
impact on other primordial shapes is also evaluated and comparisons with simulations are carried out. The final conclusion is that the evidence for ISW-tSZ-tSZ contamination on the temperature-only lensing bispectrum measurements is not very strong, but the possibility cannot be ruled out.

We also note here that another potentially important source of contamination for the local and lensing shapes is given 
by the coupling between lensing and the CIB. However, the frequency-by-frequency analysis in \cite{Hill:2018ypf} shows that in this case the expected bias is positive at all frequencies. The 
 systematically low values of $f_\mathrm{NL}^\mathrm{lens}$ observed  in temperature seem therefore to indicate that the CIB bispectrum contamination does not leak into the final component-separated maps, at least not at a level that is significant for our analysis. This is further reinforced by the fact that we do not detect any CIB signal directly in the cleaned maps (see Sect.~\ref{sec:point_sources}).

\begin{table}[htbp!]             
\begingroup
\newdimen\tblskip \tblskip=5pt
\caption{Bias in the three primordial $f_{\rm NL}$ parameters due to the
lensing signal for the four component-separation methods.}
\label{tab:lisw_bias}                            
\nointerlineskip
\vskip -6mm
\footnotesize
\setbox\tablebox=\vbox{
   \newdimen\digitwidth
   \setbox0=\hbox{\rm 0}
   \digitwidth=\wd0
   \catcode`*=\active
   \def*{\kern\digitwidth}
   \newdimen\signwidth
   \setbox0=\hbox{+}
   \signwidth=\wd0
   \catcode`!=\active
   \def!{\kern\signwidth}
   \newdimen\dotwidth
   \setbox0=\hbox{.}
   \dotwidth=\wd0
   \catcode`^=\active
   \def^{\kern\dotwidth}
\halign{\hbox to 1.1in{#\leaderfil}\tabskip 0.5em&
\hfil#\hfil\tabskip 1.0em&
\hfil#\hfil&
\hfil#\hfil&
\hfil#\hfil\tabskip 0pt\cr
\noalign{\doubleline\vskip 2pt}
\omit&\multispan4\hfil Lensing $f_{\rm NL}$ bias\hfil\cr
\noalign{\vskip -2pt}
\omit&\multispan4\hrulefill\cr
\omit\hfil Shape\hfil&\SMICA&\SEVEM&\NILC&\Commander\cr
\noalign{\vskip 4pt\hrule\vskip 6pt}
\itT\ Local&       !$*$7.3& !$*$7.2& !$*$7.3& !$*$7.3\cr
\itT\ Equilateral& *$-$0.7& *$-$0.6& *$-$0.7& *$-$0.7\cr
\itT\ Orthogonal&  $-$23^*& $-$23^*& $-$23^*& $-$23^*\cr
\noalign{\vskip 4pt\hrule\vskip 6pt}
\itE\ Local&       !$*$0.5& !$*$0.5& !$*$0.5& !$*$0.5\cr
\itE\ Equilateral& !$*$0.5& !$*$0.5& !$*$0.5& !$*$0.5\cr
\itE\ Orthogonal&  *$-$0.7& *$-$0.7& *$-$0.7& *$-$0.7\cr
\noalign{\vskip 4pt\hrule\vskip 6pt}
\itTpE\ Local&       !$*$5.0& !$*$5.0& !$*$5.0& !$*$5.0\cr
\itTpE\ Equilateral& !$*$1.0& !$*$1.1& !$*$1.1& !$*$1.0\cr
\itTpE\ Orthogonal&  *$-$9.1& *$-$9.4& *$-$9.2& *$-$9.4\cr
\noalign{\vskip 3pt\hrule\vskip 4pt}}}
\endPlancktable                    

\endgroup
\end{table}                        

In this paper our main concern with the lensing bispectrum is its
influence on the primordial shapes. The bias due to the lensing bispectrum
on the estimation of the $\fnl$ parameter of another shape $S$ is given
by the inner product of the lensing bispectrum (Eq.~\ref{LISW_redbisp_pol})
with the bispectrum of that shape $S$, divided by the inner product of
the bispectrum $S$ with itself \citepalias[see][for more details, or e.g.,
\citealt{Bucher:2015ura} for a derivation]{planck2014-a19}.
The values for the bias, as computed from theory, are given in
Table~\ref{tab:lisw_bias}. Note that
the bias values that can be read off from e.g., Table~\ref{tab:fNLall}
in Sect.~\ref{sec:Results}
can differ slightly from these, because each estimator uses values
computed using the approximations appropriate to the estimator. However,
those differences are insignificant compared to the uncertainties.
As seen already in the previous releases, for \itT-only data and for \itTpE\ the
bias is very significant for local and to a lesser extent for orthogonal NG.
For \itT-only local NG the bias is even larger than the error bars on $\fnl$.
Hence it is quite important to take this bias into account. On the other hand,
for \itE-only the effect is completely negligible.

Lastly, we would like to point out that lensing can also contribute to the covariance \citep{Babich:2004yc}. To lowest order, as was done in this analysis, the spectra in Eq.~\eqref{eq:diagcovestimator} are replaced with the lensed spectra. However, it was shown by \cite{Babich:2004yc}, and later confirmed by \cite{2011JCAP...03..018L}, that the contribution from the connected four-point function induced by lensing can become an important contribution to the covariance.  Although the analysis here has not shown the effect of lensing to be large (since tests were performed and analysis was done on lensed simulations with no obvious degradation over the Gaussian Fisher errors), it is expected that this will become an important challenge in CMB non-Gaussianity constraints beyond \Planck. Furthermore, unlike the signal contamination discussed above, this will likely be equally important for polarization. Because of the shape of the lensing-induced covariance, which tends to be largest for squeezed configurations, the local shape will be most affected. The estimates performed by \cite{Babich:2004yc} were done in the flat-sky limit and did not include polarization; in addition, computations were truncated at linear order in the potential. Although the latter seems justified, it was later shown that for the lensing power spectrum covariance the linear contribution is actually subdominant to the quadratic contribution \citep{Peloton:2016kbw}. For future CMB analysis it will be important to address these open questions. One obvious solution to the extra covariance would be to delens the maps \citep{Green:2016cjr} before applying the estimators. Interestingly, this would automatically remove some of the signal-induced lensing contributions discussed above.

\subsection{Non-Gaussianity from extragalactic point sources}
\label{sec:point_sources}

As seen in the previous releases, extragalactic point sources are a
contaminant present in the bispectrum as measured by \Planck.
They are divided into populations of unclustered and clustered sources. The former are radio and late-type infrared galaxies \citep[see e.g.,][]{Toffolatti1998,Gonzalez-Nuevo2005}, while the latter are primarily dusty star-forming galaxies constituting the cosmic infrared background \citep[CIB;][]{Lagache2005}.
For both types of point sources analytic (heuristic) bispectrum templates
have been determined, which can be fitted jointly with the primordial NG
templates to deal with the contamination.

The templates used here are the same as those used in
\citetalias{planck2014-a19} (see that
paper for more information and references). The reduced angular bispectrum
template of the unclustered sources is \citep{2001PhRvD..63f3002K}
\begin{equation}\label{Eq:unclust_template}
b_{\ell_1 \ell_2 \ell_3}^\mathrm{unclust} = b_\mathrm{PS} = \mathrm{constant.}
\end{equation}
This template is valid in polarization as well as temperature. However, we
do not detect all the same point sources in polarization as we detected
in temperature,
so that a full \itTpE\ analysis does not make sense for this template. In fact, there is no detection of unclustered point sources in the cleaned
\Planck\ polarization map, so that we do not include the \itE-only values in
the table either.
The reduced angular bispectrum template for the clustered sources, i.e.,
the CIB, is
\citep{Lacasa2014,Penin2014}
\begin{equation}\label{Eq:CIB_template}
b_{\ell_1 \ell_2 \ell_3}^\mathrm{CIB} = A_\mathrm{CIB} \left[ \frac{(1+\ell_1/\ell_\mathrm{break}) (1+\ell_2/\ell_\mathrm{break}) (1+\ell_3/\ell_\mathrm{break})}{(1+\ell_0/\ell_\mathrm{break})^3}\right]^q,
\end{equation}

where the index is $q=0.85$, the break is located at $\ell_\mathrm{break}=70$,
and $\ell_0=320$  is the pivot scale for normalization. This template is valid only for temperature; the CIB is negligibly polarized.
\begin{table}[htbp!]               
\begingroup
\newdimen\tblskip \tblskip=5pt
\caption{Joint estimates of the bispectrum amplitudes of unclustered and 
clustered point sources in the cleaned \Planck\ temperature maps, determined 
with the Binned bispectrum estimator. Uncertainties are 68\,\%~CL.}                          
\label{Table:bps_and_ACIB}                         
\nointerlineskip
\vskip -3mm
\footnotesize
\setbox\tablebox=\vbox{
   \newdimen\digitwidth
   \setbox0=\hbox{\rm 0}
   \digitwidth=\wd0
   \catcode`*=\active
   \def*{\kern\digitwidth}
   \newdimen\signwidth
   \setbox0=\hbox{+}
   \signwidth=\wd0
   \catcode`!=\active
   \def!{\kern\signwidth}
\halign{\hbox to 0.9in{#\leaderfil}\tabskip 2.2em&
\hfil#\hfil\tabskip 2.2em&
\hfil#\hfil\tabskip 0pt\cr
\noalign{\doubleline\vskip -2pt}
\omit\hfil Map\hfil& $b_\mathrm{PS}/(10^{-29})$& $A_\mathrm{CIB}/(10^{-27})$\cr
\noalign{\vskip3pt\hrule\vskip 3pt}
\SMICA&     $4.7\pm2.6$& $0.8\pm1.3$\cr
\SEVEM&     $7.0\pm2.8$& $1.4\pm1.4$\cr
\NILC&      $5.2\pm2.7$& $0.3\pm1.3$\cr
\Commander& $3.4\pm2.6$& $1.1\pm1.3$\cr
\noalign{\vskip 3pt\hrule}}}
\endPlancktable                    
\endgroup
\end{table}                        

The results for both extragalactic point source templates, as determined by the
Binned bispectrum estimator applied to the \Planck\ temperature map cleaned
with the four component-separation methods, can be found in
Table~\ref{Table:bps_and_ACIB}. Because the two templates are 
highly correlated (93\,\%), the results have been determined through a joint
analysis.
Contamination from unclustered sources is detected in all component-separated
maps, although at different levels, with \SEVEM\ having the largest
contamination. The CIB bispectrum, on the other hand, is not detected in a
joint analysis. Both point-source templates are negligibly correlated with
the primordial NG templates and the lensing template (all well below
$1\%$ for the unclustered point sources).
For this reason, and despite the detection of unclustered point sources in the
cleaned maps, it makes no difference for the primordial results in the next
sections if point sources are included in a joint analysis or completely
neglected.

\section{Results}
\label{sec:Results}

\subsection{Constraints on local, equilateral, and orthogonal $f_{\rm NL}$}
\label{fnl_loc_eq_ort_results}

We now describe our analysis of the standard local, equilateral, and orthogonal shapes. 
We employ the four bispectrum estimators described in Sects.~\ref{sec:KSW_est},
\ref{sec:modal_est}, and \ref{sec:binned_est} on the temperature 
and polarization maps generated by the \SMICA, \SEVEM, \NILC, and \Commander\ component-separation pipelines.
Further details about our data analysis setup were provided in Sect.~\ref{sec:settings}. As explained there, the main novelty, compared to the $2015$ release,
is the use of polarization multipoles in the range $4 \leq \ell < 40$, which were previously excluded.

Our final results are summarized in Table~\ref{tab:fNLall}, while data
validation tests will be presented in Sect.~\ref{sec:Sec_valid_data}. As in
2015, we show final $\fnl$ estimates for \itT-only, \itE-only, and for the
full \itTpE\ data set, with and without subtraction of the lensing bias.
When we subtract the lensing bias, we assume a theoretical prediction for the lensing 
bispectrum amplitude based on the \Planck\ best-fit
$\Lambda$CDM cosmological parameters (see Sect.~\ref{subsec:lensingISW}). We note that propagating uncertainties in the parameters 
has a negligible effect on the predicted lensing bispectrum, and the validity of the $\Lambda$CDM assumption is of course 
consistent with all \Planck\ measurements.
An alternative to the direct subtraction of the predicted lensing
bias of the primordial shapes would be performing a full
joint bispectrum analysis (accounting for the measured lensing bispectrum
amplitude and propagating the uncertainty into the final primordial NG error
budget). While the latter approach is in principle more conservative, we opt for the former 
both for simplicity and for consistency with previous analyses, 
as it turns out that the difference between
the two methods has no significant impact on our results. If, as an example, we 
perform a $T$-only joint bispectrum analysis using the \SMICA\ map and the Binned estimator, we obtain
 $\fnl^\mathrm{local} = 2.7 \pm 5.7$ and $\fnl^\mathrm{lens} = 0.60 \pm 0.34$. 
The joint $\itTpE$ analysis produces instead $\fnl^\mathrm{local} = -1.7 \pm 5.2$ and $\fnl^\mathrm{lens} =0.82 \pm 0.27$. 
Hence the uncertainties obtained from the joint analysis are practically stable with respect to those shown in Table~\ref{tab:fNLall}, while the small shift
in the local central value (due to the low value of the measured $\fnl^\mathrm{lens}$)
would not change our conclusions in any way. More details about the lensing
contribution are provided in Sect.~\ref{sec:npNG}, where we also discuss the
negligible impact of point source contamination on primordial bispectra.

Table~\ref{tab:fNLall} constitutes the most important result of this section. 
As done in $2013$ and $2015$, we select the KSW estimator and the 
\SMICA\ map to provide the final \Planck\ results for the local, equilateral, and orthogonal 
bispectra; these results are summarized in Table~\ref{Tab_KSW+SMICA}. The motivations for 
this choice are as in the past: \SMICA\ performs well in all validation tests and shows excellent stability 
across different data releases; the KSW estimator, while not able to deal with non-separable shapes or 
reconstruct the full bispectrum, can treat exactly the local, equilateral, and orthogonal templates that are analysed here.
\begin{table*}[htbp!]                 
\begingroup
\newdimen\tblskip \tblskip=5pt
\caption{Results for the $f_{\rm NL}$ parameters of the primordial local, 
equilateral, and orthogonal shapes, determined by the KSW, Binned and Modal 
estimators from the \SMICA, \SEVEM, \NILC, and \Commander\ foreground-cleaned 
maps. Results have been determined using an independent single-shape analysis
and are reported both without (first set of columns) and with (second set of columns) subtraction of the lensing bias.
Uncertainties are $68\,\%$ CL.}
\label{tab:fNLall}
\nointerlineskip
\vskip -3mm
\footnotesize
\setbox\tablebox=\vbox{
   \newdimen\digitwidth
   \setbox0=\hbox{\rm 0}
   \digitwidth=\wd0
   \catcode`*=\active
   \def*{\kern\digitwidth}
   \newdimen\signwidth
   \setbox0=\hbox{+}
   \signwidth=\wd0
   \catcode`!=\active
   \def!{\kern\signwidth}
\newdimen\dotwidth
\setbox0=\hbox{.}
\dotwidth=\wd0
\catcode`^=\active
\def^{\kern\dotwidth}
\halign{\hbox to 1.2 in{#\leaderfil}\tabskip 2.0em&
\hfil#\hfil\tabskip 0.5em&
\hfil#\hfil&
\hfil#\hfil\tabskip 2.0em&
\hfil#\hfil\tabskip 0.5em&
\hfil#\hfil&
\hfil#\hfil&
\hfil#\hfil\tabskip 0pt\cr
\noalign{\doubleline\vskip 2pt}
\omit&\multispan3\hfil Independent\hfil&
\multispan4\hfil Lensing subtracted\hfil\cr
\noalign{\vskip -4pt}
\omit&\multispan3\hrulefill& \multispan4\hrulefill\cr
\noalign{\vskip 2pt}
\omit\hfil Shape\hfil&KSW&Binned&Modal~1&**KSW&Binned&Modal~1&Modal~2\cr
\noalign{\vskip 4pt\hrule\vskip 6pt}
\omit\hfil\SMICA\ \itT\hfil&&\cr
Local&       $!*6.7\pm*5.6$& $!*6.8\pm*5.6$& $!*6.3\pm*5.8$& $*-0.5\pm*5.6$& $*-0.1\pm*5.6$& $*-0.6\pm*5.8$& $*-0.6\pm*6.4$\cr
Equilateral& $!*6^*\pm66^*$& $!27^*\pm69^*$& $!28^*\pm64^*$& $!*7^*\pm66^*$& $!26^*\pm69^*$& $!24^*\pm64^*$& $!34^*\pm67^*$\cr
Orthogonal&  $-38^*\pm36^*$& $-37^*\pm39^*$& $-29^*\pm39^*$& $-15^*\pm36^*$& $-11^*\pm39^*$& $*-4^*\pm39^*$& $-26^*\pm43^*$\cr
\noalign{\vskip 3pt}
\omit\hfil\SMICA\ \itE\hfil&&\cr
Local&       $!*48\pm*28$& $!*49\pm*26$& $!*44\pm*25$& $!*47\pm*28$& $!*48\pm*26$& $!*44\pm*25$& \cr
Equilateral& $!170\pm161$& $!170\pm140$& $!190\pm160$& $!169\pm161$& $!170\pm140$& $!200\pm160$& \cr
Orthogonal&  $-209\pm*86$& $-180\pm*83$& $-210\pm*85$& $-208\pm*86$& $-180\pm*83$& $-220\pm*85$& \cr
\noalign{\vskip 3pt}
\omit\hfil\SMICA\ \itTpE\hfil&&\cr
Local&       $!*4.1\pm*5.1$& $!*2.2\pm*5.0$& $!*4.6\pm*4.7$& $*-0.9\pm*5.1$& $*-2.5\pm*5.0$& $*-0.1\pm*4.7$& $*-2.0\pm*5.0$\cr
Equilateral& $-17^*\pm47^*$& $-16^*\pm48^*$& $*-6^*\pm48^*$& $-18^*\pm47^*$& $-19^*\pm48^*$& $*-8^*\pm48^*$& $*-4^*\pm43^*$\cr
Orthogonal&  $-46^*\pm23^*$& $-45^*\pm24^*$& $-37^*\pm24^*$& $-37^*\pm23^*$& $-34^*\pm24^*$& $-28^*\pm24^*$& $-40^*\pm24^*$\cr
\noalign{\vskip 4pt\hrule\vskip 6pt}
\omit\hfil\SEVEM\ \itT\hfil&&\cr
Local&       $!*5.0\pm*5.6$& $!*6.8\pm*5.7$& $!*6.2\pm*6.0$& $*-2.3\pm*5.6$& $!*0.0\pm*5.7$& $*-1.6\pm*6.0$& $!*0.0\pm*6.5$\cr
Equilateral& $!16^*\pm66^*$& $!45^*\pm70^*$& $!39^*\pm64^*$& $!17^*\pm66^*$& $!43^*\pm70^*$& $!33^*\pm64^*$& $!34^*\pm68^*$\cr
Orthogonal&  $!*2^*\pm37^*$& $-17^*\pm39^*$& $*-5^*\pm40^*$& $!24^*\pm37^*$& $!*9^*\pm39^*$& $!25^*\pm40^*$& $-14^*\pm43^*$\cr
\noalign{\vskip 3pt}
\omit\hfil\SEVEM\ \itE\hfil&&\cr
Local&       $!*38\pm*29$& $!*56\pm*29$& $!*55\pm*22$& $!*38\pm*29$& $!*55\pm*29$& $!*37\pm*22$& \cr
Equilateral& $!174\pm166$& $!230\pm160$& $!250\pm160$& $!173\pm166$& $!230\pm160$& $!260\pm160$& \cr
Orthogonal&  $-182\pm*88$& $-140\pm*88$& $-160\pm*87$& $-181\pm*88$& $-140\pm*88$& $-170\pm*87$& \cr
\noalign{\vskip 3pt}
\omit\hfil\SEVEM\ \itTpE\hfil&&\cr
Local&       $!*3.3\pm*5.1$& $!*3.6\pm*5.2$& $!*7.8\pm*4.7$& $*-1.7\pm*5.1$& $*-1.2\pm*5.2$& $!*1.8\pm*4.7$& $!*1.5\pm*5.1$\cr
Equilateral& $*-8^*\pm47^*$& $!*2^*\pm48^*$& $!*5^*\pm51^*$& $*-9^*\pm47^*$& $*-1^*\pm48^*$& $!*4^*\pm51^*$& $!23^*\pm45^*$\cr
Orthogonal&  $-24^*\pm23^*$& $-28^*\pm24^*$& $-19^*\pm24^*$& $-15^*\pm23^*$& $-18^*\pm24^*$& $*-6^*\pm24^*$& $-29^*\pm25^*$\cr
\noalign{\vskip 4pt\hrule\vskip 6pt}
\omit\hfil\NILC\ \itT\hfil&&\cr
Local&       $!*6.8\pm*5.7$& $!*6.8\pm*5.6$& $!*6.2\pm*5.9$& $*-0.4\pm*5.7$& $!*0.0\pm*5.6$& $*-1.7\pm*5.9$& $*-0.5\pm*6.6$\cr
Equilateral& $-10^*\pm66^*$& $!*6^*\pm69^*$& $!12^*\pm61^*$& $-10^*\pm66^*$& $!*5^*\pm69^*$& $!*6^*\pm61^*$& $!20^*\pm67^*$\cr
Orthogonal&  $-23^*\pm36^*$& $-21^*\pm39^*$& $-12^*\pm40^*$& $*-1^*\pm36^*$& $!*4^*\pm39^*$& $!17^*\pm40^*$& $-10^*\pm43^*$\cr
\noalign{\vskip 3pt}
\omit\hfil\NILC\ \itE\hfil&&\cr
Local&       $!**9\pm*30$& $!*22\pm*28$& $!*20\pm*22$& $!**9\pm*30$& $!*21\pm*28$& $!*15\pm*22$& \cr
Equilateral& $!*39\pm163$& $!*60\pm150$& $!*65\pm150$& $!*38\pm163$& $!*59\pm150$& $!*66\pm150$& \cr
Orthogonal&  $-133\pm*88$& $-120\pm*85$& $-150\pm*83$& $-133\pm*88$& $-120\pm*85$& $-160\pm*83$& \cr
\noalign{\vskip 3pt}
\omit\hfil\NILC\ \itTpE\hfil&&\cr
Local&       $!*3.9\pm*5.1$& $!*2.1\pm*5.1$& $!*3.9\pm*4.7$& $*-1.1\pm*5.1$& $*-2.7\pm*5.1$& $*-1.5\pm*4.7$& $*-2.4\pm*5.1$\cr
Equilateral& $-30^*\pm46^*$& $-29^*\pm47^*$& $-22^*\pm49^*$& $-31^*\pm46^*$& $-32^*\pm47^*$& $-24^*\pm49^*$& $-15^*\pm43^*$\cr
Orthogonal&  $-33^*\pm23^*$& $-31^*\pm23^*$& $-26^*\pm24^*$& $-24^*\pm23^*$& $-21^*\pm23^*$& $-14^*\pm24^*$& $-24^*\pm24^*$\cr
\noalign{\vskip 4pt\hrule\vskip 6pt}
\omit\hfil\Commander\ \itT\hfil&&\cr
Local&       $!*4.9\pm*5.6$& $!*5.6\pm*5.6$& $!*4.5\pm*5.9$& $*-2.3\pm*5.6$& $*-1.3\pm*5.6$& $*-3.0\pm*5.9$& $*-1.9\pm*6.6$\cr
Equilateral& $!14^*\pm66^*$& $!33^*\pm69^*$& $!33^*\pm62^*$& $!15^*\pm66^*$& $!32^*\pm69^*$& $!25^*\pm62^*$& $!36^*\pm68^*$\cr
Orthogonal&  $!*3^*\pm37^*$& $!*3^*\pm39^*$& $!14^*\pm40^*$& $!25^*\pm37^*$& $!29^*\pm39^*$& $!42^*\pm40^*$& $!*5^*\pm43^*$\cr
\noalign{\vskip 3pt}
\omit\hfil\Commander\ \itE\hfil&&\cr
Local&       $*!31\pm*29$& $*!43\pm*27$& $*!31\pm*21$& $!*31\pm*29$& $!*42\pm*27$& $!*27\pm*21$& \cr
Equilateral& $!163\pm167$& $!170\pm150$& $!190\pm160$& $!162\pm167$& $!170\pm150$& $!180\pm160$& \cr
Orthogonal&  $-179\pm*88$& $-160\pm*85$& $-180\pm*85$& $-178\pm*88$& $-160\pm*85$& $-190\pm*85$& \cr
\noalign{\vskip 3pt}
\omit\hfil\Commander\ \itTpE\hfil&&\cr
Local&       $!*3.0\pm*5.1$& $!*2.3\pm*5.1$& $!*3.4\pm*4.6$& $*-2.0\pm*5.1$& $*-2.5\pm*5.1$& $*-1.4\pm*4.6$& $*-1.7\pm*5.1$\cr
Equilateral& $*-9^*\pm47^*$& $*-6^*\pm48^*$& $!*5^*\pm50^*$& $-10^*\pm47^*$& $*-9^*\pm48^*$& $!*2^*\pm50^*$& $!35^*\pm44^*$\cr
Orthogonal&  $-23^*\pm23^*$& $-23^*\pm24^*$& $-14^*\pm24^*$& $-13^*\pm23^*$& $-12^*\pm24^*$& $*-2^*\pm24^*$& $-21^*\pm25^*$\cr
\noalign{\vskip 3pt\hrule\vskip 4pt}}}
\endPlancktablewide                 
\endgroup
\end{table*}                        
\begin{table}[htbp!]                 
	\begingroup
	\newdimen\tblskip \tblskip=5pt
	\caption{Results for the $f_{\rm NL}$ parameters of the primordial local, equilateral, and orthogonal shapes, determined by the KSW estimator from the \SMICA\ foreground-cleaned map. Both independent single-shape results and results with the lensing bias subtracted are reported; uncertainties are $68\,\%$ CL. The difference between this table and the corresponding values in the previous table is that here the equilateral and orthogonal shapes have been analysed jointly.}
\label{Tab_KSW+SMICA}
\nointerlineskip
\vskip -6mm
\footnotesize
\setbox\tablebox=\vbox{
\newdimen\digitwidth
\setbox0=\hbox{\rm 0}
\digitwidth=\wd0
\catcode`*=\active
\def*{\kern\digitwidth}
\newdimen\signwidth
\setbox0=\hbox{+}
\signwidth=\wd0
\catcode`!=\active
\def!{\kern\signwidth}
\newdimen\dotwidth
\setbox0=\hbox{.}
\dotwidth=\wd0
\catcode`^=\active
\def^{\kern\dotwidth}
\halign{\hbox to 1in{#\leaderfil}\tabskip 1em&
\hfil#\hfil\tabskip 1em&
\hfil#\hfil\tabskip 0pt\cr
\noalign{\vskip 10pt\doubleline\vskip 2pt}
\omit\hfil Shape\hfil&\hfil Independent\hfil&
\hfil Lensing subtracted\hfil\cr
\noalign{\vskip 4pt\hrule\vskip 6pt}
\multispan3\hfil\SMICA\,\, \itT\hfil\cr
Local&       $!*6.7\pm*5.6$& $*-0.5\pm*5.6$\cr
Equilateral& $!*4^*\pm67^*$& $!*5^*\pm67^*$\cr
Orthogonal&  $-38^*\pm37^*$& $-15^*\pm37^*$\cr
\noalign{\vskip 4pt\hrule\vskip 6pt}
\multispan3\hfil\SMICA\,\, \itTpE\hfil\cr
Local&       $!*4.1\pm*5.1$& $*-0.9\pm*5.1$\cr
Equilateral& $-25^*\pm47^*$& $-26^*\pm47^*$\cr
Orthogonal&  $-47^*\pm24^*$& $-38^*\pm24^*$\cr
\noalign{\vskip 3pt\hrule\vskip 4pt}}}
\endPlancktable                    
\endgroup
\end{table}                        

\begin{figure*}[htbp!]
\centering
\includegraphics[width=0.30\textwidth]{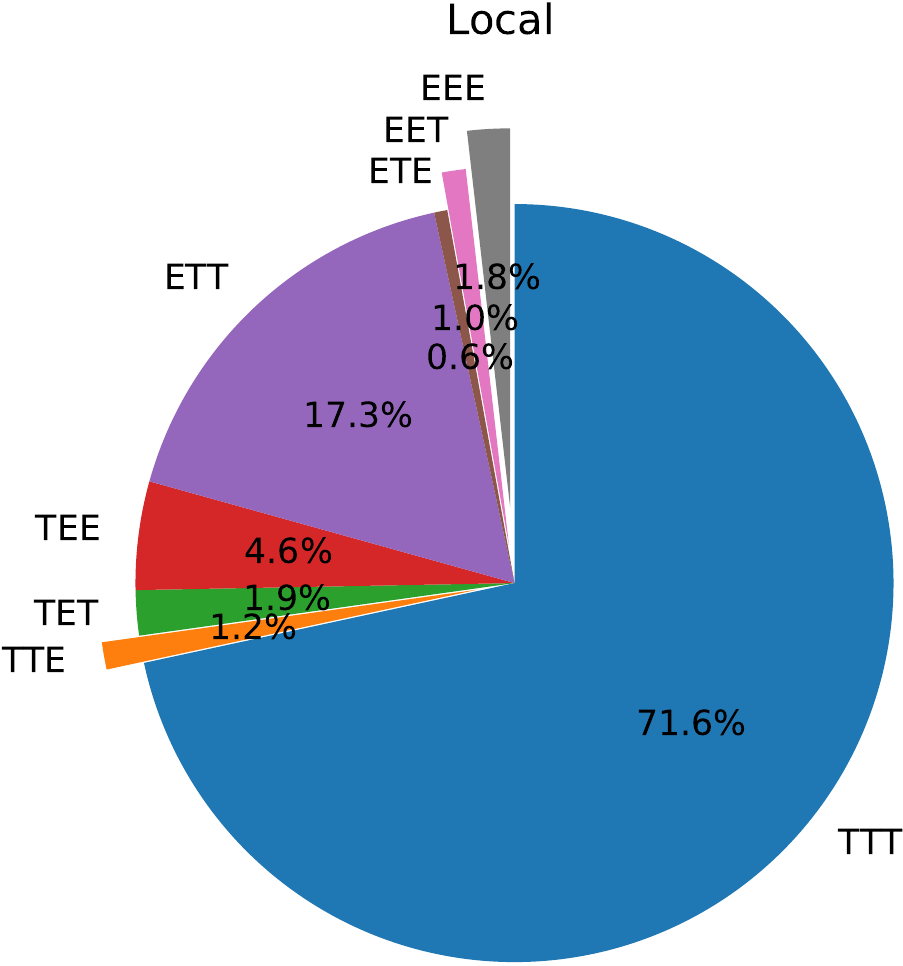}%
\includegraphics[width=0.32\textwidth]{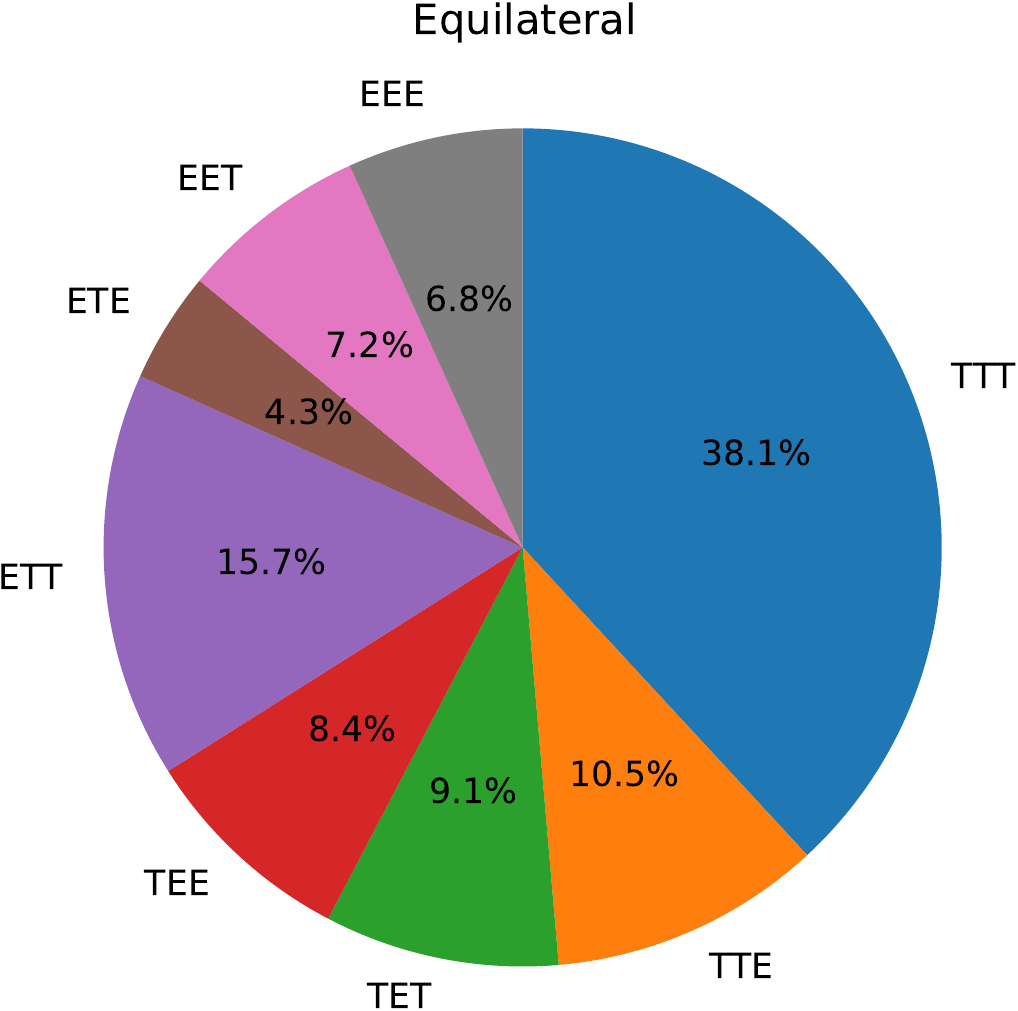}%
\includegraphics[width=0.30\textwidth]{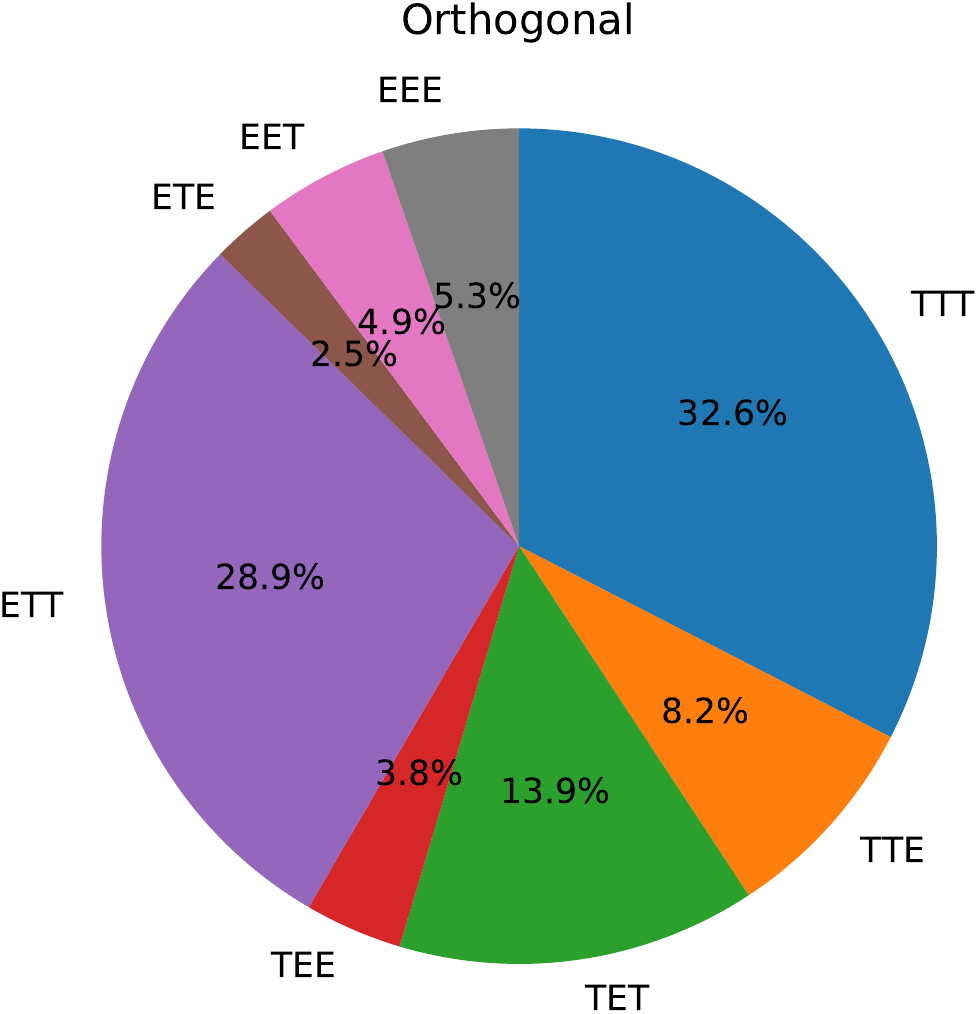}
\caption{Weights of each polarization configuration going into the total
    value of $\fnl$ for, from left to right, local, equilateral, and orthogonal
    shapes.
    Note that since we impose $\ell_1 \leq \ell_2 \leq \ell_3$, there is a
    difference between, e.g., \itT\itE\itE\ (smallest $\ell$ is temperature)
    and \itE\itE\itT\ (largest $\ell$ is temperature).}
\label{fig:contribTandEtofNL}
\end{figure*}

\begin{table*}[htbp!]                 
\begingroup
\newdimen\tblskip \tblskip=5pt
\caption{Results for $f_{\rm NL}$ for local isocurvature NG determined from the 
\SMICA\ \Planck\ 2018 map with the Binned bispectrum estimator. In each case the
adiabatic mode is considered together with one isocurvature mode
(either cold-dark-matter, neutrino-density, or neutrino-velocity isocurvature).
As explained in the text this gives six different $f_\mathrm{NL}$
parameters, indicated by the different combinations of the adiabatic (a)
and isocurvature (i) modes. Results are shown for 
both an independent and a fully joint analysis, for \itT-only, \itE-only, and 
full \itT$+$\itE\ data. In all cases the lensing bias has been subtracted.}
\label{Tab_isocurvNG}
\nointerlineskip
\vskip -3mm
\footnotesize
\setbox\tablebox=\vbox{
   \newdimen\digitwidth 
   \setbox0=\hbox{\rm 0} 
   \digitwidth=\wd0 
   \catcode`*=\active 
   \def*{\kern\digitwidth}
   \newdimen\signwidth 
   \setbox0=\hbox{+} 
   \signwidth=\wd0 
   \catcode`!=\active 
   \def!{\kern\signwidth}
   \newdimen\dotwidth 
   \setbox0=\hbox{.} 
   \dotwidth=\wd0 
   \catcode`^=\active 
   \def^{\kern\dotwidth}
\halign{\hbox to 0.8in{#\leaderfil}\tabskip 0.5em&
\hfil#\hfil\tabskip 1em&
\hfil#\hfil&
\hfil#\hfil\tabskip 2em&
\hfil#\hfil\tabskip 1em&
\hfil#\hfil&
\hfil#\hfil\tabskip 0pt\cr
\noalign{\doubleline\vskip 2pt}
\omit&\multispan3\hfil Independent\hfil& \multispan3\hfil Joint\hfil\cr
\noalign{\vskip -2pt}
\omit&\multispan3\hrulefill& \multispan3\hrulefill\cr
\noalign{\vskip 2pt}
\omit\hfil Shape\hfil& Cold dark matter& Neutrino density& Neutrino velocity& Cold dark matter& Neutrino density& Neutrino velocity\cr
\noalign{\vskip 4pt\hrule\vskip 6pt}
\itT\ a,aa& $-0.1\pm5.6$& $-0.1\pm5.6$& $-0.1\pm5.6$& $!15\pm16$& $-15\pm54$& $-29\pm49$\cr
\itT\ a,ai& $*-3\pm11$& $*-6\pm16$& $!27\pm29$& $-30\pm31$& $!*87\pm220$& $!310\pm360$\cr
\itT\ a,ii& $!440\pm950$& $-150\pm290$& $!480\pm360$& $!11000\pm8500*$& $-1800\pm4500$& $-1100\pm3800$\cr
\itT\ i,aa& $!33\pm53$& $!0.3\pm9.2$& $-0.1\pm4.9$& $!*95\pm120$& $!*46\pm110$& $-27\pm49$\cr
\itT\ i,ai& $!28\pm67$& $*-9\pm23$& $!14\pm21$& $-1300\pm1000$& $*-32\pm670$& $!*98\pm170$\cr
\itT\ i,ii& $!110\pm280$& $-130\pm250$& $!270\pm230$& $!2500\pm2000$& $*-140\pm2600$& $-1000\pm1400$\cr
\noalign{\vskip 5pt}
\itE\ a,aa& $!48\pm26$& $!48\pm26$& $!48\pm26$& $!25\pm42$& $*-60\pm12$0& $!*43\pm130$\cr
\itE\ a,ai& $!140\pm91*$& $!170\pm98*$& $!72\pm51$& $!*28\pm15$0& $!330\pm590$& $!280\pm370$\cr
\itE\ a,ii& $!*100\pm2500$& $!2000\pm1500$& $!310\pm460$& $-8100\pm4400$& $-9400\pm5500$& $-6100\pm2800$\cr
\itE\ i,aa& $!110\pm77*$& $!75\pm34$& $!31\pm21$& $!*43\pm14$0& $!*90\pm17$0& $*-94\pm120$\cr
\itE\ i,ai& $!330\pm180$& $!210\pm100$& $!43\pm35$& $!830\pm580$& $!500\pm890$& $!460\pm360$\cr
\itE\ i,ii& $!1800\pm1200$& $!1900\pm1100$& $!170\pm230$& $-1400\pm2300$& $-1300\pm4800$& $*-380\pm1000$\cr
\noalign{\vskip 5pt}
\itTpE\ a,aa& $-2.5\pm5.0$& $-2.5\pm5.0$& $-2.5\pm5.0$& $!*4\pm10$& $-53\pm28$& $!*2\pm26$\cr
\itTpE\ a,ai& $-10\pm10$& $-14\pm14$& $*-1\pm21$& $-14\pm21$& $!160\pm110$& $!250\pm110$\cr
\itTpE\ a,ii& $-450\pm520$& $-350\pm260$& $*-56\pm21$0& $-3100\pm150$0& $-4100\pm160$0& $-2100\pm920*$\cr
\itTpE\ i,aa& $!20\pm28$& $-0.5\pm8.0$& $-3.5\pm4.2$& $!96\pm52$& $!44\pm49$& $-37\pm26$\cr
\itTpE\ i,ai& $-32\pm46$& $-18\pm20$& $*-7\pm14$& $!190\pm180$& $!350\pm240$& $!23\pm77$\cr
\itTpE\ i,ii& $-290\pm210$& $-340\pm210$& $*-51\pm11$0& $-640\pm400$& $-2000\pm990*$& $!*38\pm300$\cr
\noalign{\vskip 3pt\hrule\vskip 4pt}}}
\endPlancktablewide                 
\endgroup
\end{table*}                        

As for the two previous data releases, we note that 
the agreement between the 
different estimators---for all the maps and all the shapes considered, both in temperature and polarization, as well as in 
the full \itTpE\ results---is well in line with both our theoretical 
expectations and Monte Carlo studies, where we found
an {\em average} measured $f_{\rm NL}$ scatter at the level of $\la \sigma_{f_{\rm NL}}/3$ between different pipelines (this was discussed at length in \citetalias{planck2013-p09a} and \citetalias{planck2014-a19}).
Note that the observed 
scatter between two estimators (for a given realization) can be larger than what the very small reported differences in their variance might suggest.
In an ideal, noiseless,
full-sky experiment, the observed scatter is only due to differences in the estimator weights, coming from the use of different bispectrum expansions or binning schemes. 
In such an ideal case, if two different bispectrum templates have a correlation coefficient $r$, we showed in appendix~B of \citetalias{planck2013-p09a} that the standard deviation of the expected scatter $\delta \fnl$ is given by $\sigma_{\delta \fnl} = \sigma_{\fnl} {\sqrt{(1-r^2)}/r}$. This leads to differences between $\fnl$ 
results that are a sizeable fraction of the estimator standard deviation, even for highly correlated weights.
To be more explicit, a $95\,\%$ correlation between different 
input templates leads, using the formula above, to a $\sigma_{\fnl}/3$ average scatter in $f_{\rm NL}$ estimates; on the other hand, it would produce only a $5\,\%$ difference in the final uncertainties.

Considering a more concrete example, let us focus on the current \SMICA\, KSW \itTpE\ results, which we quote as our final, recommended bounds. Let us consider, e.g., the difference between the KSW and Binned estimator for the local shape. This is relatively large, at approximately $0.3\,\sigma_{\fnl}$ after lensing-bias subtraction, compared to an error bar difference of 2\,\%. Let us now assume a correlation $r \approx 0.98$ between the weights, consistent both with what we see in simulations and with such a small difference in the errors. If we substitute this into our formula, we obtain an expected scatter of $\sigma_{\delta \fnl} = 0.2\,\sigma_{\fnl}$. The observed scatter is then $1.5\,\sigma$ away from this average and therefore fairly consistent with it. Note that the chosen example displays a relatively large difference, compared to all the other combinations that can be built in Table~\ref{tab:fNLall}; note also that the formula we are using represents an ideal case, and our validation tests on simulations in realistic conditions (e.g., masking and anisotropic noise) show, as expected, that the actual scatter between two pipelines is generally a bit larger than this ideal expectation \citepalias{planck2013-p09a,planck2014-a19}.

If, instead of comparing central values, we look at uncertainties, we see as expected that all pipelines produce nearly optimal constraints. For the local shape, we see that the Modal~1 pipeline produces $6\,\%$ smaller errors than e.g., KSW; however, this is within the expected Monte Carlo error and it seems to be just an effect of the selected simulation sample used to compute $\sigma_{\fnl}$. One should also consider that Modal~1 uses $160$ FFP10 maps to extract the standard deviation, versus $300$ maps for the other pipelines. This explanation is confirmed by our many validation tests on different sets of simulations.

A high level of internal consistency is also displayed between $\fnl$ estimates obtained from different 
component-separated maps, as well as in the comparison of current results with those from previous releases. One small 
exception is provided by the orthogonal $\fnl$ estimate obtained from \Commander\ \itT-only data, for which we notice 
both a larger fluctuation with respect to $2013$ and $2015$ results and a larger discrepancy with other foreground cleaned maps, 
in particular with the \SMICA\ one. However, we do not find this worrisome for a number of reasons. First of all, the \SMICA\ -- \Commander\ 
 difference is still at the level of the 1$\,\sigma$ orthogonal $\fnl$ uncertainty and all methods, including \Commander, show full consistency 
with $f_{\rm NL}^{\rm ortho} = 0$. Therefore, this discrepancy does not pose any problem for the theoretical interpretation of the result. 
Moreover, this fluctuation completely goes away when accounting for polarization data, the reliability of which has 
become significantly higher with respect to our previous analysis (see Sect.~\ref{sec:Sec_valid_data} for details). Finally, the discrepancy 
 is for a very specific shape and it is entirely driven by the already-noted fluctuation in the \Commander\ orthogonal result with respect to $2013$ 
and $2015$. The other
methods remain stable, in particular \SMICA, which we take as the map of choice for our final results.  
The observed fluctuation in orthogonal $\fnl$ from \Commander\ can likely be explained by the unavailability of
``detset'' (i.e., detector-subset)  maps for this release, which constituted in the past a useful input for improving the accuracy of the \Commander\ map. 
It is, however, important to stress that \Commander\ itself shows excellent agreement with other methods, when measuring $\fnl$
for all other shapes and also when correlating the bispectrum modes and bins in a model-independent fashion, both in temperature and polarization 
(again see Sect.~\ref{sec:Sec_valid_data} for a complete discussion of these tests).

Comparing the uncertainties in Table~\ref{tab:fNLall} to those in the corresponding
table in the 2015 analysis paper \citepalias{planck2014-a19}, and focussing on the ones for
the local shape, since those are most sensitive to low-$\ell$ modes, we see
the following: for \itT-only data the errors are approximately equal on average,
 slightly better for KSW, and slightly worse for the other three estimators. A
possible explanation for the slightly larger errors could be the fact that the
realism of the simulations has improved from FFP8 used in 2015 to FFP10 used
here. So the errors in 2015 might actually have been slightly underestimated.
However, the differences are small enough that they could just be random
fluctuations, especially given that not all estimators show the same effect.
For \itE-only data we see a clear improvement of the errors for all estimators.
That is as expected, since we are now including all the additional
polarization modes with $4 \leq \ell < 40$ in the analysis, and the local shape
is quite sensitive to these low-$\ell$ modes. Finally, for the full
\itTpE\ analysis, we see that all errors have remained the same, to within
fluctuations of around a few percent, at most. So one might wonder why the improvement in the
\itE-only analysis has not translated into a corresponding improvement in the
\itTpE\ analysis. The answer is relatively simple: the
\itE\itE\itE-bispectra only have a very small contribution to the final
\itTpE\ analysis, as shown in Fig.~\ref{fig:contribTandEtofNL}. 
This figure also explains why the errors for the equilateral and orthogonal
shapes improve more when going from \itT-only to the full \itTpE\ analysis
than the errors for the local shape.

\begin{figure*}[htbp!]
\centering
\includegraphics[width=0.27\textwidth]{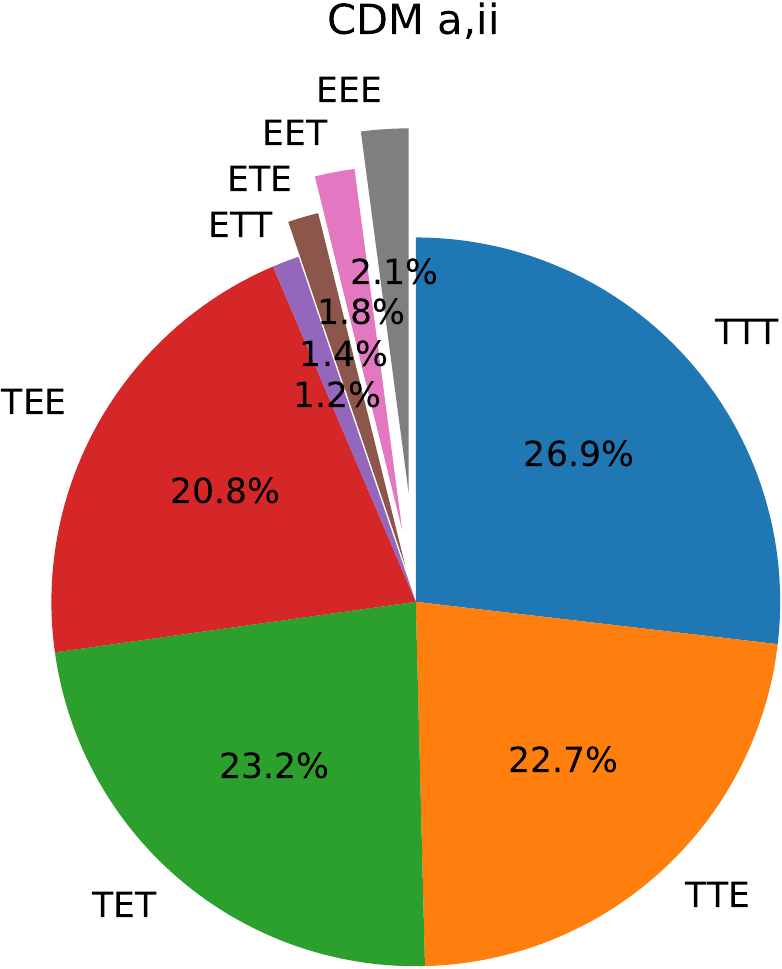}%
\includegraphics[width=0.32\textwidth]{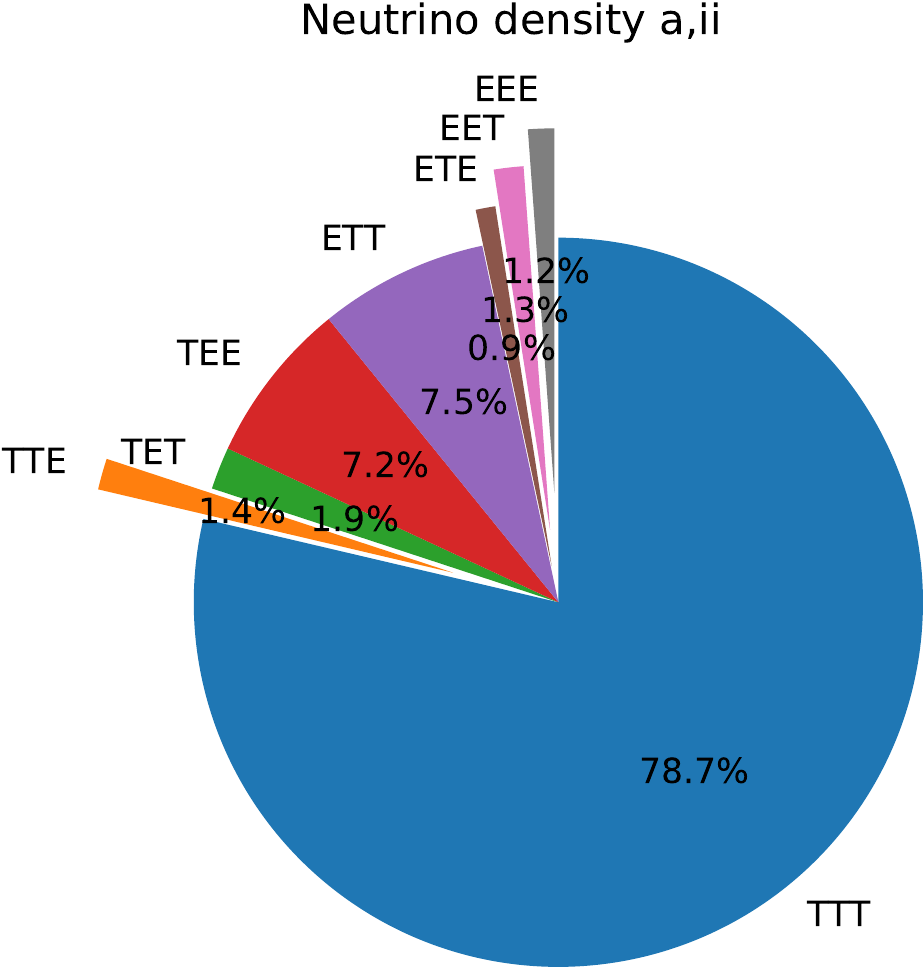}\quad%
\includegraphics[width=0.32\textwidth]{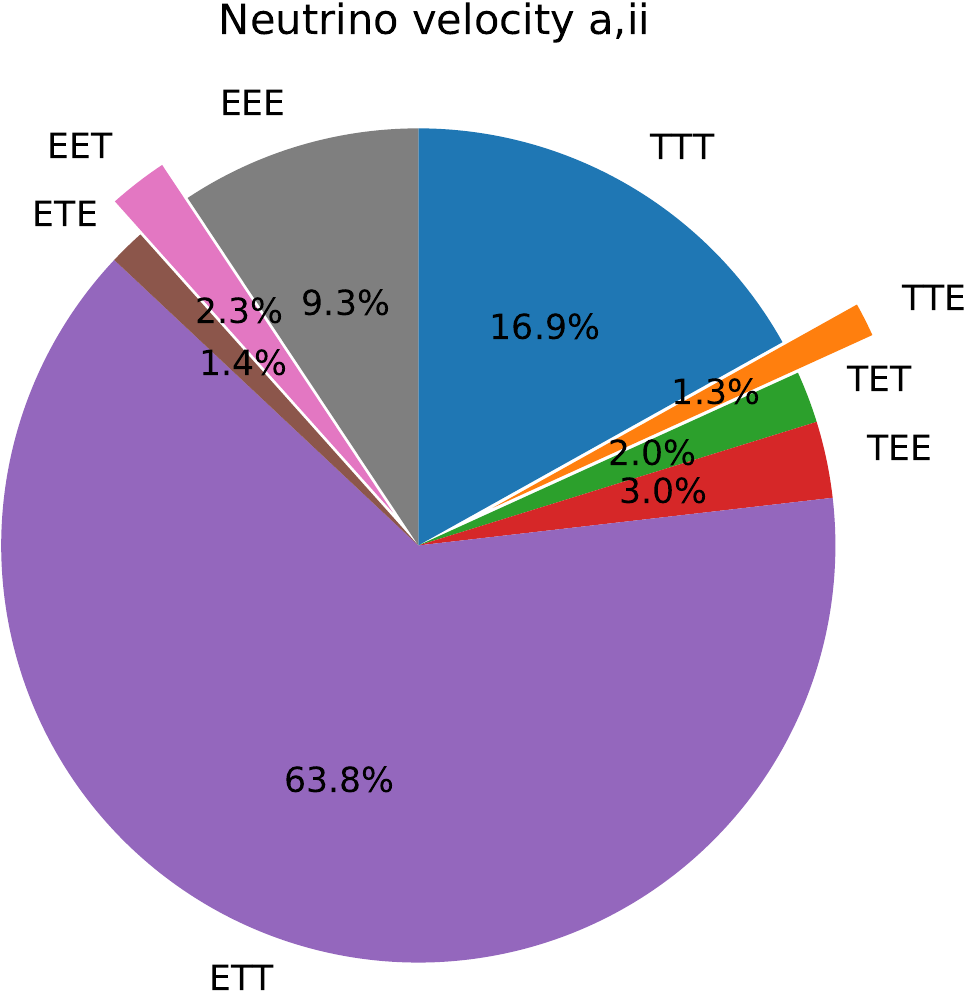}
\caption{Weights of each polarization configuration going into the total
    value of the a,ii mixed $\fnl$ parameter for, from left to right,
    CDM, neutrino-density, and neutrino-velocity isocurvature, in addition
    to the adiabatic mode.
    Note that since we impose $\ell_1 \leq \ell_2 \leq \ell_3$, there is a
    difference between, e.g., \itT\itE\itE\ (where the smallest $\ell$ is
    temperature)
    and \itE\itE\itT\ (where the largest $\ell$ is temperature).}
\label{fig:contribTandEtofNL_isocurv}
\end{figure*}

In conclusion, our current results show no evidence for non-Gaussianity of the local, equilateral, or orthogonal type and are in very good 
agreement with the previous $2013$ and $2015$ analyses.
We also show in Sect.~\ref{sec:Sec_valid_data} that the overall robustness and internal 
consistency of the polarization data set has significantly improved, as far as primordial non-Gaussian measurements are concerned. 

\begin{table*}[htbp!]                 
\begingroup
\newdimen\tblskip \tblskip=5pt
\caption{Similar to Table~\ref{Tab_isocurvNG}, except that we now assume that
the adiabatic and isocurvature modes are completely uncorrelated. Hence there
are only two $f_\mathrm{NL}$ parameters in this case, a purely adiabatic one
and a purely isocurvature one.}
\label{Tab_isocurvNG_uncorr}
\nointerlineskip
\vskip -3mm
\footnotesize
\setbox\tablebox=\vbox{
   \newdimen\digitwidth 
   \setbox0=\hbox{\rm 0} 
   \digitwidth=\wd0 
   \catcode`*=\active 
   \def*{\kern\digitwidth}
   \newdimen\signwidth 
   \setbox0=\hbox{+} 
   \signwidth=\wd0 
   \catcode`!=\active 
   \def!{\kern\signwidth}
   \newdimen\dotwidth 
   \setbox0=\hbox{.} 
   \dotwidth=\wd0 
   \catcode`^=\active 
   \def^{\kern\dotwidth}
\halign{\hbox to 0.8in{#\leaderfil}\tabskip 0.5em&
\hfil#\hfil\tabskip 1em&
\hfil#\hfil&
\hfil#\hfil\tabskip 2em&
\hfil#\hfil\tabskip 1em&
\hfil#\hfil&
\hfil#\hfil\tabskip 0pt\cr
\noalign{\doubleline\vskip 2pt}
\omit&\multispan3\hfil Independent\hfil& \multispan3\hfil Joint\hfil\cr
\noalign{\vskip -2pt}
\omit&\multispan3\hrulefill& \multispan3\hrulefill\cr
\noalign{\vskip 2pt}
\omit\hfil Shape\hfil& Cold dark matter& Neutrino density& Neutrino velocity& Cold dark matter& Neutrino density& Neutrino velocity\cr
\noalign{\vskip 4pt\hrule\vskip 6pt}
\itT\ a,aa& $-0.1\pm5.6$& $-0.1\pm5.6$& $-0.1\pm5.6$& $-0.6\pm5.7$& $!15\pm13$& $-1.3\pm5.8$\cr
\itT\ i,ii& $!110\pm280$& $-130\pm250$& $!270\pm230$& $!120\pm290$& $-710\pm590$& $!280\pm240$\cr
\noalign{\vskip 5pt}
\itE\ a,aa& $!48\pm26$& $!48\pm26$& $!48\pm26$& $!41\pm27$& $!33\pm35$& $!62\pm35$\cr
\itE\ i,ii& $!1800\pm1200$& $!1900\pm1100$& $!170\pm230$& $!1400\pm1200$& $!1000\pm1400$& $-200\pm300$\cr
\noalign{\vskip 5pt}
\itTpE\ a,aa& $-2.5\pm5.0$& $-2.5\pm5.0$& $-2.5\pm5.0$& $-1.3\pm5.1$& $!11.4\pm8.4*$& $-1.9\pm5.2$\cr
\itTpE\ i,ii& $-290\pm210$& $-340\pm210$& $-51\pm11$0& $-280\pm210$& $-710\pm360$& $*-41\pm120$\cr
\noalign{\vskip 3pt\hrule\vskip 4pt}}}
\endPlancktablewide                 
\endgroup
\end{table*}                        

\subsection{Further bispectrum shapes}

\subsubsection{Isocurvature non-Gaussianity}

In this section we present a study of the isocurvature NG in the
\Planck\ 2018 \SMICA\ map using the Binned bispectrum estimator.
This analysis is complementary to the one based on the power spectrum
presented in \cite{planck2016-l10}.
The underlying modelling approach was discussed in Sect.~\ref{Sec:isocurv_NG_intro},
and as explained there, we only investigate isocurvature NG of the local type,
and in addition always consider the adiabatic mode together with only
one isocurvature mode, i.e., we consider separately CDM-density, neutrino-density, and neutrino-velocity isocurvature. In that
case there are in general six different $f_\mathrm{NL}$ parameters: the purely
adiabatic one (a,aa) from Sect.~\ref{fnl_loc_eq_ort_results};
a purely isocurvature one (i,ii); and four mixed ones.

The results can be found in Table~\ref{Tab_isocurvNG}, both for an independent
analysis of the six parameters (i.e., assuming that only one of the six
parameters is present) and for a fully joint
analysis (i.e., assuming that all six parameters are present, which is
clearly the correct thing to do in the correlated framework described above).
Note that the reason for the very differently sized errors is a combination
of two effects. The first is a simple normalization issue, due to switching
from the more natural $\zeta$ and $S$ variables (commonly used in the inflation
literature) to $\Phi_\mathrm{adi}=3\zeta/5$ and $\Phi_\mathrm{iso}=S/5$, which
are more commonly used in the CMB literature.\footnote{Conversion factors to
obtain results based on $\zeta$ and $S$ are 6/5, 2/5, 2/15, 18/5, 6/5, and 2/5,
for the six modes, respectively.} The second is that certain parameters depend
more on the high-$\ell$ adiabatic modes (which are well-determined), while
others are dominated by the much suppressed, and hence unconstrained,
high-$\ell$ isocurvature modes.
For example, for the joint CDM \itTpE\ case, when compensating for the
normalization factor, one would find the error for a,aa and a,ai to be
around 10, and the other four errors to be around 200
\citep[see][for further discussion of these effects]{Langlois:2012tm}.

As in our analysis of the 2015 \Planck\ data
\citepalias{planck2014-a19}, we see no clear sign of any
isocurvature NG. There are a few values that deviate from zero by up to
about 2.5$\,\sigma$, but such a small deviation cannot be considered a detection,
given the large number of tests and the fact that the deviations are not
consistent between \itT-only and \itTpE\ data. For example, looking at the 300
Gaussian simulations that were used to determine the linear correction and
the uncertainties, we find that 84 of them have at least one $>2.5\,\sigma$
result in the two CDM columns of Table~\ref{Tab_isocurvNG}, while for neutrino
density and neutrino velocity the numbers are 62 and 80, respectively.

We see that many constraints are tightened considerably when 
including polarization, by up to the predicted factor of about 5--6 for
the CDM a,ii, i,ai, and i,ii modes in the joint analysis
(e.g., $-1300 \pm 1000$ for $T$ only, decreasing to $190 \pm 180$ for \itTpE\ in the CDM i,ai case).
Focussing now on the independent results, where it is easier to understand
their behaviour (as things are not mixed together), we see that the
uncertainties of some of the CDM and neutrino-velocity modes improve by a
factor of about 2 when going from the \itT-only to the full \itTpE\ analysis
(e.g., neutrino velocity i,ii changes from $270 \pm 230$ to $-51 \pm 110$),
while the improvements
for the neutrino-density modes are much smaller, of the order of what we see
for the pure adiabatic mode. This can be explained if we look at the
contribution of the various polarization modes to the total $\fnl$ parameters,
as indicated in Fig.~\ref{fig:contribTandEtofNL_isocurv} for the a,ii modes
(to give an example).
While the diagram for the neutrino-density mode very much resembles the one for the pure
adiabatic local case in Fig.~\ref{fig:contribTandEtofNL}, those
for CDM and neutrino velocity are quite different and have a much larger
contribution from polarization modes (but quite different between the two
cases). Hence isocurvature NG also provides a good test of the quality of the
polarization maps, and a clear motivation for extending our analyses to
include polarization in addition to temperature.

Comparing the results in Table~\ref{Tab_isocurvNG} to those of the
corresponding table in our 2015 analysis \citepalias{planck2014-a19} we see
in the first place that the \itT-only results are mostly very stable, generally
shifting by much less than $1\,\sigma$ (e.g., joint neutrino velocity i,ii is
$-970 \pm 1400$ in 2015 versus $-1000 \pm 1400$ now). Secondly, for the
\itE-only results we see that all the errors have significantly decreased
(e.g., taking again joint neutrino velocity i,ii, we had $2200 \pm 1600$ in
2015 versus $-380 \pm 1000$ now), in line with the fact that we are now using
the additional $4 \leq \ell < 40$ $E$ modes. We also generally see larger
changes in the central values, with several shifts of $1\,\sigma$ or larger;
nevertheless, the results are consistent with zero. Because we have added new polarization
data to the analysis compared to 2015, and the quality of the polarization
data has improved overall, these larger shifts are not unexpected.
Finally, looking at the full
\itTpE\ analysis, we see that the central values shift a bit more than
for $T$ only, but remain consistent with zero. For CDM and neutrino density
the errors are marginally larger than in 2015, similarly to what we see
with the Binned estimator for the local adiabatic shape (see
Sect.~\ref{fnl_loc_eq_ort_results}). As shown in
Fig.~\ref{fig:contribTandEtofNL_isocurv}, both of these depend very little
on low-$\ell$ polarization, and so are likely mostly driven by the marginal
increase in the \itT-only errors. For the neutrino-velocity mode, on the
other hand, some of the errors do decrease (e.g., $480 \pm 430$ in 2015
versus $38 \pm 300$ now for joint neutrino velocity i,ii). As seen in
Fig.~\ref{fig:contribTandEtofNL_isocurv}, this mode depends more strongly
on the \itE\itT\itT\ combination, which does involve low-$\ell$ polarization.

In the results so far we looked at the most general case, having a possible
correlation between the isocurvature and adiabatic modes. However, if we
assume that the adiabatic and the isocurvature modes have a cross-power
spectrum of zero and are completely uncorrelated, then there are only 
two free $f_\mathrm{NL}$ parameters, the a,aa and the i,ii ones. In 
Table~\ref{Tab_isocurvNG_uncorr} we present the results for this uncorrelated 
case. The independent results are the same as in the previous table and have
been repeated for convenience. The significant increase in the \itT-only and
\itTpE\ errors for the neutrino density case when going from the
independent to the joint analysis clearly illustrates the fact that its
bispectrum template has a large overlap with the adiabatic one, something
that also explains the similarity of Figs.~\ref{fig:contribTandEtofNL}
and \ref{fig:contribTandEtofNL_isocurv} for that case.
The CDM and neutrino-velocity modes, on the other hand, have templates that are
very different from the adiabatic one and their errors hardly increase
(except for neutrino velocity for $E$-only data).
Again there is no evidence for any isocurvature NG: we do not consider
the almost $2\,\sigma$ result for the neutrino-density isocurvature mode
in the \itTpE\ joint analysis to be significant, although it will
be interesting to keep an eye on this in future CMB experiments with
even better polarization measurements.

\subsubsection{Running non-Gaussianity}
\label{running_analysis}

\begin{figure}[htbp!]
    \centering
    \includegraphics[width=0.5\textwidth]{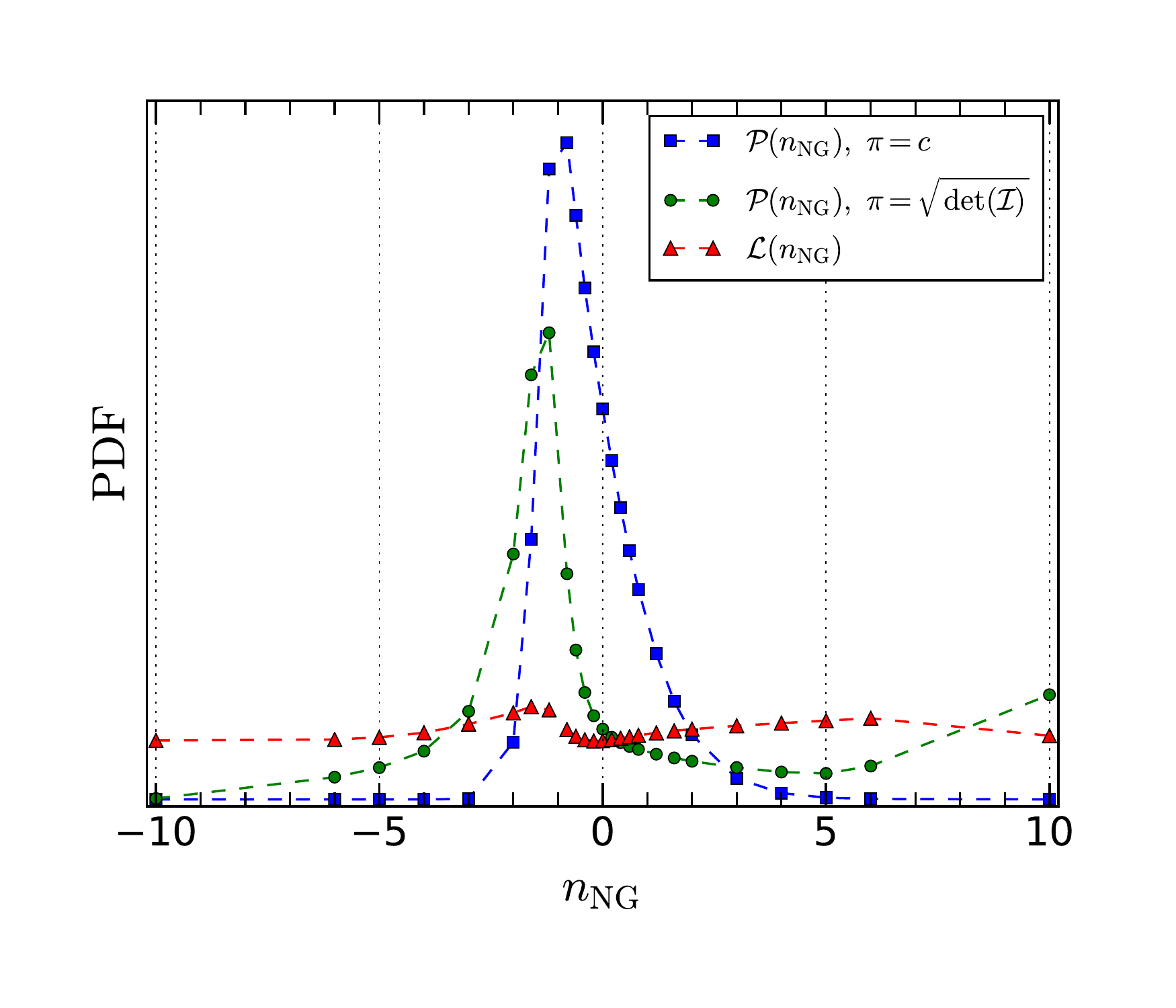}\\
\vspace{-0.8cm}
    \includegraphics[width=0.5\textwidth]{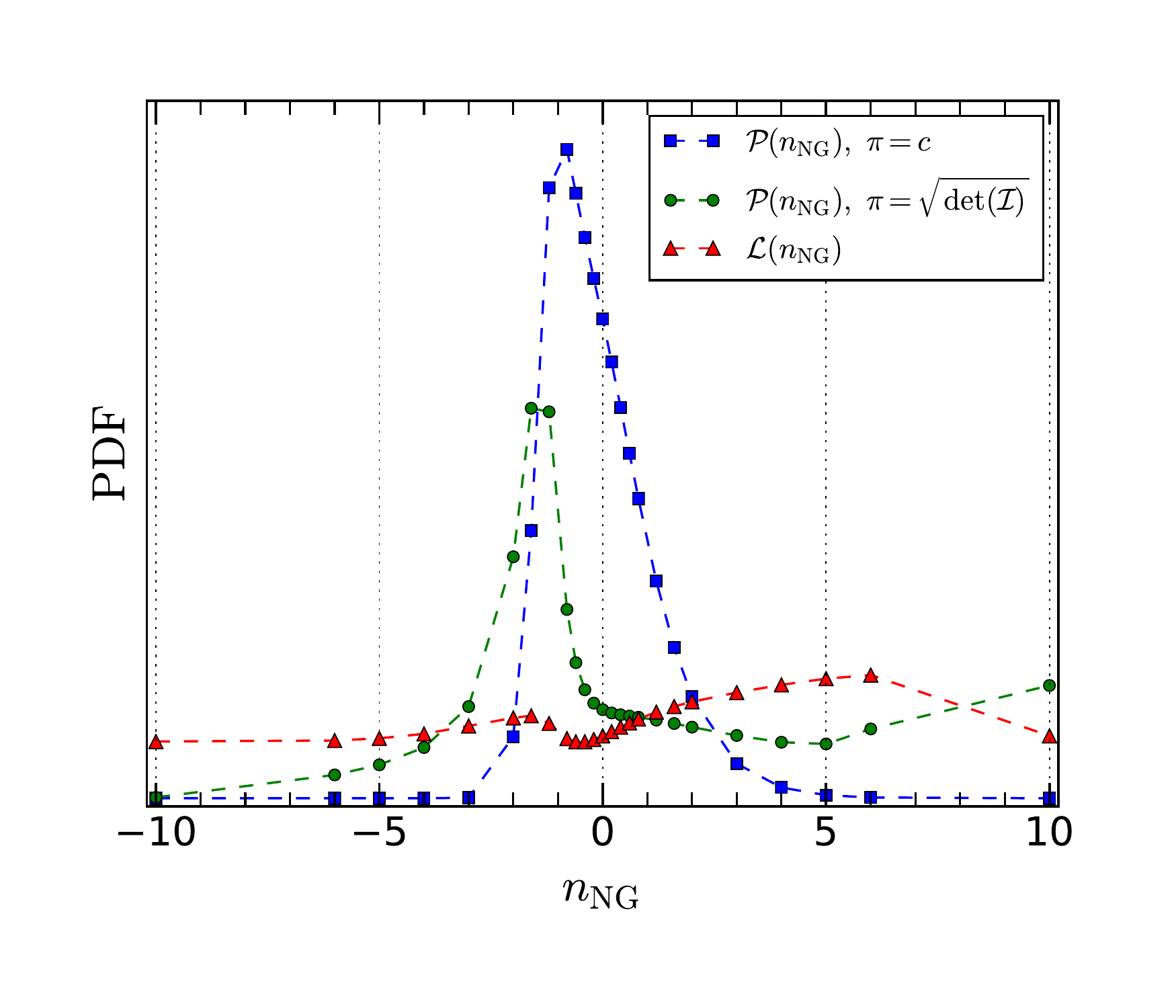}\\
\vspace{-0.5cm}
    \caption{PDF of the running parameter $n_{\rm NG}$ for the one-field local model. {\it Top}: \SMICA\ map. {\it Bottom}: \Commander\ map. Blue squares give the marginalized posterior assuming a constant prior, green circles are the posterior assuming a Jeffreys prior, and red triangles are the profiled Likelihood.}
    \label{fig:onf}
\end{figure}

\begin{figure}[htbp!]
    \centering
    \includegraphics[width=0.50\textwidth]{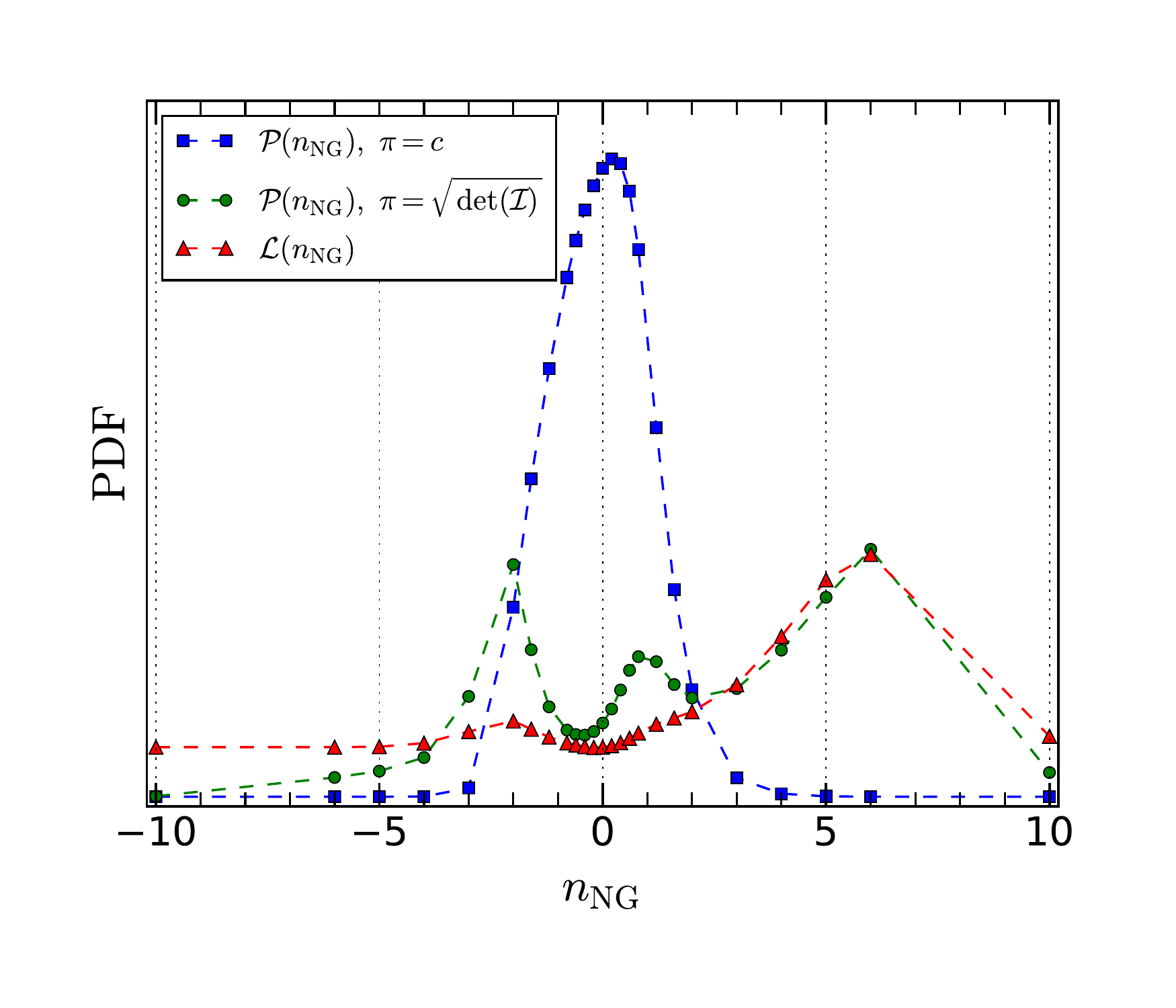}\\
\vspace{-0.8cm}
    \includegraphics[width=0.50\textwidth]{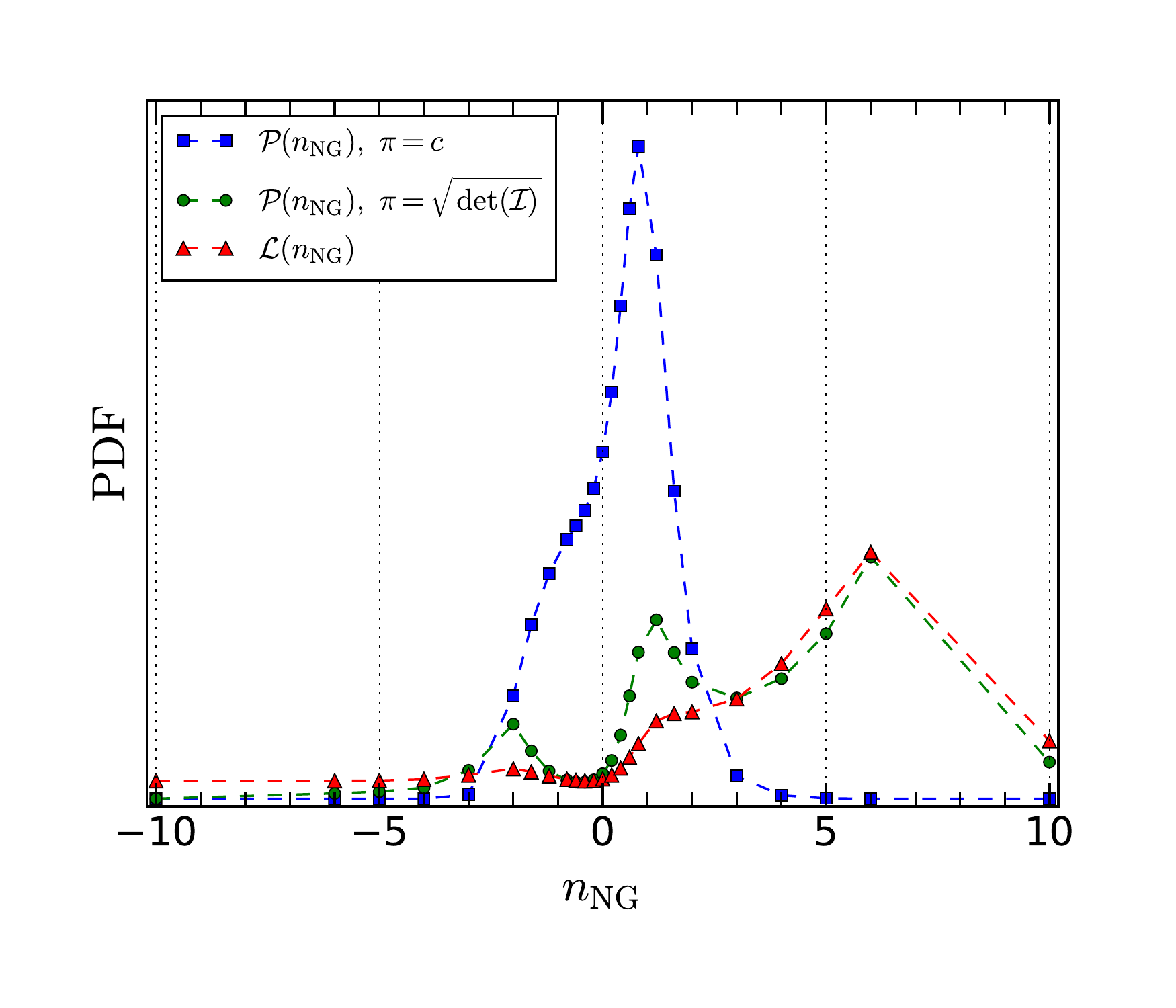}\\
\vspace{-0.5cm}
    \caption{PDF of the running parameter $n_{\rm NG}$ for the two-field local model.  {\it Top}: \SMICA\ map. {\it Bottom}: \Commander\ map. Blue squares give the marginalized posterior assuming a constant prior, green circles are the posterior assuming a Jeffreys prior, and red triangles are the profiled Likelihood.}
    \label{fig:twf}
\end{figure}

\begin{figure}[htbp!]
    \centering
    \includegraphics[width=0.50\textwidth]{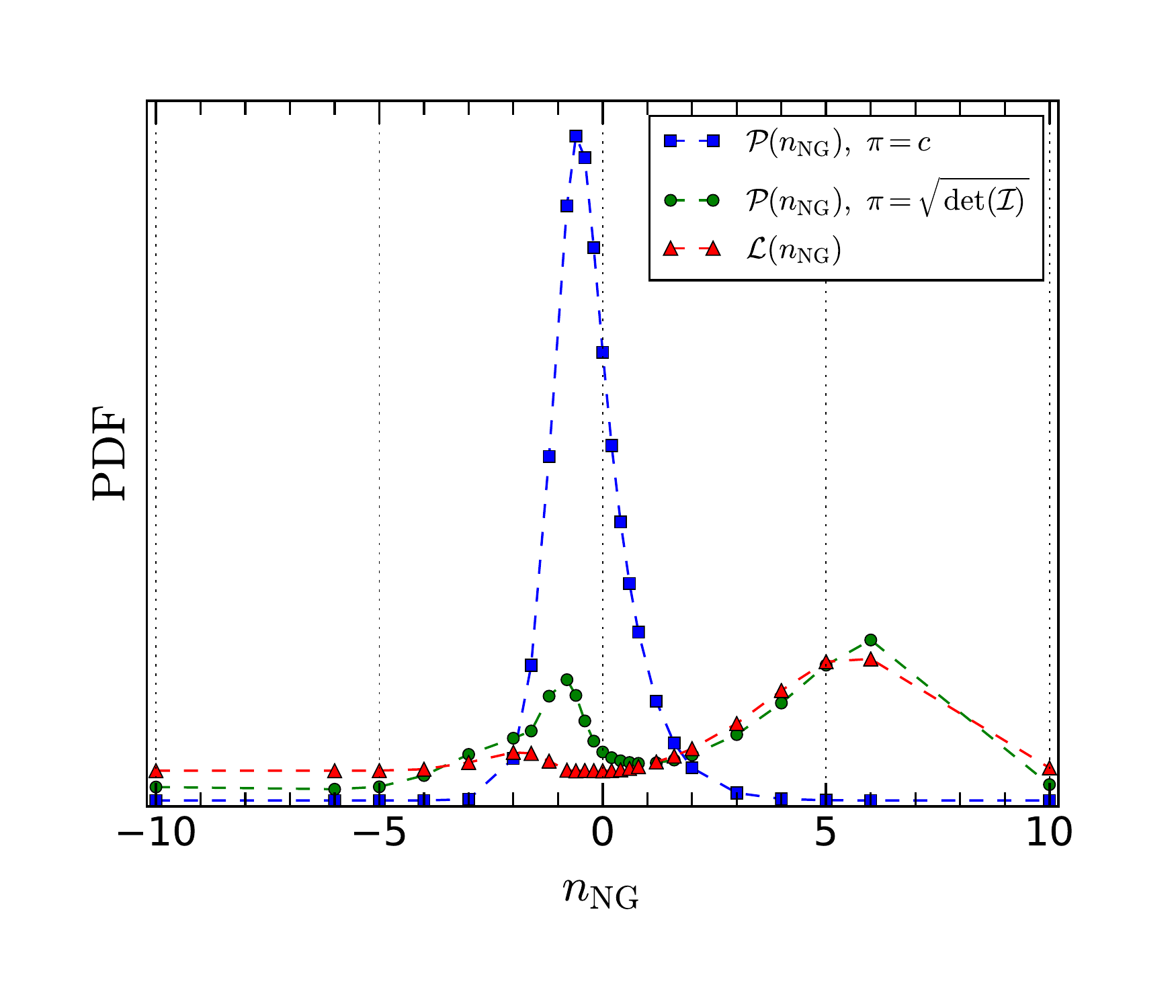}\\
\vspace{-0.8cm}
    \includegraphics[width=0.50\textwidth]{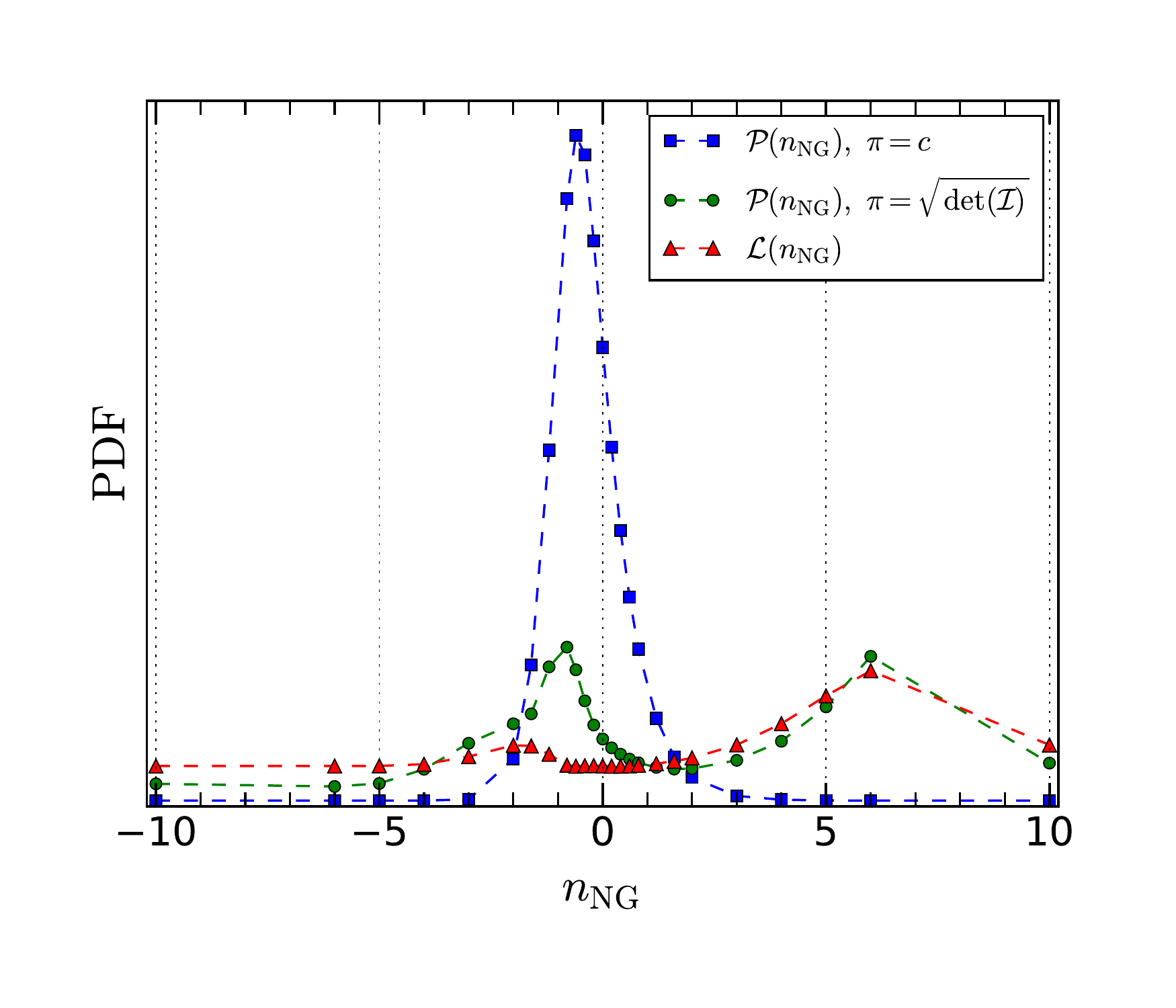}\\
\vspace{-0.5cm}
    \caption{PDF of the running parameter $n_{\rm NG}$ for the geometric mean equilateral parametrization. {\it Top}: \SMICA\ map. {\it Bottom}: \Commander\ map. Blue squares give the marginalized posterior assuming a constant prior, green circles are the posterior assuming a Jeffreys prior, and red triangles are the profiled Likelihood.}
    \label{fig:gme}
\end{figure}
In this section we present our analysis of the scale-dependent bispectrum shapes described in Sect.~\ref{Sec:Running}; we obtain these results following the pipeline described in Sect.~\ref{runningmethod}.
In Figs.~\ref{fig:onf}, \ref{fig:twf}, and \ref{fig:gme} we show the results, respectively, for the one-field local model (Eq.~\ref{running1}), the two-field local model (Eq.~\ref{running2}), and the geometric-mean equilateral model, where $f_{\rm NL}^*$ is parametrized as in Eq.~\eqref{eq:fnlgm}.  Here we present results for both the \SMICA\ and \Commander\ temperature maps. 
Since the $E$-mode polarization maps are not expected to significantly improve the constraints on $n_{\rm NG}$, while requiring a significant growth in the computational cost of the estimation pipeline, we do not include polarization in this particular analysis.
We show the PDF inferred from all three of the methodologies described.
Each point is derived from a KSW estimate of the amplitude for the corresponding scale-dependent template with the given running. 
The KSW pipeline and the map processing steps are the same as applied to the estimation of the local, equilateral, and orthogonal shapes.  We include in this analysis the multipole range from 2 to 2000 and the results are corrected for the lensing bias.  All curves in Figs.~\ref{fig:onf}--\ref{fig:gme}
are normalized to integrate to one.
We consider possible values of the running in the interval $n_{\rm NG}=[-10,10] $. 
This interval is two orders of magnitude wider than the theoretical expectation of the models, which are valid in the regime of mild scale dependence, i.e., $n_{\rm NG}\approx0.1$.

 The effects of the choice of prior are very obvious: if we adopt a constant prior (blue squares) we can always identify a peak in the distribution and define proper constraints; however, this is not the case for the other priors. 
If we implement 
an uninformative prior (green circles), the shape of the distribution becomes complex, showing multiple peaks or even diverging on the boundaries, making it impossible to define constraints. 
A similar behaviour appears in the profiled likelihood approach (red triangles; see Eq.~\ref{eq:marglik}). 
We used the likelihood to also perform a likelihood ratio test between its maximum value and the value for $n_{\rm NG}=0$. 
Notice that in the case of zero running, these models reduce to the usual local or equilateral shapes. We assume an acceptance threshold of $\alpha=0.01$
As expected, we do not find any evidence in favour of scale dependent models.

\subsubsection{Resonance and axion monodromy}
\begin{figure*}[htbp!]
\centering
\includegraphics[width=.48\linewidth]{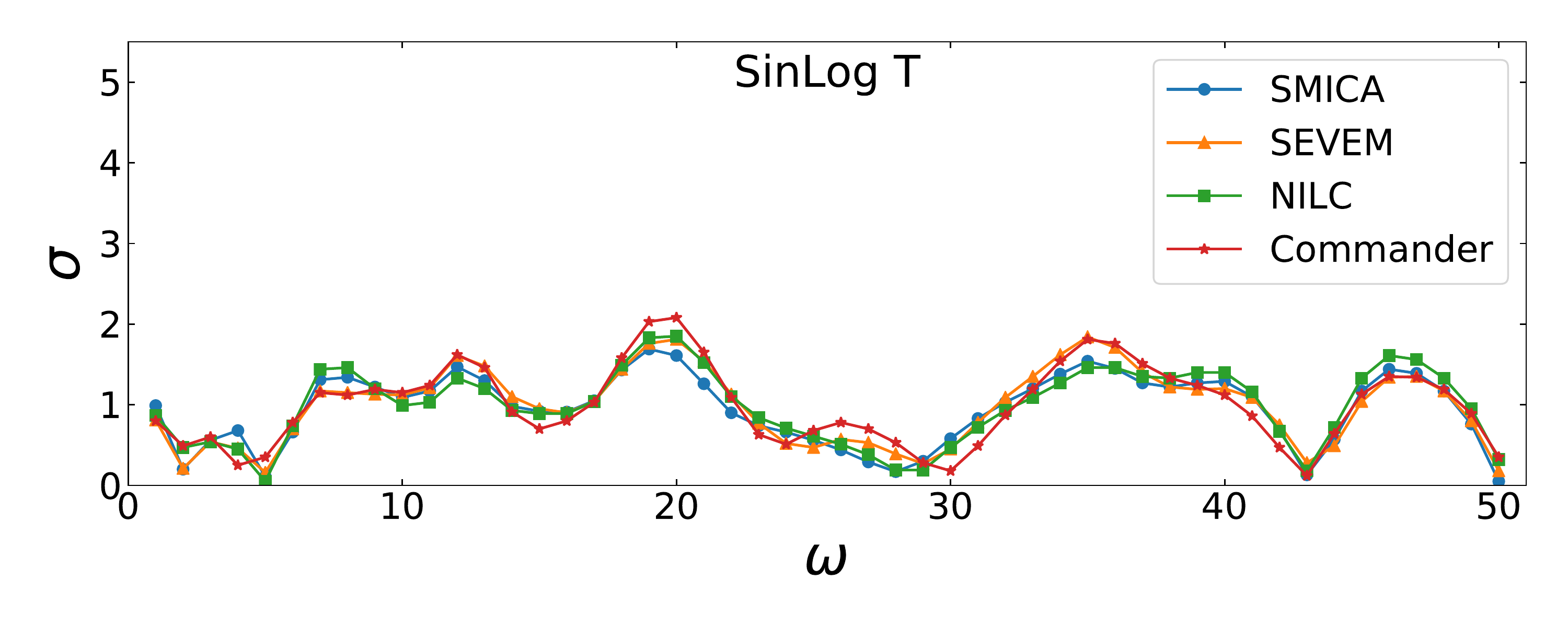} \hskip0.1in\includegraphics[width=.48\linewidth]{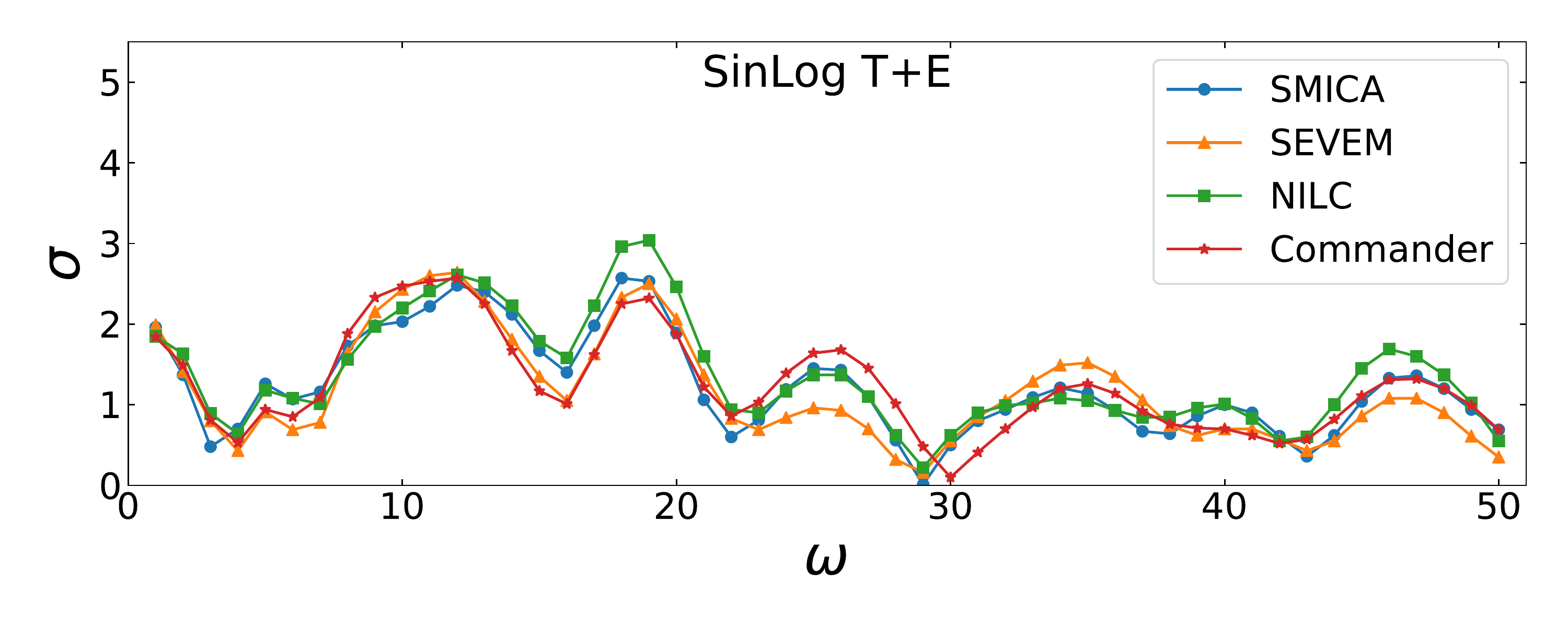}
\includegraphics[width=.48\linewidth]{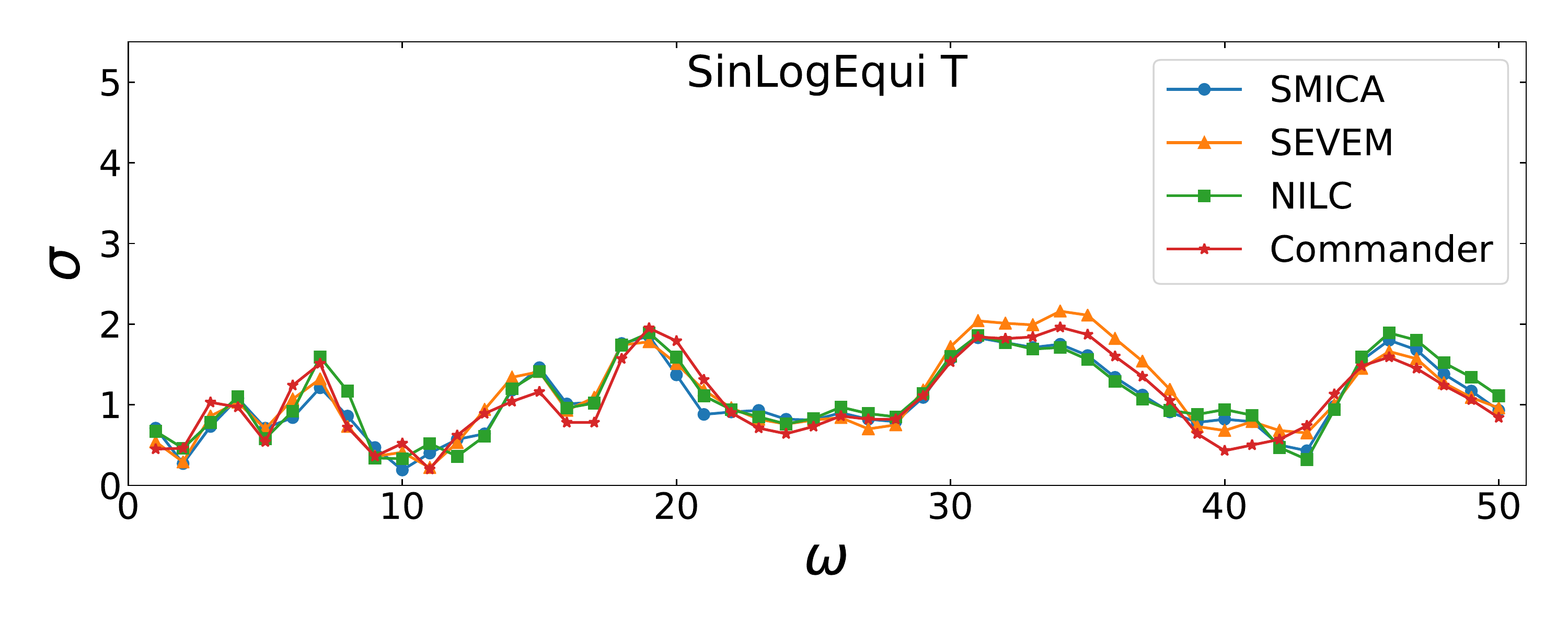} \hskip0.1in\includegraphics[width=.48\linewidth]{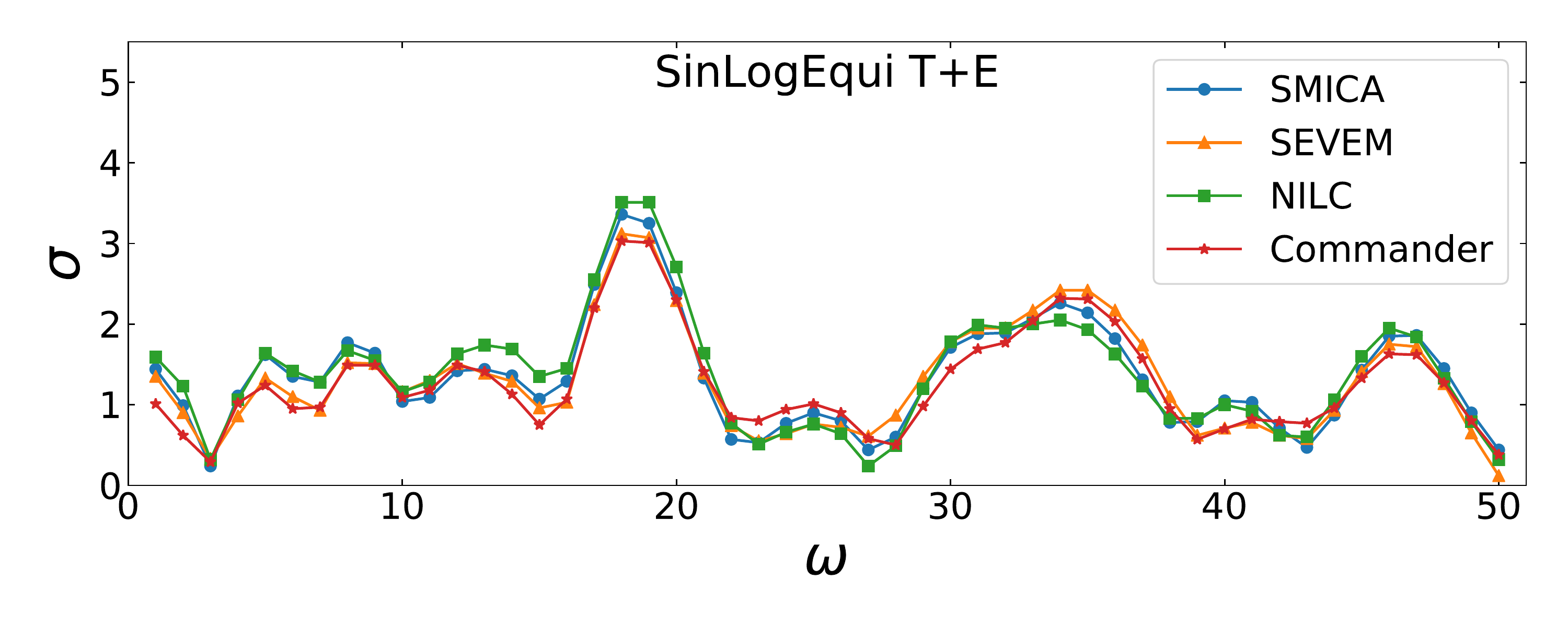}
\includegraphics[width=.48\linewidth]{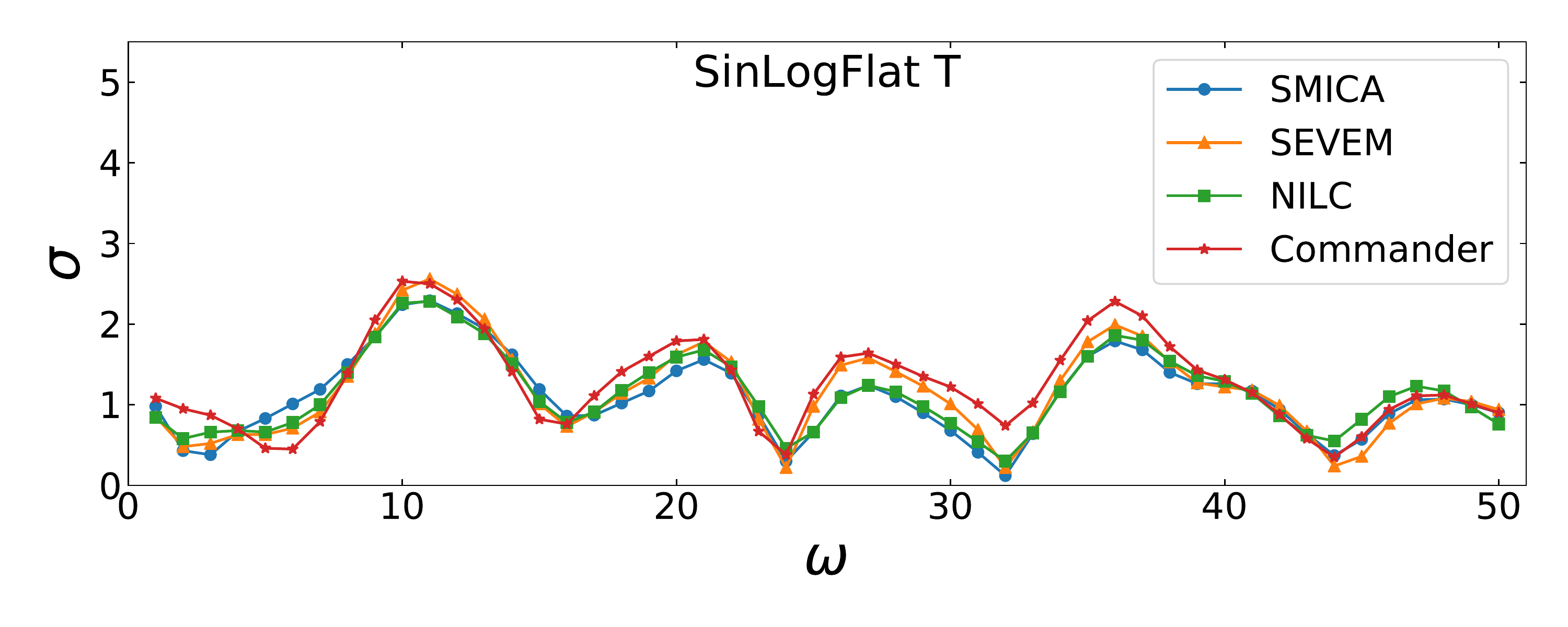} \hskip0.1in\includegraphics[width=.48\linewidth]{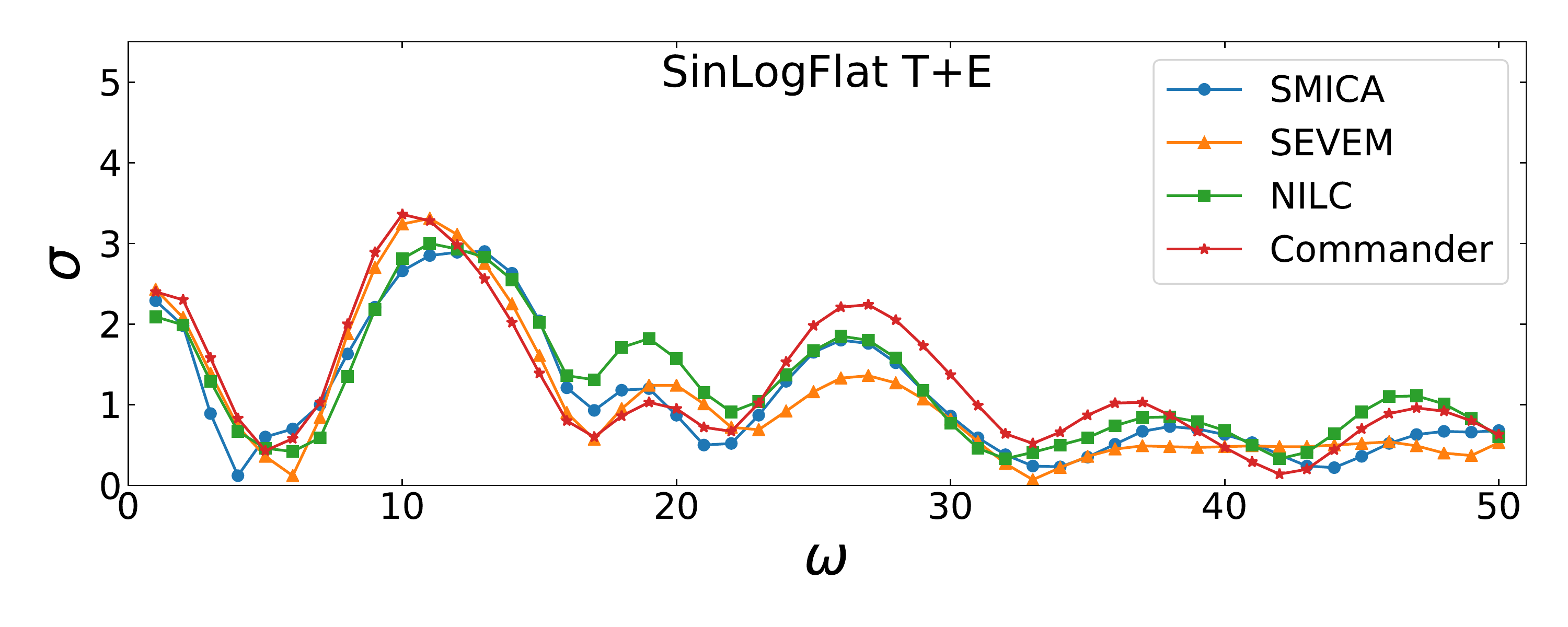}
\includegraphics[width=.48\linewidth]{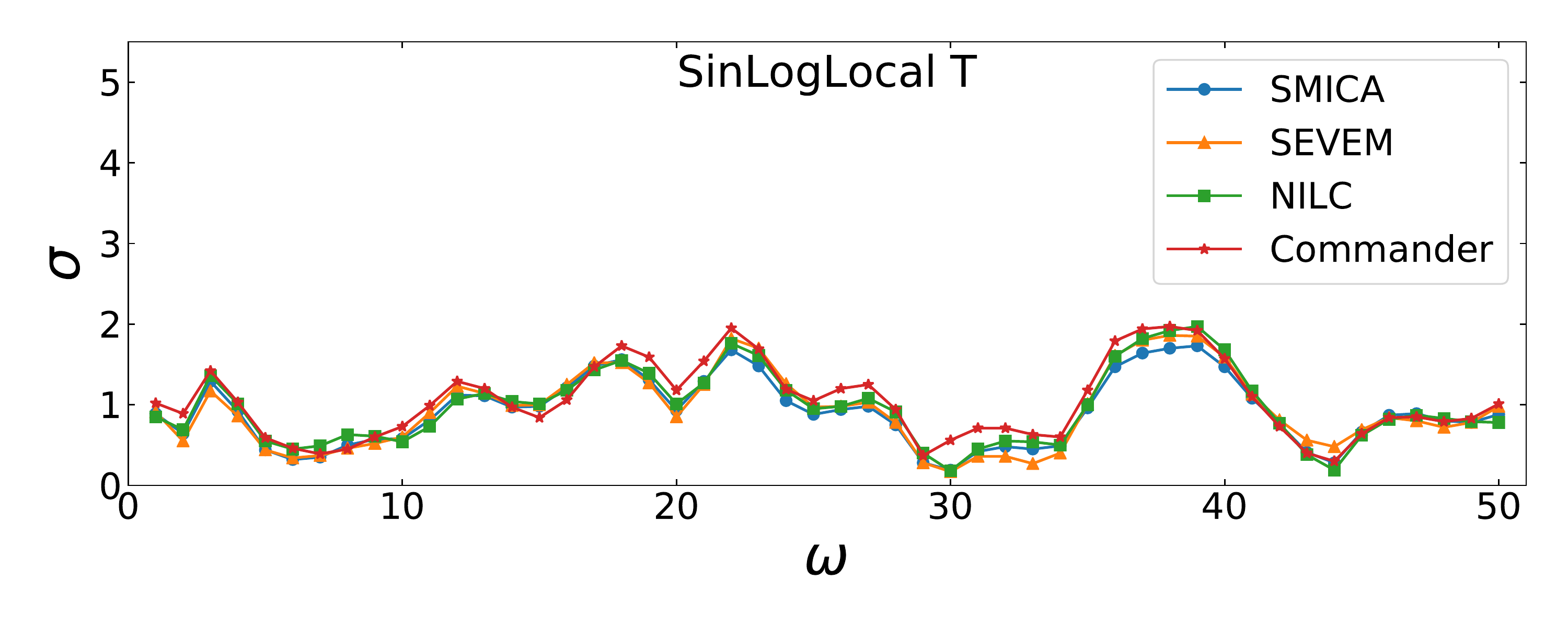}\hskip0.1in \includegraphics[width=.48\linewidth]{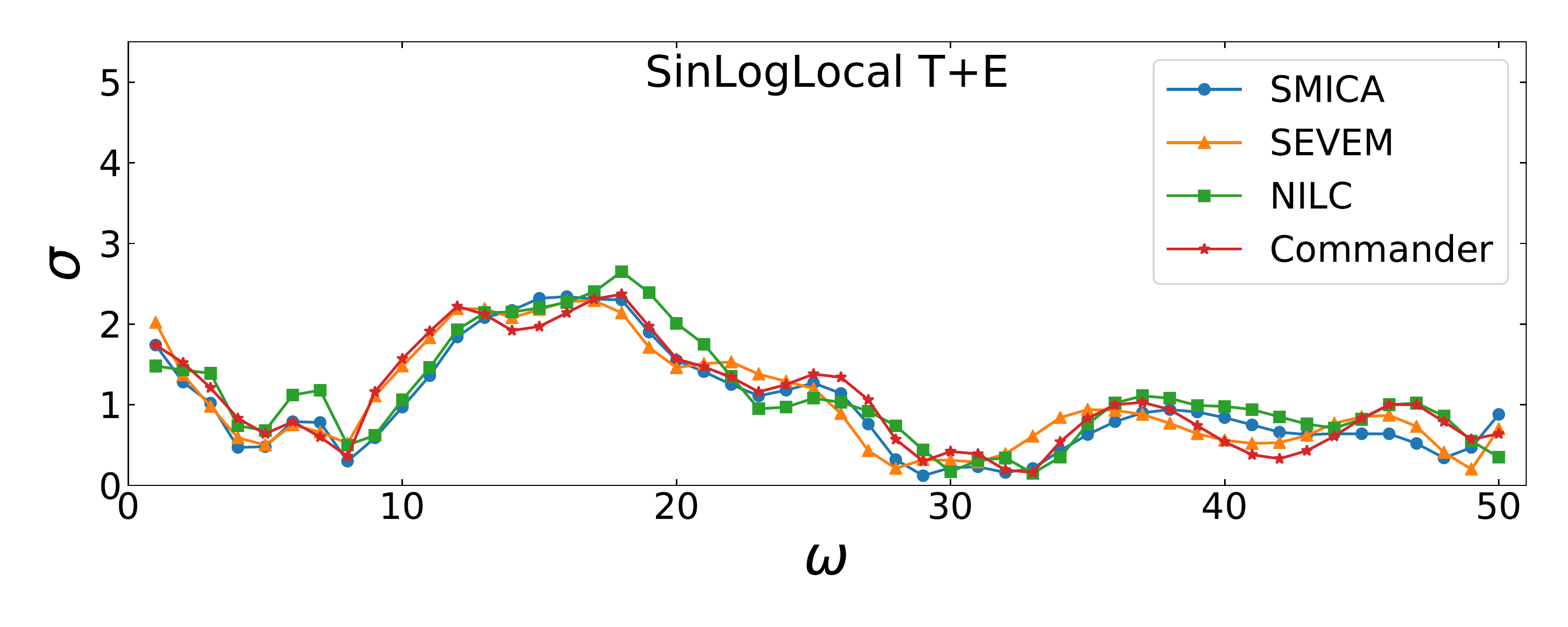}
\caption[]{Generalized resonance-model significance surveyed over the Modal~2 frequency range with the uppermost panels for the constant resonance model (Eq.~\ref{eq:resBprim}), showing \itT-only (left) and \textit{T+E} (right) results, then below this the equilateral resonance model (Eq.~\ref{eq:equilBprim}), followed by the flattened model (Eq.~\ref{eq:resflatBprim}) with the bottom panels, representing the local resonance model with a squeezed envelope.  These models have been investigated using \Planck\ temperature data to $\ell_{\rm max} =2000$ and polarization data to $\ell_{\rm max} =1500$, with the \Planck\ component-separation methods \SMICA\ (blue), \SEVEM\ (orange), \NILC\ (green), and \Commander\ (red).   These results are broadly consistent with those found previously \citepalias{planck2014-a19}, with some broad peaks of moderate significance emerging in both the equilateral and flattened resonance models, somewhat enhanced by including polarization data  (upper middle and lower middle right panels). }
\label{fig:resonance_models}
\end{figure*}

Now we present results from the Modal~2 and from an adapted KSW-type estimator for the broad class of resonance-type models.  These models can be characterized by a bispectrum template having logarithmic oscillations with a scale, as introduced in Sect.~\ref{sec:resonance}.  For the Modal~2 analysis we examine templates with frequency in the range $0 < \omega \le 50$ and with a range of possible cross-sectional behaviour covering constant, flat, local, and equilateral type templates to span a broader range of possible models.  The raw results have been maximized over phase and are presented in Fig.~\ref{fig:resonance_models} for the four component-separation methods.  Since the errors grow with frequency due to increasing suppression by the both transfer functions and projection effects, we only plot the raw significance.  The results are consistent across component-separation methods and are comparable with those previously presented in \citetalias{planck2014-a19}, but with slight reductions in significance.  Since we are surveying a large range of frequencies, we must correct our results for the ``look-elsewhere'' effect to asses their true significance.  This is done using the optimal methods first proposed in \cite{Fergusson:2014hya}, which assess the true significance of both single peaks and also clusters of multiple peaks (which may indicate a model in this class but with a waveform that only partially correlates with the templates used).  The results are presented in Table~\ref{tab:peakstatres}, with the largest look-elsewhere-corrected result being for the equilateral cross-section with around $2\,\sigma$ for a single peak.  This is not on its own significant, but may warrant further investigation with more realistic exact templates near this frequency.

\begin{table*}[htbp!]
\begingroup
\newdimen\tblskip \tblskip=5pt
\caption{Peak statistics, as defined in \cite{Fergusson:2014hya}, for the resonance models, showing the maximum ``Raw'' peak significance, the ``Single'' peak significance after accounting for the parameter survey look-elsewhere effect, and the ``Multi''-peak statistic integrating across all peaks (also accounting for the look-elsewhere correction). This table does not include the results of the high-frequency resonance-model estimator, whose significance was assessed independently and presented in subsection \ref{sec:HFFest}.}
\label{tab:peakstatres}
\nointerlineskip
\vskip -3mm
\footnotesize
\setbox\tablebox=\vbox{
  \newdimen\digitwidth 
  \setbox0=\hbox{\rm 0} 
  \digitwidth=\wd0 
  \catcode`*=\active 
  \def*{\kern\digitwidth}
  \newdimen\signwidth 
  \setbox0=\hbox{+} 
  \signwidth=\wd0 
  \catcode`!=\active 
  \def!{\kern\signwidth}
\halign{\hbox to 1.75in{#\leaderfil}\tabskip 1em&
\hfil#\hfil\tabskip=1em& \hfil#\hfil& \hfil#\hfil\tabskip=2em&
\hfil#\hfil\tabskip=1em& \hfil#\hfil& \hfil#\hfil\tabskip=2em&
\hfil#\hfil\tabskip=1em& \hfil#\hfil& \hfil#\hfil\tabskip=2em&
\hfil#\hfil\tabskip=1em& \hfil#\hfil& \hfil#\hfil\tabskip=0pt\cr 
\noalign{\doubleline}
\omit& \multispan3\hfil\SMICA\hfil& \multispan3\hfil\SEVEM\hfil&
 \multispan3\hfil\NILC\hfil& \multispan3\hfil\Commander\hfil\cr
\noalign{\vskip -4pt}
\omit& \multispan3\hrulefill& \multispan3\hrulefill&
 \multispan3\hrulefill& \multispan3\hrulefill\cr
\omit& Raw& Single& Multi& Raw& Single& Multi& Raw& Single& Multi& Raw& Single& Multi\cr							
\noalign{\vskip 3pt\hrule\vskip 5pt}
Sin(log) constant       \textit{T} only& 1.7& 0.0& 0.0& 1.8& 0.1& 0.1& 1.8& 0.1& 0.1& 2.1& 0.3& 0.2\cr
Sin(log) constant          \textit{T+E}& 2.6& 0.9& 0.9& 2.6& 1.0& 1.0& 3.0& 1.6& 1.4& 2.6& 0.9& 0.9\cr
Sin(log) equilateral    \textit{T} only& 1.8& 0.1& 0.1& 2.2& 0.4& 0.3& 1.9& 0.1& 0.1& 2.0& 0.2& 0.2\cr
Sin(log) equilateral       \textit{T+E}& 3.4& 2.0& 1.5& 3.1& 1.7& 1.4& 3.5& 2.2& 1.8& 3.0& 1.6& 1.2\cr
Sin(log) flattened      \textit{T} only& 2.3& 0.5& 0.4& 2.6& 0.9& 0.7& 2.3& 0.5& 0.4& 2.5& 0.9& 0.7\cr
Sin(log) flattened         \textit{T+E}& 2.9& 1.4& 1.4& 3.3& 2.0& 1.8& 3.0& 1.5& 1.5& 3.4& 2.0& 1.9\cr
Sin(log) local          \textit{T} only& 1.7& 0.1& 0.0& 1.9& 0.1& 0.1& 2.0& 0.2& 0.1& 2.0& 0.2& 0.2\cr
Sin(log) local             \textit{T+E}& 2.3& 0.6& 0.6& 2.3& 0.5& 0.6& 2.7& 1.0& 0.8& 2.4& 0.6& 0.6\cr	
\noalign{\vskip 5pt\hrule\vskip 3pt}}}
\endPlancktablewide                    
\endgroup
\end{table*}

\subsubsection{Scale-dependent oscillatory features}
\begin{figure*}[htbp!]
\centering
\includegraphics[width=.48\linewidth]{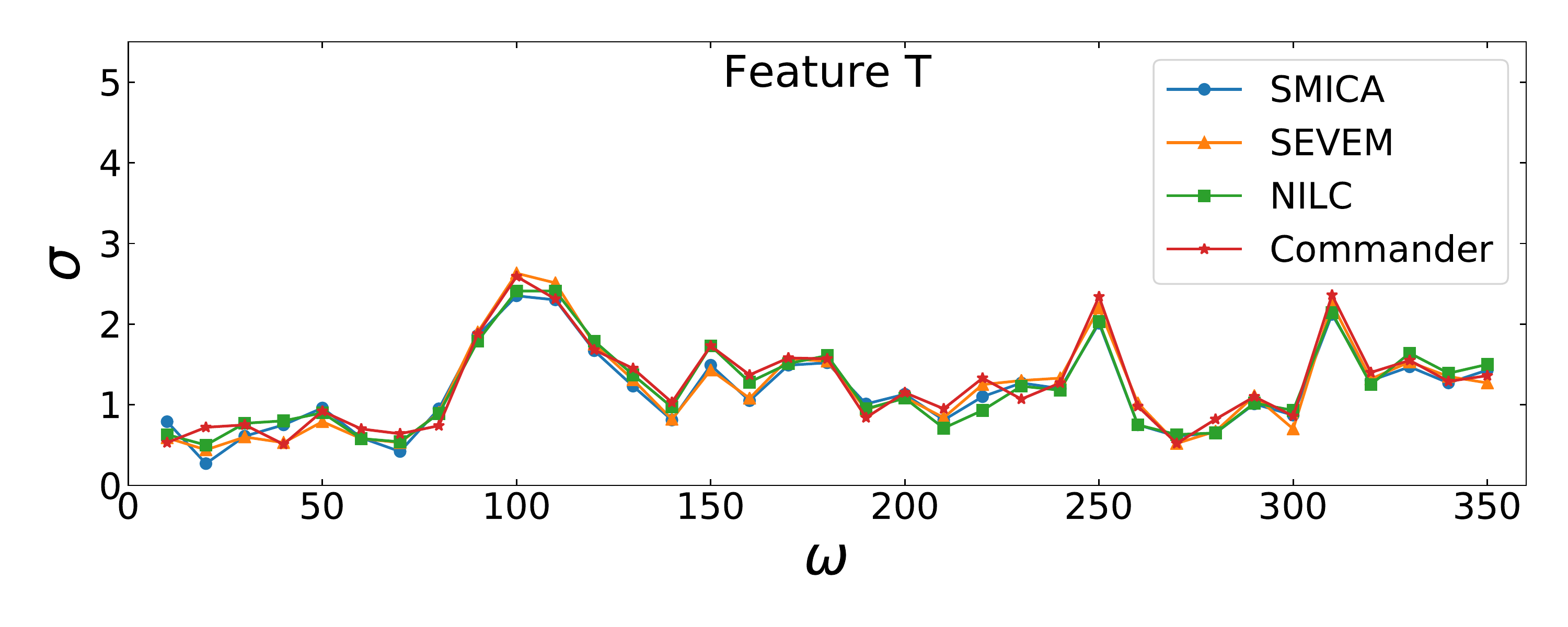} \hskip0.1in\includegraphics[width=.48\linewidth]{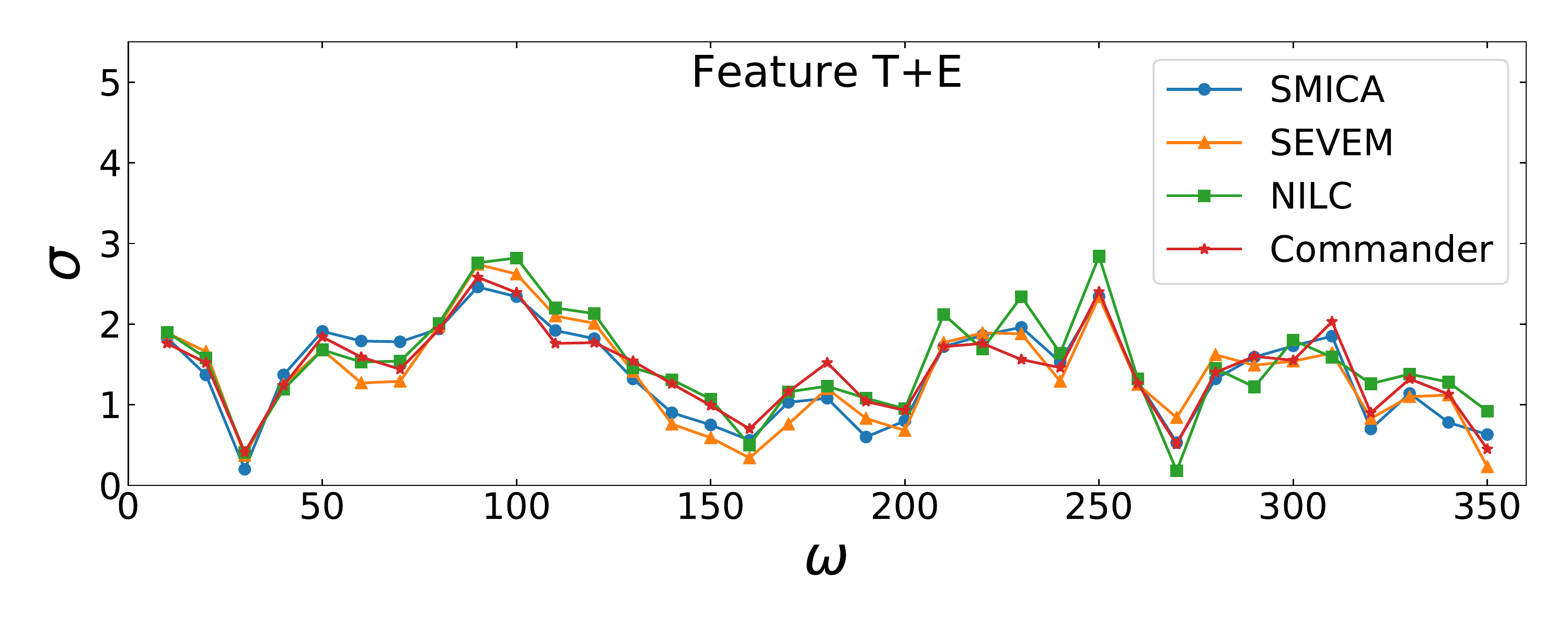}
\includegraphics[width=.48\linewidth]{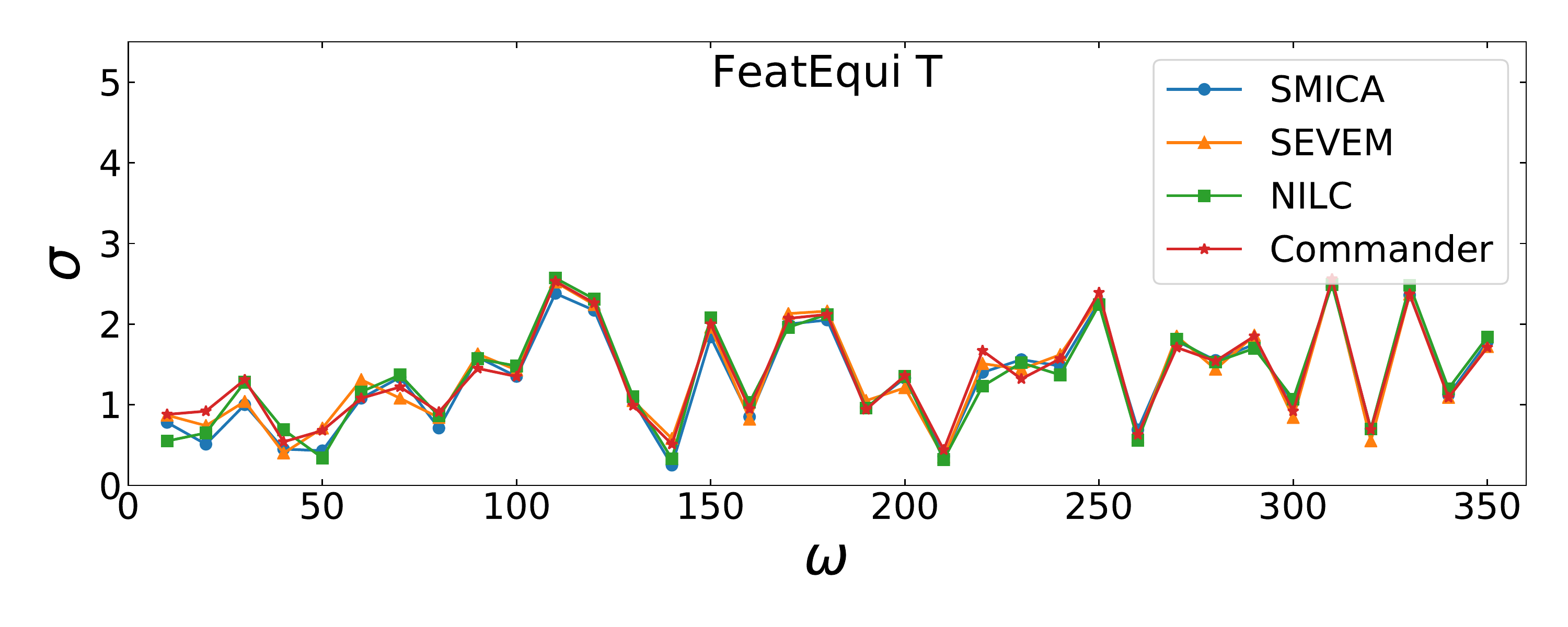} \hskip0.1in\includegraphics[width=.48\linewidth]{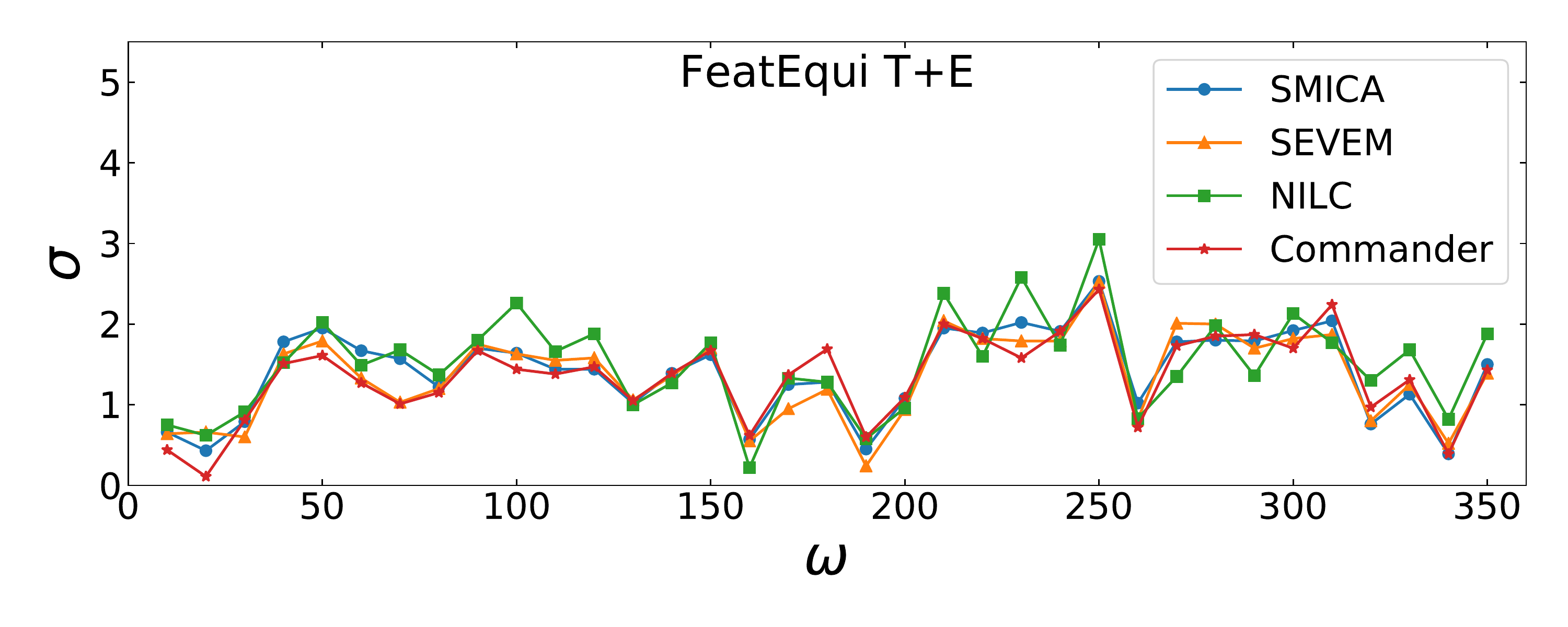}
\includegraphics[width=.48\linewidth]{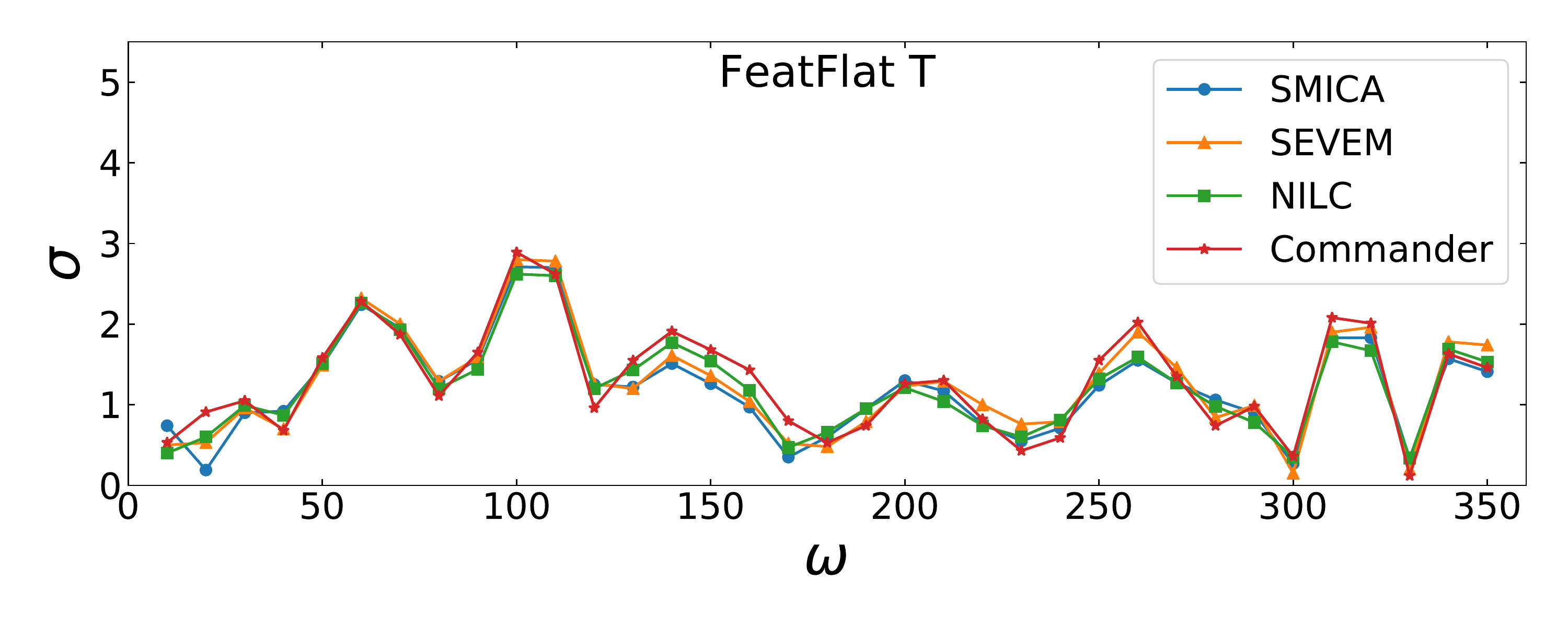} \hskip0.1in\includegraphics[width=.48\linewidth]{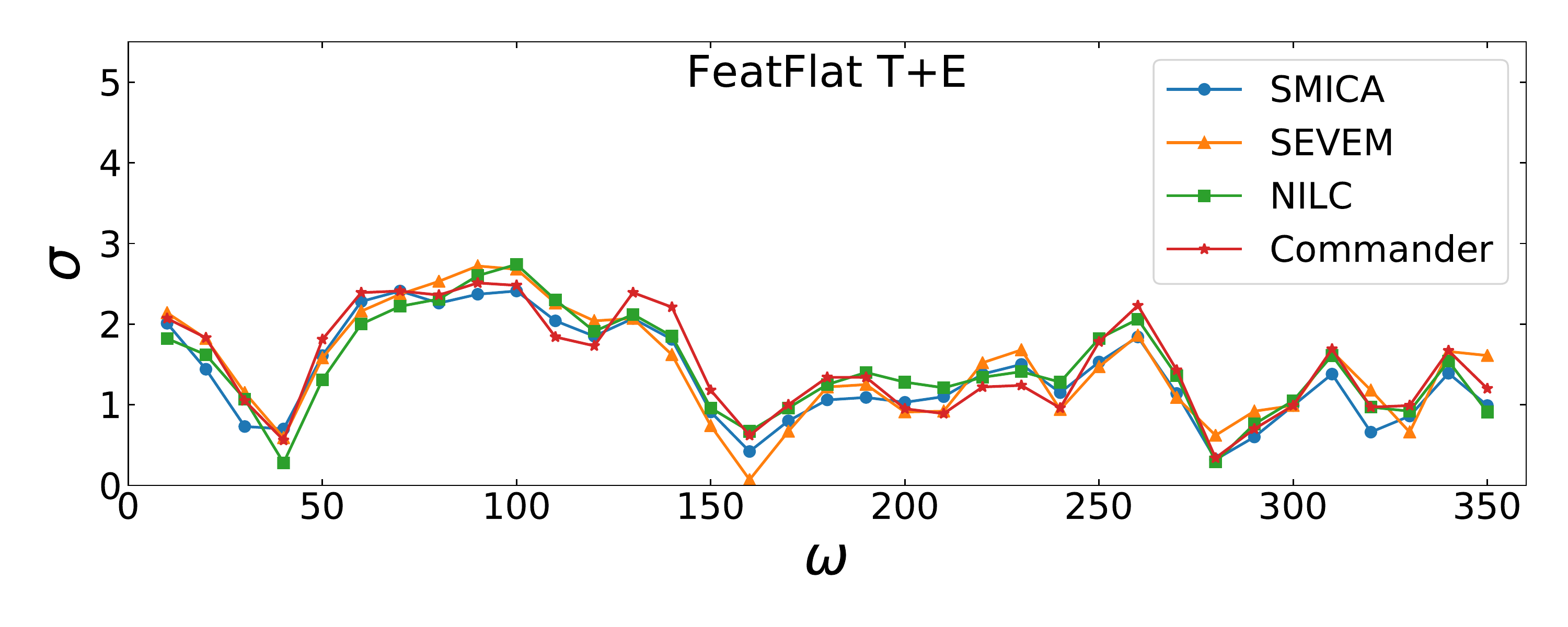}
\caption[]{Generalized feature model significance surveyed over the Modal~2 frequency range $0< \omega< 350$, after marginalizing over the phase $\phi$.  The top panels show results for the constant feature model (Eq.~\ref{eq:featureBprim}) with $T$ only (left) and \textit{T+E} (right), the middle panels for the equilateral feature model (Eq.~\ref{eq:featequilBprim}), and the bottom panels for the flattened feature model (Eq.~\ref{eq:featflatBprim}).  The same conventions apply as in Fig.~\ref{fig:resonance_models}.   These feature model results generally have lower significance than obtained previously \citepalias{planck2014-a19}, with polarization data not tending to reinforce apparent peaks found using temperature data only. }
\label{fig:feature_models}
\end{figure*}

\begin{figure*}[htbp!]
\centering
\includegraphics[width=.48\linewidth]{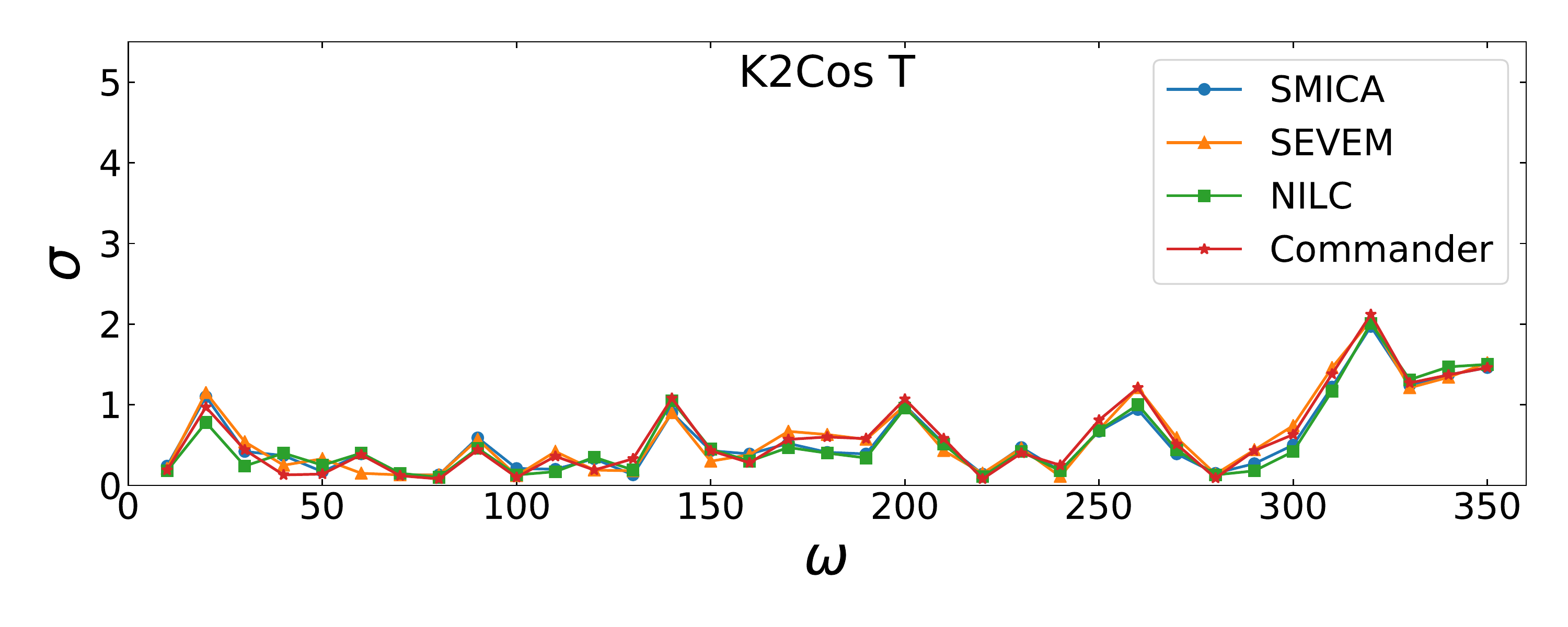} \hskip0.1in\includegraphics[width=.48\linewidth]{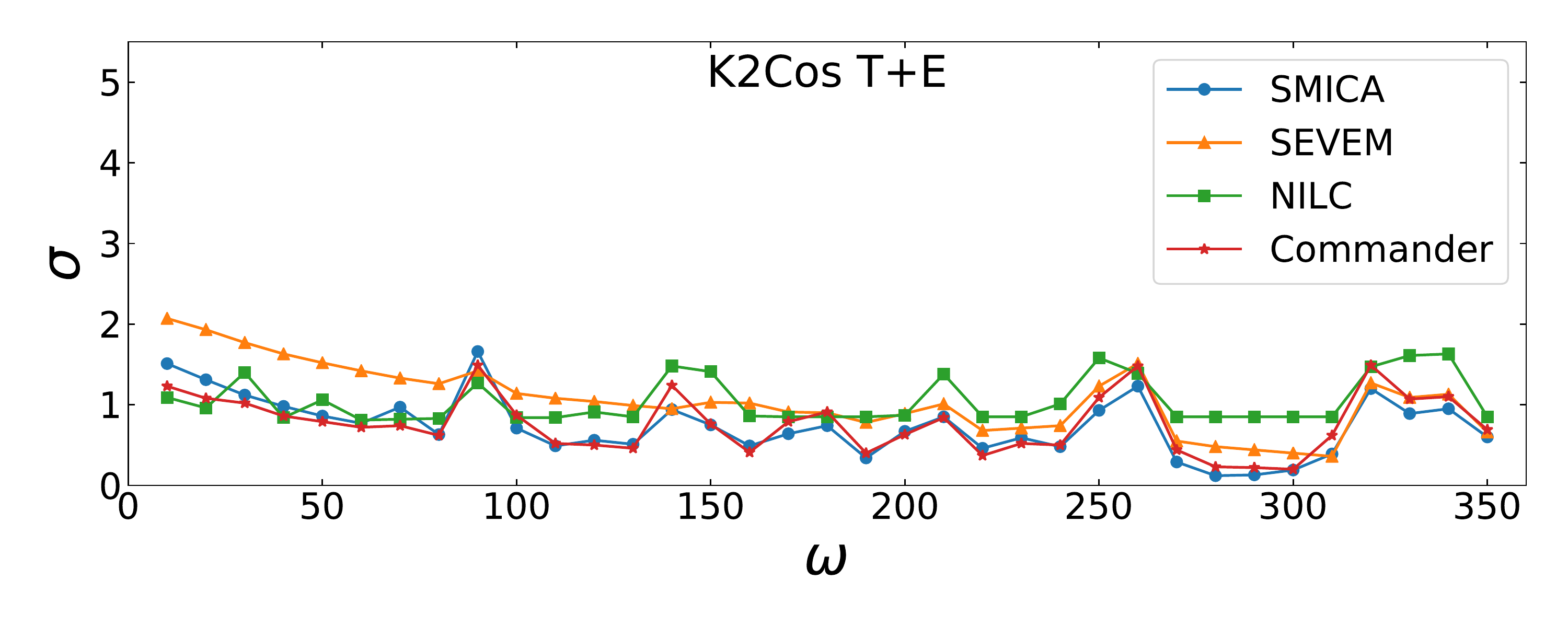}
\includegraphics[width=.48\linewidth]{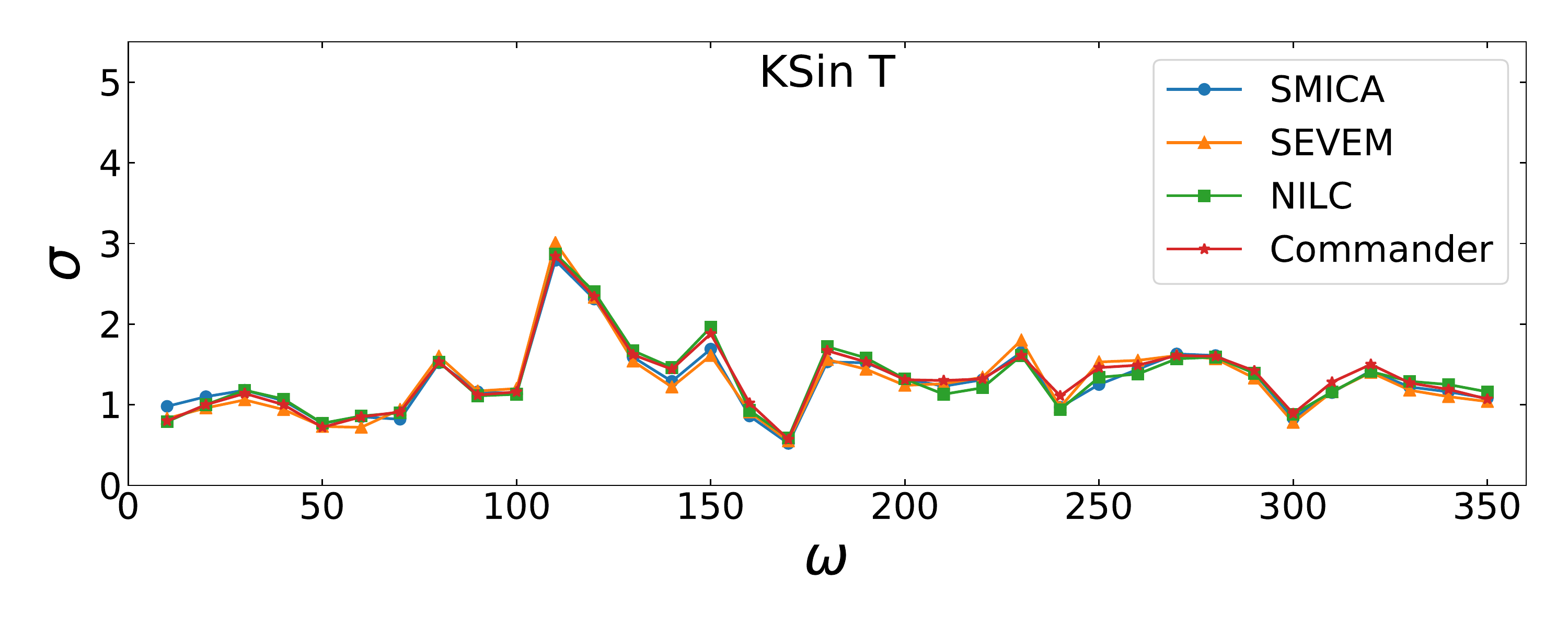} \hskip0.1in\includegraphics[width=.48\linewidth]{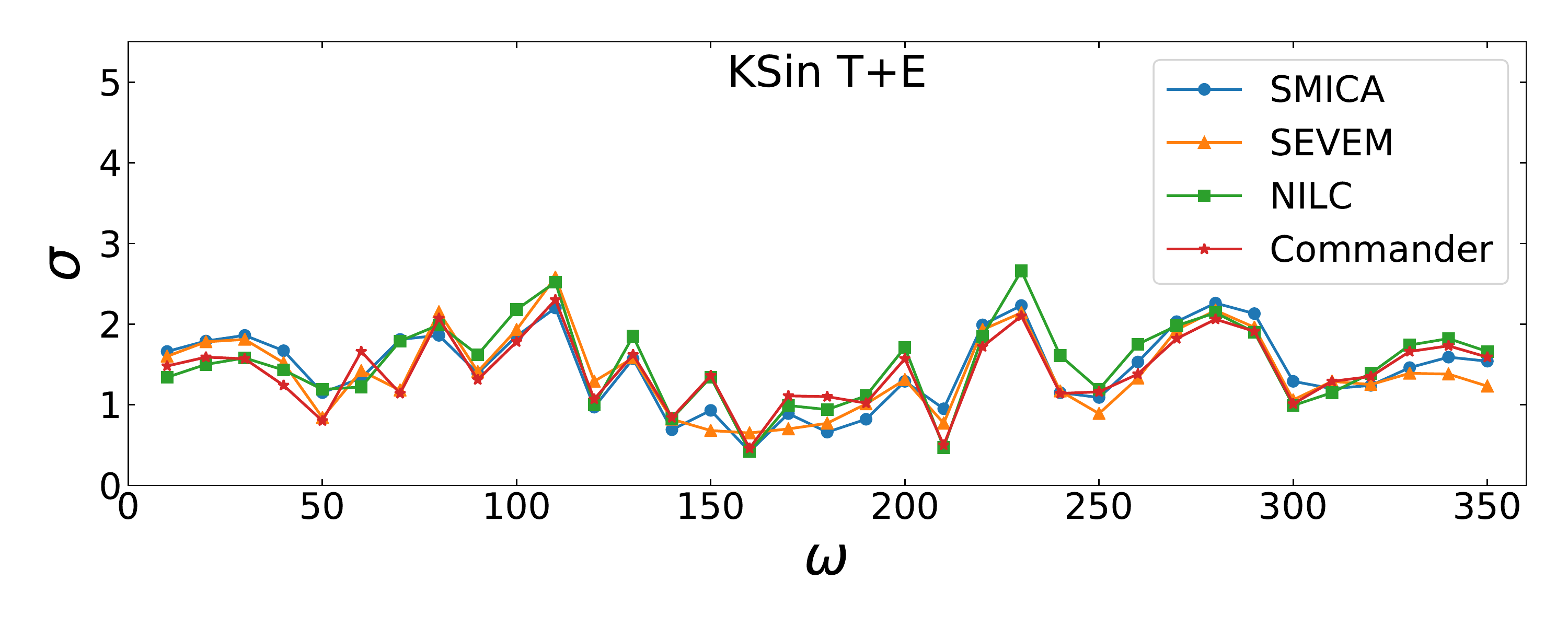}
\caption[]{{\it Top}: Significance of single-field models for the potential
feature case with a $K^2 \cos \omega K$ scaling dependence
(Eq.~\ref{eq:Adsetal1}).  {\it Bottom}: significance for rapidly varying sound
speed with a $K\sin \omega K$ scaling (Eq.~\ref{eq:Adsetal2}).  Left panels
are for $T$ only and left panels are for $T+E$.
These results have been marginalized over the envelope parameter $\alpha$ (determined by feature width and height) from $\alpha = 0$ to $\alpha\omega = 90$. There appears to be no evidence for these very specific signatures in this frequency range when polarization data are included.}
\label{fig:feat_singlefields}
\end{figure*}

In this section we show the broad range of oscillatory models sourced through features in the inflationary potential or sound speed, described in Sect.~\ref{sec:feature}, which can be described by a sinusoid multiplied by either a cross-sectional template or a scaling function.  The results are plotted after maximization over phase in Figs.~\ref{fig:feature_models} and \ref{fig:feat_singlefields} with look-elsewhere-adjusted results presented in Table~\ref{tab:feat_peak_stats}.  Here all results are consistent with Gaussianity for all models in this class.

\begin{table*}[htbp!]
\begingroup
\newdimen\tblskip \tblskip=5pt
\caption{Peak statistics, as defined in \cite{Fergusson:2014hya}, for the different feature models, showing the ``Raw'' peak maximum significance (for the given Modal~2 survey domain), the corrected significance of this ``Single'' maximum peak after accounting for the parameter survey size (the look-elsewhere effect) and the ``Multi''-peak statistic which integrates across the adjusted significance of all peaks to determine consistency with Gaussianity. This table does not include the results of the high-frequency resonance-model estimator, whose significance was assessed independently and presented in subsection \ref{sec:HFFest}.}
\label{tab:feat_peak_stats}
\nointerlineskip
\vskip -3mm
\footnotesize
\setbox\tablebox=\vbox{
  \newdimen\digitwidth 
  \setbox0=\hbox{\rm 0} 
  \digitwidth=\wd0 
  \catcode`*=\active 
  \def*{\kern\digitwidth}
  \newdimen\signwidth 
  \setbox0=\hbox{+} 
  \signwidth=\wd0 
  \catcode`!=\active 
  \def!{\kern\signwidth}
\halign{\hbox to 1.75in{#\leaderfil}\tabskip 1em&
\hfil#\hfil\tabskip=1em& \hfil#\hfil& \hfil#\hfil\tabskip=2em&
\hfil#\hfil\tabskip=1em& \hfil#\hfil& \hfil#\hfil\tabskip=2em&
\hfil#\hfil\tabskip=1em& \hfil#\hfil& \hfil#\hfil\tabskip=2em&
\hfil#\hfil\tabskip=1em& \hfil#\hfil& \hfil#\hfil\tabskip=0pt\cr 
\noalign{\doubleline}
\omit& \multispan3\hfil\SMICA\hfil& \multispan3\hfil\SEVEM\hfil&
 \multispan3\hfil\NILC\hfil& \multispan3\hfil\Commander\hfil\cr
\noalign{\vskip -4pt}
\omit& \multispan3\hrulefill& \multispan3\hrulefill&
 \multispan3\hrulefill& \multispan3\hrulefill\cr
\omit& Raw& Single& Multi& Raw& Single& Multi& Raw& Single& Multi& Raw& Single& Multi\cr
\noalign{\vskip 3pt\hrule\vskip 5pt}
Features constant    \textit{T} only& 2.4& 0.2& 0.2& 2.6& 0.5& 0.5& 2.4& 0.2& 0.3& 2.6& 0.4& 0.5\cr
Features constant    \textit{T+E}   & 2.5& 0.3& 0.3& 2.7& 0.6& 0.7& 2.8& 0.8& 1.2& 2.6& 0.4& 0.5\cr
Features equilateral \textit{T} only& 2.5& 0.3& 0.4& 2.5& 0.3& 0.5& 2.6& 0.4& 0.6& 2.6& 0.4& 0.6\cr
Features equilateral \textit{T+E}   & 2.5& 0.3& 0.3& 2.5& 0.3& 0.3& 3.0& 1.1& 1.1& 2.4& 0.2& 0.2\cr
Features flattened   \textit{T} only& 2.7& 0.6& 0.8& 2.8& 0.7& 1.0& 2.6& 0.5& 0.6& 2.9& 0.9& 0.9\cr
Features flattened   \textit{T+E}   & 2.4& 0.2& 0.4& 2.7& 0.6& 0.9& 2.7& 0.6& 0.7& 2.5& 0.3& 0.5\cr
$K^2 \cos$ features  \textit{T} only& 2.0& 0.0& 0.0& 2.0& 0.0& 0.0& 2.0& 0.0& 0.0& 2.1& 0.0& 0.0\cr
$K^2 \cos$ features  \textit{T+E}   & 1.7& 0.0& 0.0& 2.1& 0.0& 0.0& 1.6& 0.0& 0.0& 1.5& 0.0& 0.0\cr
$K \sin$ features    \textit{T} only& 2.8& 0.7& 0.7& 3.0& 1.0& 1.0& 2.9& 0.8& 0.8& 2.8& 0.8& 0.8\cr
$K \sin$ features    \textit{T+E}   & 2.3& 0.1& 0.1& 2.6& 0.4& 0.4& 2.7& 0.5& 0.6& 2.3& 0.1& 0.1\cr
\noalign{\vskip 5pt\hrule\vskip 3pt}}}
\endPlancktablewide                    
\endgroup
\end{table*}

\subsubsection{High-frequency feature and resonant-model estimator}\label{sec:HFFest}

\begin{figure*}[htbp!]
\centering
\includegraphics[width=1.0\linewidth]{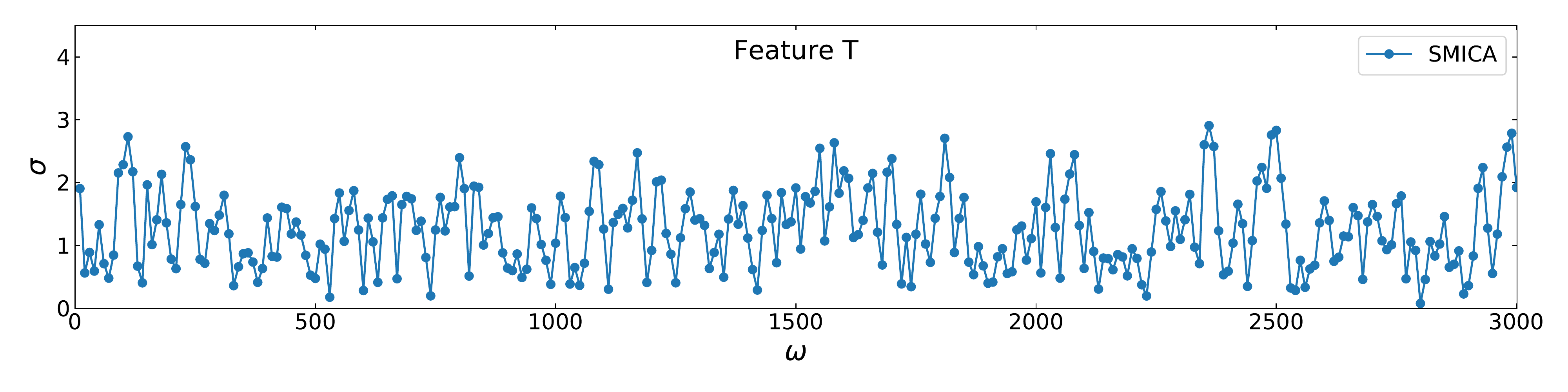} 
\includegraphics[width=1.0\linewidth]{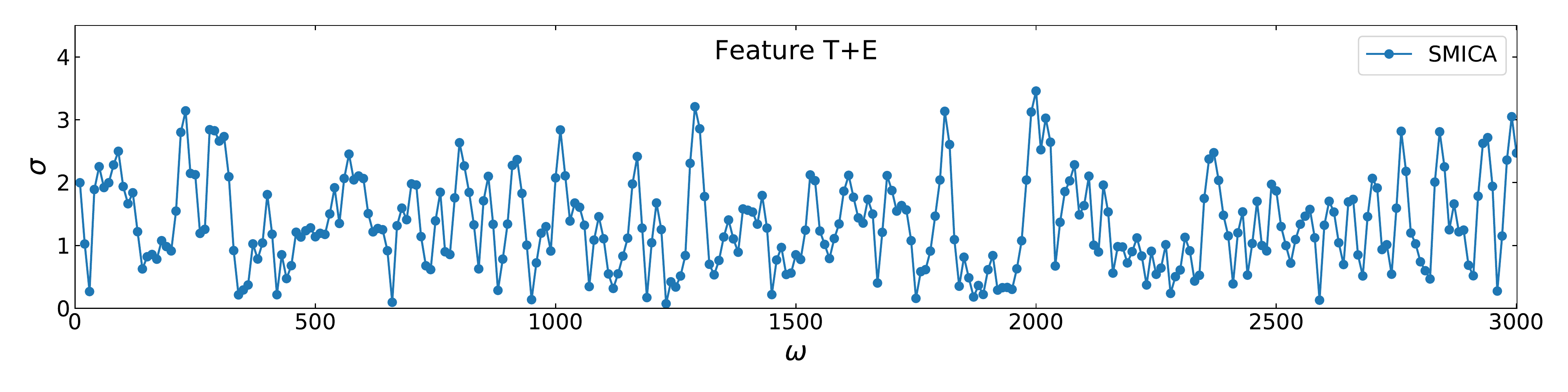} 
\includegraphics[width=1.0\linewidth]{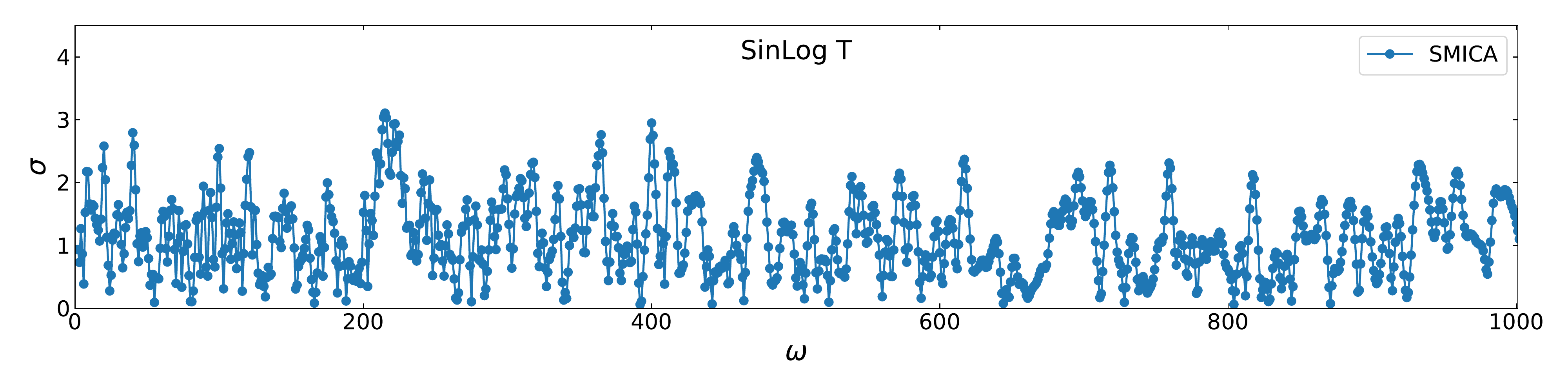} 
\includegraphics[width=1.0\linewidth]{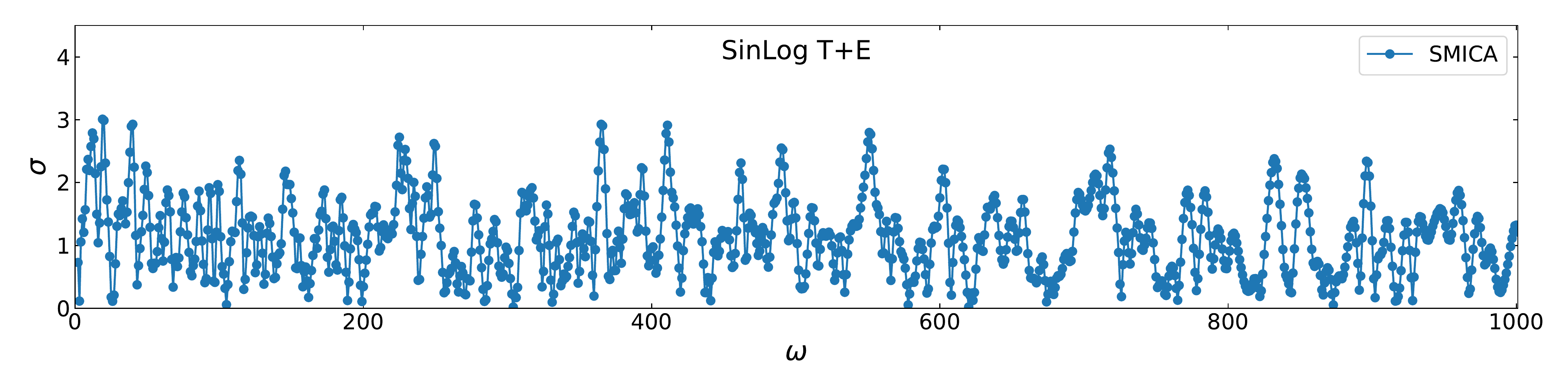} 
\caption[]{High-frequency estimator results for feature and resonance models.
The upper two panels show significance for the constant feature model
(Eq.~\ref{eq:featureBprim}) surveyed over the frequency range $0< \omega< 3000$
after marginalizing over the phase $\phi$, for $T$ and $T+E$ \SMICA\ maps.
The bottom two panels show significance for the constant resonance model
(Eq.~\ref{eq:resBprim}) surveyed over the frequency range $0< \omega< 1000$
after marginalizing over the phase $\phi$, for $T$ and $T+E$ \SMICA\ maps.
As for the Modal expansion case, the high-frequency results generally have lower significance than obtained previously \citepalias{planck2014-a19}, with polarization data not tending to reinforce apparent peaks found using temperature data only.}
\label{fig:high_feature_models}
\end{figure*}

As in \citetalias{planck2014-a19}, we have extended the frequency range of the constant feature model and the constant resonance model with a second set of estimators. The constant feature bispectrum in Eq.~\eqref{eq:featureBprim} is separable and thus allows for the construction of a KSW estimator \citep{Munchmeyer:2014nqa} for direct bispectrum estimation at any given frequency. We use this method to probe frequencies up to $\omega=3000$. In the overlapping frequency range we also confirmed that the results are compatible with those of the Modal pipeline. The results are shown in the upper two panels of Fig.~\ref{fig:high_feature_models}. Due to the computational demands of this estimator we have only analysed \SMICA\ maps, since it was already shown in \citetalias{planck2014-a19} that all component-separations methods agree well for this estimator. No statistically significant peak is found, and the distribution of peaks is consistent with the 2015 analysis. The highest peak is 2.9$\,\sigma$ in $T$ only and 3.5$\,\sigma$ in $T+E$, compared to the Gaussian expectation of $3.1(\pm 0.3)\,\sigma$.\footnote{The Gaussian expectation is not zero because it is the average value of the most significant fluctuation, maximized over the frequency and phase parameters of the shape. The expectation value of a single amplitude with a fixed frequency and phase is of course zero for a Gaussian map.}

For the constant resonance model, we use the method of \cite{Munchmeyer:2014cca}, which expands the logarithmic oscillations in terms of separable linear oscillations. Here we use 800 sine and cosine modes, to cover the frequency range $0<\omega<1000$. The results are shown in Fig.~\ref{fig:feature_models} (lower two panels) for \SMICA\ data. The highest peak is 3.1$\,\sigma$ for $T$ only and 3.0$\,\sigma$ for $T+E$, compared to the Gaussian expectation of $3.4\,\sigma \pm 0.4\,\sigma$. The Gaussian expectation was obtained from the covariance matrix of the estimators using the method of \cite{Meerburg:2015owa}. In summary we do not find evidence for non-Gaussianity in the high-frequency feature and resonance-model analysis.

We used the results obtained here to perform a joint power and bispectrum analysis, which was presented in the accompanying paper \cite{planck2016-l10} in a search for correlated features. No evidence was found for such correlated features in either the power spectrum or the bispectrum. 

\subsubsection{Equilateral-type models and the effective field theory of inflation}
Many physically well-motivated inflationary models produce non-Gaussianity of the equilateral type. While the equilateral template accurately approximates models in this class, it is nevertheless interesting to constrain the exact templates for two reasons.  Firstly, while the equilateral template correlates well with the true models, they are are only normalized in the equilateral limit, which neglects the rest of configuration space and so is only approximate.  This is demonstrated by the fact that the uncertainties for the true templates differ by up to a factor of $2$ compared with the equilateral limit. Secondly, even small deviations in correlation can result in much larger shifts in measurements, for example models that are $95\,\%$ correlated should produce measurements that are on average $1/3\,\sigma$ apart. Here we constrain the exact templates for all inflationary bispectra that fit into the broad equilateral class; for details on these exact templates we refer the reader to the paper \citetalias{planck2014-a19}.  The results for all models in this class are presented in Table~\ref{tab:equilmodels} and are all below $1\,\sigma$, as expected due to their correlation with the equilateral template, but with significant spread due to the small differences in correlation between templates.

\begin{table*}[htbp!]
\begingroup
\newdimen\tblskip \tblskip=5pt
\caption{Constraints on models with equilateral-type NG covering the shapes predicted by the effective field theory of inflation, together with constraints on specific non-canonical inflation models, such as DBI inflation; see section~2 of \citetalias{planck2013-p09a} for detailed explanation of these specific models, with further implications discussed in Sect.~\ref{sec:Implications}. }
\label{tab:equilmodels}
\nointerlineskip
\vskip -3mm
\footnotesize
\setbox\tablebox=\vbox{
 \newdimen\digitwidth 
   \setbox0=\hbox{\rm 0} 
   \digitwidth=\wd0 
   \catcode`*=\active 
   \def*{\kern\digitwidth}
   \newdimen\signwidth 
   \setbox0=\hbox{+} 
   \signwidth=\wd0 
   \catcode`!=\active 
   \def!{\kern\signwidth}
   \newdimen\dotwidth 
   \setbox0=\hbox{.} 
   \dotwidth=\wd0 
   \catcode`^=\active 
   \def^{\kern\dotwidth}
\halign{\hbox to 1.5in{#\leaderfil}\tabskip 0.5em&
\hfil#\hfil\tabskip=1em& \hfil#\hfil\tabskip=2em&
\hfil#\hfil\tabskip=1em& \hfil#\hfil\tabskip=2em&
\hfil#\hfil\tabskip=1em& \hfil#\hfil\tabskip=2em&
\hfil#\hfil\tabskip=1em& \hfil#\hfil\tabskip0pt\cr 
\noalign{\doubleline}
\omit& \multispan2\hfil\SMICA\hfil& \multispan2\hfil\SEVEM\hfil& \multispan2\hfil\NILC\hfil& \multispan2\hfil\Commander\hfil\cr 
\noalign{\vskip -5pt}
\omit& \multispan2\hrulefill& \multispan2\hrulefill& \multispan2\hrulefill& \multispan2\hrulefill\cr 
\omit\hfil Equilateral-type model\hfil& $!*A\pm\sigma_A$& !S/N& $!*A\pm\sigma_A$& !S/N& $!*A\pm\sigma_A$& !S/N& $!*A\pm\sigma_A$& !S/N\cr
\noalign{\vskip 3pt\hrule\vskip 5pt}
Constant        \textit{T} only& !$31\pm38$& !0.8& !$24\pm38$& !0.6& !$15\pm38$& !0.4& !$15\pm38$& !0.4\cr
Constant           \textit{T+E}& !$21\pm24$& !0.9& !$18\pm25$& !0.7& !$*7\pm24$&   !0.3&   !$14\pm25$&   !0.6\cr
Equilateral     \textit{T} only& !$34\pm67$& !0.5& !$34\pm68$& !0.5& !$20\pm67$& !0.3& !$36\pm68$& !0.5\cr
Equilateral     \textit{T+E}& *$-5\pm43$&  $-$0.1& $!*2\pm45$& !0.1& $-16\pm43$& $-$0.4& $!*4\pm44$&   !0.1\cr
EFT shape 1     \textit{T} only& !$43\pm61$& !0.7& !$41\pm63$& !0.6& !$26\pm61$& !0.4& !$39\pm62$& !0.6\cr
EFT shape 1     \textit{T+E}   & $!*8\pm40$& !0.2& !$10\pm41$& !0.2& *$-7\pm40$& $-$0.2& !$10\pm41$&   !0.3\cr
EFT shape 2     \textit{T} only& !$51\pm47$& !1.1& !$45\pm48$& !1.0& !$33\pm46$& !0.7& !$37\pm47$& !0.8\cr
EFT shape 2     \textit{T+E}   & !$28\pm30$& !0.9& !$23\pm31$& !0.8& !$11\pm30$& !0.4& !$21\pm31$& !0.7\cr
DBI inflation   \textit{T} only& !$46\pm58$& !0.8& !$43\pm60$& !0.7& !$29\pm58$& !0.5& !$39\pm59$& !0.7\cr
DBI inflation   \textit{T+E}   & !$14\pm38$& !0.4& !$14\pm39$& !0.4& *$-2\pm38$& $-$0.1& !$14\pm39$& !0.4\cr
Ghost inflation \textit{T} only& !$*6\pm81$& !0.1& !$13\pm83$& !0.2& *$-0\pm81$& $-$0.0& !$25\pm82$& !0.3\cr
Ghost inflation \textit{T+E}   & $-48\pm52$& $-$0.9& $-30\pm54$& $-$0.6& $-50\pm52$& $-$1.0& $-25\pm53$& $-$0.5\cr
Inverse decay   \textit{T} only& !$38\pm41$& !0.9& !$32\pm42$& !0.8& $!21\pm41$& !0.5& $!23\pm42$& !0.5\cr
Inverse decay   \textit{T+E}   & !$24\pm27$& !0.9& !$19\pm27$& !0.7& $!*8\pm26$& !0.3& $!17\pm27$& !0.6\cr 
\noalign{\vskip 3pt\hrule\vskip 3pt}}}
\endPlancktablewide                    
\endgroup
\end{table*}

\subsubsection{Models with excited initial states (non-Bunch-Davies vacua)}
Inflationary models that modify the initial vacuum for the inflaton generally produce shapes in the flattened class, meaning that they peak in triangle configurations with zero area. There is a wide variety of models of this type proposed in the literature, which are described in detail in Sect.~\ref{Sec:NBD_models}.  For simplicity we also include here the results for the warm inflation template, described in Sect.~\ref{Sec:Multifield_models}, since it exhibits similar behaviour once projected, despite the mechanism behind it being quite different. The results for this class are presented in Table~\ref{tab:fnlnonstandard}. The most significant measurement is for the NBD sin template, which produces results around $2\,\sigma$.  However, it is important to note that this result involves a marginalization over a frequency type parameter so there is a look-elsewhere effect that has not yet been taken into account.  Because of this we expect the true significance will be lower, so we can safely claim that our results are consistent with Gaussianity, while noting that it may be interesting to revisit oscillatory NBD templates, like the NBD sin model, using future data sets.

\begin{table*}[htbp!]
\begingroup
\newdimen\tblskip \tblskip=5pt
\caption{Constraints on models with excited initial states (non-Bunch-Davies models), as well as warm inflation; see Sect. 2 for further explanation and the labelling of these classes of NBD models.  Note that the NBD, NBD1, and NBD2 models contain free parameters, so here we quote the maximum significance found over the entire parameter range.  Note that the location of the maximum for \itT\ and \itTpE\ can occur for different parameter values for the model, and so the results with and without polarization cannot be directly compared; however, the model parameters are held fixed across different component-separation methods.}
\label{tab:fnlnonstandard}
\nointerlineskip
\vskip -3mm
\footnotesize
\setbox\tablebox=\vbox{
\setbox0=\hbox{\rm 0} 
   \digitwidth=\wd0 
   \catcode`*=\active 
   \def*{\kern\digitwidth}
   \newdimen\signwidth 
   \setbox0=\hbox{+} 
   \signwidth=\wd0 
   \catcode`!=\active 
   \def!{\kern\signwidth}
   \newdimen\dotwidth 
   \setbox0=\hbox{.} 
   \dotwidth=\wd0 
   \catcode`^=\active 
   \def^{\kern\dotwidth}
\halign{\hbox to 2.1in{#\leaderfil}\tabskip 0.25em&
\hfil#\hfil\tabskip=0.5em& \hfil#\hfil\tabskip=1em&
\hfil#\hfil\tabskip=0.5em& \hfil#\hfil\tabskip=1em&
\hfil#\hfil\tabskip=0.5em& \hfil#\hfil\tabskip=1em&
\hfil#\hfil\tabskip=0.5em& \hfil#\hfil\tabskip0pt\cr 
\noalign{\doubleline}
\omit& \multispan2\hfil\SMICA\hfil& \multispan2\hfil\SEVEM\hfil& \multispan2\hfil\NILC\hfil& \multispan2\hfil\Commander\hfil\cr 
\noalign{\vskip -5pt}
\omit& \multispan2\hrulefill& \multispan2\hrulefill& \multispan2\hrulefill& \multispan2\hrulefill\cr 
\omit\hfil Flattened-type model\hfil& $!*A\pm\sigma_A^$& !S/N& $!*A\pm\sigma_A^$& !S/N& $!*A\pm\sigma_A^$& !S/N& $!*A\pm\sigma_A^$& !S/N\cr
\noalign{\vskip 3pt\hrule\vskip 5pt}
Flat model                        \textit{T} only&    !*56^* $\pm$ *72^*&   !0.8&   !*37^* $\pm$ *72^*&    !0.5&    !*23^* $\pm$ *71^*&   !0.3&    !**7^* $\pm$ *72^*&   !0.1\cr
Flat model                        \textit{T+E}   &    !*60^* $\pm$ *43^*&   !1.4&   !*44^* $\pm$ *44^*&    !1.0&    !*29^* $\pm$ *43^*&   !0.7&    !*33^* $\pm$ *44^*&   !0.8\cr
Non-Bunch-Davies                  \textit{T} only&    !*13^* $\pm$ *89^*&   !0.1&   !**6^* $\pm$ *90^*&    !0.1&   *$-$25^* $\pm$ *88^*& $-$0.3&  *$-$17^* $\pm$ *90^*& $-$0.2\cr
Non-Bunch-Davies                  \textit{T+E}   &    !*60^* $\pm$ *54^*&   !1.1&   !*38^* $\pm$ *56^*&    !0.7&    !*32^* $\pm$ *53^*&   !0.6&    !*44^* $\pm$ *55^*&   !0.8\cr
NBD sin                           \textit{T} only&  $-$630^* $\pm$ 445^*& $-$1.4& $-$673^* $\pm$ 450^*&  $-$1.5&  $-$629^* $\pm$ 444^*& $-$1.4&  $-$747^* $\pm$ 445^*& $-$1.7\cr
NBD sin                           \textit{T+E}   &  $-$496^* $\pm$ 247^*& $-$2.0& $-$483^* $\pm$ 255^*&  $-$1.9&  $-$547^* $\pm$ 249^*& $-$2.2&  $-$505^* $\pm$ 253^*& $-$2.0\cr
NBD1 cos flattened                \textit{T} only&    !**6^* $\pm$ *21^*&   !0.3&    !**8^* $\pm$ *21^*&    !0.4&    !**3^* $\pm$ *21^*&   !0.1&    !**5^* $\pm$ *22^*&   !0.2\cr
NBD1 cos flattened                \textit{T+E}   &    !**2^* $\pm$ *15^*&   !0.1&    !**6^* $\pm$ *16^*&    !0.4&  **$-$7^* $\pm$ *15^*& $-$0.5&    !**1^* $\pm$ *16^*&   !0.0\cr
NBD2 cos squeezed                 \textit{T} only&    !*48^* $\pm$ 167^*&   !0.3&    !*41^* $\pm$ 170^*&    !0.2&  *$-$14^* $\pm$ 166^*& $-$0.1&    !*11^* $\pm$ 170^*&   !0.1\cr
NBD2 cos squeezed                 \textit{T+E}   &  **$-$5.0 $\pm$ **6.0& $-$0.8& **$-$6.5 $\pm$ **6.2&  $-$1.1&  **$-$6.7 $\pm$ **6.0& $-$1.1&  **$-$5.6 $\pm$ **6.0& $-$0.9\cr
NBD1 sin flattened                \textit{T} only&  *$-$22^* $\pm$ *26^*& $-$0.8& *$-$33^* $\pm$ *26^*&  $-$1.3&  *$-$27^* $\pm$ *27^*& $-$1.0&  *$-$38^* $\pm$ *27^*& $-$1.4\cr
NBD1 sin flattened                \textit{T+E}   &  *$-$14^* $\pm$ *20^*& $-$0.7& *$-$22^* $\pm$ *20^*&  $-$1.1&  *$-$19^* $\pm$ *20^*& $-$1.0&  *$-$24^* $\pm$ *20^*& $-$1.2\cr
NBD2 sin squeezed                 \textit{T} only&  **$-$0.5 $\pm$ **0.6& $-$0.7& **$-$0.3 $\pm$ **0.5&  $-$0.6&  **$-$2.9 $\pm$ **2.6& $-$1.1&  **$-$0.8 $\pm$ **0.7& $-$1.1\cr
NBD2 sin squeezed                 \textit{T+E}   &  **$-$0.3 $\pm$ **0.4& $-$0.7& **$-$0.2 $\pm$ **0.4&  $-$0.5&  **$-$0.4 $\pm$ **0.4& $-$0.9&  **$-$0.4 $\pm$ **0.4& $-$1.0\cr
NBD3 non-canonical ($\times1000$) \textit{T} only&  **$-$4.4 $\pm$ **7.8& $-$0.6& **$-$5.6 $\pm$ **7.9&  $-$0.7&  **$-$4.6 $\pm$ **8.0& $-$0.6&  **$-$5.5 $\pm$ **7.9& $-$0.7\cr
NBD3 non-canonical ($\times1000$) \textit{T+E}   &  **$-$6.4 $\pm$ **5.8& $-$1.1& **$-$7.5 $\pm$ **5.9&  $-$1.3&  **$-$5.8 $\pm$ **5.9& $-$1.0&  **$-$6.2 $\pm$ **5.9& $-$1.0\cr
WarmS inflation                   \textit{T} only&  *$-$39^* $\pm$ *44^*& $-$0.9& *$-$35^* $\pm$ *44^*&  $-$0.8&  *$-$34^* $\pm$ *44^*& $-$0.8&  *$-$18^* $\pm$ *44^*& $-$0.4\cr
WarmS inflation                   \textit{T+E}   &  *$-$48^* $\pm$ *27^*& $-$1.8& *$-$37^* $\pm$ *28^*&  $-$1.3&  *$-$41^* $\pm$ *28^*& $-$1.5&  *$-$27^* $\pm$ *28^*& $-$1.0\cr
\noalign{\vskip 5pt\hrule\vskip 3pt}}}
\endPlancktablewide                    
\endgroup
\end{table*}

\subsubsection{Direction-dependent primordial non-Gaussianity}
Here we present results for inflationary models where gauge fields induce a direction dependance to a ``local'' bispectrum, as described in Sect.~\ref{sec:vector_models}. Results for the $L=1$ and $L=2$ templates are presented in Table~\ref{tab:dirdep} from the Modal~2 pipeline. Due to the complicated behaviour in the squeezed limit, the convergence for these models is lower than for some others, but the Modal correlation remains above $93\,\%$ so a perfect estimator may see shifts of order $0.5\,\sigma$.  The largest result is $1.9\,\sigma$ for $L=1$ and \SMICA, but this is not seen consistently across all component-separation methods so cannot be considered robust. We can then conclude that our results are consistent with Gaussianity.

\begin{table*}[htbp!]
\begingroup
\newdimen\tblskip \tblskip=5pt
\caption{Direction-dependent NG results for both the $L=1$ and $L=2$ models from the Modal~2 pipeline.}
\label{tab:dirdep}
\nointerlineskip
\vskip -3mm
\setbox\tablebox=\vbox{
\setbox0=\hbox{\rm 0} 
   \digitwidth=\wd0 
   \catcode`*=\active 
   \def*{\kern\digitwidth}
   \newdimen\signwidth 
   \setbox0=\hbox{+} 
   \signwidth=\wd0 
   \catcode`!=\active 
   \def!{\kern\signwidth}
   \newdimen\dotwidth 
   \setbox0=\hbox{.} 
   \dotwidth=\wd0 
   \catcode`^=\active 
   \def^{\kern\dotwidth}
\halign{\hbox to 1.5in{#\leaderfil}\tabskip 0.5em&
\hfil#\hfil\tabskip=1em& \hfil#\hfil\tabskip=2em&
\hfil#\hfil\tabskip=1em& \hfil#\hfil\tabskip=2em&
\hfil#\hfil\tabskip=1em& \hfil#\hfil\tabskip=2em&
\hfil#\hfil\tabskip=1em& \hfil#\hfil\tabskip0pt\cr 
\noalign{\doubleline}
\omit& \multispan2\hfil\SMICA\hfil& \multispan2\hfil\SEVEM\hfil& \multispan2\hfil\NILC\hfil& \multispan2\hfil\Commander\hfil\cr 
\noalign{\vskip -5pt}
\omit& \multispan2\hrulefill& \multispan2\hrulefill& \multispan2\hrulefill& \multispan2\hrulefill\cr 
\omit\hfil Equilateral-type model\hfil& $!*A\pm\sigma_A$& !S/N& $!*A\pm\sigma_A$& !S/N& $!*A\pm\sigma_A$& !S/N& $!*A\pm\sigma_A$& !S/N\cr
\noalign{\vskip 3pt\hrule\vskip 5pt}
\omit\hfil $L = 1$\hfil& &&&&&&&\cr
Modal~2 \textit{T} only& $-51^*\pm51^*$& $-$1.0& $-40^*\pm51^*$& $-$0.8&   $-41^*\pm52^*$&  $-$0.8& $-18^*\pm52^*$& $-$0.3\cr 
Modal~2 \textit{T+E}   & $-57^*\pm30^*$& $-$1.9& $-42^*\pm31^*$& $-$1.4&   $-42^*\pm30^*$&  $-$1.4& $-31^*\pm31^*$& $-$1.0\cr 
\noalign{\vskip 5pt\hrule\vskip 3pt}
\omit\hfil $L = 2$\hfil&&&&&&&&\cr
Modal~2 \textit{T} only&   $!*1.2\pm*3.0$&   !0.4&   $!*1.7\pm*3.1$&   !0.5&     $!*1.3\pm*3.1$&    !0.4&   $!*0.5\pm*3.1$&   !0.1\cr 
Modal~2 \textit{T+E}   &   $!*0.9\pm*2.4$&   !0.4&   $!*1.8\pm*2.4$&   !0.7&   *$-0.2\pm*2.4$&  $-$0.1& *$-0.2\pm*2.4$& $-$0.1\cr  
\noalign{\vskip 5pt\hrule\vskip 3pt}}}
\endPlancktablewide                    
\endgroup
\end{table*}

\subsubsection{Parity-violating tensor non-Gaussianity motivated by pseudo-scalars} 

In this section, we report constraints on the tensor nonlinearity parameter $f_{\rm NL}^{\rm \,tens}$ (Eq.~\ref{eq:fnltens}) obtained from temperature and $E$-mode polarization maps. As in our 2015 analysis, we examine even and odd $\ell_1 + \ell_2 + \ell_3$ multipole domains, employing the original parity-even Modal estimator, $\hat{f}_{\rm NL}^{\rm \,even}$ \citep{2010PhRvD..82b3502F,2012JCAP...12..032F,2019arXiv190402599S}, and its parity-odd version, $\hat{f}_{\rm NL}^{\rm \,odd}$ \citep{2014JCAP...05..008S,2015JCAP...01..007S,2019arXiv190402599S}, respectively. The constraints obtained from both domains are also combined by computing 
\begin{equation}
  \hat{f}_{\rm NL}^{\rm \,all} = \frac{F^{\rm even} \hat{f}_{\rm NL}^{\rm \,even} + F^{\rm odd} \hat{f}_{\rm NL}^{\rm \,odd}}{F^{\rm even} + F^{\rm odd}} ,
\end{equation}
where $F^{\rm even / odd}$ is the Fisher matrix from $\ell_1 + \ell_2 + \ell_3 = {\rm even / odd}$. This analysis is performed for under the multipole ranges $2 \leq \ell \leq 500$ in temperature and $4 \leq \ell \leq 500$ in polarization.  Here, the use of the first 40 multipoles of the $E$-mode polarization data, which were disregarded in the 2015 analysis, boosts contributions of $TTE$, $TEE$, and $EEE$ to constraining $f_{\rm NL}^{\rm \,tens}$. Although the data, simulations (used for the computation of the linear terms and error bars) and analysis details (e.g., masks, beams, and noise distributions) are somewhat different, other settings for the $f_{\rm NL}^{\rm \,tens}$ estimation are basically the same as in the 2015 analysis \citepalias{planck2014-a19}.

The results from four different component-separated maps are summarized in Table~\ref{tab:fnltens}. We confirm there that the sizes of errors in the parity-odd {\it T+E} analysis reduce to almost the same level as the parity-even counterparts. This is because the low-$\ell$ signal of $TTE$, which dominates the signal-to-noise ratio from $\ell_1 + \ell_2 + \ell_3 = {\rm odd}$ \citep{2013JCAP...11..051S}, is now taken into account. Although the errors for the parity-even case are as large as the 2015 ones, owing to the sensitivity improvement of the parity-odd part, the whole-domain constraints become more stringent.

Regardless of some updates, we find no ${>}\,2\,\sigma$ signal, which is consistent with the conclusion of the 2015 analysis. This indicates no parity violation in the primordial Universe and accordingly gives constraints on some axion inflationary models (see Sect.~\ref{subsec:nonstand}).

\begin{table}[htbp!]                 
\begingroup
\newdimen\tblskip \tblskip=5pt
\caption{Results for the tensor nonlinearity parameter $f_{\rm NL}^{\rm \,tens} / 10^2$ obtained from the \SMICA, \SEVEM, \NILC, and \Commander\ temperature and polarization maps. The central values and the errors ($68\,\%$ CL) extracted from $\ell_1 + \ell_2 + \ell_3 = {\rm even}$ (``Even''), $\ell_1 + \ell_2 + \ell_3 = {\rm odd}$ (``Odd''), and their whole domain (``All'') are separately described. One can see that all {\it T}-only results are in good agreement with both the \Planck\ 2015 ones \citepalias{planck2014-a19} and WMAP ones \citep{2015JCAP...01..007S}.
}
\label{tab:fnltens}
\nointerlineskip
\vskip -6mm
\footnotesize
\setbox\tablebox=\vbox{
   \newdimen\digitwidth
   \setbox0=\hbox{\rm 0}
   \digitwidth=\wd0
   \catcode`*=\active
   \def*{\kern\digitwidth}
   \newdimen\signwidth
   \setbox0=\hbox{+}
   \signwidth=\wd0
   \catcode`!=\active
   \def!{\kern\signwidth}
\newdimen\dotwidth
\setbox0=\hbox{.}
\dotwidth=\wd0
\catcode`^=\active
\def^{\kern\dotwidth}
\halign{\hbox to 1in{#\leaderfil}\tabskip 1em&
\hfil#\hfil\tabskip 2em&
\hfil#\hfil&
\hfil#\hfil\tabskip 0pt\cr
\noalign{\vskip 10pt\doubleline}
\omit& 
\hfil !Even \hfil& 
\hfil !Odd \hfil& 
\hfil !All \hfil\cr
\noalign{\vskip 5pt\hrule\vskip 3pt}
\omit\hfil \SMICA\hfil&&\cr
\textit{T}  & $!*4\pm17$& $*!100\pm100$& $!*6\pm16$\cr
\textit{E}  & $!33\pm67$& $*-570\pm720$& $!29\pm67$\cr
\textit{T+E}& $!11\pm14$& $!***1\pm*18$& $!*8\pm11$\cr
\noalign{\vskip 5pt\hrule\vskip 3pt}
\omit\hfil \SEVEM\hfil&&\cr
\textit{T}  & $!*4\pm17$& $!**90\pm100$& $!*6\pm16$\cr
\textit{E}  & $!75\pm75$& $*-790\pm830$& $!70\pm75$\cr
\textit{T+E}& $!16\pm14$& $!***2\pm*20$& $!13\pm12$\cr
\noalign{\vskip 5pt\hrule\vskip 3pt}
\omit\hfil \NILC\hfil&&\cr
\textit{T}  & $!*4\pm17$& $!**90\pm100$& $!*6\pm16$\cr
\textit{E}  & $-16\pm81$& $*-540\pm820$& $-19\pm80$\cr
\textit{T+E}& $!*6\pm14$& $!***3\pm*21$& $!*5\pm11$\cr
\noalign{\vskip 5pt\hrule\vskip 3pt}
\omit\hfil \Commander\hfil&&\cr
\textit{T}  & $!*5\pm17$& $!**90\pm100$& $!*6\pm16$\cr
\textit{E}  & $!21\pm69$& $-1200\pm700$& $!13\pm69$\cr
\textit{T+E}& $!10\pm14$& $***-2\pm*19$& $!*7\pm11$\cr
\noalign{\vskip 5pt\hrule\vskip 3pt}}}
\endPlancktable                    
\endgroup
\end{table}                        

\subsection{Bispectrum reconstruction}

\subsubsection{Modal bispectrum reconstruction}

\begin{figure*}[htbp!]
\centering
\includegraphics[width=.5\linewidth]{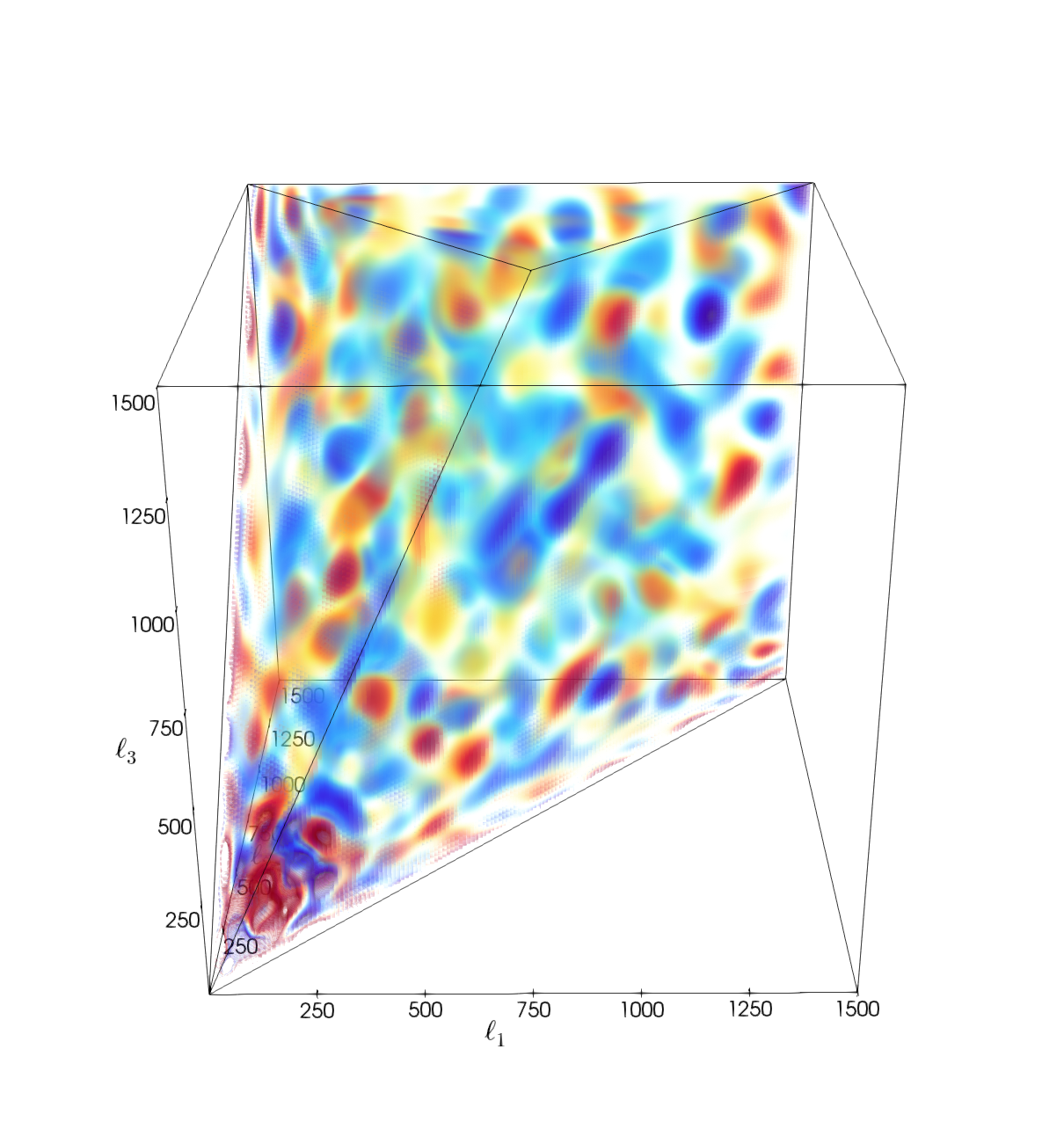}\includegraphics[width=.5\linewidth]{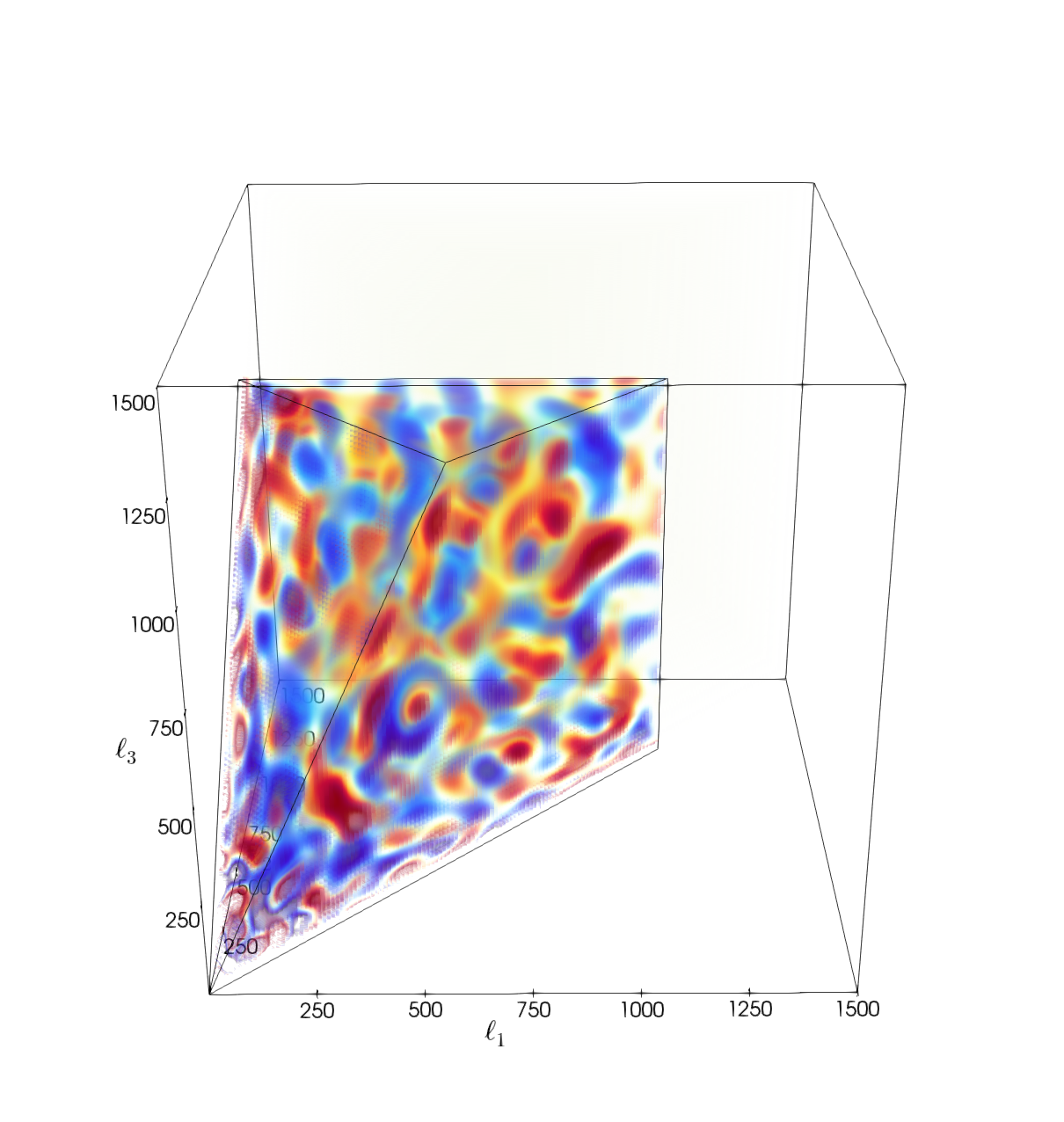}
\includegraphics[width=.5\linewidth]{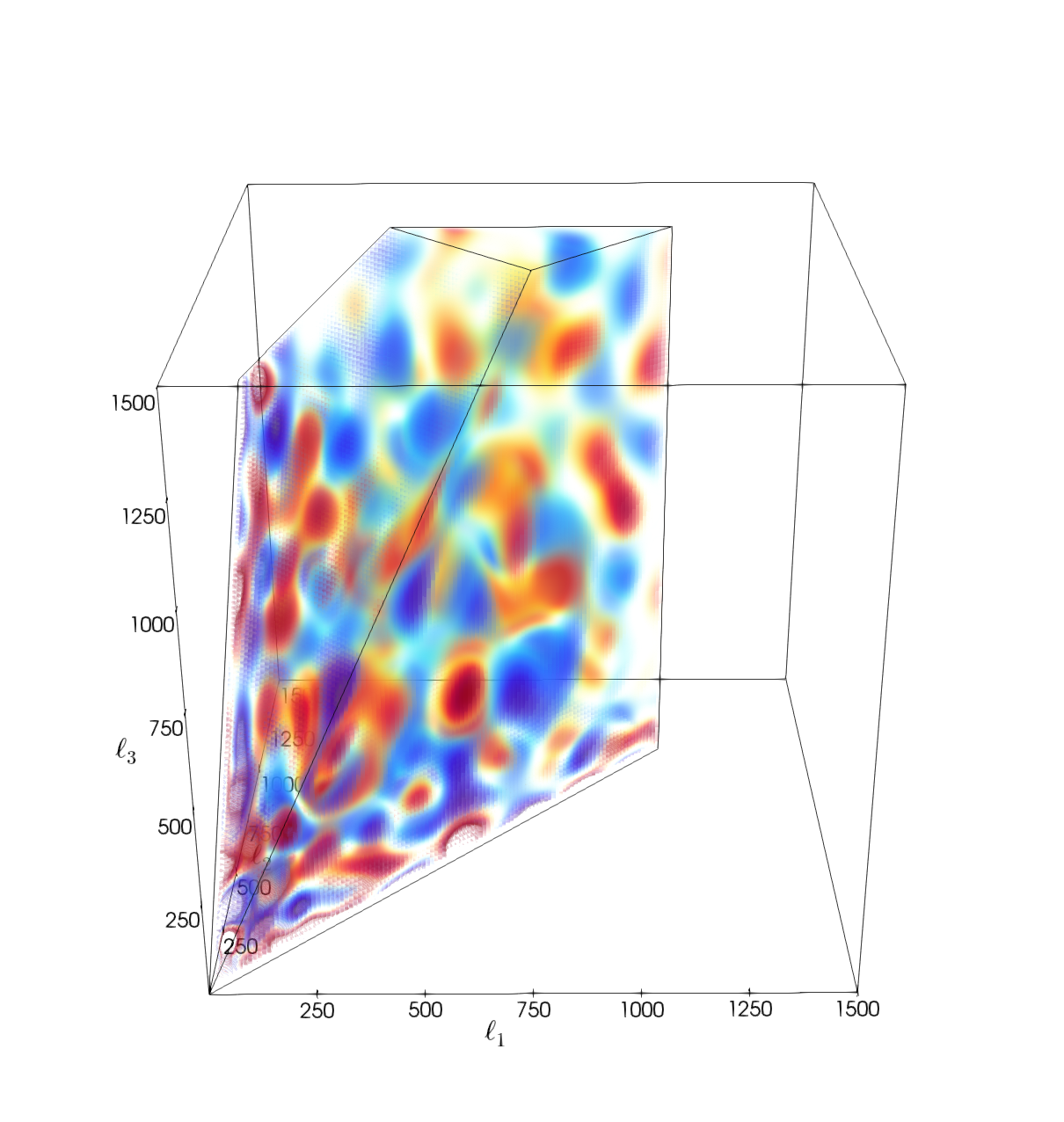}\includegraphics[width=.5\linewidth]{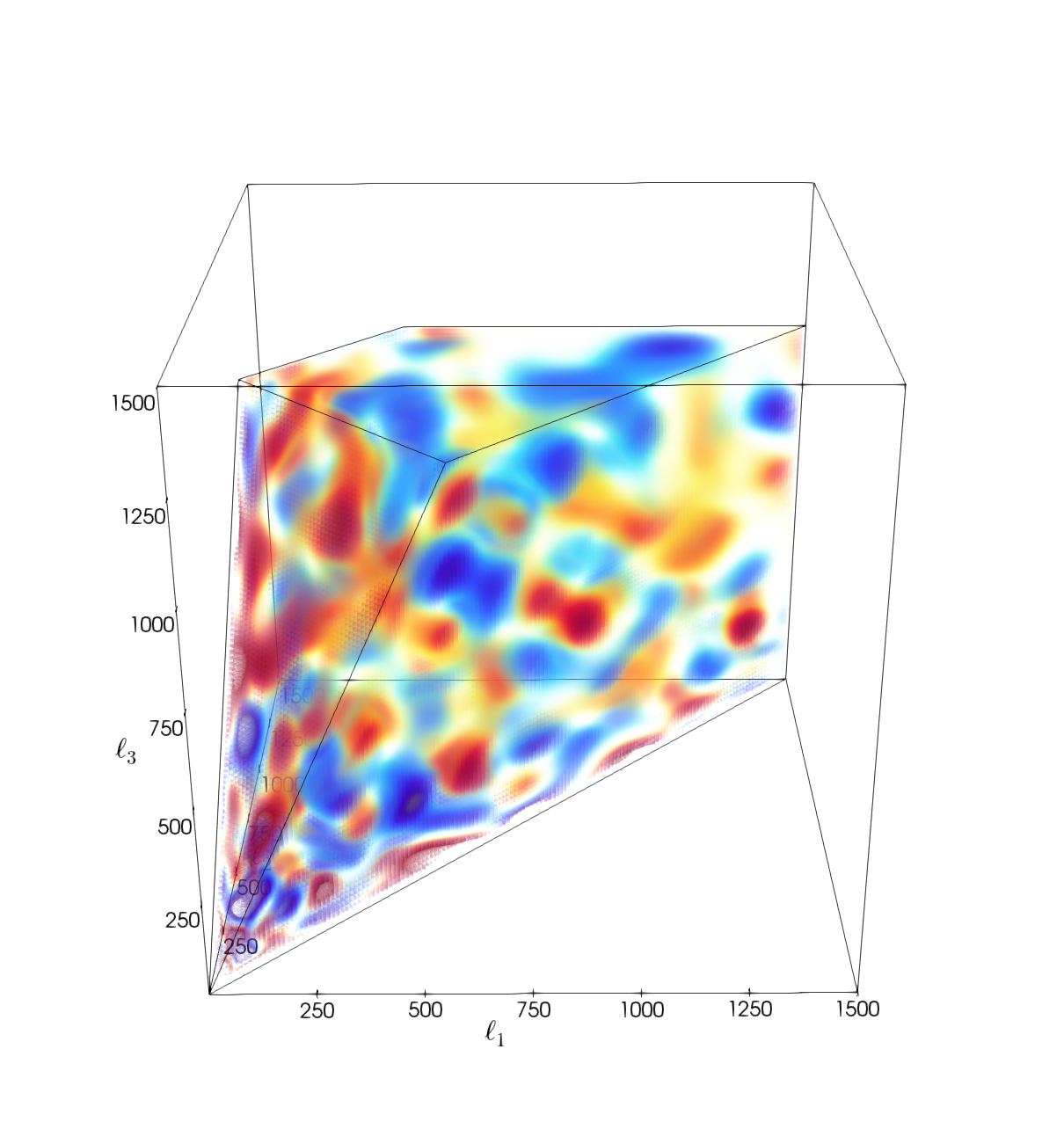}
\caption[]{Signal-to-noise-weighted temperature and polarization bispectra obtained from \Planck\ \SMICA\ maps using the Modal reconstruction with  $n_{\rm max} = 2001$ polynomial modes. The bispecra are (clockwise from top left) the  temperature \textit{TTT} for $\ell \le 1500$,  the \itE-mode polarization \textit{EEE} for $ \ell \le 1100$, the mixed temperature/polarization \textit{TEE}  (with \itT\ multipoles in the $z$-direction), and lastly \textit{TTE} (with \itE\ multipoles in the $z$-direction).  The S/N thresholds are the same between all plots.  }
\label{fig:reconstructTE}
\end{figure*}

The Modal bispectrum estimator filters the \Planck\ foreground-removed CMB maps, i.e., \SMICA, \SEVEM, \NILC, and \Commander, using $n_{\rm max} = 2001$ polynomial modes to obtain model coefficients $\beta_n$.  This procedure is undertaken to obtain all auto- and cross-correlations between the temperature and polarization components \textit{TTT, TTE, TEE}, and \textit{EEE}.  We can then use the $\beta_n$ coefficients with the polynomial modes to obtain a full 3D reconstruction of the \Planck\ temperature and polarization bispectra and this is shown in Fig.~\ref{fig:reconstructTE}.  These bispectra are in close agreement with those published in the paper analysing the \Planck\ 2015 results \citepalias{planck2014-a19} when comparison is made in the signal-dominated regime.

\subsubsection{Binned bispectrum reconstruction}

The Binned bispectrum estimator can also be used to study the
bispectrum itself, in addition to determining the $\fnl$ amplitudes
of specific templates, as in the previous sections. In particular
we can investigate if any non-Gaussianity beyond that of the explicit
models tested can be found in the cleaned CMB maps.
Since we want to detect specific bin-triplets
standing out from the noise, we work with the linear-term-corrected
signal-to-noise-ratio bispectrum. Except for the high-$\ell$ bins,
where a point-source signal is present, the bin-triplets are noise
dominated. Instead of rebinning, we can use a Gaussian kernel to smooth
the bispectrum, so that structure localized in harmonic
space stands out from the noise.  In this process, we mask out a few
bin-triplets that have non-Gaussian noise (due to the fact that they contain
very few valid $\ell$-triplets). We also have to take into account
edge effects from the non-trivial domain of definition of the
bispectrum. The method has been described in \citetalias{planck2014-a19}
and more extensively in \cite{Bucher:2015ura}.

\begin{figure*}[htbp!]
\centering
\begin{tabular}{>{\centering\arraybackslash}m{0.2in}
 >{\centering\arraybackslash}m{1.5in}
 >{\centering\arraybackslash}m{1.5in}
 >{\centering\arraybackslash}m{1.5in}
 >{\centering\arraybackslash}m{1.5in}
}
 & \SMICA&  \SEVEM& \NILC& \Commander\\

$\mathcal{B}^{TTT}$& \includegraphics[trim=0.15cm 0.22cm 0.28cm 0.1cm, clip,width=1.6in]{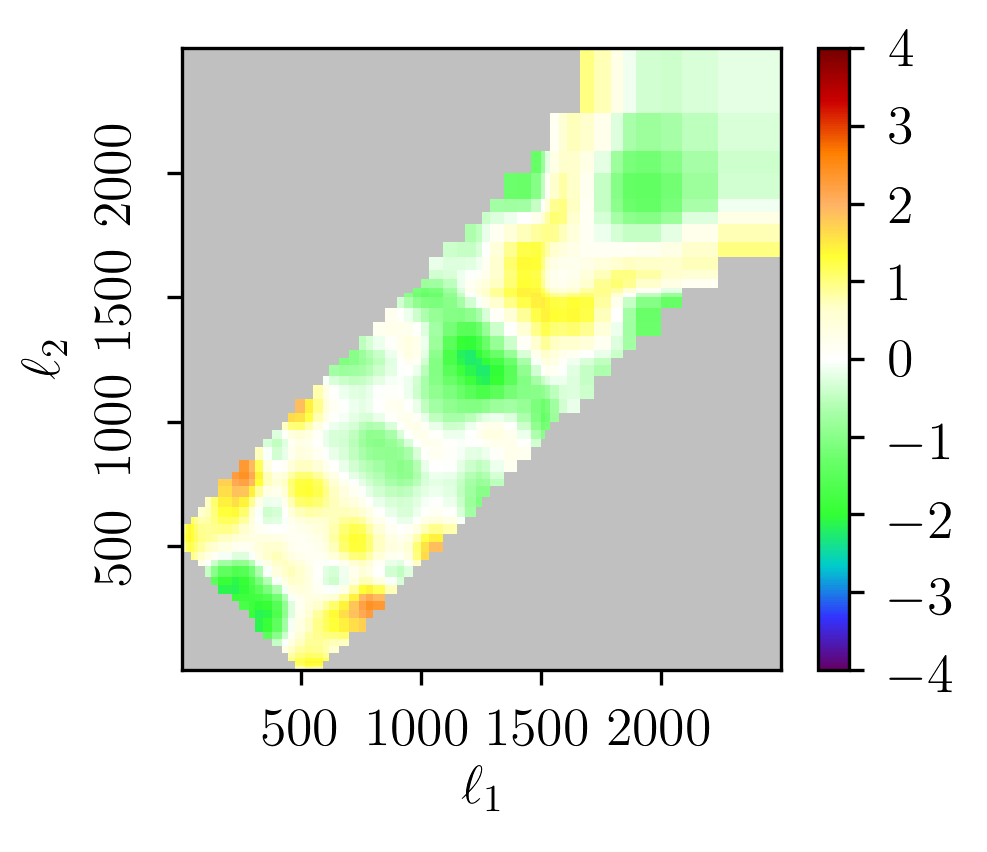}& \includegraphics[trim=0.15cm 0.22cm 0.28cm 0.1cm, clip,width=1.6in]{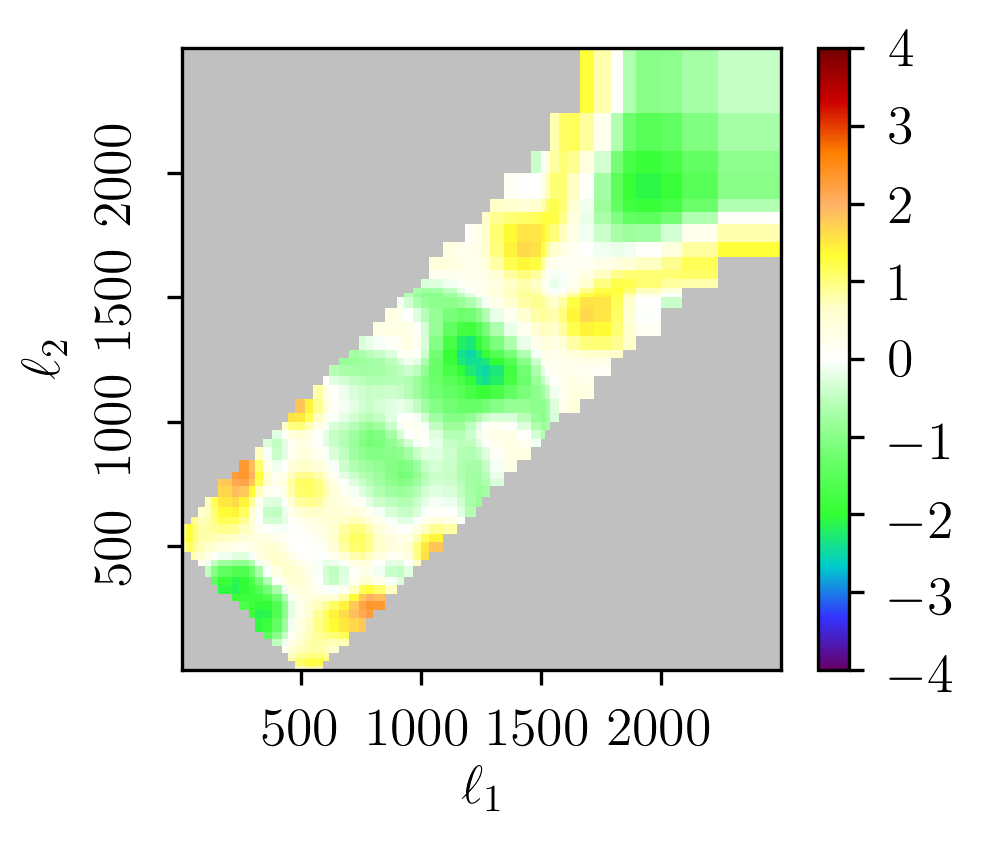}& \includegraphics[trim=0.15cm 0.22cm 0.28cm 0.1cm, clip,width=1.6in]{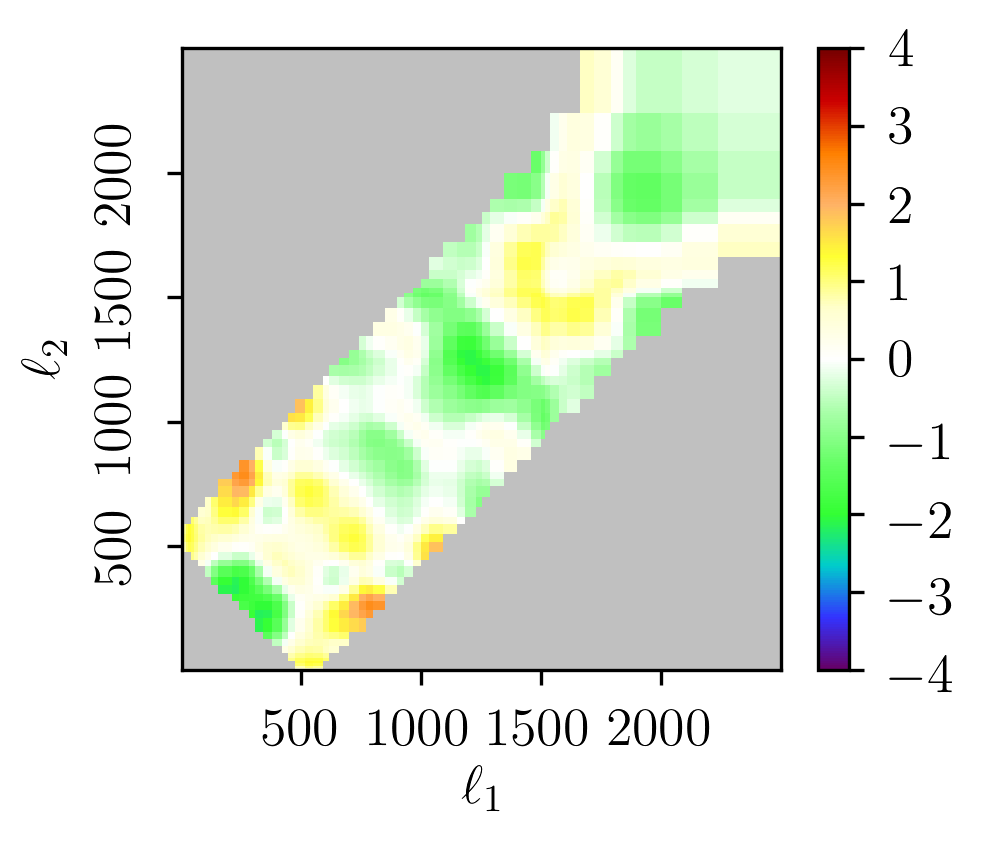}& \includegraphics[trim=0.15cm 0.22cm 0.28cm 0.1cm, clip,width=1.6in]{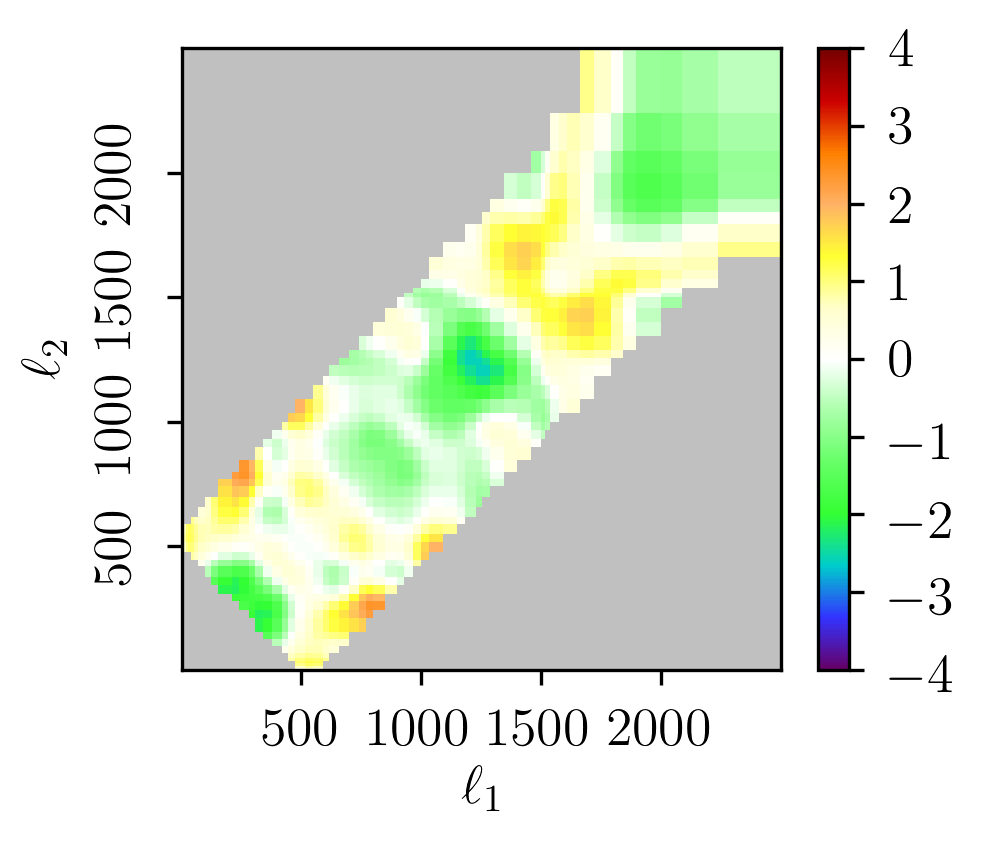} \\
$\mathcal{B}^{TTT}$ no PS& \includegraphics[trim=0.15cm 0.22cm 0.28cm 0.1cm, clip,width=1.6in]{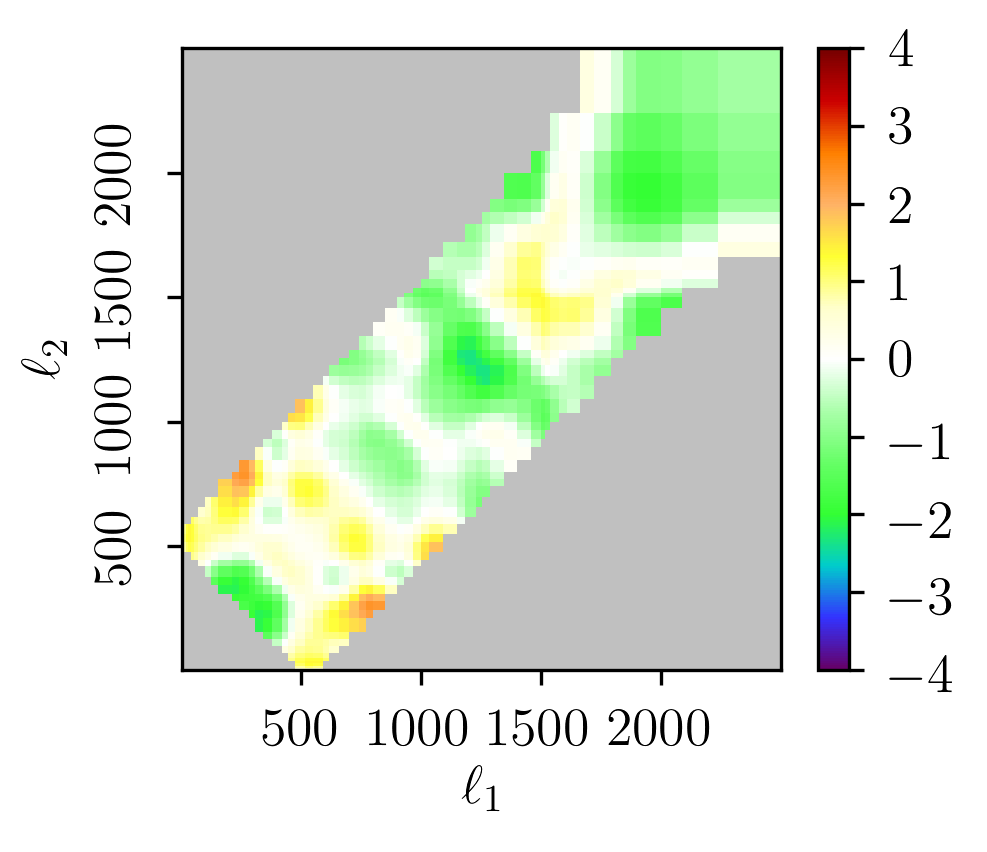}& \includegraphics[trim=0.15cm 0.22cm 0.28cm 0.1cm, clip,width=1.6in]{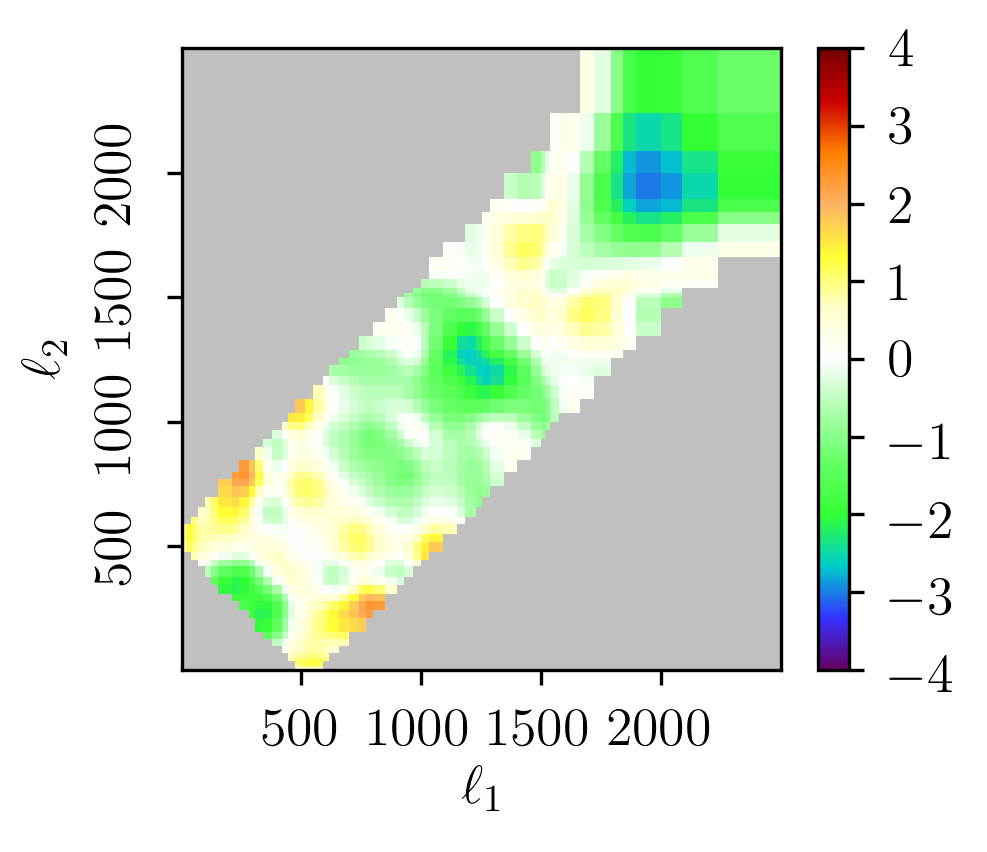}& \includegraphics[trim=0.15cm 0.22cm 0.28cm 0.1cm, clip,width=1.6in]{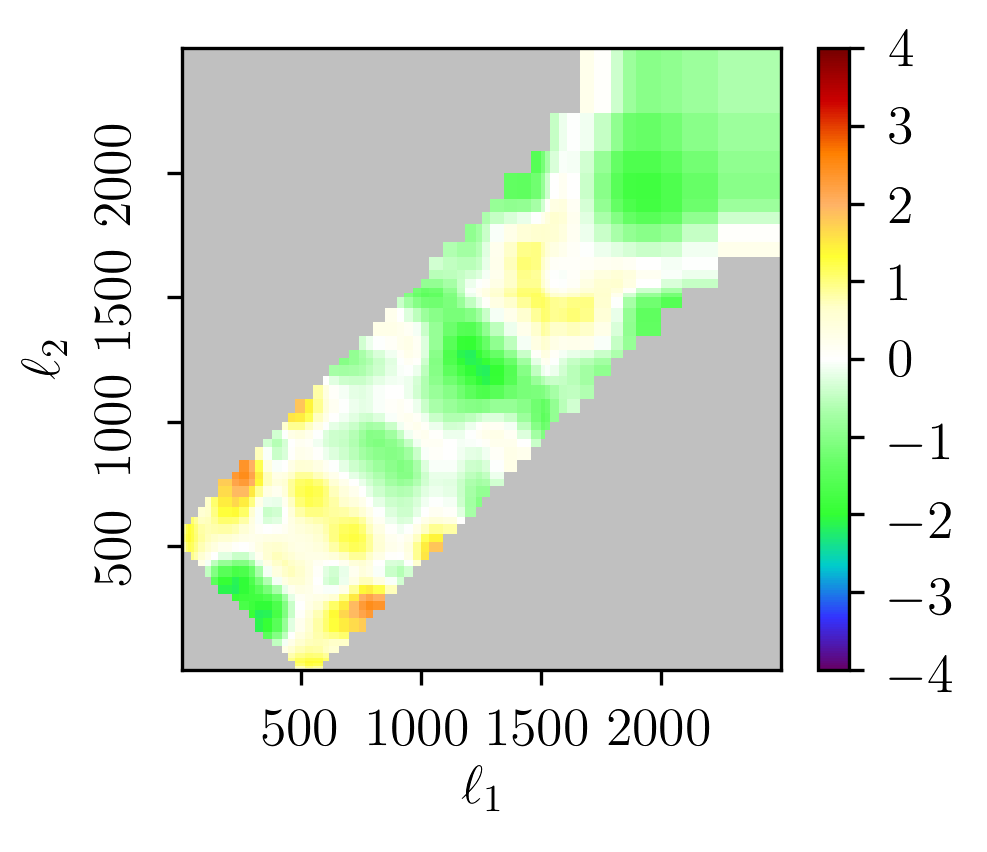}& \includegraphics[trim=0.15cm 0.22cm 0.28cm 0.1cm, clip,width=1.6in]{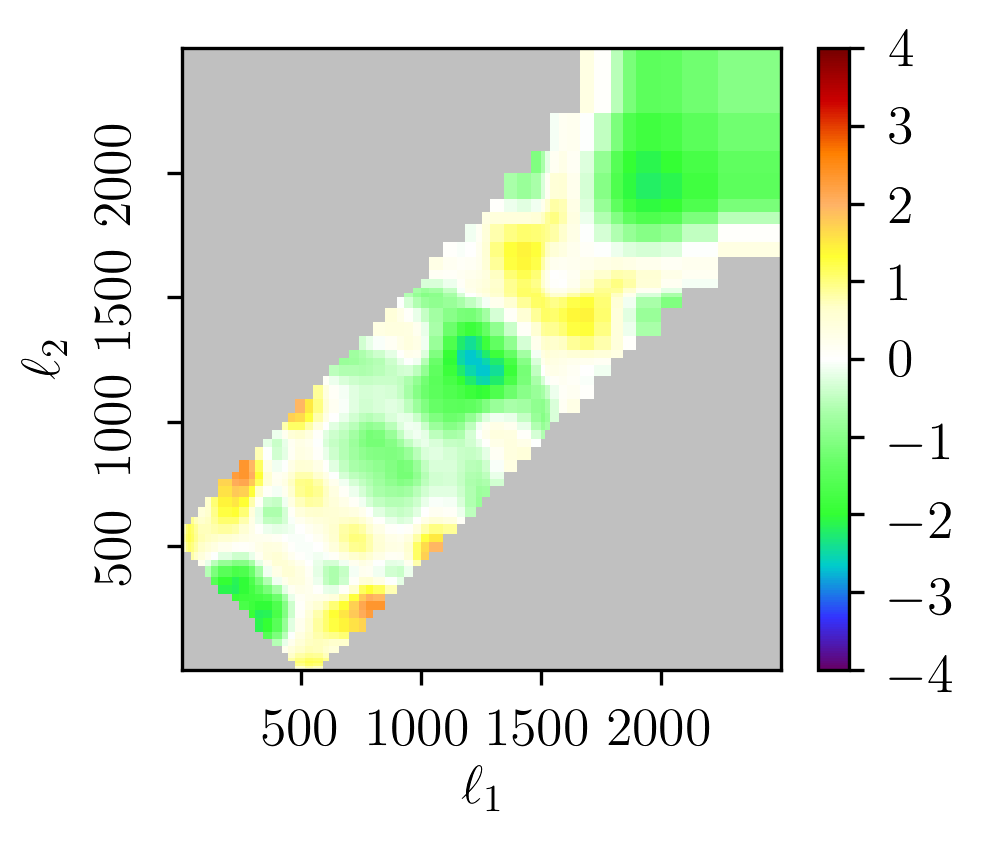} \\
$\mathcal{B}^{T2E}$& \includegraphics[trim=0.15cm 0.22cm 0.28cm 0.1cm, clip,width=1.6in]{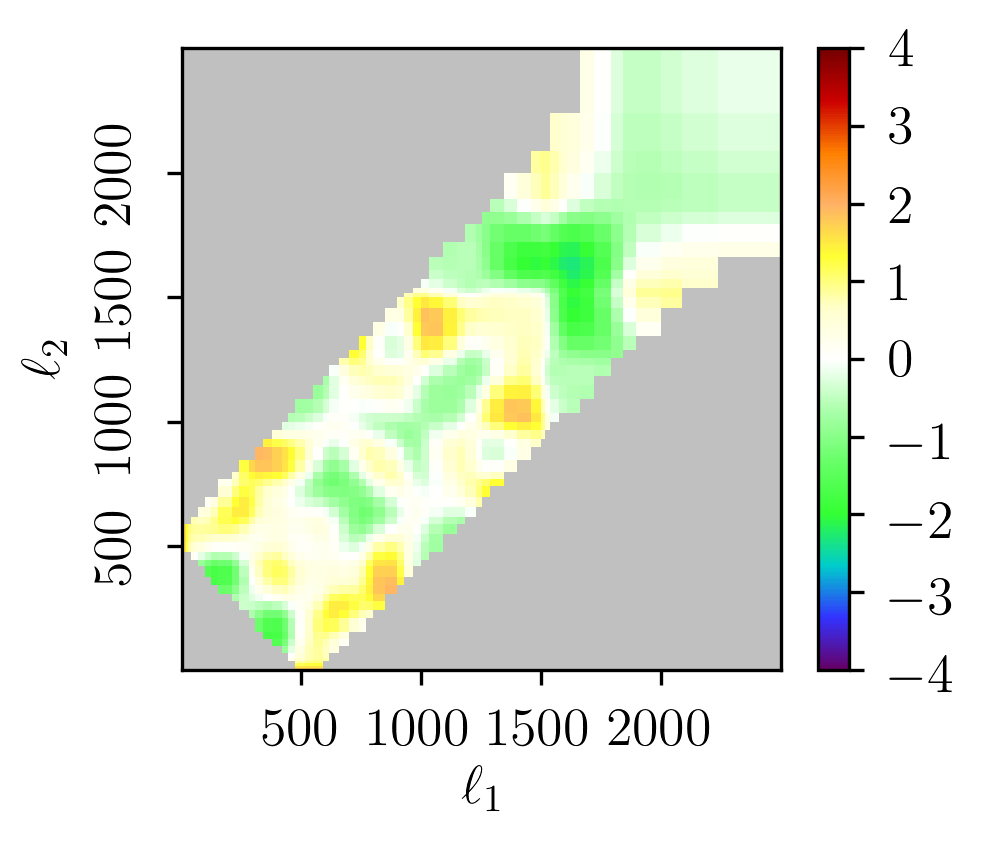}& \includegraphics[trim=0.15cm 0.22cm 0.28cm 0.1cm, clip,width=1.6in]{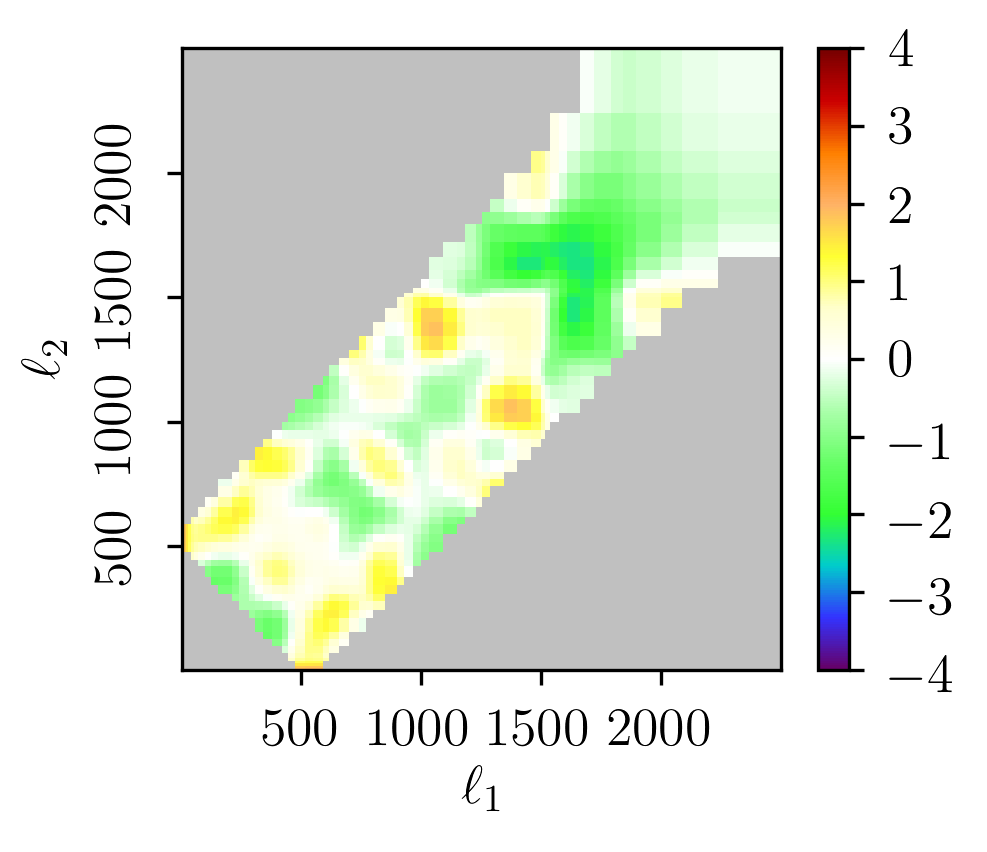}& \includegraphics[trim=0.15cm 0.22cm 0.28cm 0.1cm, clip,width=1.6in]{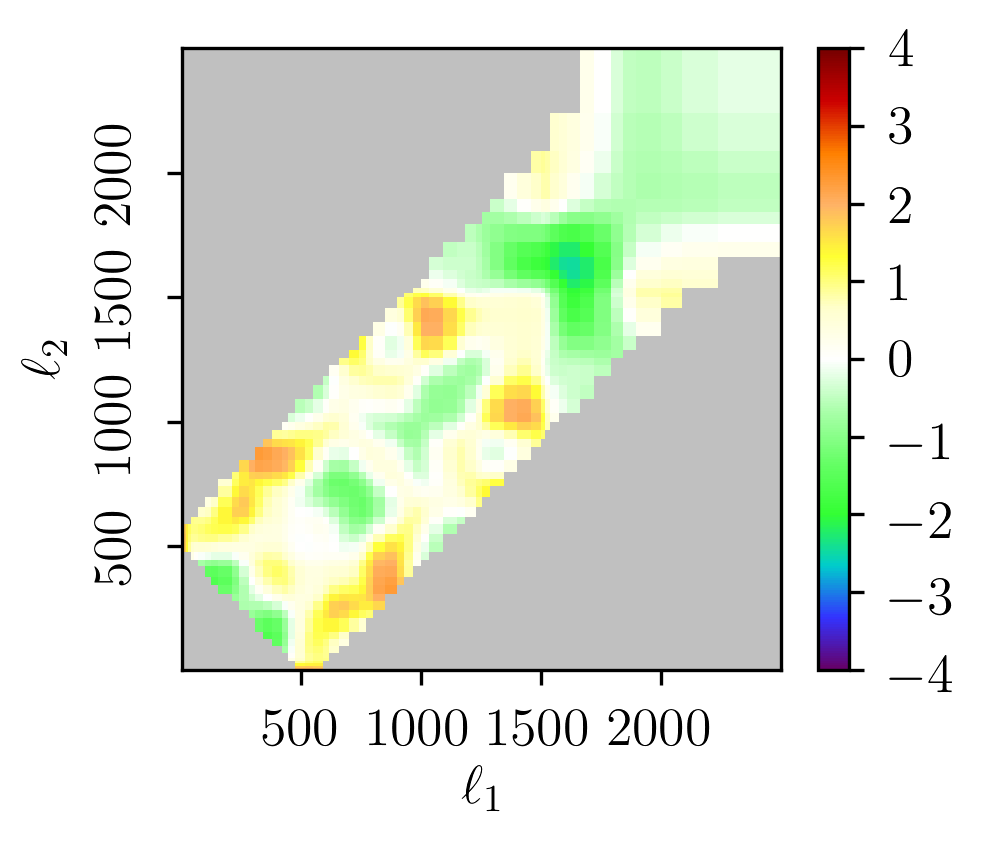}& \includegraphics[trim=0.15cm 0.22cm 0.28cm 0.1cm, clip,width=1.6in]{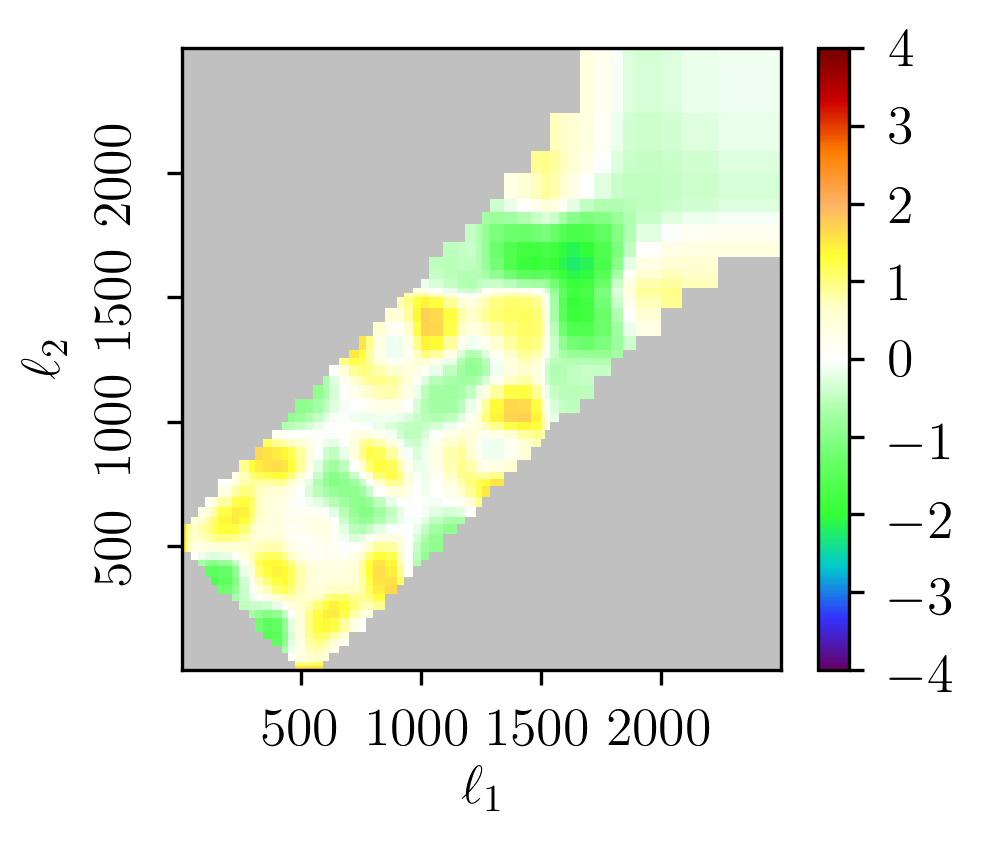} \\
$\mathcal{B}^{TE2}$& \includegraphics[trim=0.15cm 0.22cm 0.28cm 0.1cm, clip,width=1.6in]{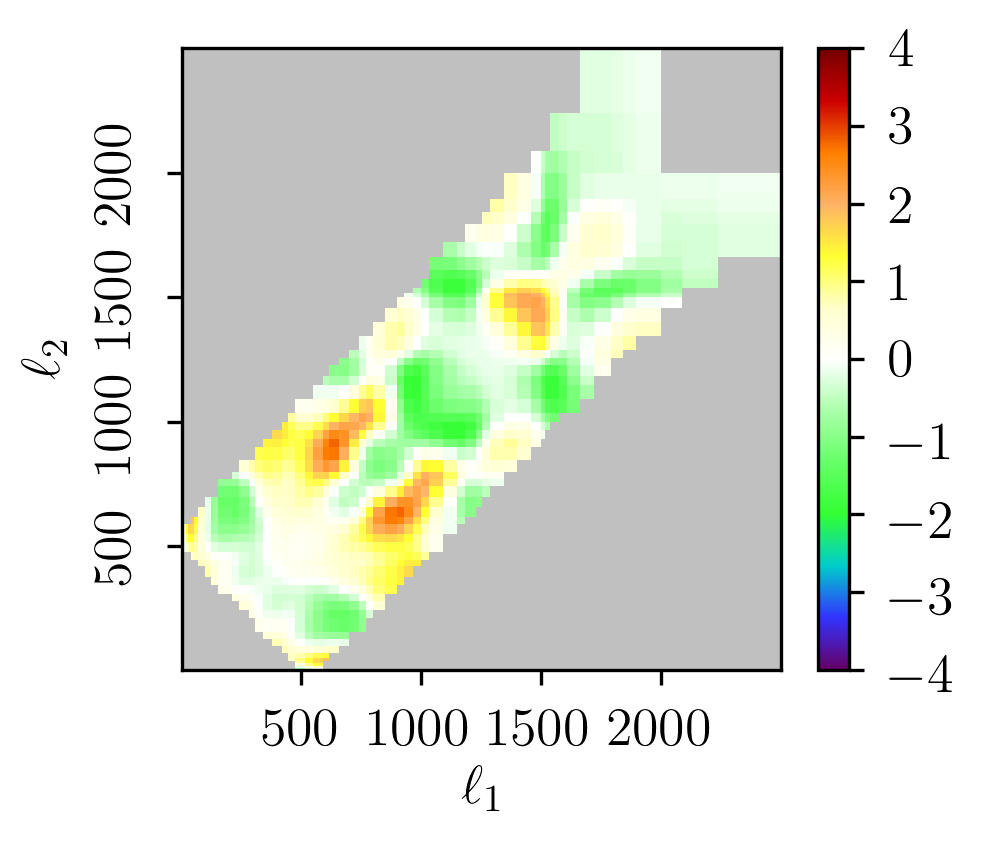}& \includegraphics[trim=0.15cm 0.22cm 0.28cm 0.1cm, clip,width=1.6in]{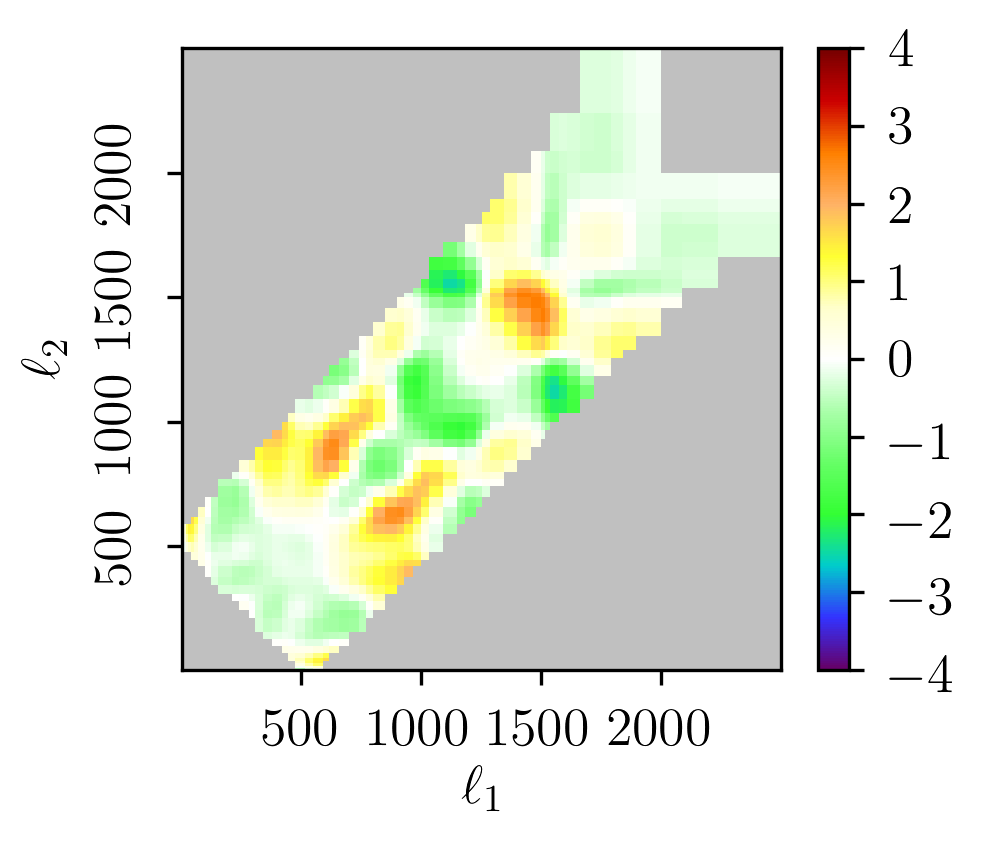}& \includegraphics[trim=0.15cm 0.22cm 0.28cm 0.1cm, clip,width=1.6in]{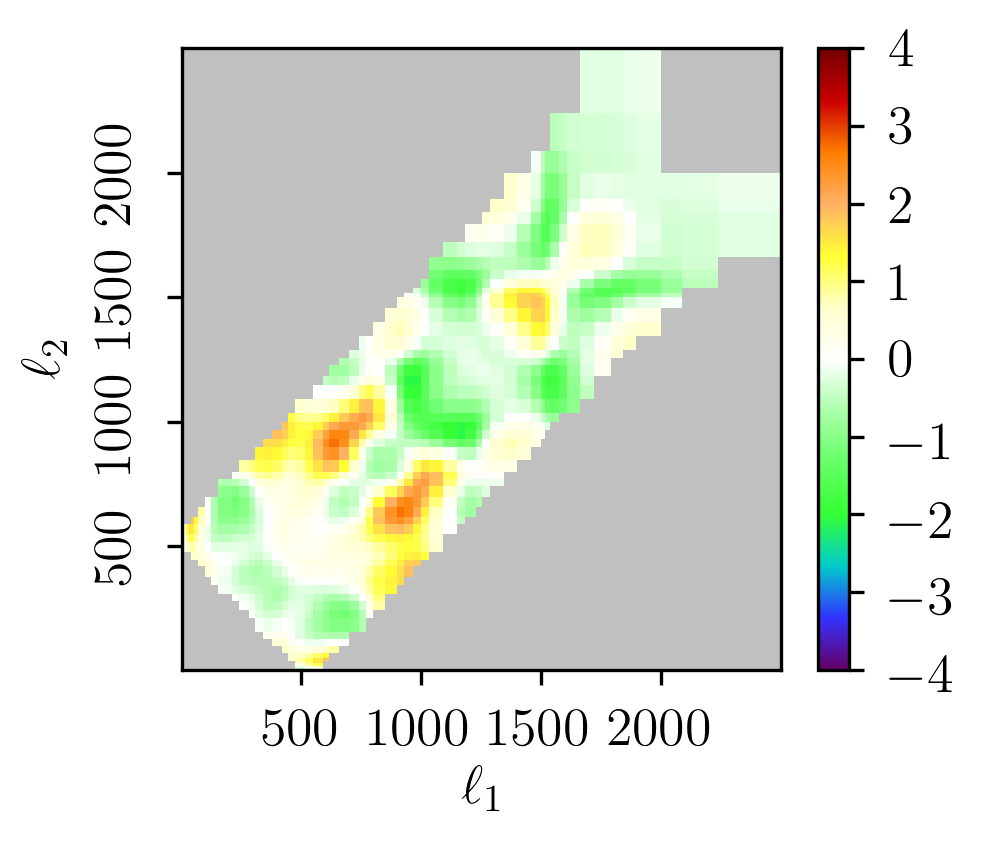}& \includegraphics[trim=0.15cm 0.22cm 0.28cm 0.1cm, clip,width=1.6in]{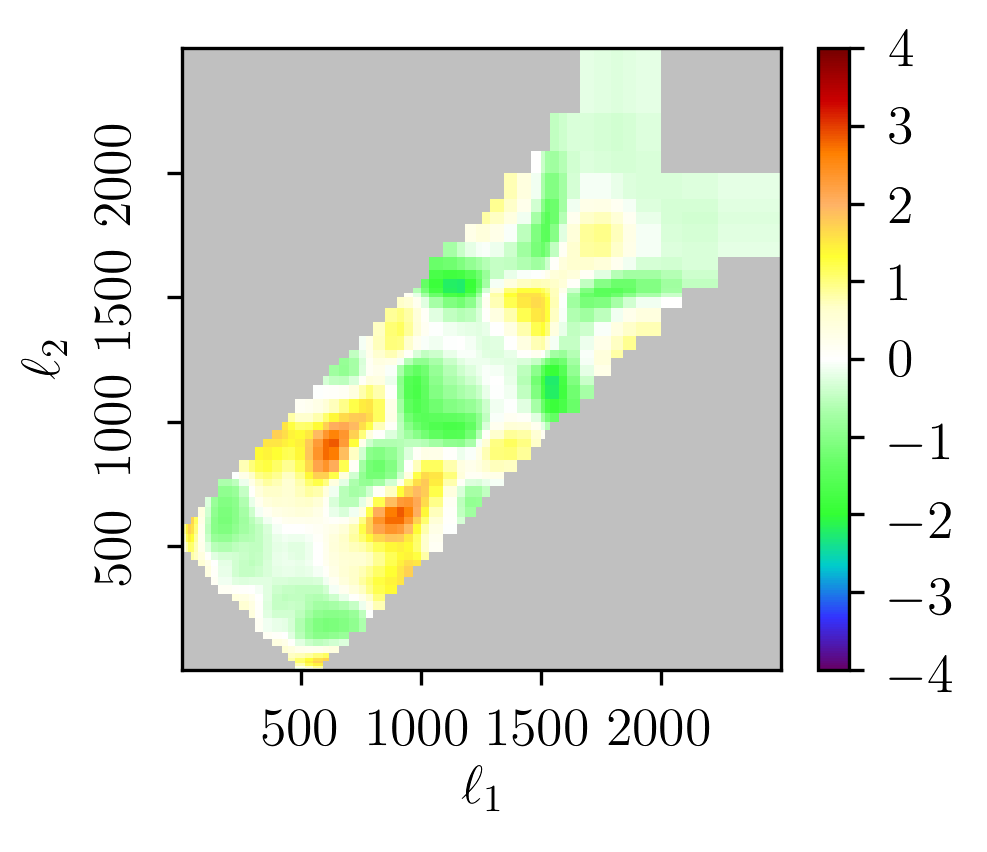} \\
$\mathcal{B}^{EEE}$& \includegraphics[trim=0.15cm 0.22cm 0.28cm 0.1cm, clip,width=1.6in]{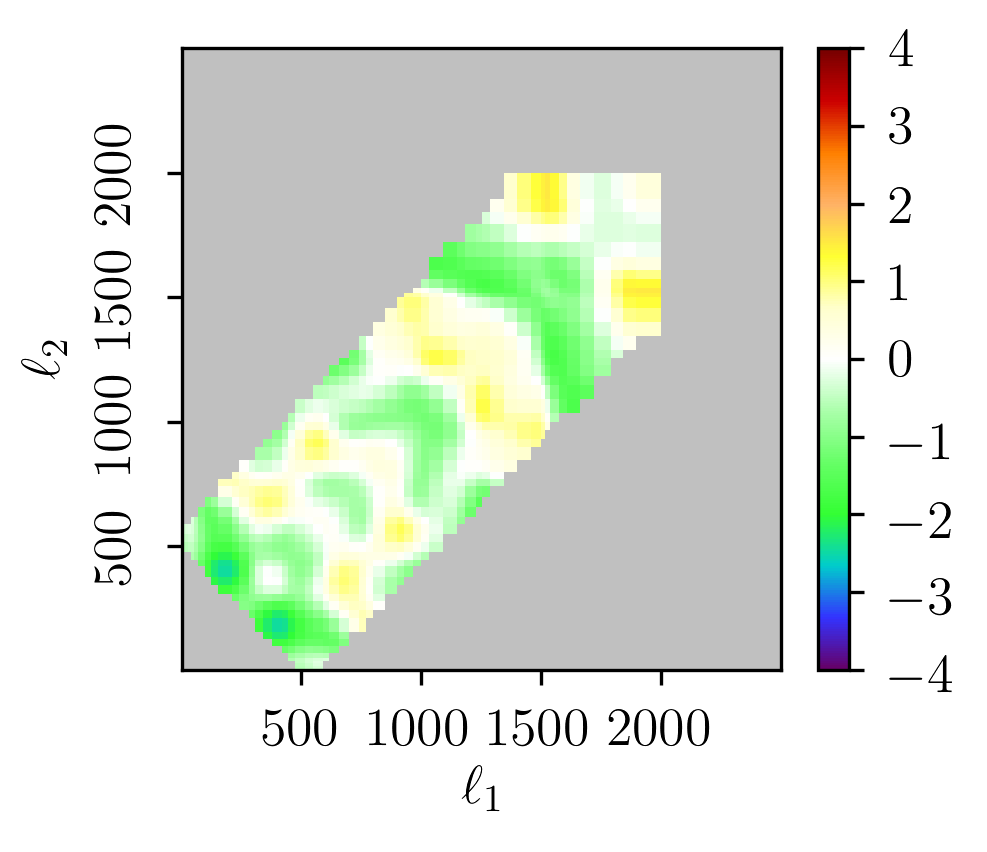}& \includegraphics[trim=0.15cm 0.22cm 0.28cm 0.1cm, clip,width=1.6in]{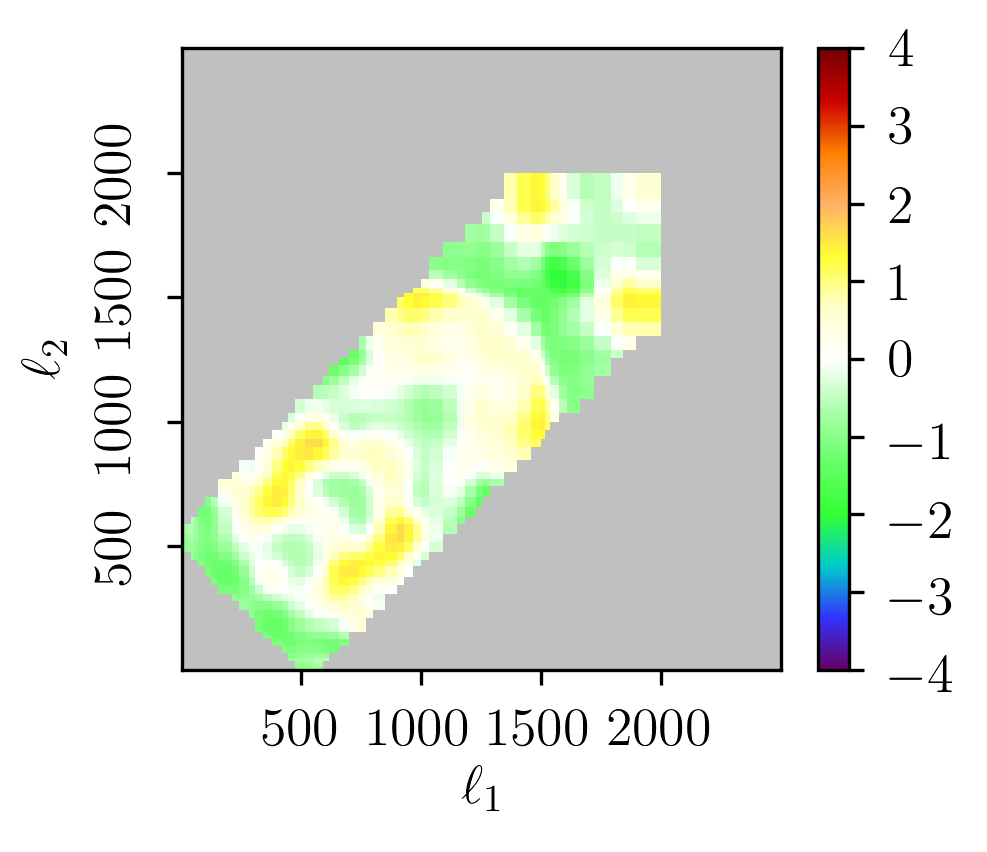}& \includegraphics[trim=0.15cm 0.22cm 0.28cm 0.1cm, clip,width=1.6in]{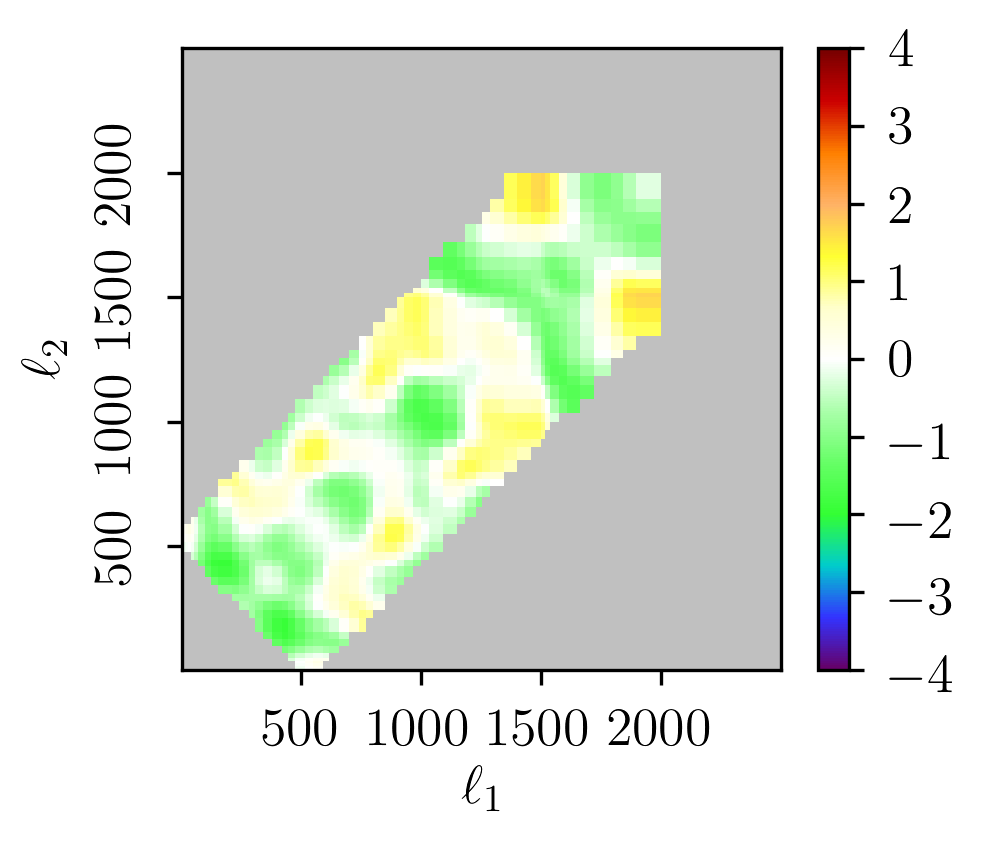}& \includegraphics[trim=0.15cm 0.22cm 0.28cm 0.1cm, clip,width=1.6in]{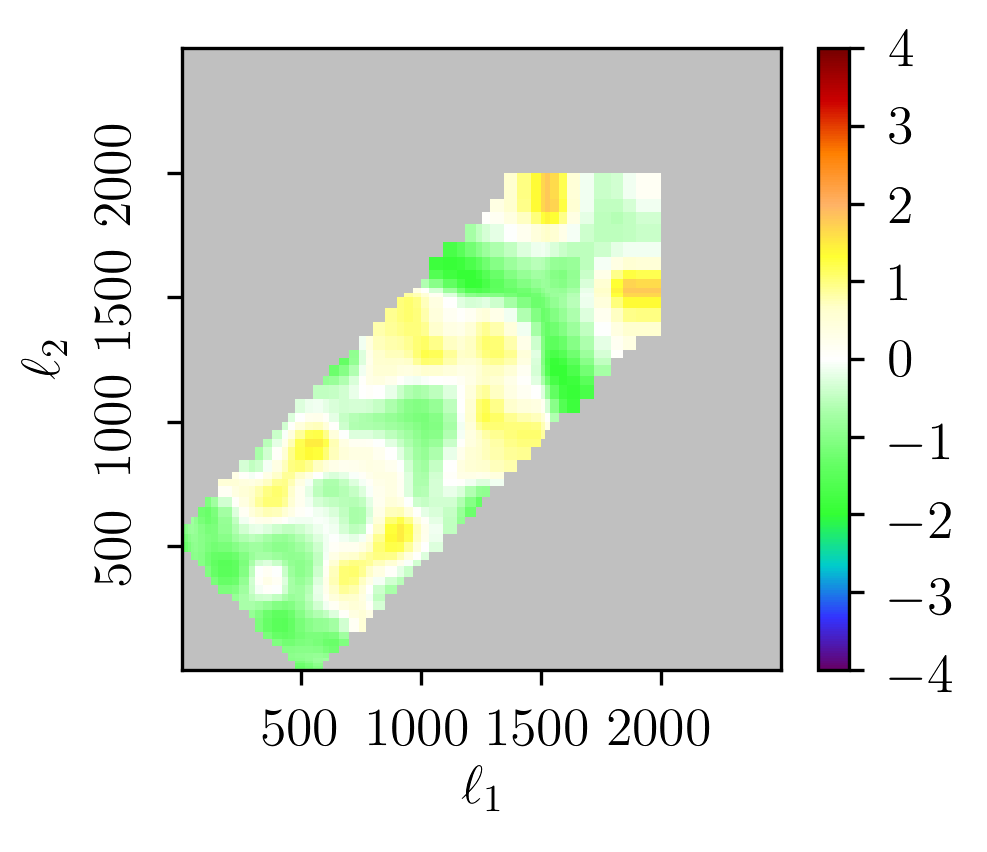} \\
\end{tabular}
\caption{Smoothed binned signal-to-noise bispectra $\mathcal{B}$ for the
\Planck\ 2018 cleaned sky maps. We show slices as a function of $\ell_1$ and $\ell_2$ for a fixed $\ell_3$-bin [518, 548]. 
From left to right results are shown for the
four component-separation methods \SMICA, \SEVEM, \NILC, and \Commander.
  From top to bottom we show: \textit{TTT; TTT} cleaned of clustered and unclustered point sources; \textit{T2E; TE2}; and \textit{EEE}. The colour
 range shows signal-to-noise from $-4$ to $+4$. 
  The light grey regions are where the bispectrum is not defined, either
because it lies outside the triangle inequality or because of the cut
$\ell_\mathrm{max}^E = 2000$.}
\label{fig:smoothed_1}
\end{figure*}

\begin{figure*}[htbp!]
\centering
\begin{tabular}{>{\centering\arraybackslash}m{0.2in}
 >{\centering\arraybackslash}m{1.5in}
 >{\centering\arraybackslash}m{1.5in}
 >{\centering\arraybackslash}m{1.5in}
 >{\centering\arraybackslash}m{1.5in}
}
 & \SMICA& \SEVEM& \NILC& \Commander \\

$\mathcal{B}^{TTT}$& \includegraphics[trim=0.15cm 0.22cm 0.28cm 0.1cm, clip,width=1.6in]{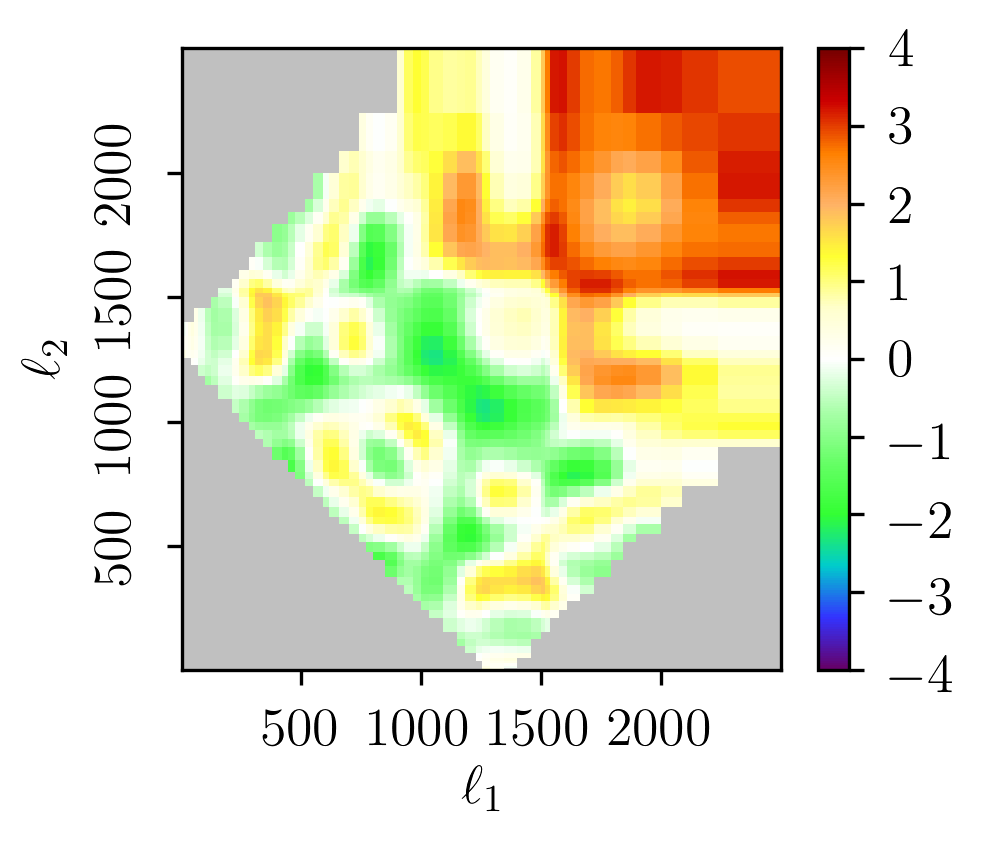}& \includegraphics[trim=0.15cm 0.22cm 0.28cm 0.1cm, clip,width=1.6in]{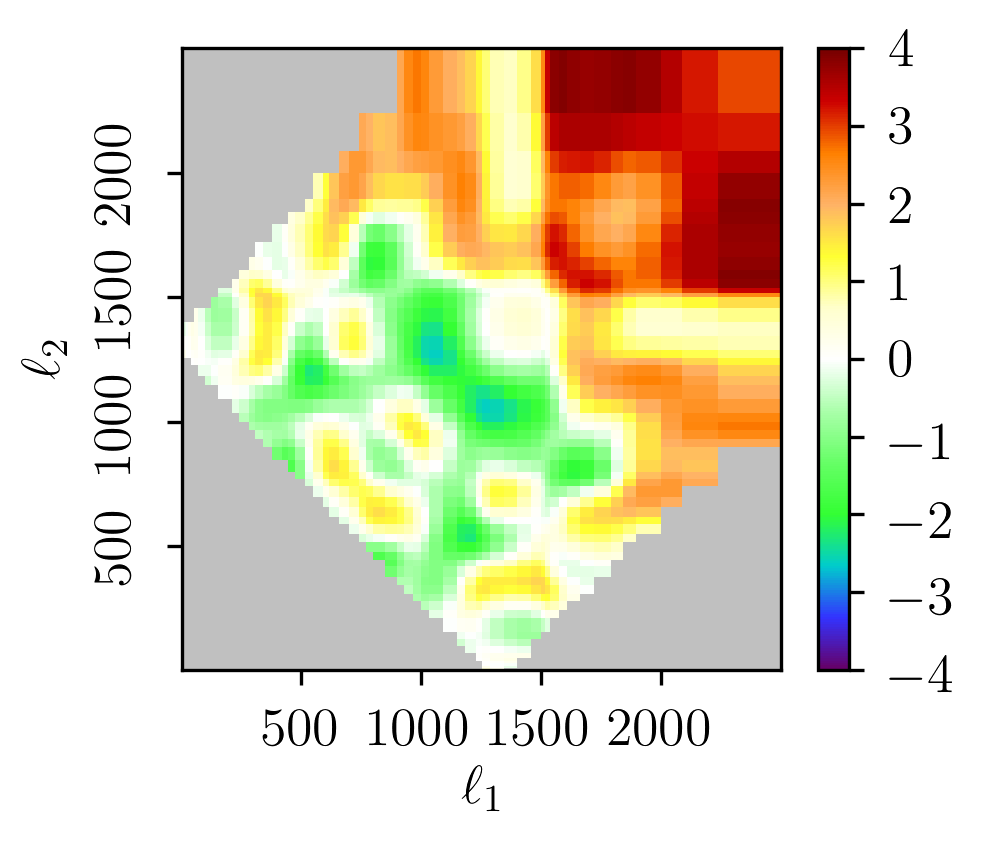}& \includegraphics[trim=0.15cm 0.22cm 0.28cm 0.1cm, clip,width=1.6in]{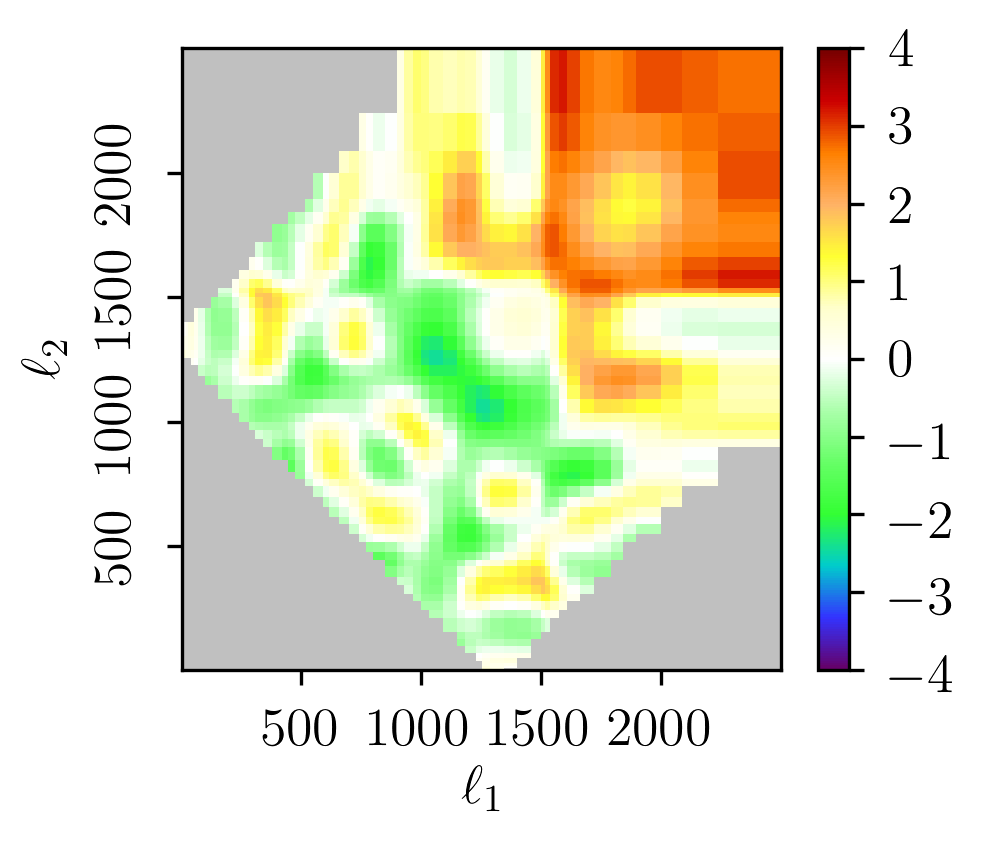}& \includegraphics[trim=0.15cm 0.22cm 0.28cm 0.1cm, clip,width=1.6in]{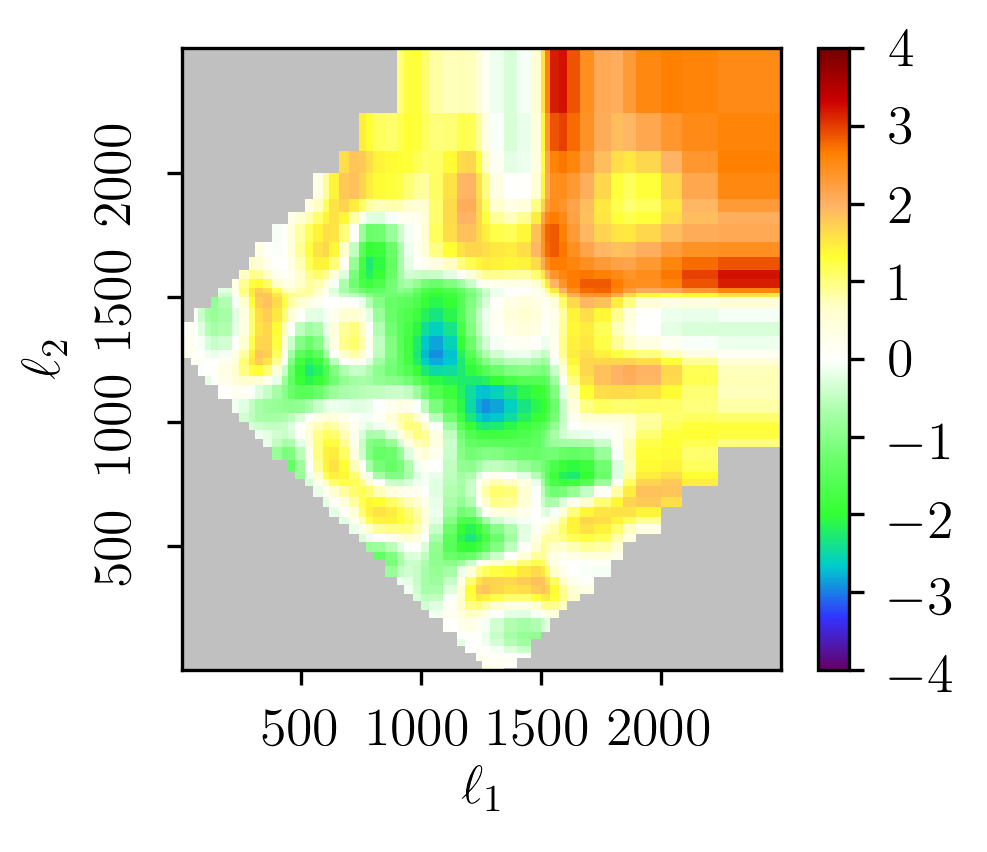} \\
$\mathcal{B}^{TTT}$ no PS& \includegraphics[trim=0.15cm 0.22cm 0.28cm 0.1cm, clip,width=1.6in]{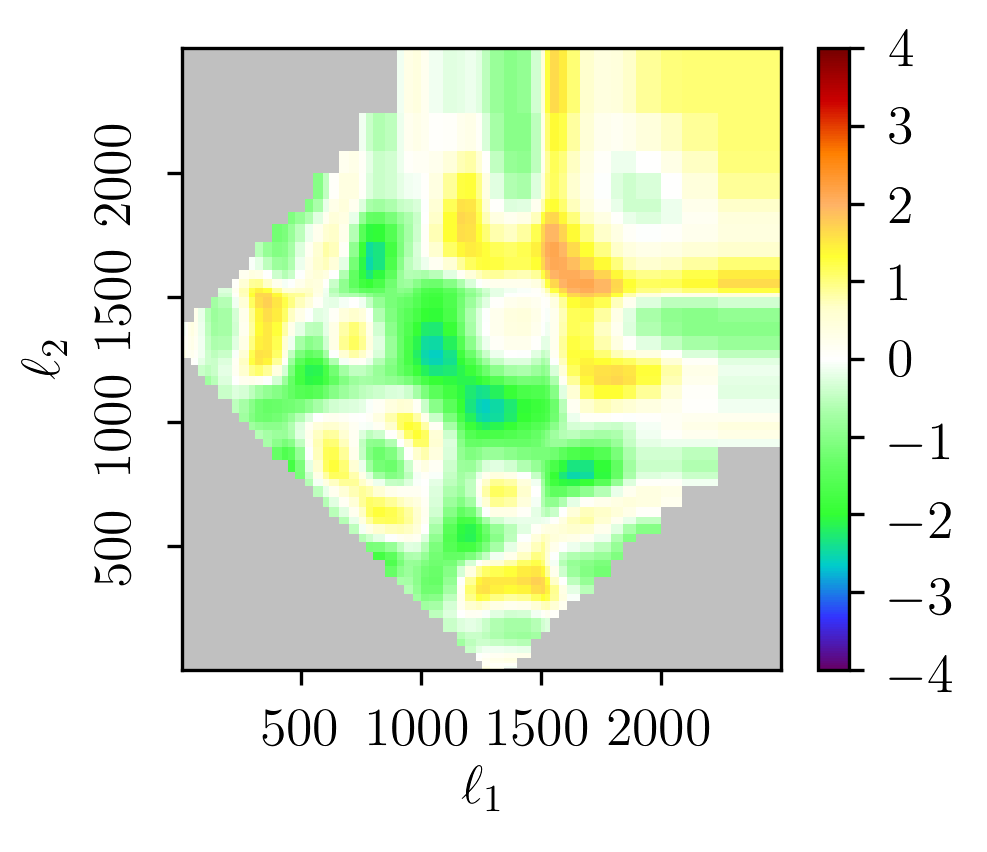}& \includegraphics[trim=0.15cm 0.22cm 0.28cm 0.1cm, clip,width=1.6in]{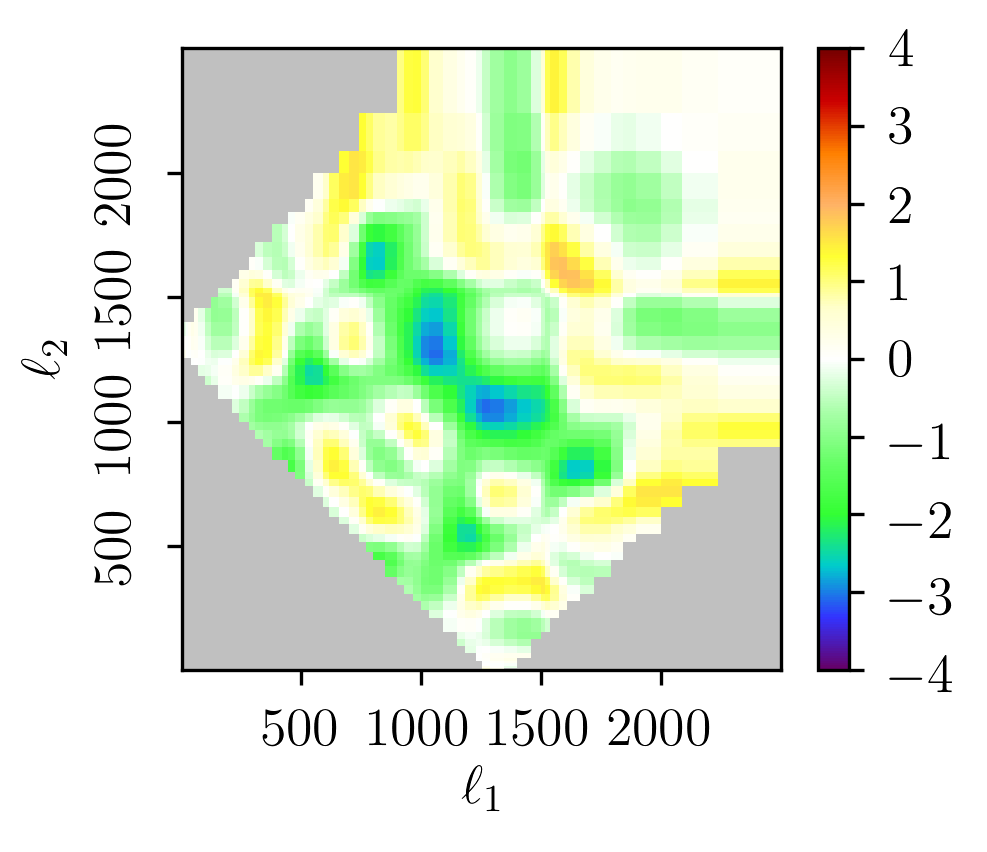}& \includegraphics[trim=0.15cm 0.22cm 0.28cm 0.1cm, clip,width=1.6in]{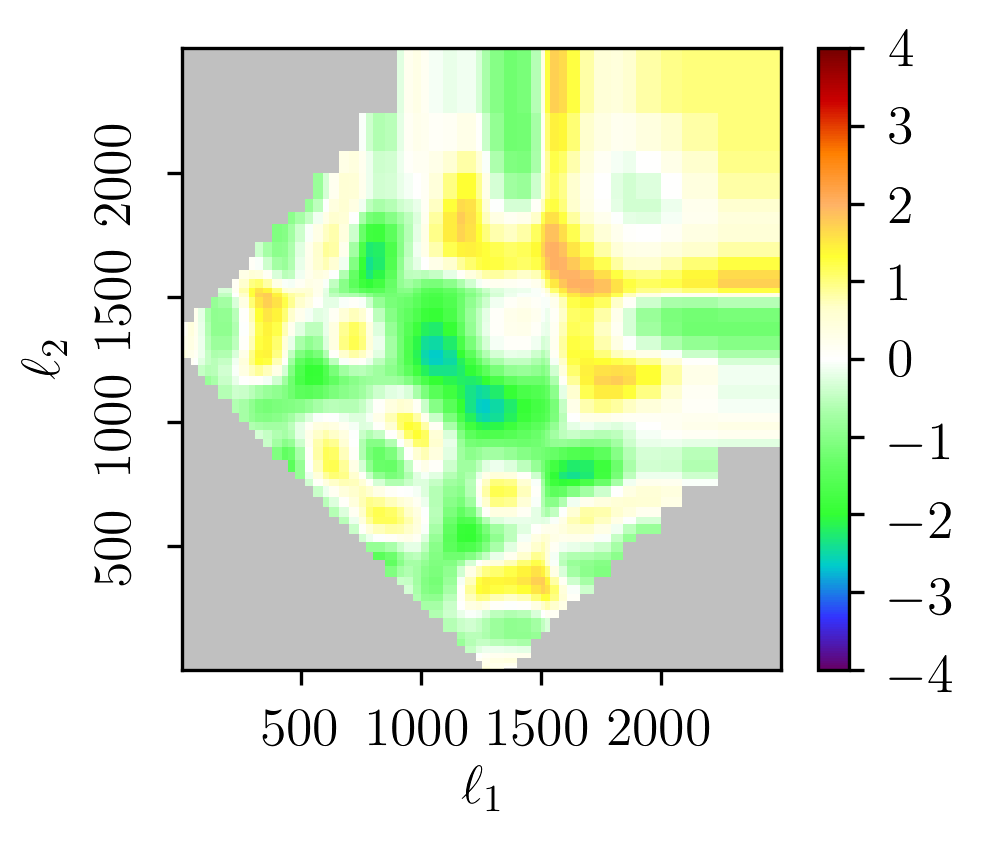}& \includegraphics[trim=0.15cm 0.22cm 0.28cm 0.1cm, clip,width=1.6in]{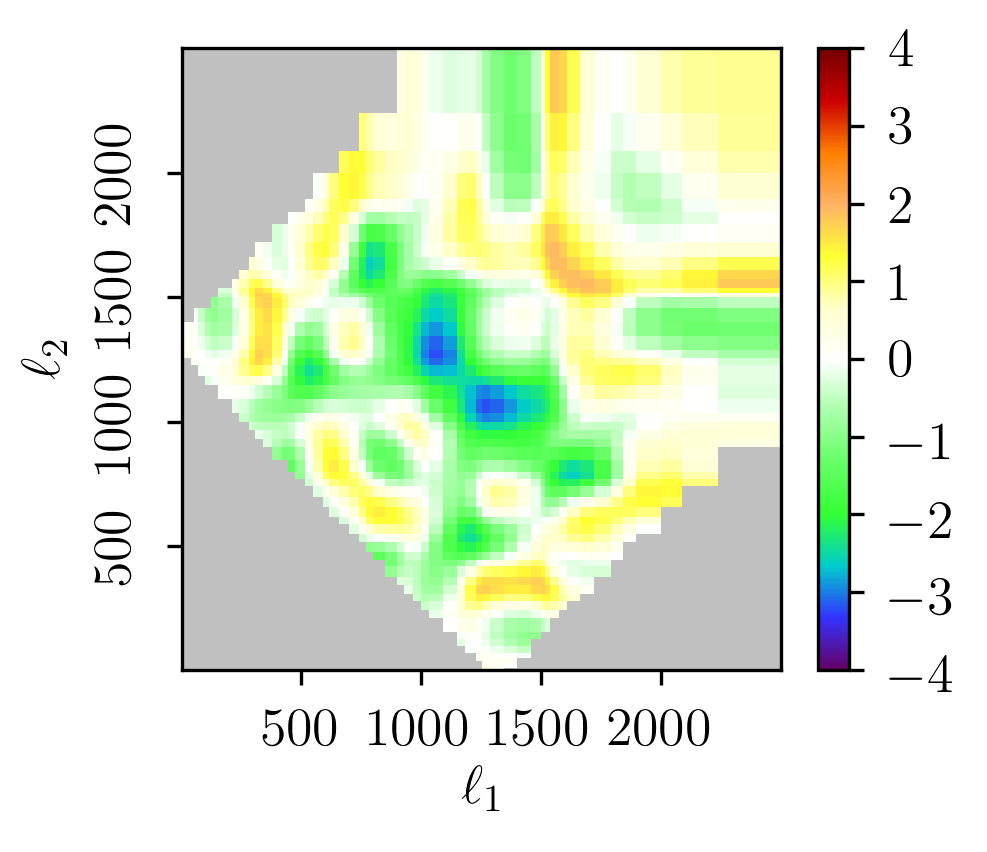} \\
$\mathcal{B}^{T2E}$& \includegraphics[trim=0.15cm 0.22cm 0.28cm 0.1cm, clip,width=1.6in]{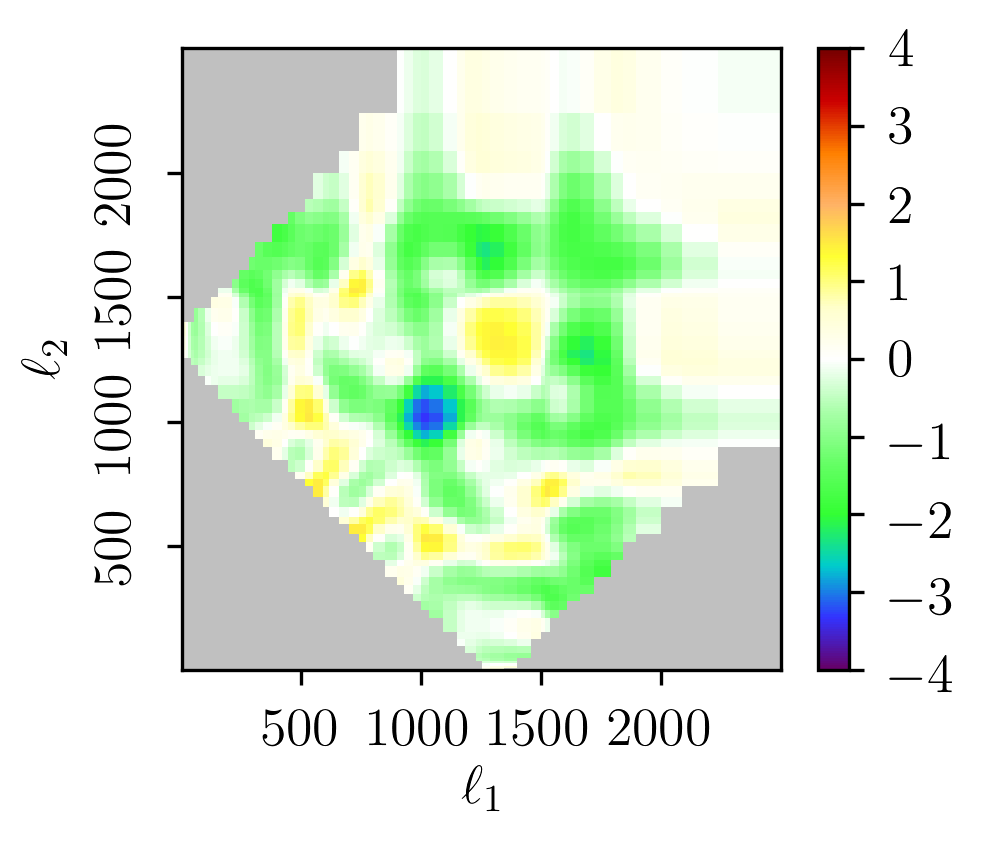}& \includegraphics[trim=0.15cm 0.22cm 0.28cm 0.1cm, clip,width=1.6in]{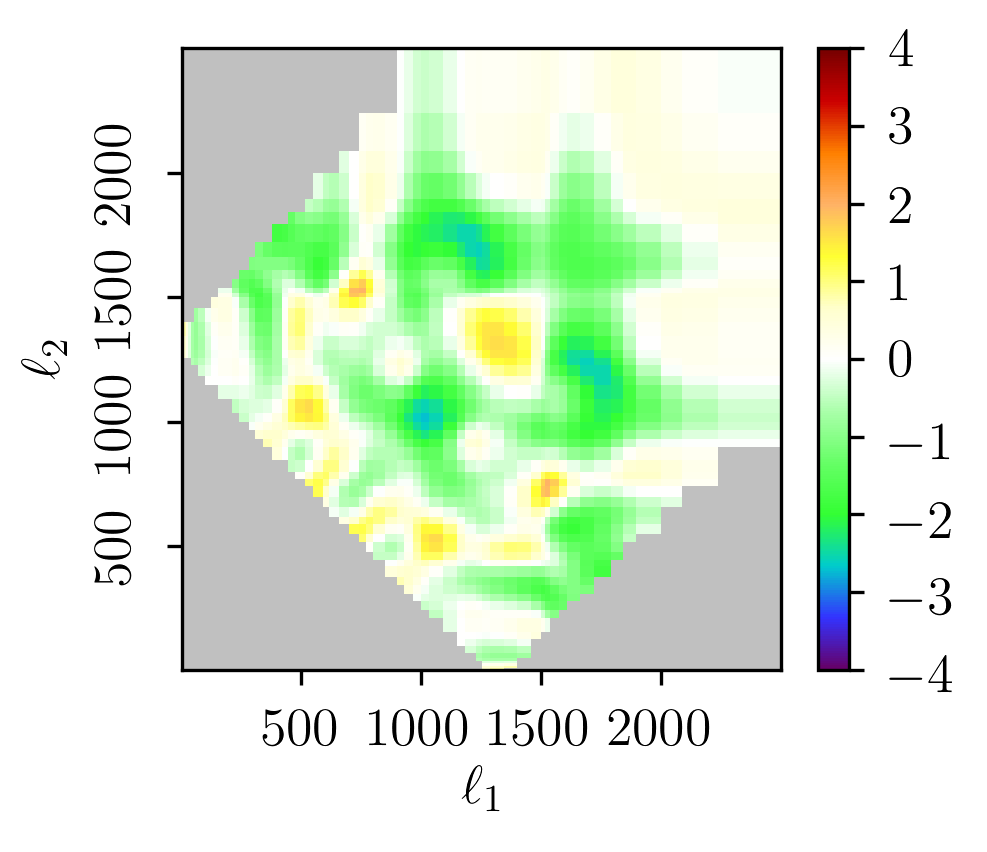}& \includegraphics[trim=0.15cm 0.22cm 0.28cm 0.1cm, clip,width=1.6in]{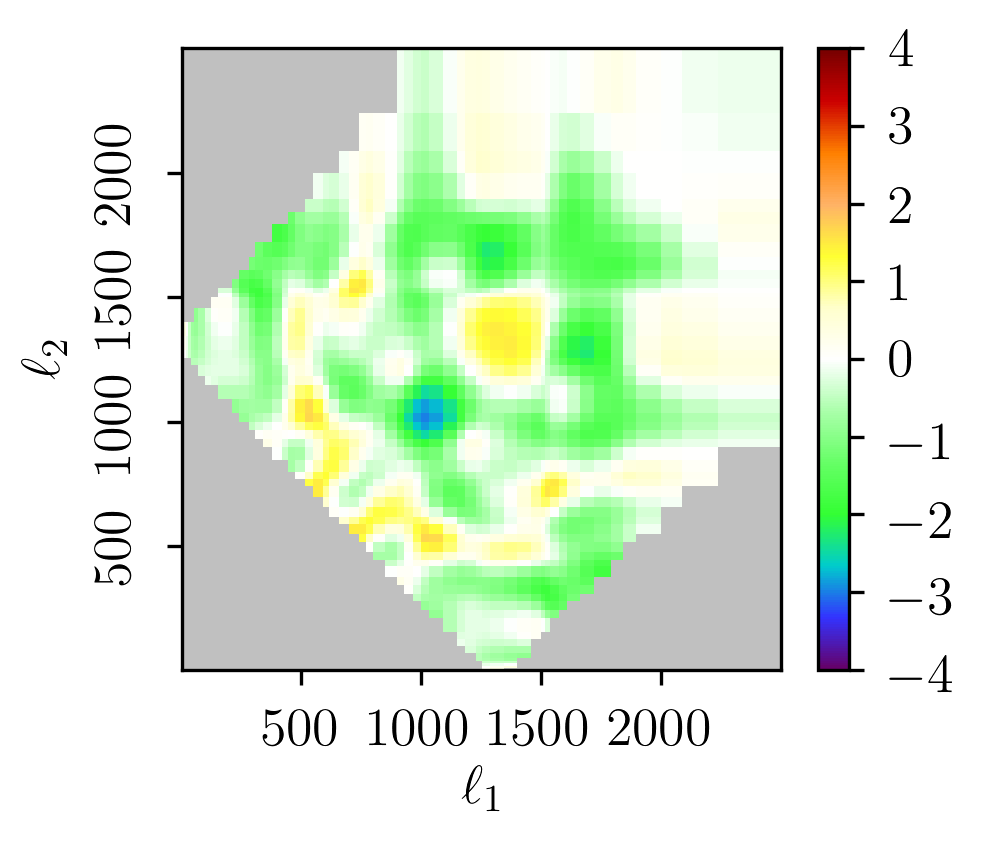}& \includegraphics[trim=0.15cm 0.22cm 0.28cm 0.1cm, clip,width=1.6in]{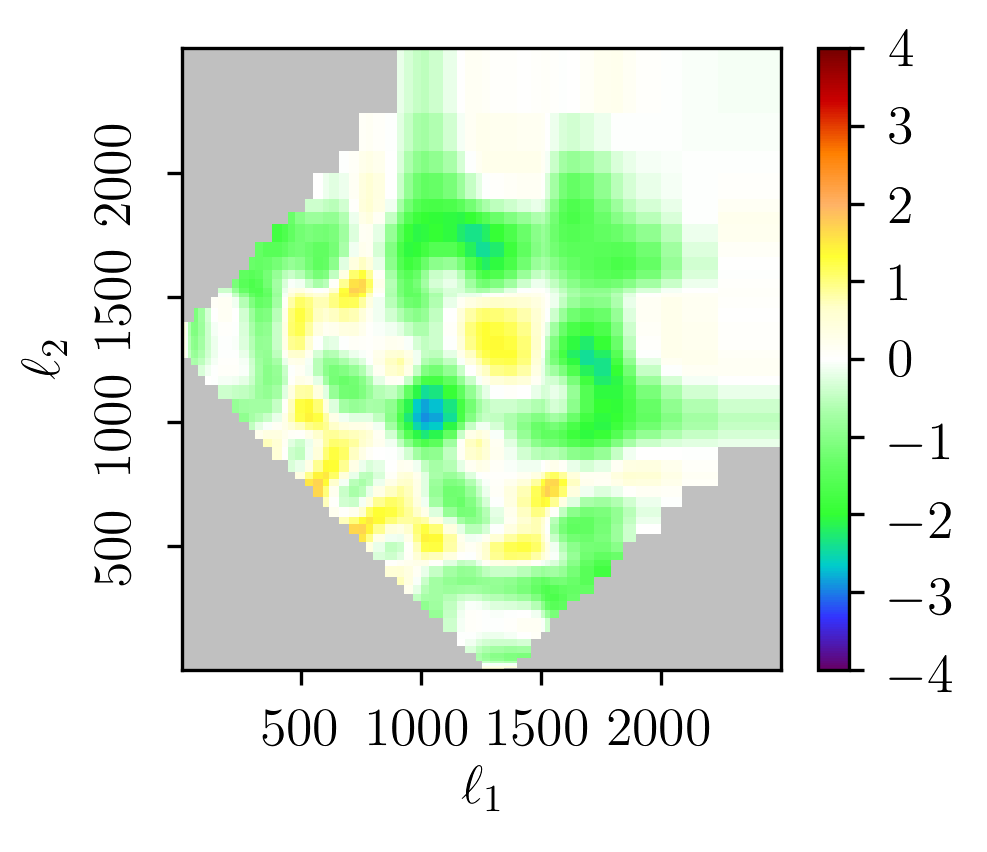} \\
$\mathcal{B}^{TE2}$& \includegraphics[trim=0.15cm 0.22cm 0.28cm 0.1cm, clip,width=1.6in]{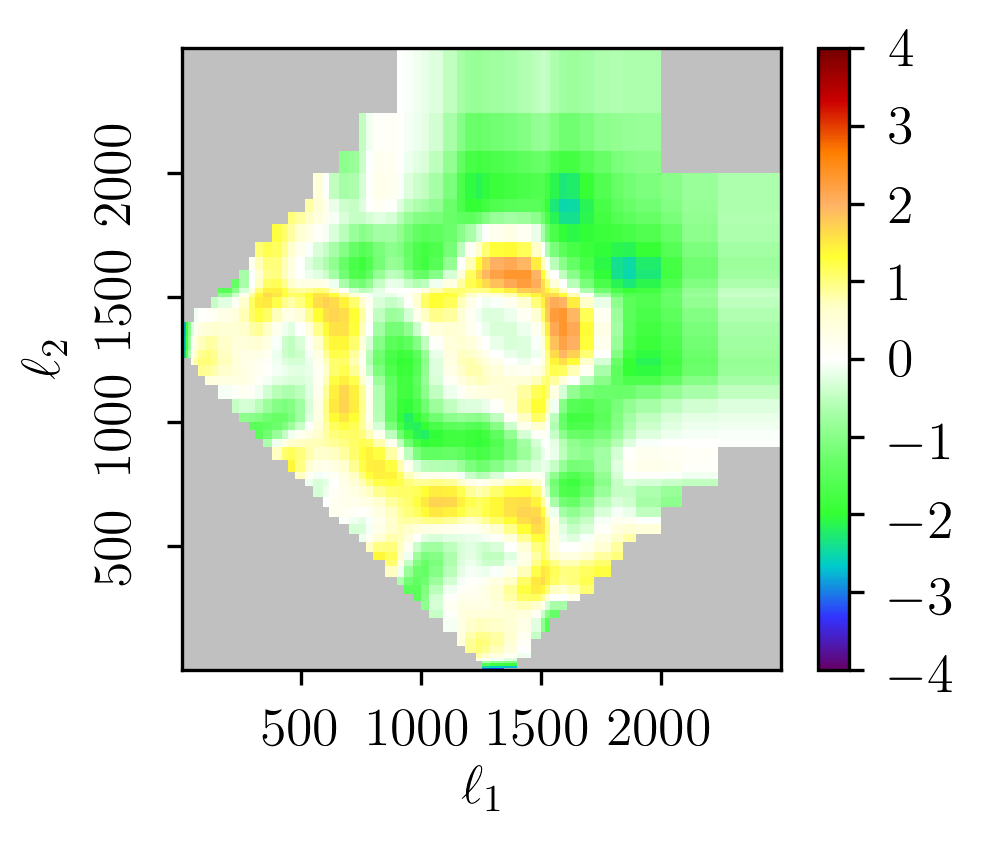}& \includegraphics[trim=0.15cm 0.22cm 0.28cm 0.1cm, clip,width=1.6in]{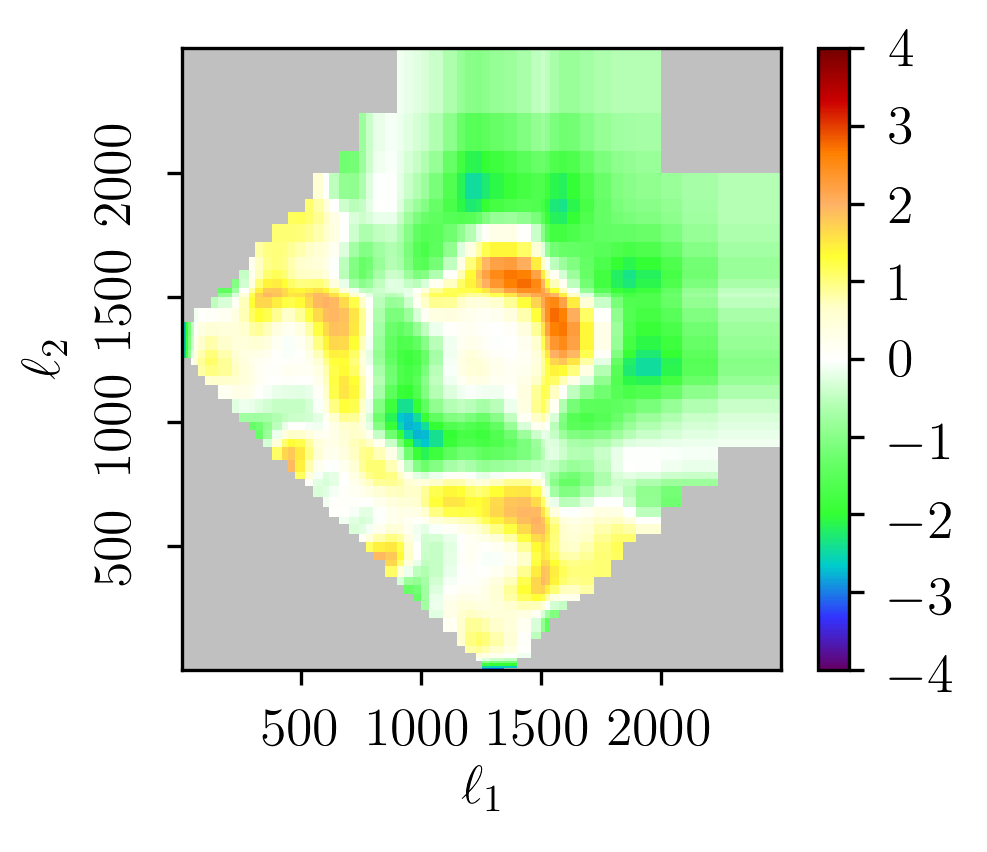}& \includegraphics[trim=0.15cm 0.22cm 0.28cm 0.1cm, clip,width=1.6in]{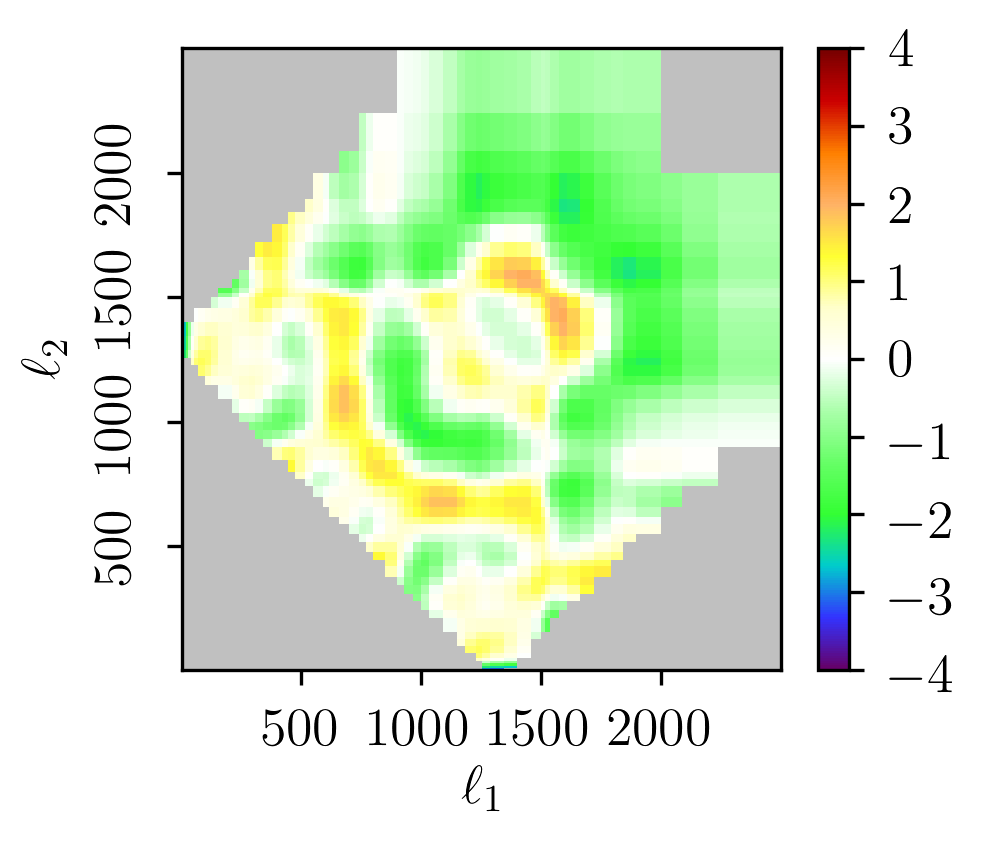}& \includegraphics[trim=0.15cm 0.22cm 0.28cm 0.1cm, clip,width=1.6in]{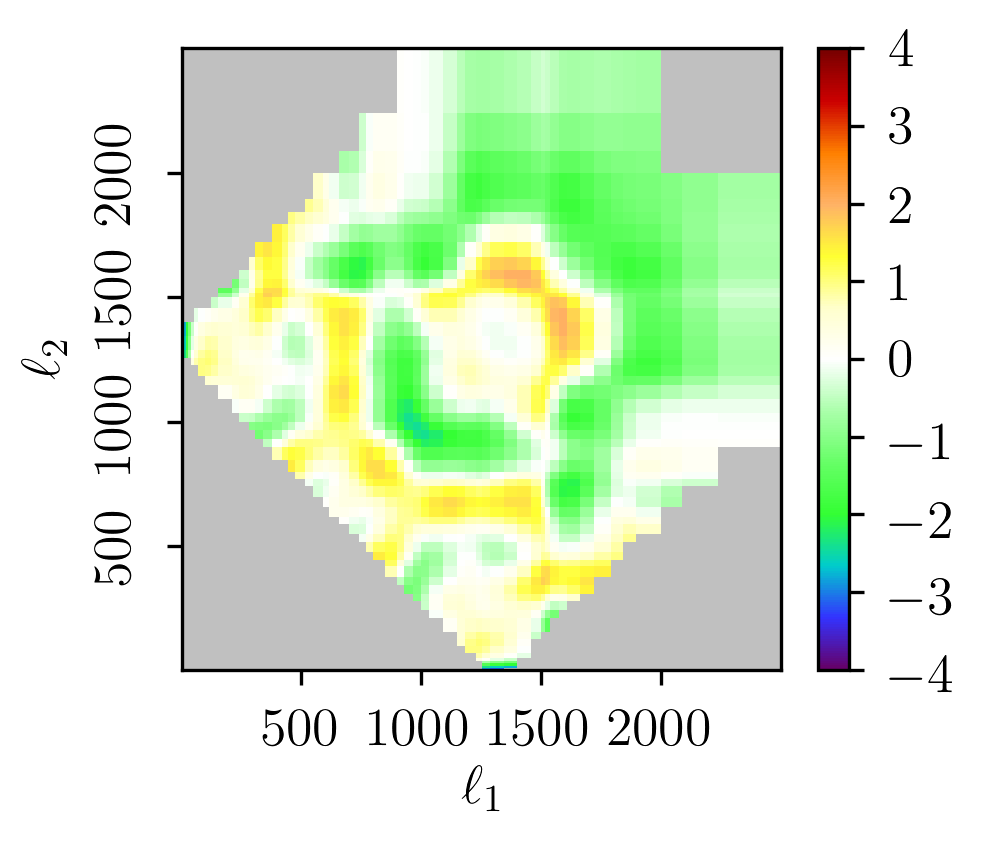} \\
$\mathcal{B}^{EEE}$& \includegraphics[trim=0.15cm 0.22cm 0.28cm 0.1cm, clip,width=1.6in]{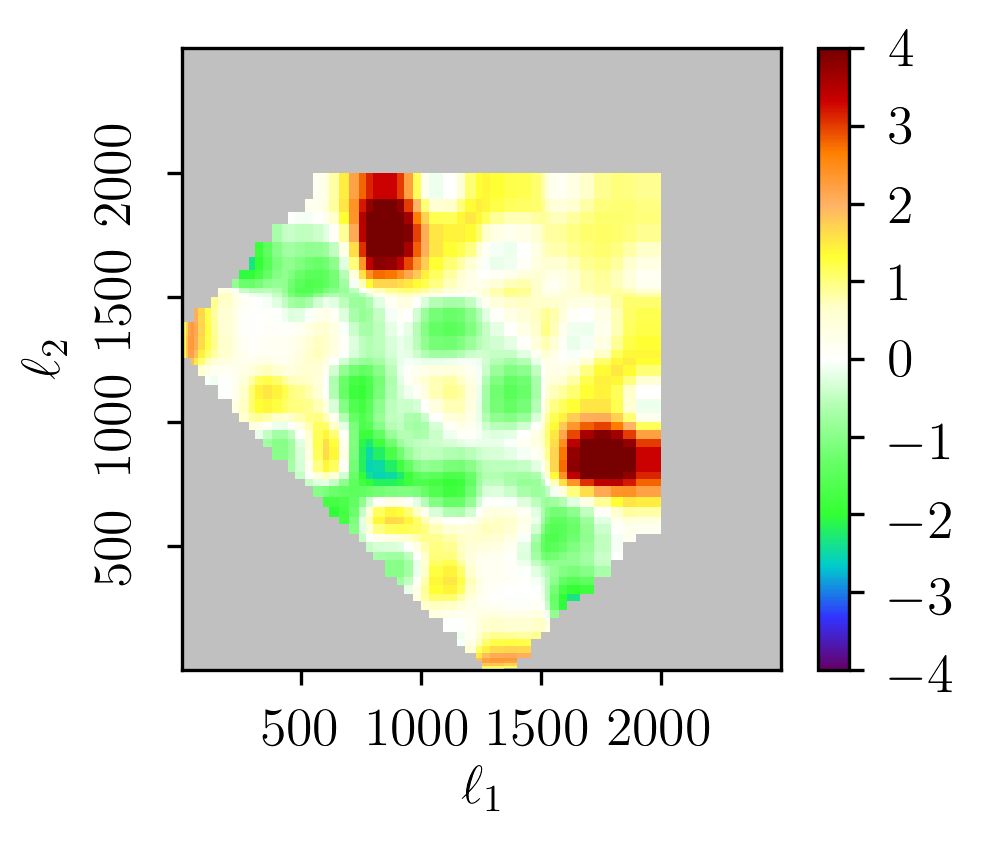}& \includegraphics[trim=0.15cm 0.22cm 0.28cm 0.1cm, clip,width=1.6in]{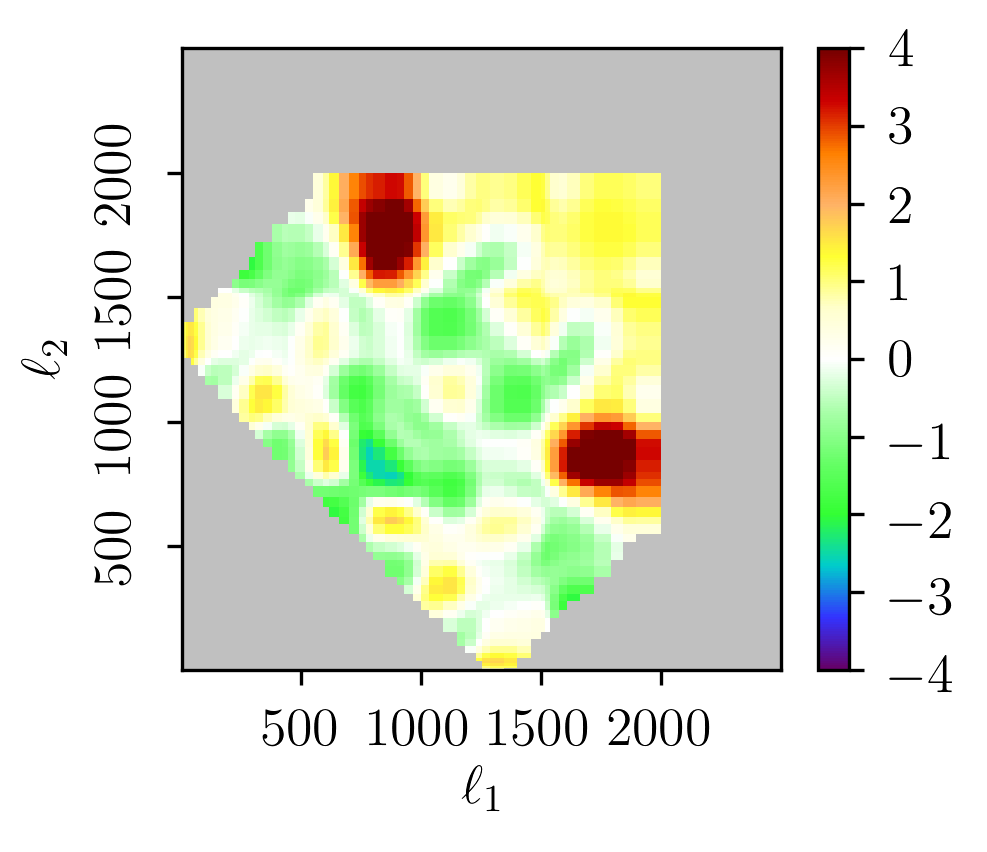}& \includegraphics[trim=0.15cm 0.22cm 0.28cm 0.1cm, clip,width=1.6in]{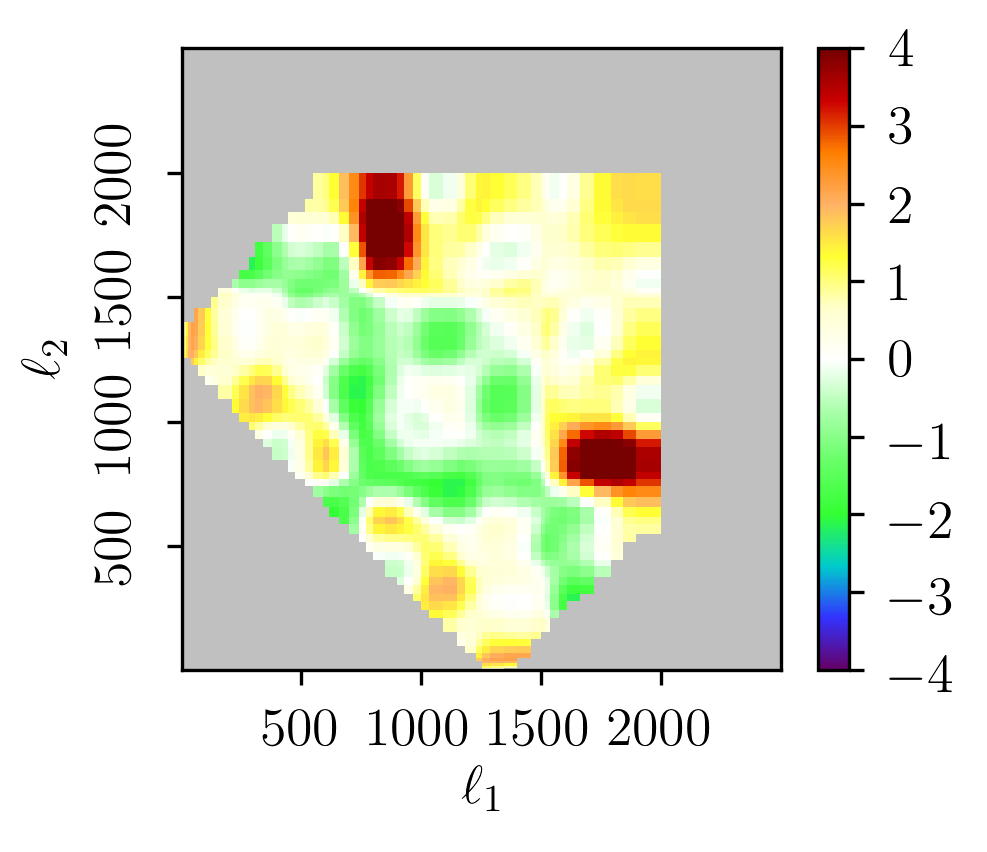}& \includegraphics[trim=0.15cm 0.22cm 0.28cm 0.1cm, clip,width=1.6in]{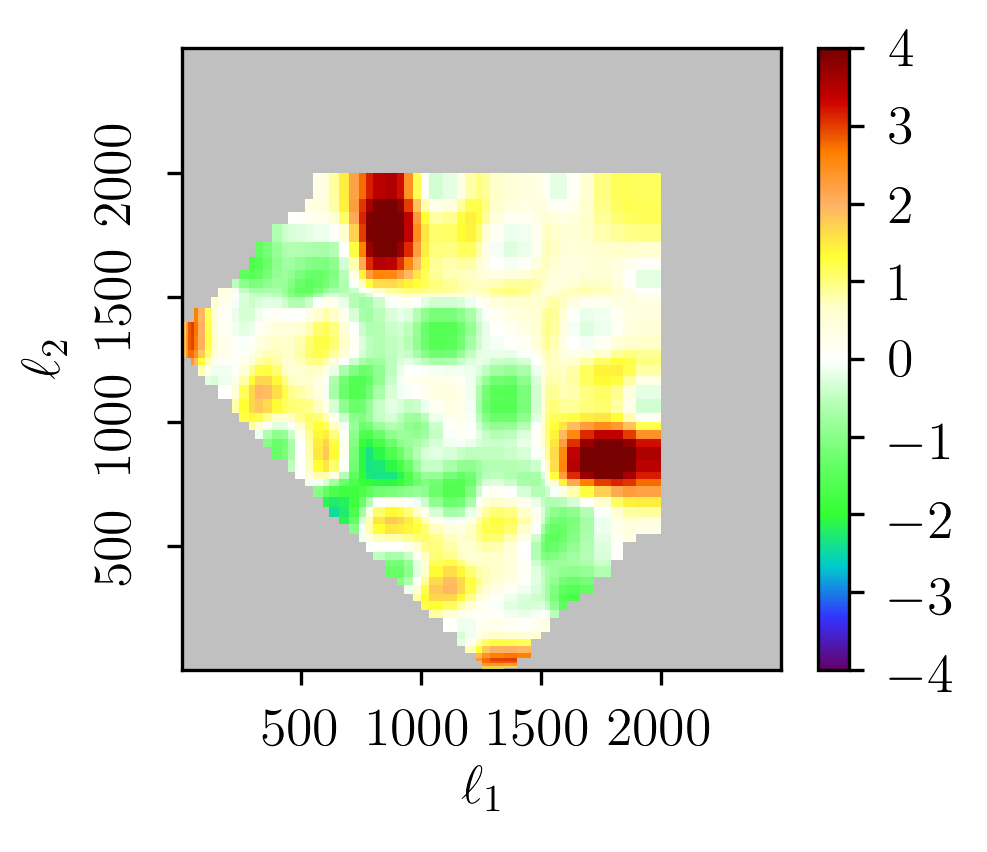} \\
\end{tabular}
\caption{Similar to Fig.~\ref{fig:smoothed_1}
but for the $\ell_3$-bin [1291, 1345].}
\label{fig:smoothed_2}
\end{figure*}

Slices of the smoothed binned signal-to-noise
bispectrum $\mathcal{B}_{i_1 i_2 i_3}$, with a Gaussian smoothing of 
$\sigma_{\rm bin}\,{=}\,2$, are shown in Figs.~\ref{fig:smoothed_1} and
\ref{fig:smoothed_2}. These are slices for the 20th and 40th $\ell_3$-bin as a function of $\ell_1$ and $\ell_2$. For the cross-bispectra mixing $T$ and $E$ modes, we defined $B_{i_1 i_2 i_3}^{T2E} \equiv
B_{i_1 i_2 i_3}^{TTE} + B_{i_1 i_2 i_3}^{TET} + B_{i_1 i_2 i_3}^{ETT}$; and 
$B_{i_1 i_2 i_3}^{TE2} \equiv B_{i_1 i_2 i_3}^{TEE} + B_{i_1 i_2 i_3}^{ETE}
+ B_{i_1 i_2 i_3}^{EET}$, with corresponding variances Var($B^{T2E}$) = 
Var($TTE$) + Var($TET$) + Var($ETT$) + 2 Cov($TTE,TET$) + 2 Cov($TTE,ETT$)
+ 2 Cov($TET,ETT$), and similarly for Var($B^{TE2}$), where we have omitted the bin indices for clarity. 
The red and blue regions correspond to a significant NG, whereas grey areas in Figs.~\ref{fig:smoothed_1} and
\ref{fig:smoothed_2} show regions where the bispectrum is not defined. 
Results are shown for the four component-separation methods \SMICA,
\SEVEM, \NILC, and \Commander, and for \textit{TTT, T2E, TE2}, and
\textit{EEE}. We also show the \textit{TTT} bispectra after we remove
the best joint-fit unclustered and clustered point-source
contribution (see Table~\ref{Table:bps_and_ACIB}). In the top row
of Fig.~\ref{fig:smoothed_2}, we can clearly see this significant point-source
signal at high $\ell$. The \textit{T2E} and \textit{TE2} bispectra do not have
any obvious signals standing out, but we see some stronger NG in the
\textit{EEE} combination. Removing the best joint-fit contribution from all
shapes (Tables~\ref{Table:bps_and_ACIB} and \ref{tab:fNLall}) does not reduce
this region (a case we do not show). The main difference with the previous
release is the better qualitative agreement between the four
component-separation methods in polarization, as was already the case for
temperature.  \Commander\ and \SEVEM\ now show similar structures
as the \NILC\ and \SMICA\ bispectra, which have remained quite stable.

\begin{table*}[htbp!]
\begingroup
\newdimen\tblskip \tblskip=5pt
\caption{Two-tailed $p$-values of the maxima and the minima of the smoothed bispectra. We report three smoothing scales ($\sigma_{\rm bin}=1,\sigma_{\rm bin}=2,\sigma_{\rm bin}=3$), for five cases: $TTT$; $TTT$ cleaned of clustered and unclustered point sources; $T2E$; $TE2$; and $EEE$.}
\label{tab:smoothedbisp}
\nointerlineskip
\vskip -3mm
\footnotesize
\setbox\tablebox=\vbox{
  \newdimen\digitwidth 
  \setbox0=\hbox{\rm 0} 
  \digitwidth=\wd0 
  \catcode`*=\active 
  \def*{\kern\digitwidth}
  \newdimen\signwidth 
  \setbox0=\hbox{+} 
  \signwidth=\wd0 
  \catcode`!=\active 
  \def!{\kern\signwidth}
\halign{\hbox to 0.75in{#\leaderfil}\tabskip=0.5em& \hfil#\hfil\tabskip=1em&
\hfil#\hfil& \hfil#\hfil&
\hfil#\hfil& \hfil#\hfil\tabskip=0pt\cr 
\noalign{\doubleline}
\omit& $ \mathcal{B}^{\rm TTT}$& $ \mathcal{B}^{\rm TTT} $ (no PS)& $ \mathcal{B}^{\rm T2E}$& $ \mathcal{B}^{\rm TE2}$& $ \mathcal{B}^{\rm EEE}$\cr 
\noalign{\vskip 3pt\hrule\vskip 5pt}
\multispan6\hfil Maximum $p$-values\hfil\cr
\noalign{\vskip 3pt}
    \SMICA& (0.08, 0.04, 1.8 $\times 10^{-3}$)& (0.84, 0.80, 0.95)& (0.21, 0.11, 0.72)& (0.82, 0.36, 0.20)& (72, 0.34, 0.22) $\times 10^{-2}$\cr
    \SEVEM& (1.2 $\times 10^{-4}$ , $<10^{-6}$, $<10^{-6}$)&  (0.27, 0.03, 0.03)& (0.32, 0.18, 0.82)& (0.37, 0.50, 0.89)&  (44, 0.32, 0.14) $\times 10^{-2}$\cr
     \NILC& (0.18, 0.03, 3.9 $\times 10^{-3}$)& (0.59, 0.38, 0.68)& (0.74, 0.33, 0.82)& (0.94, 0.74, 0.49)& (31, 0.21, 0.25) $\times 10^{-2}$\cr
\Commander& (70, 1.4, 0.3) $\times 10^{-3}$& (12, 0.66, 0.61) $\times 10^{-2}$&  (0.12, 0.19, 0.71)& (0.70, 0.40, 0.19)& (88, 2.0, 0.92) $\times 10^{-2}$\cr
\noalign{\vskip 3pt\hrule\vskip 5pt}
\multispan6\hfil Minimum $p$-values\hfil\cr
\noalign{\vskip 3pt}
    \SMICA& (0.02, 0.14, 0.20)& (0.25, 0.23, 0.62)& (0.47, 0.21, 0.54)& (0.59, 0.72, 0.86)& (0.27, 0.94, 0.68)\cr
    \SEVEM& (0.32, 0.43, 0.98)& (0.96, 0.84, 0.45)& (0.08, 0.75, 0.81)& (0.18, 0.70, 0.64)& (0.62, 0.68, 0.53)\cr
     \NILC& (8.1 $\times 10^{-3}$, 0.05, 0.27)& (0.30, 0.21, 0.67)& (0.32, 0.54, 0.93)& (0.73, 0.73, 0.95)& (0.89, 0.74, 0.72)\cr
\Commander&  (0.21, 0.42, 0.76)& (0.74, 0.99, 0.37)& (0.25, 0.71, 0.63)& (0.73, 0.69, 0.77)& (4.6 $\times 10^{-3}$, 0.25, 0.19)\cr
\noalign{\vskip 3pt\hrule\vskip 5pt}}}
\endPlancktablewide                    
\endgroup
\end{table*}

\begin{figure*}[htbp!]
 \centering
        \begin{tabular}{>{\centering\arraybackslash}m{0.2in}
 >{\centering\arraybackslash}m{1.5in}
 >{\centering\arraybackslash}m{1.5in}
 >{\centering\arraybackslash}m{1.5in}
 >{\centering\arraybackslash}m{1.5in}
}

 & \SMICA& \SEVEM& \NILC& \Commander\\

$\mathcal{B}^{TTT}$ no PS& \includegraphics[trim=0.15cm 0.22cm 0.28cm 0.1cm, clip,width=1.6in]{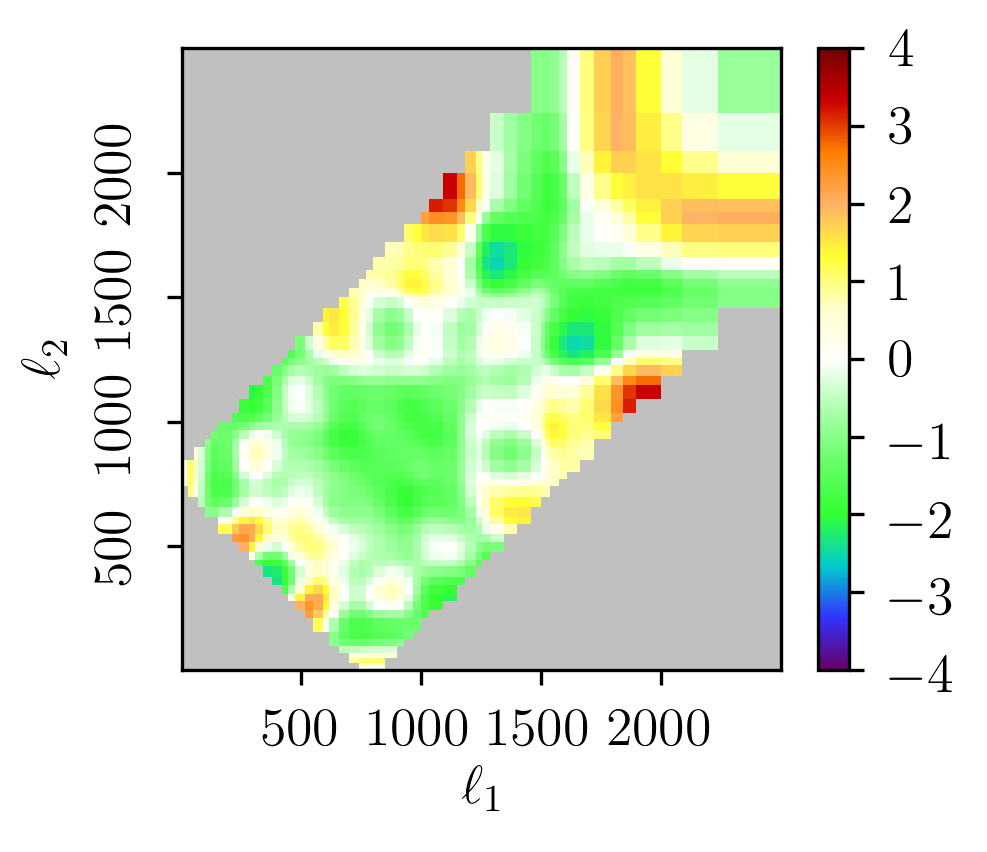}&\includegraphics[trim=0.15cm 0.22cm 0.28cm 0.1cm, clip,width=1.6in]{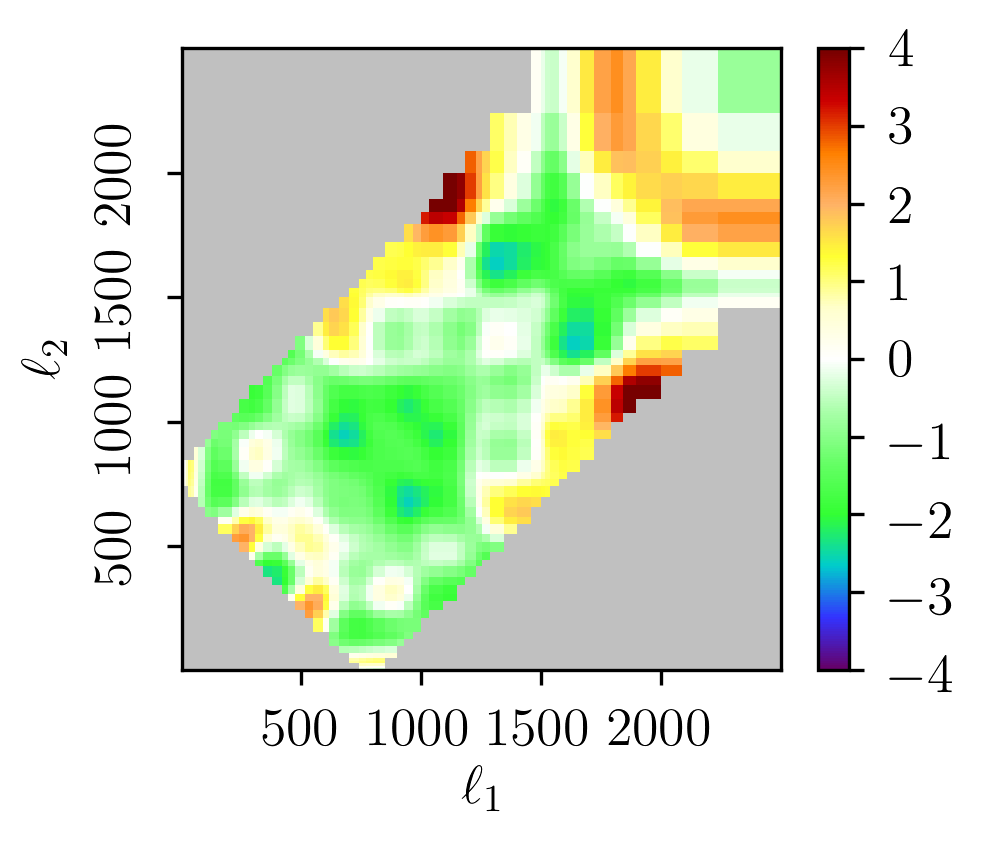}&\includegraphics[trim=0.15cm 0.22cm 0.28cm 0.1cm, clip,width=1.6in]{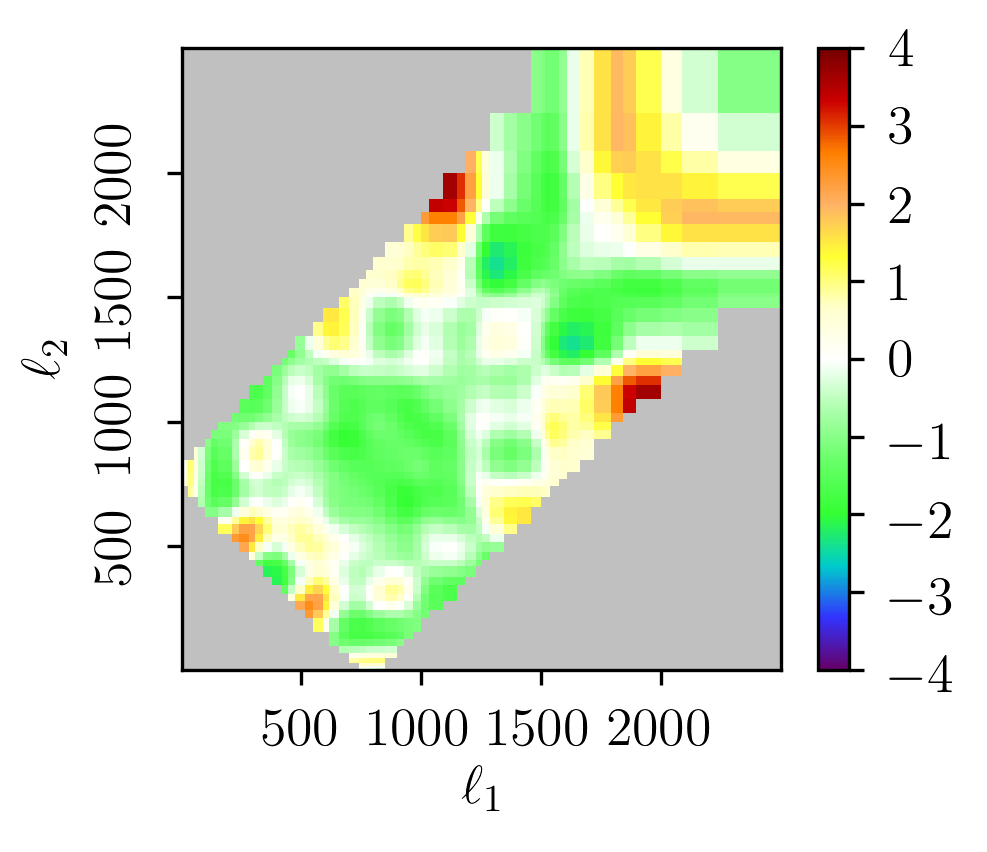}&\includegraphics[trim=0.15cm 0.22cm 0.28cm 0.1cm clip,width=1.6in]{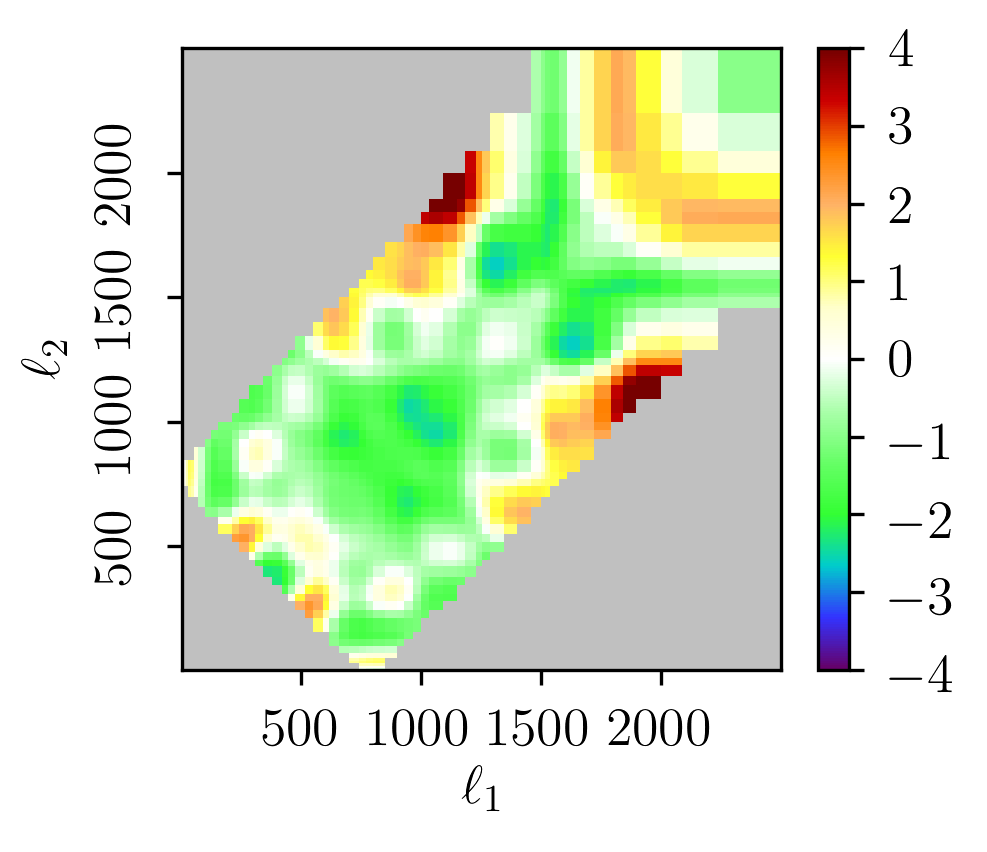}\\
       \end{tabular}
\caption{Similar to Fig.~\ref{fig:smoothed_1} but for the $\ell_3$-bin [771,799], and only for \textit{TTT} cleaned of clustered and unclustered point sources.}
\label{fig:smoothed_3}
\end{figure*}

In order to quantify the possible residual non-Gaussianity in these smoothed bispectra, we focus on the minimum and maximum of our bin-triplets. In the case of statistically independent Gaussian numbers, we can calculate the probability distribution of the extreme value statistics. However, once we introduce correlations due to the smoothing with non-trivial boundary conditions, we do not have an analytical formula. We can instead generate Monte Carlo simulations of Gaussian random numbers with the same boundary conditions, apply the smoothing as for the data signal-to-noise bispectrum, and compute the $p$-values of the observed extremum statistics as the fraction of simulations having a more extreme extremum than our data. This requires many simulations to study the very unlikely events. In the current analysis, $10^6$ Monte Carlo simulations turn out to be sufficient and so we do not need to use the semi-analytical Ansatz introduced in \cite{Bucher:2015ura}.

In Table~\ref{tab:smoothedbisp}, we report for the smoothing lengths $\sigma_{\rm bin} = 1$, 2, and 3, the two-tailed $p$-value\footnote{While a one-tailed $p$-value quantifies how likely it is for the maxima (a similar description holds for the minima) of the Monte Carlo simulations to be higher than the maximum of the data, the two-tailed $p$-value also considers how likely it is for it to be lower than that of the data. This means that a maximum that is too low will yield a low $p$-value too.} of the maximum and of the minimum, defined as $p = 2 \,{\rm min}\bigl[{\rm Prob}(X_{\rm MC} \leq X_{\rm data}), {\rm Prob}(X_{\rm MC}\geq X_{\rm data})\bigr]$, where $X$ is either the minimum or the maximum of the distribution. As expected, we detect a highly significant departure from Gaussian statistics in the maxima for \textit{TTT} when we do not correct for the contribution from point sources, and the signal stands out more when increasing the smoothing kernel size. When looking at the \SEVEM\ data for $\sigma_{\rm bin} = 2$ and 3, we find no simulation with a higher maximum, but for all the other cases our analysis should be robust. Most bispectra seem to be compatible with a simple Gaussian distribution, except for the \textit{EEE} bispectrum in the region shown in Fig.~\ref{fig:smoothed_2}, for multipole triplets around [900, 1300, 1800]. We also see some significantly high maxima for the \textit{TTT} bispectrum of \SEVEM\ and \Commander\ (even after correcting for point sources), located around [800, 1100, 2000] (shown in Fig.~\ref{fig:smoothed_3}). The origin of these signals clustered in multipole space is not understood.

\section{Validation of \textit{Planck} results}
\label{sec:Sec_valid_data}

Two important potential sources of systematic effects in $\fnl$ estimation are foreground residuals (which can also be related to the choice of mask) and an imperfect modelling of instrumental noise, either of which can lead to miscalibration of the linear correction term in the estimator. In this section, we perform a battery of tests aimed at testing the impact of these systematics. We will typically choose only one of the the KSW, Binned, or Modal pipelines for each of the tests described below. This is possible because of their excellent and well-verified agreement on both data and simulations.

\subsection{Dependence on foreground-cleaning method}
\label{sec:dep_methods}

\subsubsection{Comparison between $\fnl$ measurements}\label{sec:fnldiff}

For this test, we consider $160$ FFP10-based simulations with realistic beams and noise levels. A starting set of single-frequency Gaussian maps is processed through the different component-separation pipelines. Each pipeline combines the various frequency maps by adopting exactly the same coadding and filtering approach as done for the data. At the end, we have four sets of $160$ frequency coadded realizations (the \SMICA\ set, the \SEVEM\ set, the \NILC\ set and the \Commander\ set).
For each realization, in any of the four sets, we measure $\fnl$ for the local, equilateral, and orthogonal shapes, using the Modal~1 pipeline. Then, for each shape and pair of methods, we build the difference between the two results and call it $\Delta \fnl$. 
To give an example, let us consider \SMICA, \SEVEM, and the local shape. For each of the $160$ realizations, we measure $f_{\rm NL}^{\rm local}$ from the \SMICA\ map 
($\equiv f_{\rm NL}^{\SMICA}$), repeat the operation with the \SEVEM\ map ($\equiv f_{\rm NL}^{\SEVEM}$), and build the difference $\Delta \fnl \equiv f_{\rm NL}^{\SMICA} - f_{\rm NL}^{\SEVEM}$. 
By repeating this for each possible pair of methods, we obtain six sets of $160$ differences $\Delta \fnl$ for a given shape.

We then extract the standard deviation of each set, which we call $\sigma_{\Delta \fnl}$ (not to be confused with $\sigma_{\fnl}$; $\sigma_{\Delta \fnl}$ is the standard deviation of the scatter between two cleaning methods, whereas $\sigma_{\fnl}$ is simply the $\fnl$ uncertainty for a given method).
The simulations we use do not include any foreground component; therefore, the extracted $\sigma_{\Delta \fnl}$ is only due to different frequency weighting and filtering schemes, and to further operations during NG estimation, such as inpainting. We can use these quantities to verify the consistency between the $\fnl$ differences obtained from data and from simulations. A large $\Delta \fnl$ scatter observed in the data, for a given pair of cleaned maps, compared to the corresponding expectation $\sigma_{\Delta \fnl}$, would signal potential residual foreground contamination in at least one of the two maps.

Our results for this test are summarized in Table~\ref{tab:methodsdiff}. For each pair of component-separation methods, and the three standard shapes, we show the measured scatter $\Delta \fnl$ and the ratio $ {\Delta \fnl / \sigma_{\Delta}}$. In order to assess the statistical significance of the result, we need of course to take into account that a multiplicity correction is necessary, since we are considering six pairs of methods. In Table~\ref{tab:multiplicity} we report the fraction of starting simulations for which, after the component-separation processing, ${\Delta \fnl / \sigma_{\Delta}}$ is larger than $1$, $2$ or $3$, for {\em any} pair of methods. We also report the largest value of ${\Delta \fnl / \sigma_{\Delta}}$, measured for all combinations and simulations. We see from this table that measured values of  ${\Delta \fnl / \sigma_{\Delta}}$ from the data (Table~\ref{tab:methodsdiff}) are not particularly unusual up to ${\Delta \fnl / \sigma_{\Delta}} \approx 3$.

This leads us to a first interesting observation, namely that polarization-only $\fnl$ results show no significant discrepancy among different cleaned maps. This is a large improvement with respect to our $2015$ analysis, where in a similar test we found large differences in \itE\itE\itE\ bispectra for several combinations. Such discrepancies, together with other anomalies, led us in the previous release to warn the reader that all polarization-based $\fnl$ measurements were less robust than \itT\itT\itT\ estimates, and had to be considered as preliminary. This is no longer the case, since both this and other tests (see next section) show that the polarization data are now fully reliable for primordial NG studies. Achieving such reliability was indeed one of the main goals for our analysis in this data release.

Despite consistency in polarization,
somewhat surprisingly we now find some relevant discrepancies in \itT-only results, even though limited to specific methods and the orthogonal shape only. The most striking example is a large difference between the orthogonal \itT-only $\fnl$ measurements obtained with \SMICA\ and \Commander\ ($ {\Delta \fnl / \sigma_{\Delta}} \approx 10$). Smaller but non-negligible are also the orthogonal \itT-only discrepancies for the \NILC\ -- \Commander\ and \SMICA\ -- \SEVEM\ pairs (both with $ {\Delta \fnl / \sigma_{\Delta}} \approx 4$).
These results have been cross-checked using the Binned pipeline, finding agreement between estimators: the Binned estimator finds
${\Delta \fnl / \sigma_{\Delta}}$ of approximately
8, 4, and 4, for these three cases, respectively.

As anticipated in Sect.~\ref{fnl_loc_eq_ort_results}, a closer inspection shows that these differences are less worrisome that they might appear at first glance. First of all, a comparison with our $2013$ and $2015$ results shows that \SMICA\ and \NILC\ orthogonal \itT\itT\itT\ measurements have remained very stable, while \SEVEM\ and especially \Commander\ display significant changes in this data release. Considering that {\em all} pipelines agreed very well and displayed robustness to a large number of validation tests in temperature in both previous data releases, we conclude that the latter two methods can be identified as the sources of the current orthogonal \itT-only discrepancy. This is good news, since the main component-separation method that we have focused on for $\fnl$ analysis (including in \citetalias{planck2013-p09a} and \citetalias{planck2014-a19}) is \SMICA. 

It is also important to stress that all significant discrepancies are specifically confined to the orthogonal \itT\itT\itT\ case. The other shapes and also a mode-by-mode or bin-by-bin correlation analysis over the full bispectrum domain (see next section) show no other signs of anomalies for any component-separation method.
Even considering the largest discrepancy, arising from the \SMICA\ -- \Commander\ pair, this still amounts to just a $1\,\sigma$ deviation in $\fnl^{\rm ortho}$. This is due to the fact that the level of agreement displayed by different cleaning methods on simulations typically amounts to a small fraction of the $\fnl$ errors. The implications for inflationary constraints of shifting $\fnl^{\rm ortho}$ by $1\,\sigma$  are essentially negligible. As mentioned earlier, the changes in orthogonal $\fnl$ coming from \Commander\ can be explained through the unavailability of detector-set maps for the current release. 

The main conclusion to be drawn from the validation tests described in this section is that the reliability of our final \itTpE\ $\fnl$ results has significantly increased with respect to the previous release, thanks to a clear improvement in the robustness of the polarization data. The issues we find in the temperature data are confined to the orthogonal shape and to specific component-separation pipelines, not affecting the final \SMICA\ measurements used for inflationary constraints. 

Of course, any considerations in this section that might lead us to a preference for specific component-separation methods, apply only to primordial NG analysis, and not to other cosmological or astrophysical analyses. There is no generally preferred cleaning method for all applications, and separate assessments should be conducted case by case.

\begin{table*}[htbp!]                 
\begingroup
\newdimen\tblskip \tblskip=5pt
\caption{Comparison between local, equilateral, and orthogonal $\fnl$ results, obtained using  
the four different component-separation pipelines and the Modal~1 bispectrum estimator. We calculate the difference $\Delta$ in the sense
$f_{\rm NL}({\rm method~1}) - f_{\rm NL}({\rm method~2})$.
For each pair of cleaned maps, we start by considering actual data and 
we compute the scatter $\Delta$ in our estimates of local, equilateral, and orthogonal $\fnl$. We then  
compute the ratio of $\Delta$ over its standard deviation, obtained using $160$ FFP10 simulations. 
These ratios have to be compared with the benchmarks provided in Table~\ref{tab:multiplicity}.
}\label{tab:methodsdiff}
\nointerlineskip
\vskip -3mm
\footnotesize
\setbox\tablebox=\vbox{
   \newdimen\digitwidth
   \setbox0=\hbox{\rm 0}
   \digitwidth=\wd0
   \catcode`*=\active
   \def*{\kern\digitwidth}
   \newdimen\signwidth
   \setbox0=\hbox{+}
   \signwidth=\wd0
   \catcode`!=\active
   \def!{\kern\signwidth}
\newdimen\dotwidth
\setbox0=\hbox{.}
\dotwidth=\wd0
\catcode`^=\active
\def^{\kern\dotwidth}
\halign{\hbox to 1.25in{#\leaderfil}\tabskip1em&
\hfil#\hfil\tabskip 1em&
\hfil#\hfil\tabskip 2em&
\hfil#\hfil\tabskip 1em&
\hfil#\hfil\tabskip 2em&
\hfil#\hfil\tabskip 1em&
\hfil#\hfil\tabskip 0pt\cr
\noalign{\doubleline\vskip 2pt}
\omit&\multispan2\hfil *Local\hfil&\multispan2\hfil *Equilateral\hfil& \multispan2\hfil *Orthogonal\hfil\cr
\noalign{\vskip -4pt}
\omit&\multispan2\hrulefill&\multispan2\hrulefill& \multispan2\hrulefill\cr
\omit\hfil Methods\hfil& $\Delta$& ${\Delta/\sigma}$& $\Delta$& ${\Delta/\sigma}$& $\Delta$& ${\Delta/\sigma}$\cr
\noalign{\vskip 3pt\hrule\vskip 5pt}
\omit\hfil\SMICA\ -- \SEVEM\hfil&&\cr
\itT&   $*-0.1$& $!0.0$& $*-11^*$& $-1.2$& $-24^*$& $*-4.1$\cr
\itE&   $*-5.4$& $-0.4$& $*-59^*$& $-1.1$& $*-5.0$& $*-1.8$\cr
\itTpE& $*-1.1$& $-0.6$& $*-10^*$& $-1.0$& $-19^*$& $*-3.8$\cr
\noalign{\vskip 3pt\hrule\vskip 5pt}
\omit\hfil\SMICA\ -- \NILC\hfil&&\cr
\itT&   $!*0.1$& $!0.1$& $!*16^*$& $!2.4$& $-17^*$& $*-3.0$\cr
\itE&   $!25^*$& $!1.8$& $!123^*$& $!2.1$& $-66^*$& $*-2.4$\cr
\itTpE& $!*0.3$& $!0.2$& $!*17^*$& $!1.8$& $-12^*$& $*-2.4$\cr
\noalign{\vskip 3pt\hrule\vskip 5pt}
\omit\hfil\SMICA\ -- \Commander\hfil&&\cr
\itT&   $!*1.7$& $!2.1$& $**-4.5$& $-0.5$& $-43^*$& $-10.0$\cr
\itE&   $*-0.2$& $!0.0$& $!**4.9$& $!0.1$& $-32^*$& $*-1.4$\cr
\itTpE& $!*0.9$& $!0.9$& $*-10^*$& $-1.1$& $-23^*$& $*-5.2$\cr
\noalign{\vskip 3pt\hrule\vskip 5pt}
\omit\hfil\SEVEM\ -- \NILC\hfil&&\cr
\itT&   $!*0.2$& $!0.2$& $!*27^*$& $!2.6$& $!*7.6$& $*!1.2$\cr
\itE&   $!30^*$& $!2.2$& $!183^*$& $!2.7$& $*-6.9$& $*-0.2$\cr
\itTpE& $!*1.4$& $!1.2$& $!*27^*$& $!2.1$& $!*7.7$& $*!1.3$\cr
\noalign{\vskip 3pt\hrule\vskip 5pt}
\omit\hfil \SEVEM\ -- \Commander\hfil&&\cr
\itT&   $!*1.8$& $!0.8$& $!**6.5$& $!0.6$& $-19^*$& $*-2.6$\cr
\itE&   $!*5.2$& $!0.4$& $!*64^*$& $!1.1$& $!26^*$& $*!0.8$\cr
\itTpE& $!*1.9$& $!0.9$& $**-0.4$& $!0.0$& $*-4.4$& $*-0.7$\cr
\noalign{\vskip 3pt\hrule\vskip 5pt}
\omit\hfil \NILC\ -- \Commander\hfil&&\cr
\itT&   $!*1.6$& $!0.7$& $*-20^*$& $-2.0$& $-26^*$& $*-3.8$\cr
\itE&   $-25^*$& $-1.8$& $-119^*$& $-1.6$& $!33^*$& $*!1.0$\cr
\itTpE& $!*0.6$& $!0.3$& $*-27^*$& $-2.1$& $-12^*$& $*-2.0$\cr
\noalign{\vskip 3pt\hrule\vskip 4pt}
}}
\endPlancktablewide                 
\endgroup
\end{table*}                        

\begin{table*}[htbp!]
\begingroup
\newdimen\tblskip \tblskip=5pt
\caption{Fraction of simulations for which the differences between pairs of
component-separation methods are above various levels of $\sigma$.
We use 160 FFP10 simulations (without foreground residuals) to compute the standard deviation $\sigma_{\Delta \fnl}$ of the measured $\fnl$ scatter, obtained by processing a given realization through the different component-separation pipelines.
We report the largest measured significance ${\Delta \fnl/\sigma_{\Delta \fnl}}$, across all method pairs and simulations. We then report the fraction of the total number of simulations for which at least one method pair returns a value $\Delta_{\fnl}$
larger than $1\,\sigma_{\Delta \fnl}$, $2\,\sigma_{\Delta \fnl}$, or $3\,\sigma_{\Delta \fnl}$. These numbers provide a benchmark against which to assess the significance of the scatter measured on the data (Table~\ref{tab:methodsdiff}), taking into account the multiple comparisons.}\label{tab:multiplicity}
\nointerlineskip
\vskip -3mm
\footnotesize
\setbox\tablebox=\vbox{
 \newdimen\digitwidth 
 \setbox0=\hbox{\rm 0} 
 \digitwidth=\wd0 
 \catcode`*=\active 
 \def*{\kern\digitwidth}
 \newdimen\signwidth 
 \setbox0=\hbox{+} 
 \signwidth=\wd0 
 \catcode`!=\active 
 \def!{\kern\signwidth}
\halign{\hbox to 0.75in{#\leaderfil}\tabskip=1em&
\hfil#\hfil\tabskip=1em& \hfil#\hfil& \hfil#\hfil& \hfil#\hfil\tabskip=2em&
\hfil#\hfil\tabskip=1em& \hfil#\hfil& \hfil#\hfil& \hfil#\hfil\tabskip=2em&
\hfil#\hfil\tabskip=1em& \hfil#\hfil& \hfil#\hfil& \hfil#\hfil\tabskip=0pt\cr  
\noalign{\doubleline}
\omit& \multispan4\hfil Local\hfil& \multispan4\hfil Equilateral\hfil& \multispan4\hfil  Orthogonal\hfil\cr
\noalign{\vskip -4pt} 		
\omit&\multispan4\hrulefill&\multispan4\hrulefill&\multispan4\hrulefill\cr
\omit\hfil Data set\hfil& ${\Delta f_{\rm NL}/\sigma_{\Delta}} ^{\rm max}$& $N_{\rm sims}^{> 1\,\sigma}$& $N_{\rm sims}^{> 2\,\sigma}$& $N_{\rm sims}^{> 3\,\sigma}$& ${\Delta f_{\rm NL}/\sigma_{\Delta}} ^{\rm max}$& $N_{\rm sims}^{> 1\,\sigma}$& $N_{\rm sims}^{> 2\,\sigma}$& $N_{\rm sims}^{> 3\,\sigma}$&${\Delta f_{\rm NL}/\sigma_{\Delta}} ^{\rm max}$&  $N_{\rm sims}^{> 1\,\sigma}$& $N_{\rm sims}^{> 2\,\sigma}$& $N_{\rm sims}^{> 3\,\sigma}$\cr
\noalign{\vskip 3pt\hrule\vskip 5pt} 		
\itT& $3.7$& $0.66$& $0.18$& $0.006$& $3.3$& $0.74$& $0.18$& $0.006$& $3.5$& $0.72$& $0.16$& $0.019$\cr
\itE& $3.6$& $0.72$& $0.16$& $0.025$& $3.2$& $0.71$& $0.16$& $0.013$& $3.3$& $0.76$& $0.19$& $0.019$\cr
\itTpE& $2.9$& $0.66$& $0.13$& $0$& $3.3$& $0.64$& $0.15$& $0.013$& $3.7$& $0.65$& $0.16$& $0.019$\cr
\noalign{\vskip 5pt\hrule\vskip 3pt}
}}
\endPlancktablewide                    
\endgroup
\end{table*}

\begin{figure*}[htbp!]
\centering
\includegraphics[width=0.49\linewidth, height=.3\textheight]{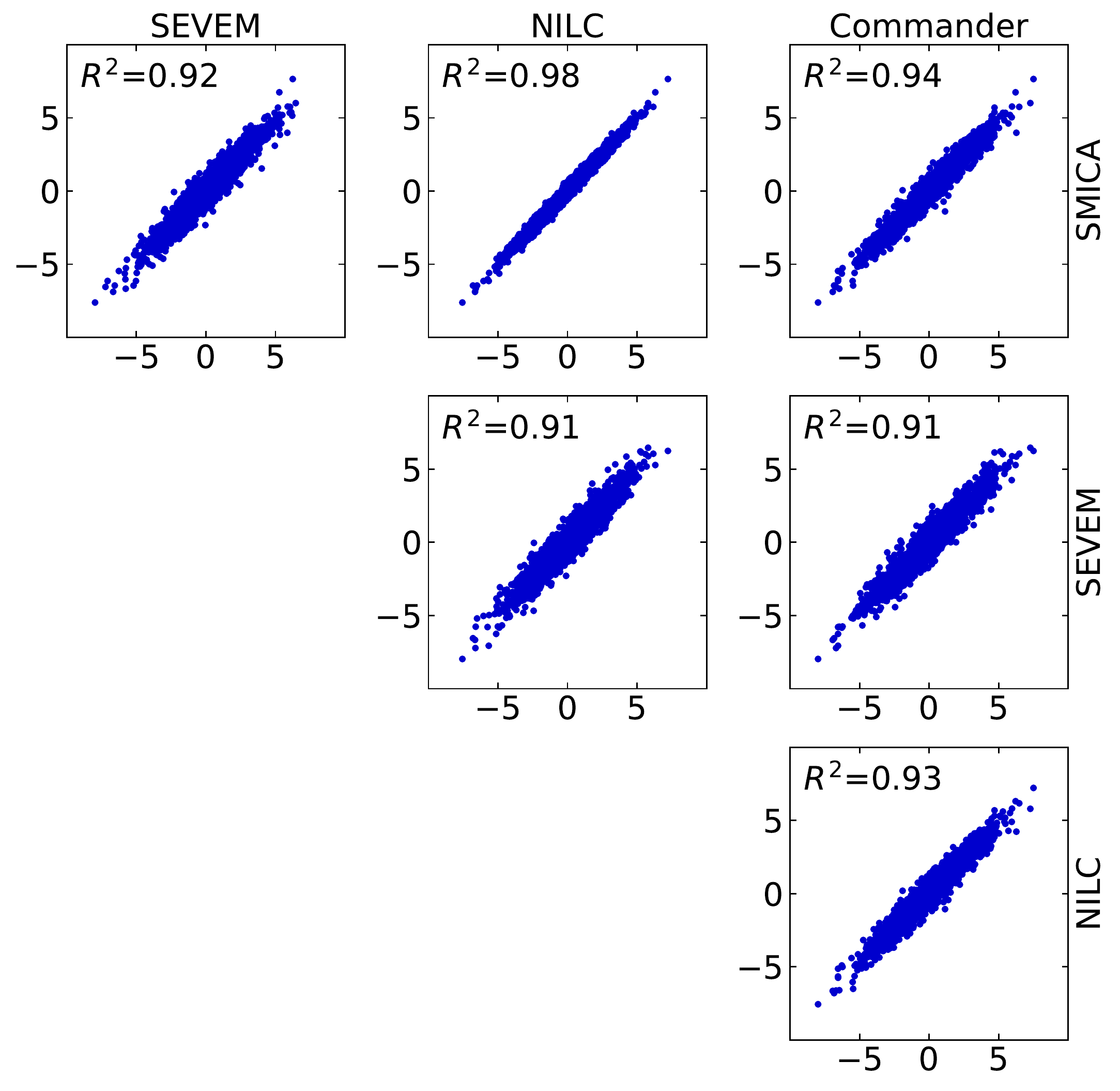}
\includegraphics[width=0.49\linewidth, height=.3\textheight]{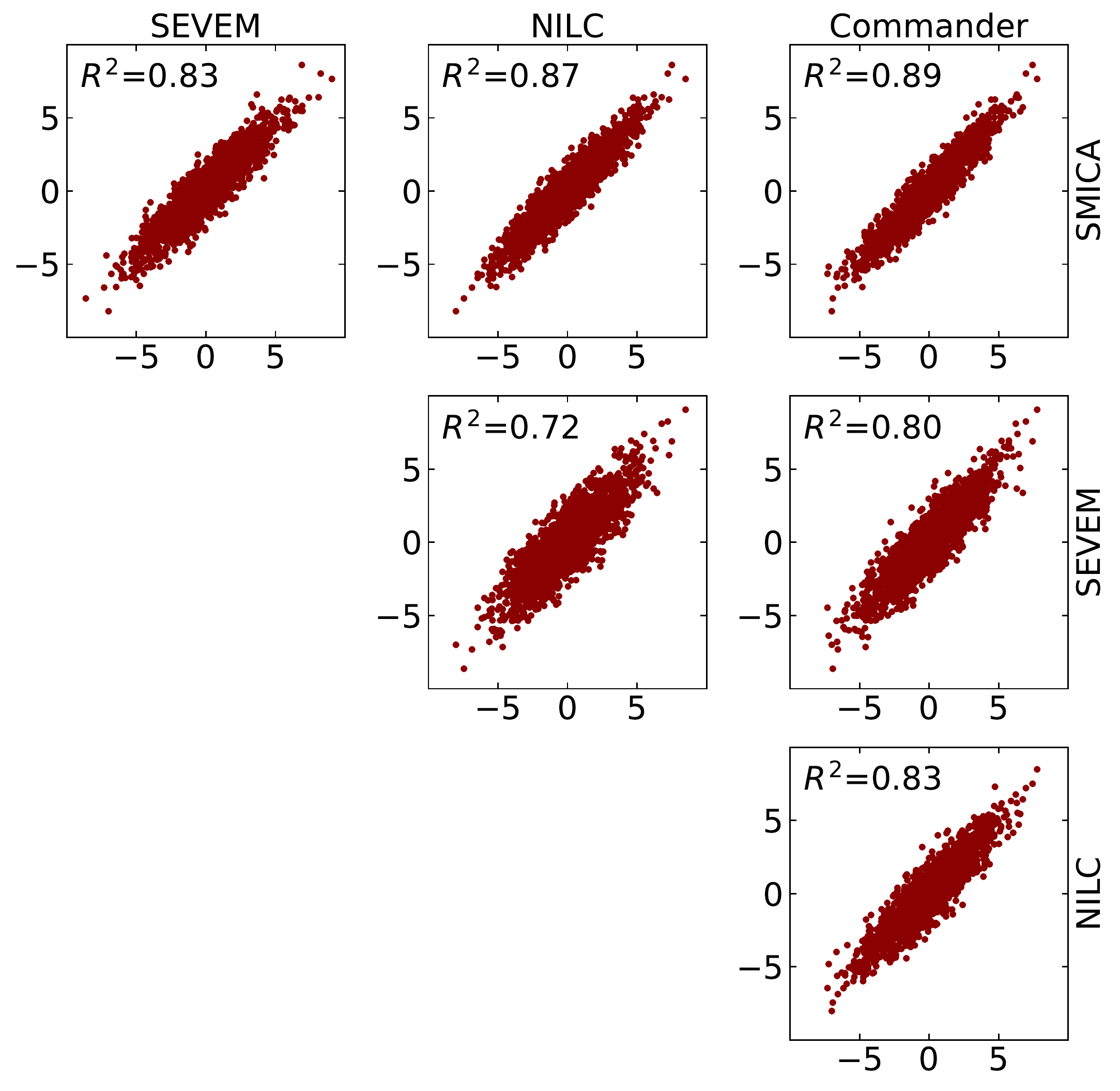}
\caption[]{Scatter plots of the 2001 Modal~2 coefficients for each combination of component-separation methods, labelled with the $R^2$ coefficient of determination. The figures on the left are for the temperature modes and those on the right for pure polarization modes (the Modal~2 pipeline reconstructs the component of the \itE\itE\itE\ bispectrum that is orthogonal to \itT\itT\itT, so this is not exactly the same as the \itE\itE\itE\ bispectrum of the other estimators).}
\label{fig:scatterplots_modal}
\end{figure*}

\begin{figure*}[htbp!]
\centering
\includegraphics[width=.49\linewidth, height=.3\textheight]{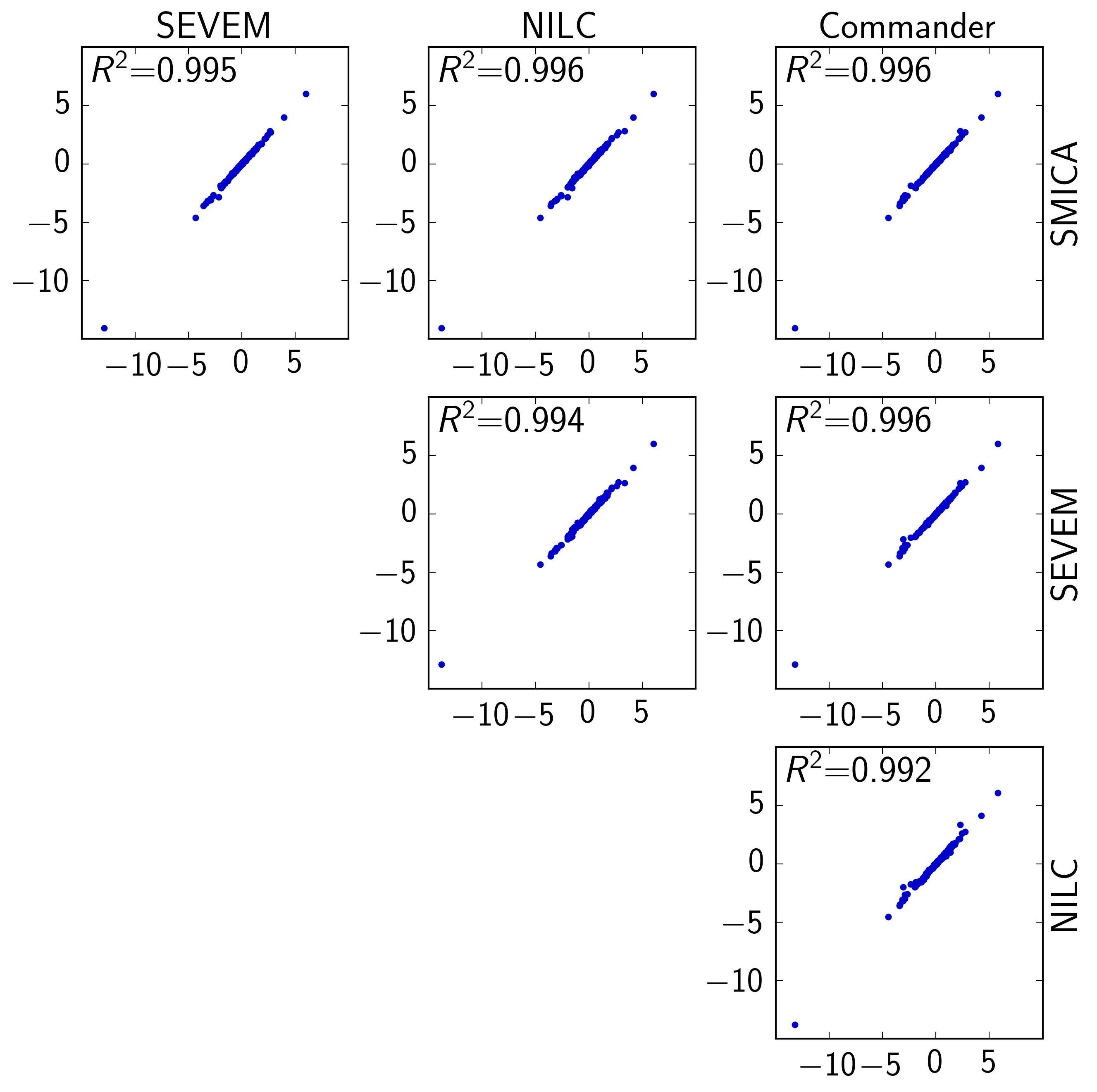}
\includegraphics[width=.49\linewidth, height=.3\textheight]{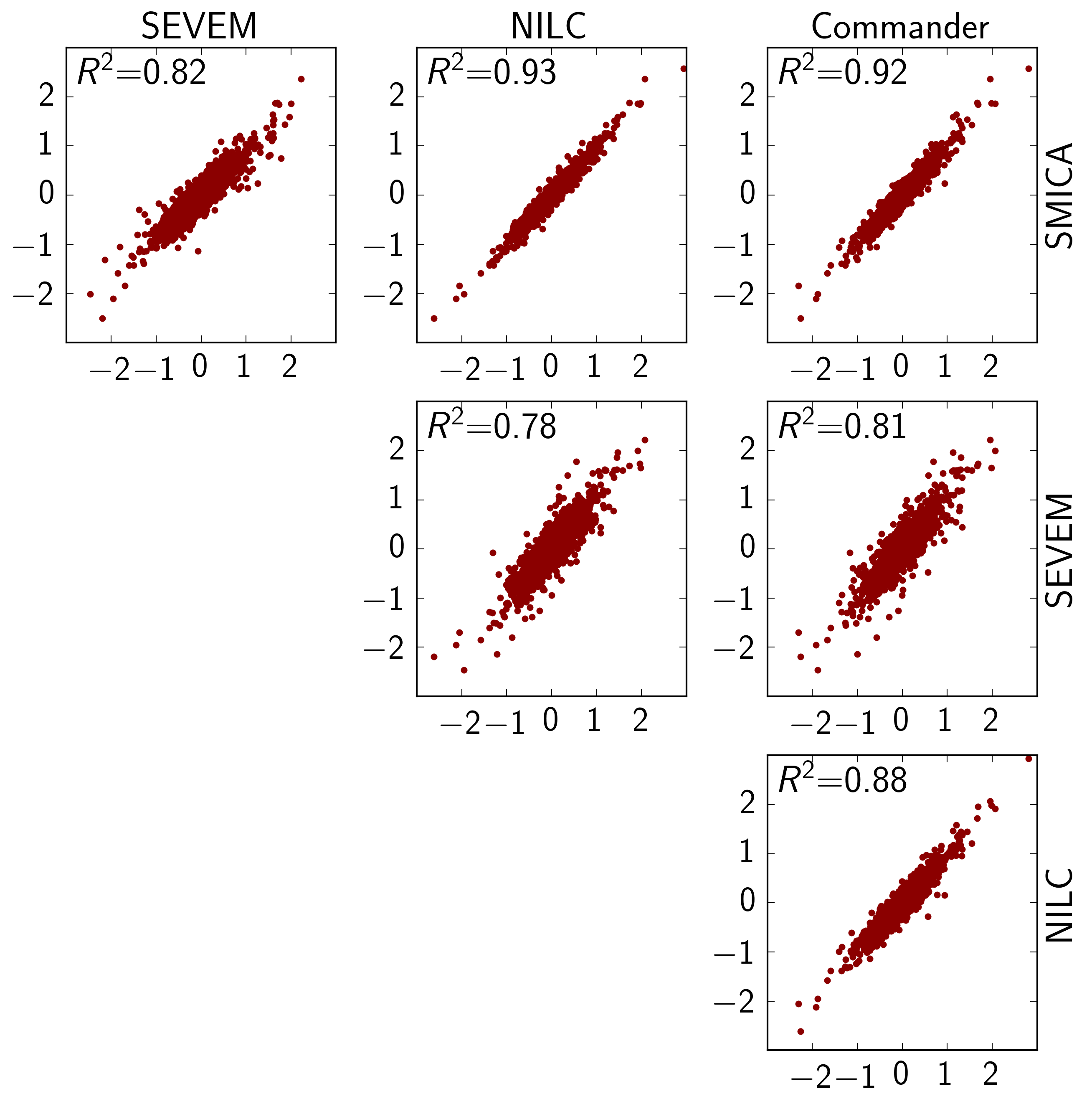}
\caption[]{Scatter plots of the bispectrum values in each bin-triplet (13020 for \itT\itT\itT, all of which have been multiplied by $10^{16}$, and 11221 for \itE\itE\itE, all of which have been multiplied by $10^{20}$) for all combinations of component-separation methods, with the $R^2$ coefficient of determination. The figures on the left are for \itT\itT\itT\ and those on the right for \itE\itE\itE.}
\label{fig:scatterplots_binned}
\end{figure*}

\subsubsection{Comparison between reconstructed bispectra}\label{sec:bispcomp}

The local, equilateral, and orthogonal directions already cover a significant part of the entire bispectrum domain, and deserve special attention, 
since they are crucial for inflationary constraints. It is nevertheless useful to also check the agreement between component-separation pipelines in a more general, model-independent fashion. 
For this purpose we calculate the coefficient of determination \citep[see][]{coffofdet}, denoted by  $R^2$, from the mode or bin amplitudes extracted from different 
foreground-cleaned maps.
The coefficient of determination is a standard statistical measure of what proportion of the variance of one variable can be predicted from another. For example in our case a score of $0.9$ for \SMICA\ -- \SEVEM\  would mean that 90\% of the mode or bin amplitudes measured in \SMICA\  could be explained by the mode or bin amplitudes measured in \SEVEM\  and 10\% of the amplitudes would be unexplained.

Scatter plots of modes are shown in Fig.~\ref{fig:scatterplots_modal}, while
those for bins are shown in Fig.~\ref{fig:scatterplots_binned}.
For \itT\itT\itT\ the lowest Modal coefficient of determination is $0.91$
(for \SEVEM\ -- \NILC), while for the Binned bispectrum values all \itT\itT\itT\
coefficients of determination are larger than $0.99$. For \itE\itE\itE\ the
lowest Modal value is $0.71$, again for \SEVEM\ -- \NILC, while the corresponding
Binned value is $0.78$. The lowest Binned value overall is $0.71$, for
\SMICA\ -- \SEVEM\ \itE\itT\itT\ (not shown in the figure). Note that it is
also possible to see the impact of excluding the lowest $[2,3]$ bin from the
analysis for \itE; if it were included that value would drop from $0.71$ to $0.59$.

These results show a very good level of agreement between all methods. Again, we see a large improvement in the polarization maps, by comparing with a similar test performed in \citetalias{planck2014-a19}.  Also interesting is the fact that no anomalies show up in temperature data for any specific mode or bin. The issues discussed in the previous section, which affected only specific component-separation pipelines, seem also completely confined to the combination of modes/bins that selects the orthogonal shape region of the bispectrum domain, in such a way
as to be invisible in this more general analysis.

\subsection{Testing noise mismatch}\label{sec:noisetest}

Accurate tests of the FFP10 maps have found some level of mismatch between the noise model adopted in the simulations and the actual noise levels in the data \citep{planck2016-l03,planck2016-l04}. In temperature this is roughly at the $3\,\%$ level in the noise power spectrum at $\ell \approx 2000$. Percent level differences are also seen in polarization.
This issue raises some concern for our estimators, since we use simulations to calibrate them. We thus decided to perform some simulation-based tests to check to what extent this noise mismatch may affect our Monte Carlo errors (note that, as long as the noise is Gaussian, noise mismatches of this kind cannot bias the estimators). For all the work in this section we consider \SMICA\ maps only. 

Uncertainties in non-Gaussianity parameters might be affected by two effects. One is the suboptimality of the estimator weights in the cubic term, if the power spectra extracted from simulations do not match the data. The other is an imperfect Monte Carlo calibration of the linear correction term, leading to the inability to fully correct for anisotropic and correlated noise features. Given that we are considering percent-level corrections, we do not expect the former of the two effects to be of particular significance; this is also confirmed by past analyses, in which the cosmological parameters have been updated several times, leading to percent changes in the fiducial power spectrum, without any appreciable difference in the $\fnl$ results. The effect on the linear term is harder to assess, and to some extent it depends on the specific noise correlation properties in the data, which are possibly not fully captured in the simulations. In general, it is reasonable to expect a power spectrum mismatch of a few percent to have only a small impact, unless significant spatial correlations between large and small scales are present in the data and not captured by the simulation noise model. Note also that the linear term correction is generally dominated by the mask, in the mostly signal-dominated regime we consider for our analysis.

As a first test, we generate ``extra-noise'' multipoles, drawing them from a zero-mean Gaussian distribution having as power spectrum the difference between the simulation and data noise power spectra. We then add these realizations to the original noise simulations and apply our estimators to this extra-noise mock data set. However, we calibrate our linear term, estimator normalization, and weights using the original noise maps. We check the effect of the mis-calibration through a comparison between the $\fnl$ measurements obtained in this way and those extracted from the original realizations, without extra noise included. As a figure of merit, we consider the local $\fnl$ results, since they are generally most sensitive to noise features. At the end of this analysis, we verify that the impact of the noise mismatch is very small: the uncertainties change by a negligible amount with respect to the exactly calibrated case, while the scatter between $\fnl$ measured with and without extra noise in the maps is zero on average, as expected, and has a standard deviation $\sigma_{\Delta \fnl} \approx \sigma_{\fnl} / 10$.

\begin{figure}[htbp!]
    \centering
    \includegraphics[width=0.5\textwidth]{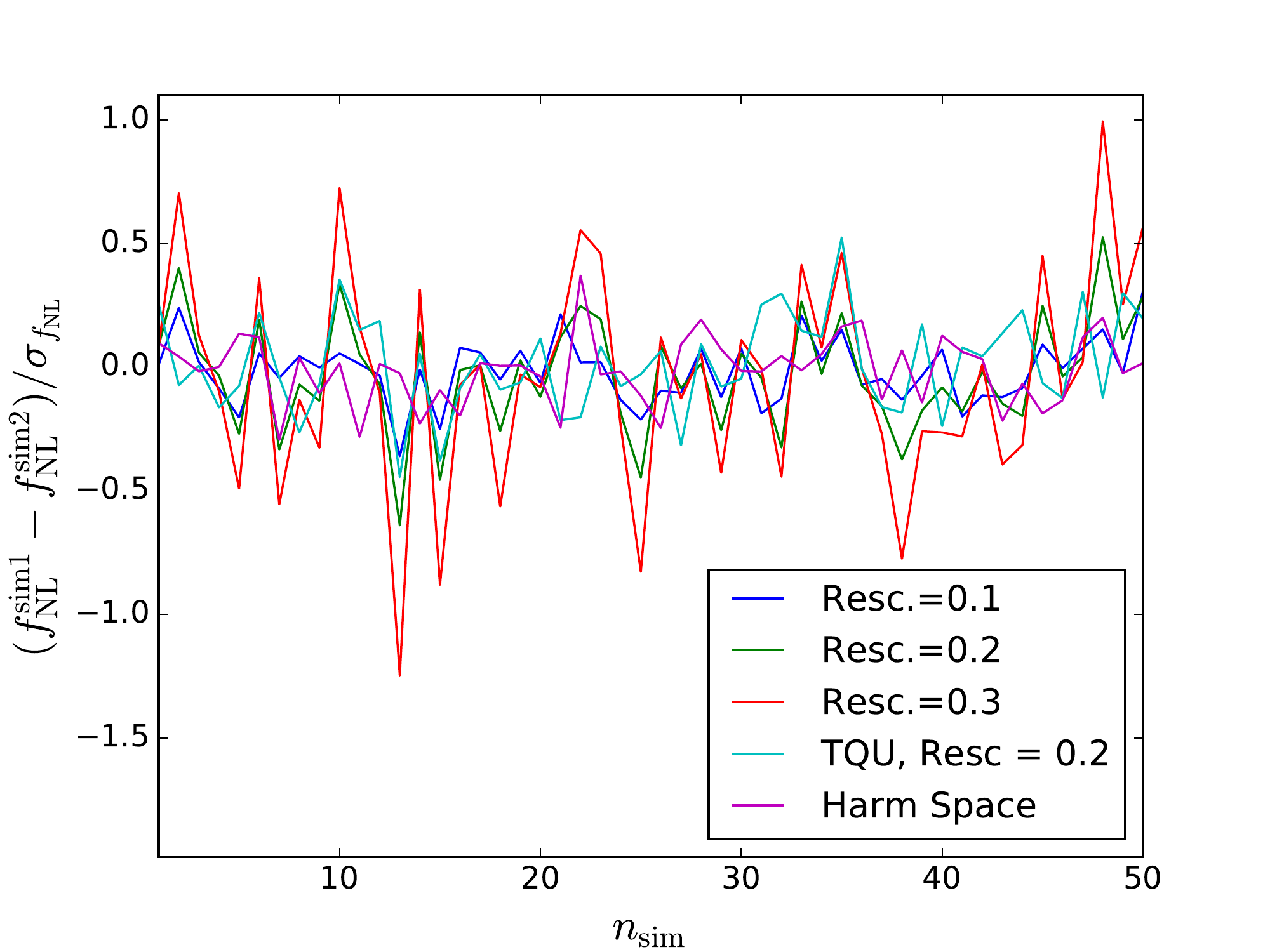}
    \includegraphics[width=0.5\textwidth]{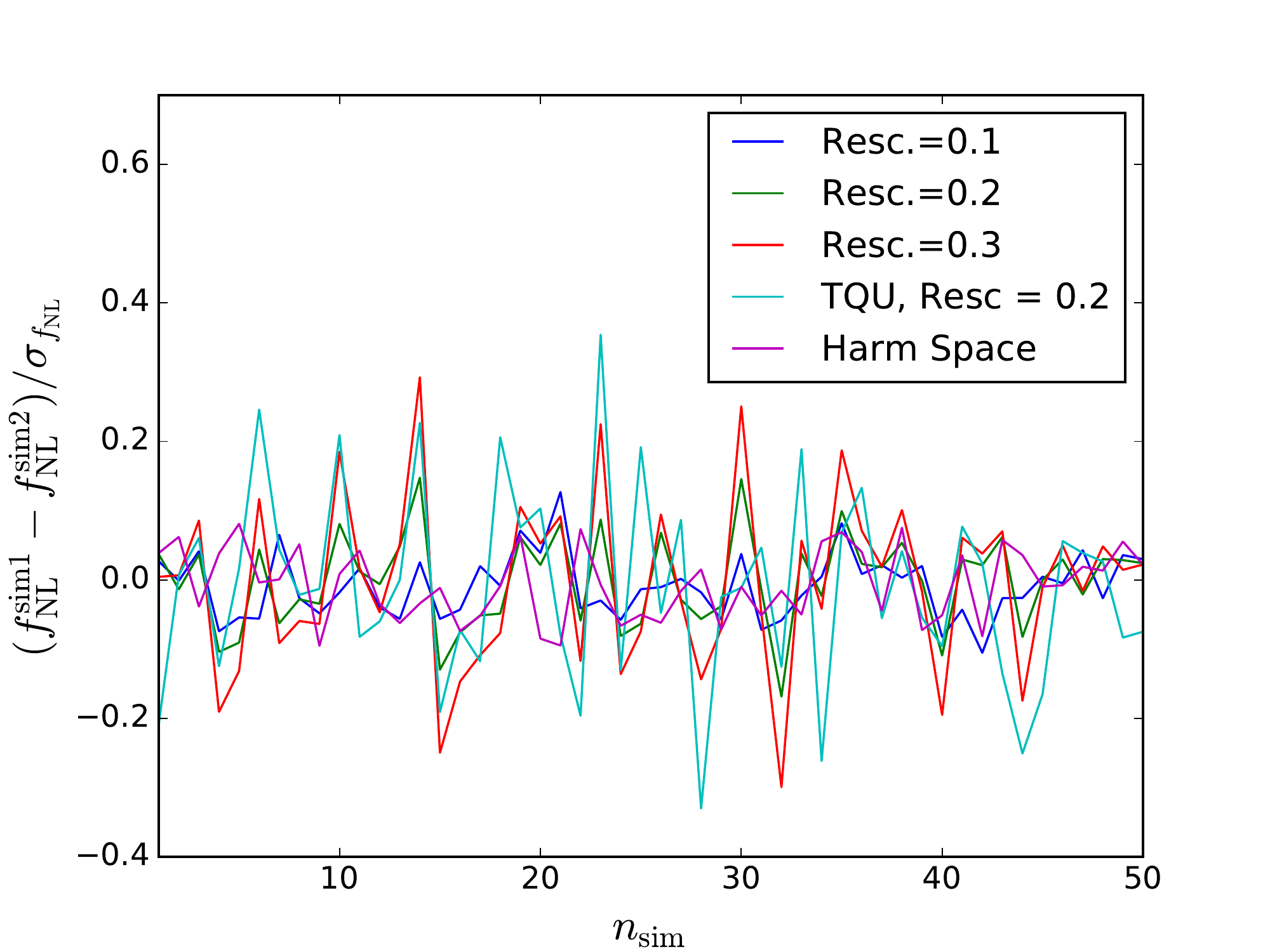}
    \caption{Effects on $\fnl$ scatter, simulation-by-simulation, of different levels of noise mismatch, as described in Sect.~\ref{sec:noisetest}. {\it Top}: \itE\itE\itE-only results. {\it Bottom}: combined \itT$+$\itE~results. The magenta line represents the case in which we generate an extra-noise component in harmonic space, both in temperature and polarization, with variance equal to the difference between the noise spectra in the data and the simulations. All other lines represent cases in which we generate extra-noise maps in pixel space, with the rms per pixel extracted from simulations and rescaled by different factors, as specified in the legend; the label ``TQU'' further specifies the pixel-space test in which extra noise is added to both temperature and polarization data, unlike in the other three pixel-space cases, in which we include only a polarization extra-noise component.}
    \label{fig:noisetest}
\end{figure}

Our results are shown in detail in the first row of Table~\ref{tab:noisetest} and in Fig.~\ref{fig:noisetest}. The former reports the $\fnl$ error bars and the standard deviation of map-by-map $\fnl$ differences, while the latter shows the map-by-map $\fnl$ scatter for $50$ realizations. These very small deviations were largely expected, given the small change in the amplitude of the noise power spectrum we are considering.

In the test we have just described we generate uncorrelated extra-noise multipoles (with a non-white spectrum). On the other hand, we know that the estimator is most sensitive to couplings between large and small scales, which require a properly 
calibrated linear term correction.
Therefore, we decided to also test the impact of a possible linear term miscalibration of this type, by generating and studying an extra-noise component directly in pixel space. 
In this case, $\ell$-space correlations between large and small scales 
arise due the spatially anisotropic distribution of the noise.
We proceed as follows.  After extracting the noise rms of the \SMICA\ FFP10 polarization simulations, we rescale it by a fixed factor $A$, 
and we use this rescaled rms map to generate new Gaussian ``extra-noise" realizations in pixel space, which we add to the original
noise maps. We consider different cases. Firstly, we take a rescaling factor $A_{TQU}=0.2$ for both temperature and polarization noise maps. 
We then perform a more detailed study of the effect on polarization maps only. For this purpose, we leave the temperature noise unchanged and rescale the 
polarization rms noise by factors ranging from $A=0.1$ up to $A=0.3$. 

We see again from the summary of results reported in Table~\ref{tab:noisetest} that the $\fnl$ error change is always very small. 
The same can be said of the standard deviation of the $\fnl$ scatter between realizations with and without extra noise, which reaches at most 
a value $\sigma_{\Delta \fnl} \approx \sigma_{\fnl} / 3$, for a large $A=0.3$. 
These results provide a good indication that a noise mismatch between simulations and data is not a concern for primordial NG estimation, unless 
the mismatch itself is well above the estimated percent level in the noise power spectrum and produces large correlations between small and large scales. 
It is also worth mentioning that, in the very early stages of our primordial NG analysis of \Planck\ data, 
when accurate simulations were not yet available, we calibrated our estimators using very simple noise models (e.g., we initially generated 
noise in harmonic space, with a non-flat power spectrum consistent with the data, but neglecting any correlations between different scales; we then went to pixel space and modulated the noise map with the hit-count map to anisotropize it). Despite the simplicity of this approach, we were able to
verify later, using FFP simulations, that such simple models already produced an accurate linear-term correction. This further reinforces our confidence in the robustness of 
$\fnl$ estimators to imperfect modelling of the noise.

\begin{table}[htbp!]                 
\begingroup
\newdimen\tblskip \tblskip=5pt
\caption{Results of the noise mismatch test described in Sect.~\ref{sec:noisetest}. We consider FFP10 simulations and add extra noise to them. We then measure
 $f_{\rm NL}^{\rm local}$ using the Modal~1 pipeline, normalized and calibrated without the extra-noise component. We verify how this affects the local $\fnl$ errors ($\sigma_{f_{\rm NL}}$ columns) 
and measure the standard deviation of the scatter $\fnl^{1,i} - \fnl^{2,i}$, where $\fnl^{2,i}$ and $\fnl^{1,i}$ are the measurements extracted with and without noise mismatch, respectively, for the $i$th simulation ($\sigma_{\Delta f_{\rm NL}}$ columns).
Results in the first row are obtained by generating extra noise in harmonic space, while in the second row we generate extra noise in pixel space, both in temperature and polarization. In the following rows, we still generate noise in pixel space, but leave temperature noise maps unchanged, while scaling the polarization rms by the given factors.} 
\label{tab:noisetest}
\nointerlineskip
\vskip -3mm
\footnotesize
\setbox\tablebox=\vbox{
   \newdimen\digitwidth 
   \setbox0=\hbox{\rm 0} 
   \digitwidth=\wd0 
   \catcode`*=\active 
   \def*{\kern\digitwidth}
   \newdimen\signwidth 
   \setbox0=\hbox{+} 
   \signwidth=\wd0 
   \catcode`!=\active 
   \def!{\kern\signwidth}
   \newdimen\signwidth 
   \setbox0=\hbox{.} 
   \signwidth=\wd0 
   \catcode`^=\active 
   \def^{\kern\signwidth}
\halign{\hbox to 0.9in{#\leaderfil}\tabskip1em& \hfil#\hfil\tabskip=1em&
\hfil#\hfil\tabskip 2em& \hfil#\hfil\tabskip=1em&
\hfil#\hfil\tabskip=2em& \hfil#\hfil\tabskip=1em&
\hfil#\hfil\tabskip 0pt\cr
\noalign{\doubleline}
\omit&\multispan2\hfil\itT\ only\hfil& \multispan2\hfil\itE\ only\hfil& \multispan2\hfil\itTpE\hfil\cr
\noalign{\vskip -4pt}
\omit&\multispan2\hrulefill& \multispan2\hrulefill& \multispan2\hrulefill\cr
\omit\hfil Method\hfil& $\sigma_{f_{\rm NL}}$&$\sigma_{\Delta f_{\rm NL}}$&$\sigma_{f_{\rm NL}}$&$\sigma_{\Delta f_{\rm NL}}$& $\sigma_{f_{\rm NL}}$& $\sigma_{\Delta f_{\rm NL}}$\cr
\noalign{\vskip 3pt\hrule\vskip 5pt}
$\ell$ space&   $5.8$& 0.1& 26& 3.5& 4.7& 0.2\cr 
$A_{\rm all} = 0.2$& $5.8$& $0.6$& $26$& $5.1$& $4.8$& $0.6$\cr
$A = 0.1$&&& $26$& $3.3$& $4.7$& $0.2$\cr
$A = 0.2$&&& 28& 5.8& 4.7& 0.3\cr
$A = 0.3$&&& 29& 11.0& 4.7& 0.6\cr
\noalign{\vskip 3pt\hrule\vskip 5pt}
}}
\endPlancktable                 
\endgroup
\end{table}                        

\subsection{Effects of foregrounds}\label{sec:foregroundtest}

Here we look at two non-primordial contributions to the \Planck\ bispectrum, namely Galactic dust and the Sunyaev-Zeldovich effect, which
could potentially be present in the cleaned maps, but which we do not
detect in the end.\footnote{Another potential contaminant for both
primordial and lensing bispectrum results is the ``intrinsic bispectrum,'' induced in the CMB by weak (second-order) non-linearities from gravity in 
general relativity and by non-linearities in the recombination physics. This would set the minimal level of CMB NG present even for Gaussian initial conditions of the primordial curvature perturbation.
However, this is of no particular concern to us, since
for the \Planck\ data set its expected impact is very small both in temperature
\citep{2004JCAP...01..003B,2004PhRvD..70h3532C,2005JCAP...08..010B,2009JCAP...08..029B,2009JCAP...05..014N,2009JCAP...09..038S,2009PhRvD..79b3501K,2009JCAP...03..017B,2010PhRvD..81j3518K,2010AdAst2010E..75B,2011JCAP...11..025C,2012JCAP...02..017B,2012arXiv1212.3573H,2012arXiv1212.6968S,2014PhRvD..90j3010P} and polarization~\citep{2012JCAP...06..023L,2014PhRvD..90j3010P}.} Non-primordial contributions that we \textit{do} detect (lensing and extragalactic point sources) were discussed in Sect.~\ref{sec:npNG}.

\subsubsection{Non-Gaussianity of the thermal dust emission}\label{sec:dust}

In principle, the foreground-cleaned maps should not contain any noticeable NG of Galactic origin. However, in raw observations the strongest contamination to the primordial NG is due to the thermal dust emission, which induces a large negative bias in the measurements of $f_\mathrm{NL}^\mathrm{local}$. Therefore, it is important to verify that this contamination has been removed entirely through the different component-separation methods.

There is no analytical template for the dust bispectrum, unlike the extragalactic templates discussed in Sect.~\ref{sec:point_sources}. A simple method using instead numerical templates for the different Galactic foregrounds has been described in \cite{Jung:2018rgf} \citep[see also][]{2019arXiv190104515C}. First, using the Binned bispectrum estimator, the dust bispectral shape is computed from the thermal dust emission map at 143\,GHz (the dominant frequency channel in the cleaned CMB maps) produced by the \Commander\ technique. Note that this is actually
a map determined at higher frequencies, where the dust dominates, and then
rescaled to 143\,GHz. Since there was no improvement for the temperature map of this foreground in the latest release, we use the \Planck\ 2015 temperature map here. Then, this numerical dust bispectrum can be used as a theoretical template in the analysis of any other data map with the Binned bispectrum estimator. The only condition is that the same mask, beam, and binning are used for both the determination of the dust bispectrum and the analysis itself. 

An illustration of the large bias induced by the presence of dust in the map is given in Table~\ref{tab:raw_sky}. It shows the $f_\mathrm{NL}$ parameters of the primordial local shape and the thermal dust bispectrum for both an independent and a joint analysis of the raw 2018 143-GHz \Planck\ temperature map. As a reminder, in an independent analysis we assume that only one of
the templates is present in the data (and we repeat the analysis for each individual template), while in
a fully joint analysis we assume that all templates are present, so that we have
to take their correlations into account. In the independent case, there is a strong detection of local NG (at about the 5$\,\sigma$ level), while the dust is observed at the expected level (within the 1$\,\sigma$ interval centred on the expected value: $f_\mathrm{NL}^\mathrm{dust}=1$). The joint analysis shows that indeed the large negative $f_\mathrm{NL}^\mathrm{local}$ of the independent case is entirely due to the presence of dust in the map, while the dust is still detected at the expected amount. 

\begin{table}[htbp!]               
\begingroup
\newdimen\tblskip \tblskip=5pt
\caption{Independent and joint estimates (see the main text for a definition)
of the $\fnl$ parameters of the
  primordial local shape and the thermal dust bispectrum 
  in the raw \Planck\ 143-GHz temperature map, determined using the
  Binned bispectrum estimator. Uncertainties are 68\,\%~CL.$^{\rm a}$}
\label{tab:raw_sky}
\nointerlineskip
\vskip -3mm
\footnotesize
\setbox\tablebox=\vbox{
   \newdimen\digitwidth
   \setbox0=\hbox{\rm 0}
   \digitwidth=\wd0
   \catcode`*=\active
   \def*{\kern\digitwidth}
   \newdimen\signwidth
   \setbox0=\hbox{+}
   \signwidth=\wd0
   \catcode`!=\active
   \def!{\kern\signwidth}
\halign{\hbox to 0.9in{#\leaderfil}\tabskip 2.2em&
\hfil#\hfil\tabskip 2.2em&
\hfil#\hfil\tabskip 0pt\cr
\noalign{\doubleline}
\omit\hfil Shape\hfil& *Independent& Joint\cr
\noalign{\vskip3pt\hrule\vskip 3pt}
Local& $-64\pm13 $& $*8\pm20$\cr
Dust& $!1.21\pm0.35$& $1.32\pm0.52$\cr
\noalign{\vskip 3pt\hrule\vskip 5pt}}}
\endPlancktable                    
\tablenote {{\rm a}} Uncertainties in this table only are Fisher forecasts multiplied by a factor larger than 1, which depends on the shape, due to the small breaking of the weak non-Gaussianity approximation in this map. Here, we use the same factors as in \cite{Jung:2018rgf}, which were determined by comparing the observed errors and Fisher forecasts in the analysis of 100 Gaussian CMB maps to which the dust map was added.\par
\endgroup
\end{table}                        

\begin{table*}[htbp!]                 
\begingroup
\newdimen\tblskip \tblskip=5pt
\caption{Independent and joint estimates of the $\fnl$ parameters of the
  indicated shapes, including in particular the dust template, for the
  cleaned maps produced by the four component-separation methods, as determined 
  with the Binned bispectrum estimator.}
\label{tab:cmb}
\nointerlineskip
\vskip -3mm
\footnotesize
\setbox\tablebox=\vbox{
   \newdimen\digitwidth 
   \setbox0=\hbox{\rm 0} 
   \digitwidth=\wd0 
   \catcode`*=\active 
   \def*{\kern\digitwidth}
   \newdimen\signwidth 
   \setbox0=\hbox{+} 
   \signwidth=\wd0 
   \catcode`!=\active 
   \def!{\kern\signwidth}
\halign{\hbox to 1.0in{#\leaderfil}\tabskip 0.5em&
\hfil#\hfil\tabskip 1em& \hfil#\hfil\tabskip 2em&
\hfil#\hfil\tabskip 1em& \hfil#\hfil\tabskip 2em&
\hfil#\hfil\tabskip 1em& \hfil#\hfil\tabskip 2em&
\hfil#\hfil\tabskip 1em& \hfil#\hfil\tabskip 0pt\cr
\noalign{\doubleline}
\omit&\multispan2\hfil\SMICA\hfil& \multispan2\hfil\SEVEM\hfil&
\multispan2\hfil\NILC\hfil& \multispan2\hfil\Commander\hfil\cr
\noalign{\vskip -4pt}
\omit& \multispan2\hrulefill& \multispan2\hrulefill&
\multispan2\hrulefill& \multispan2\hrulefill\cr
\noalign{\vskip 2pt}
\omit\hfil Shape\hfil& *Independent& *Joint& Independent& Joint& Independent& *Joint& *Independent& Joint\cr
\noalign{\vskip 4pt\hrule\vskip 5pt}
$\fnl^\mathrm{local}$& $-0.1\pm5.6$& $!5.0\pm8.4$& $0.0\pm5.7$& $1.7\pm8.7$& $0.0\pm5.6$& $!5.2\pm8.5$& $-1.3\pm5.6$& $3.1\pm8.3$\cr
$\fnl^\mathrm{equil}$& $!26\pm69$& $*!5\pm73$& $43\pm70$& $30\pm74$& $*5\pm69$& $-12\pm73$& $!32\pm69$& $20\pm73$\cr
$\fnl^\mathrm{ortho}$& $-11\pm39$& $*-5\pm44$& $*8\pm39$& $13\pm45$& $*4\pm39$& $!13\pm45$& $!29\pm39$& $35\pm44$\cr
$b_\mathrm{PS} / (10^{-29})$& $!6.3\pm1.0$& $!5.0\pm2.7$& $9.7\pm1.1$& $7.1\pm2.9$& $5.7\pm1.1$& $!5.4\pm2.7$& $!5.4\pm1.0$& $3.6\pm2.6$\cr
$A_\mathrm{CIB} / (10^{-27})$& $!3.0\pm0.5$& $!0.6\pm1.3$& $4.6\pm0.5$& $1.3\pm1.4$& $2.6\pm0.5$& $!0.1\pm1.3$& $!2.6\pm0.5$& $0.9\pm1.3$\cr
$\fnl^\mathrm{dust} / (10^{-2})$& $!6.6\pm4.4$& $!6.7\pm5.9$& $4.8\pm4.6$& $1.9\pm6.1$& $4.8\pm4.4$& $!5.1\pm5.9$& $!4.4\pm4.3$& $3.1\pm5.7$\cr
\noalign{\vskip 3pt\hrule\vskip 4pt}}}
\endPlancktablewide                 
\endgroup
\end{table*}                        

The analysis of the \itT-only maps produced by the four component-separation methods is the main result of this subsection. Table~\ref{tab:cmb} gives the values of $f_\mathrm{NL}$ for several primordial NG shapes (local, equilateral, and orthogonal), the amplitudes of some extragalactic foreground bispectra (unclustered point sources and the CIB, as defined in Sect.~\ref{sec:point_sources}) and the $\fnl$ of the thermal dust emission bispectrum, after subtracting the lensing bias, in both an independent and a fully joint analysis. There is no significant detection of dust in any of the four maps (the worst case being \SMICA\ with slightly more than a 1$\,\sigma$ deviation).\footnote{While no dust is detected in
the cleaned maps using the template determined
from the dust map, it should be pointed out that there is no guarantee that
the dust residuals (or negative dust residuals in the case of an
oversubtraction), after passing through the component-separation
pipelines, have exactly the same form as the original dust
bispectrum. However, it seems reasonable to assume that the resulting
shape would still be highly correlated with the original dust template, so
that this remains a meaningful test.}
Given the small size of the $\fnl^\mathrm{dust}$ errors, this non-detection
means that there is at most a few percent of dust contamination in the cleaned
maps (outside the mask). However, the errors of the local shape increase
significantly in the joint analysis because the dust and the local shapes are
quite correlated (more than 60\,\%).

\subsubsection{Impact of the tSZ effect}\label{sec:noSZ}

The \SMICA\ component-separation method also produces a foreground-cleaned
temperature map, ``\SMICA\ no-SZ,'' where, in addition to the usual foregrounds, contamination
by the thermal Sunyaev-Zeldovich (tSZ) effect has been subtracted as well \citep[see][]{planck2016-l04}. This allows us to test if the tSZ contamination has
any significant impact on our primordial results (we already saw in
Sect.~\ref{subsec:lensingISW} that it does seem to have an
impact on the determination of the lensing NG). This is an
important test, since there have been recent claims in the literature
\citep{Hill:2018ypf} that it might have an effect.

The results of the analysis can be found in Table~\ref{tab:fNLnoSZ}.
Because this effect is only important in temperature,
we restrict ourselves to a \itT-only
analysis. Results have been determined with the KSW estimator, with the
Binned estimator (this time using only 150 maps for the linear correction
and the error bars instead of 300, which is enough for the purposes of this
test), and with the Modal~1 estimator. The mask used is the
same as for the main analysis. The table also contains the difference with
the result determined from the normal \SMICA\ temperature map (without
tSZ removal) and the uncertainties on this difference.

We see a shift of about $1\,\sigma_{\Delta \fnl}$ (hence insignificant) in the local shape result,
which together with orthogonal is the shape predicted to be most contaminated
by the tSZ effect \citep[see][]{Hill:2018ypf}. We actually see a larger shift
in the equilateral result, which is supposed to be almost unaffected by this effect, while
the orthogonal shift is the largest, at more than $2\,\sigma_{\Delta \fnl}$ for all
estimators. However, such a marginal effect in the orthogonal shape without a
corresponding effect in the local shape leads us to conclude that we do not detect any
significant impact of contamination of the usual foreground-corrected maps by the tSZ
effect on our primordial NG results. In other words, while some tSZ contamination, 
peaking in the squeezed limit, is expected to be present in the standard 
temperature maps, this is too small to be clearly disentangled from the
statistical fluctuations in the $f_{\rm NL}$ results; in other words the tSZ contamination is not bigger than effects due to the
different processing of the data when tSZ is included in the foreground
components for the \SMICA\ analysis.
Furthermore, all shifts discussed here are much smaller
than the uncertainties on the $\fnl$ values themselves, due to the fact that 
the $f_{\rm NL}$ scatter between different cleaned maps ($\sigma_{\Delta f_{\rm NL}}$)
is much smaller than the $f_{\rm NL}$ error ($\sigma_{f_{\rm NL}}$).

Finally, it should also be pointed out that, using the same criteria, we cannot
strictly call the shift in $\fnl^\mathrm{lens}$ discussed in
Sect.~\ref{subsec:lensingISW} significant either. The observation that all
\itT\itT\itT\ measurements are below the expected value $\fnl^\mathrm{lens} = 1$
and that adding polarization systematically shifts them up, does point to some tSZ
contamination in \itT-only results for the lensing bispectrum.
However, the analysis discussed in this section finds again only a
$2\,\sigma_{\Delta \fnl}$ effect in this case; this is slightly larger or smaller than the
significance of the orthogonal $f_{\rm NL}$ shift, depending on the estimator.
Again, intrinsic statistical uncertainties make it hard to detect this
systematic effect.

\begin{table*}[htbp!]                 
\begingroup
\newdimen\tblskip \tblskip=5pt
\caption{Impact of the thermal Sunyaev-Zeldovich (tSZ) effect on $\fnl$ estimators for temperature data.  First three columns: results for the $f_{\rm NL}$ parameters of the primordial
local, equilateral, and orthogonal shapes, determined by the indicated
estimators from the \SMICA\ foreground-cleaned temperature
map with additional removal of the tSZ contamination.
Last three columns: difference with the result from the normal \SMICA\ map (see
Table~\ref{tab:fNLall}), where the error bar is the standard deviation of the
differences in $\fnl$ for the Gaussian FFP10 simulations when processed
through the two different \SMICA\ foreground-cleaning procedures.
Results have been determined using an independent single-shape analysis
and are reported with subtraction of the lensing bias; 
error bars are $68\,\%$ CL.}
\label{tab:fNLnoSZ}
\nointerlineskip
\vskip -3mm
\footnotesize
\setbox\tablebox=\vbox{
   \newdimen\digitwidth
   \setbox0=\hbox{\rm 0}
   \digitwidth=\wd0
   \catcode`*=\active
   \def*{\kern\digitwidth}
   \newdimen\signwidth
   \setbox0=\hbox{+}
   \signwidth=\wd0
   \catcode`!=\active
   \def!{\kern\signwidth}
\newdimen\dotwidth
\setbox0=\hbox{.}
\dotwidth=\wd0
\catcode`^=\active
\def^{\kern\dotwidth}
\halign{\hbox to 1.0 in{#\leaderfil}\tabskip 1em&
\hfil#\hfil\tabskip 1em& \hfil#\hfil& \hfil#\hfil\tabskip 2em&
\hfil#\hfil\tabskip 1em& \hfil#\hfil& \hfil#\hfil\tabskip 0pt\cr
\noalign{\doubleline}
\omit&\multispan3\hfil\SMICA\ no-SZ\hfil& \multispan3\hfil Difference\hfil\cr
\noalign{\vskip -4pt}
\omit&\multispan3\hrulefill& \multispan3\hrulefill\cr
\noalign{\vskip 2pt}
\omit\hfil Shape\hfil& !KSW& !Binned& !Modal~1& !KSW& !Binned& !Modal~1\cr
\noalign{\vskip 4pt\hrule\vskip 6pt}
Local& $!1.3\pm5.6$& $!1.1\pm5.9$& $!0.9\pm5.8$& $-1.8\pm1.4$& $-1.3\pm1.6$& $-1.5\pm1.3$\cr
Equilateral& $*-6\pm67$& $*-6\pm67$& $!*5\pm64$& $!14\pm14$& $!32\pm17$& $!19\pm14$\cr
Orthogonal& $-36\pm37$& $-31\pm42$& $-23\pm39$& $!21\pm8*$& $!20\pm9*$& $!19\pm7*$\cr
\noalign{\vskip 3pt\hrule\vskip 4pt}}}
\endPlancktablewide                 
\endgroup
\end{table*}                        

\subsection{Dependence on sky coverage}\label{sec:depmask}

The temperature and polarization mask we are using have been determined to be
the optimal masks for use on the
maps produced by the component-separation pipelines, according to criteria
explained in \cite{planck2016-l04}, and hence are used for all \Planck\ analyses
on those maps. However, the choice of mask can have an impact on the results
for $\fnl$. In the first place the sky fraction of the mask will have a direct
effect on the size of the uncertainties. Potentially more important, however, is
the effect the mask might have on the amount of foreground residuals. Hence we
judge it important to investigate the impact of the choice of mask on our
results by comparing results for several different masks. All tests are performed on
the \SMICA\ maps (with one exception, detailed below) using the Binned bispectrum
estimator, using 150 maps for the linear correction and the errors.

As a first test, performed on the temperature map only, we take the union
of the mask used in this paper ($f_\mathrm{sky}=0.78$) with the mask we used
in our 2015 analysis ($f_\mathrm{sky}=0.76$), leading to a mask that leaves
a fraction $f_\mathrm{sky}=0.72$ of the sky uncovered. The results for the
standard primordial shapes are given in Table~\ref{tab:fNLcombinedmask}, as
well as their differences with the results using the standard mask (see
Table~\ref{tab:fNLall}). The errors on the differences have been
determined from the scatter among 150 simulations when analysed with the
two different masks. This particular test is performed both on the \SMICA\ and the
\Commander\ maps, since its initial purpose was a further check on the discrepancy
between those two component-separation methods regarding the \itT-only orthogonal result, discussed
in detail in Sect.~\ref{sec:fnldiff}.
We see, however, that the orthogonal result is very stable for both methods, while the local result
shifts by about $1\,\sigma_{\Delta \fnl}$ and the equilateral result moves by about
$2\,\sigma_{\Delta \fnl}$,
which corresponds to about $(2/3)\,\sigma_{\fnl}$. Checking the 150 Gaussian
simulations used for determining the errors and the linear correction,
we see that there are 20 that have at least one fluctuation larger than
$2\,\sigma_{\Delta \fnl}$ in at least one of the three shapes, which corresponds
to a 13.3\,\% probability. This is large enough that we
consider the shift consistent with a statistical fluctuation. In any case,
all results remain consistent with zero. It is interesting to note that this
result for the equilateral shape is much closer to the one determined for the 2015 \Planck\ data.

\begin{table}[htbp!]                 
\begingroup
\newdimen\tblskip \tblskip=5pt
\caption{Tests of dependence on choice of mask.  The first column gives results for the $f_{\rm NL}$ parameters of the primordial
local, equilateral, and orthogonal shapes, determined using the Binned
estimator on the \SMICA\ and \Commander\ foreground-cleaned temperature
maps, using as a mask the union of the 2018 and 2015 common masks.
The second column gives the differences from the results using the 2018 common mask
(see Table~\ref{tab:fNLall}), where the
error is the standard deviation of the differences in $\fnl$ for 150
simulations when analysed with the two masks.
Results have been determined using an independent single-shape analysis
and are reported with subtraction of the lensing bias.}
\label{tab:fNLcombinedmask}
\nointerlineskip
\vskip -3mm
\footnotesize
\setbox\tablebox=\vbox{
   \newdimen\digitwidth
   \setbox0=\hbox{\rm 0}
   \digitwidth=\wd0
   \catcode`*=\active
   \def*{\kern\digitwidth}
   \newdimen\signwidth
   \setbox0=\hbox{+}
   \signwidth=\wd0
   \catcode`!=\active
   \def!{\kern\signwidth}
\newdimen\dotwidth
\setbox0=\hbox{.}
\dotwidth=\wd0
\catcode`^=\active
\def^{\kern\dotwidth}
\halign{\hbox to 1.0in{#\leaderfil}\tabskip 1em&
\hfil#\hfil\tabskip 2em&
\hfil#\hfil\tabskip 0pt\cr
\noalign{\doubleline}
\omit\hfil Shape\hfil& Combined mask& Difference\cr
\noalign{\vskip 3pt\hrule\vskip 5pt}
\multispan3\hfil \SMICA\ \itT\hfil\cr
\noalign{\vskip 2pt}
Local& $-1.8\pm6.0$& $-1.7\pm1.6$\cr
Equilateral& $-23\pm71$& $-49\pm24$\cr
Orthogonal& $-13\pm42$& $*-1\pm12$\cr
\noalign{\vskip 3pt\hrule\vskip 5pt}
\multispan3\hfil \Commander\ \itT\hfil\cr
\noalign{\vskip 2pt}
Local& $-3.2\pm6.0$& $-1.9\pm1.6$\cr
Equilateral& $-11\pm72$& $-42\pm24$\cr
Orthogonal& $!26\pm43$& $*-3\pm12$\cr
\noalign{\vskip 3pt\hrule\vskip 4pt}}}
\endPlancktable                    
\endgroup
\end{table}                        
\begin{table}[htbp!]                 
\begingroup
\newdimen\tblskip \tblskip=5pt
\caption{Tests of growing the polarization mask size.  The first column gives results for the $f_{\rm NL}$ parameters of the primordial
local, equilateral, and orthogonal shapes, determined using the Binned
estimator on the \SMICA\ foreground-cleaned maps. The standard temperature
mask is used, but the polarization mask has been enlarged by surrounding
every hole by a region either 20 or 40 pixels in width (``Extra~20''
or ``Extra~40'').
The second column gives the difference with the result using the 2018 common mask
(see Table~\ref{tab:fNLall}), where the
error is the standard deviation of the differences in $\fnl$ for 150
simulations when analysed with the two masks.
Results have been determined using an independent single-shape analysis
and are reported with subtraction of the lensing bias.}
\label{tab:fNLpolextramask}
\nointerlineskip
\vskip -3mm
\footnotesize
\setbox\tablebox=\vbox{
   \newdimen\digitwidth
   \setbox0=\hbox{\rm 0}
   \digitwidth=\wd0
   \catcode`*=\active
   \def*{\kern\digitwidth}
   \newdimen\signwidth
   \setbox0=\hbox{+}
   \signwidth=\wd0
   \catcode`!=\active
   \def!{\kern\signwidth}
\newdimen\dotwidth
\setbox0=\hbox{.}
\dotwidth=\wd0
\catcode`^=\active
\def^{\kern\dotwidth}
\halign{\hbox to 1.0in{#\leaderfil}\tabskip 1em&
\hfil#\hfil\tabskip 2em&
\hfil#\hfil\tabskip 0pt\cr
\noalign{\doubleline}
\omit\hfil Shape\hfil& ``Extra'' mask& Difference\cr
\noalign{\vskip 3pt\hrule\vskip 5pt}
\multispan3\hfil Extra~20, \itE\hfil\cr
\noalign{\vskip 2pt}
Local& $*!36\pm28*$& $*-12\pm17$\cr
Equilateral& $*!70\pm150$& $*-96\pm65$\cr
Orthogonal& $-180\pm91*$& $**-3\pm43$\cr
\noalign{\vskip 3pt\hrule\vskip 5pt}
\multispan3\hfil Extra~20, \itTpE\hfil\cr
\noalign{\vskip 2pt}
Local& $-5.0\pm5.2$& $*-2.4\pm1.8$\cr
Equilateral& $-31\pm48$& $*-12\pm13$\cr
Orthogonal& $-23\pm25$& $!*11\pm7*$\cr
\noalign{\vskip 3pt\hrule\vskip 5pt}
\multispan3\hfil Extra~40, \itE\hfil\cr
\noalign{\vskip 2pt}
Local& $!*41\pm33*$& $**-7\pm22$\cr
Equilateral& $!*16\pm170$& $-150\pm85$\cr
Orthogonal& $-160\pm100$& $!*20\pm61$\cr
\noalign{\vskip 3pt\hrule\vskip 5pt}
\multispan3\hfil Extra~40, \itTpE\hfil\cr
\noalign{\vskip 2pt}
Local& $-2.4\pm5.2$& $!*0.2\pm2.2$\cr
Equilateral& $-30\pm50$& $*-11\pm17$\cr
Orthogonal& $-35\pm26$& $**-1\pm10$\cr
\noalign{\vskip 3pt\hrule\vskip 4pt}}}
\endPlancktable                    
\endgroup
\end{table}                        
\begin{figure*}[htbp!]
\centering
\includegraphics[width=.32\linewidth]{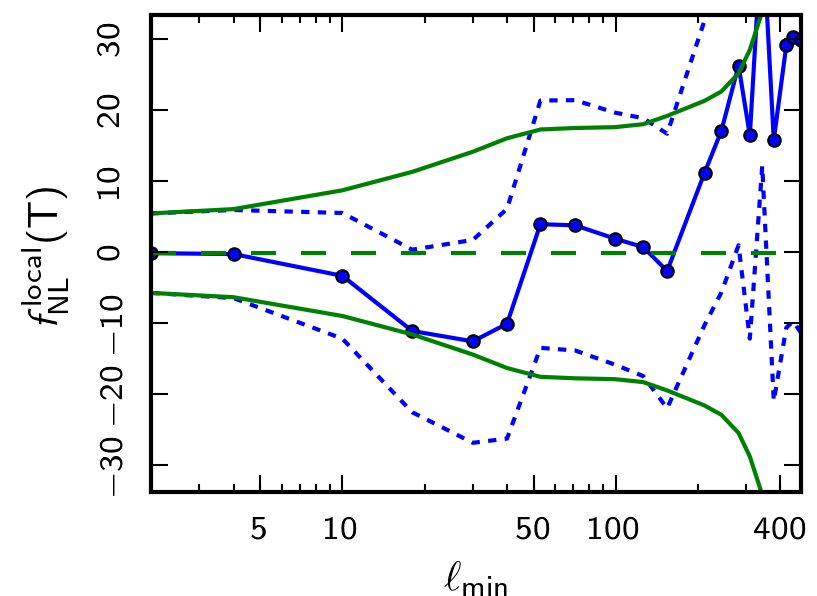}
\includegraphics[width=.32\linewidth]{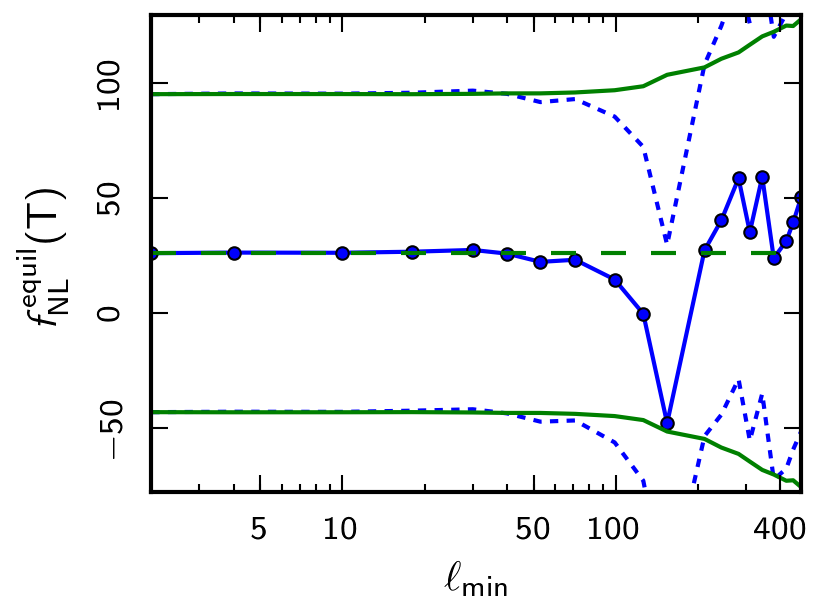}
\includegraphics[width=.32\linewidth]{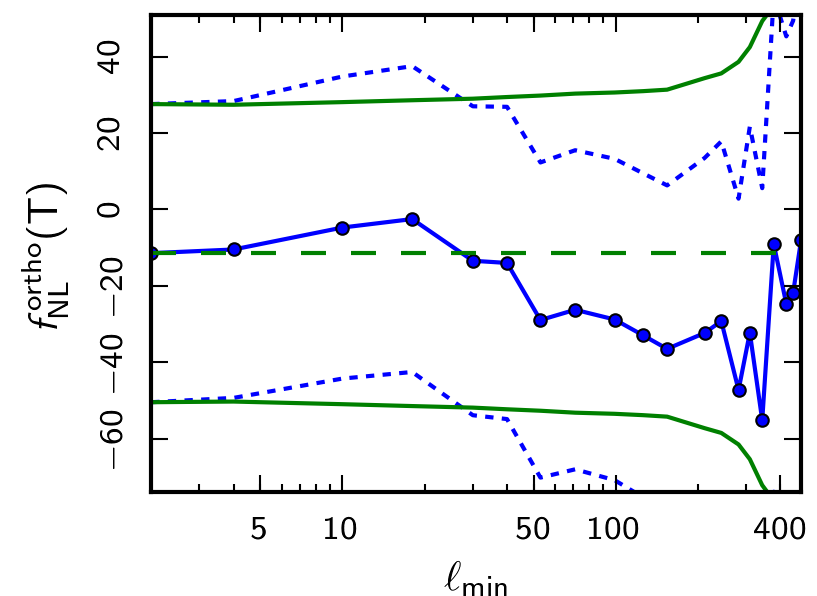}

\vspace{0.2cm}

\includegraphics[width=.32\linewidth]{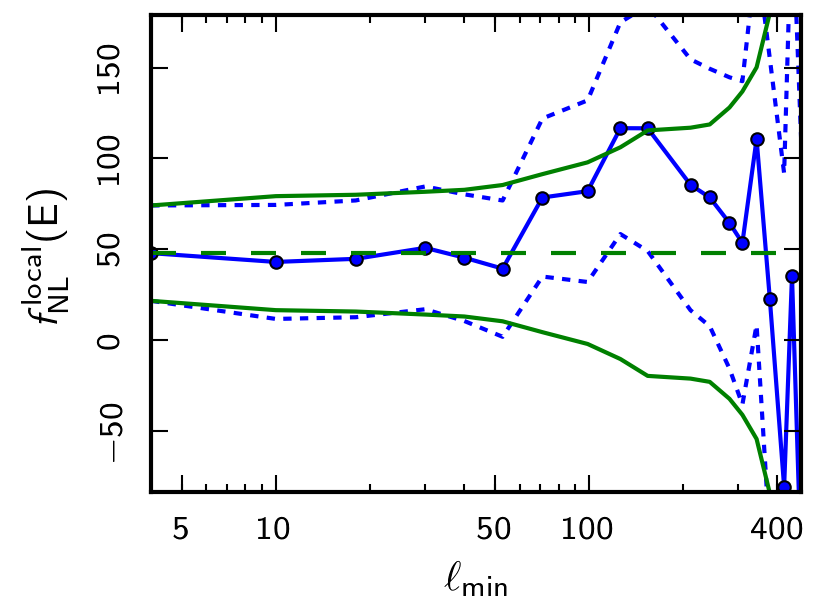}
\includegraphics[width=.32\linewidth]{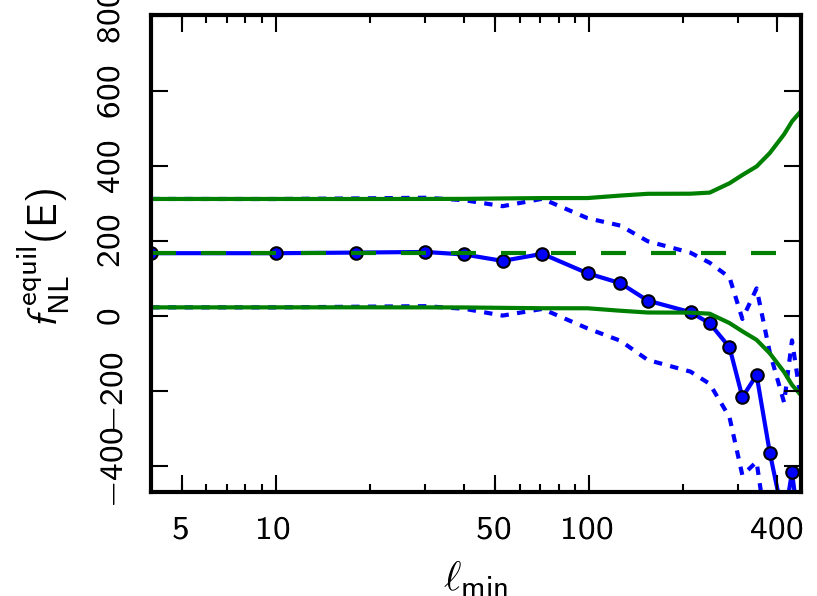}
\includegraphics[width=.32\linewidth]{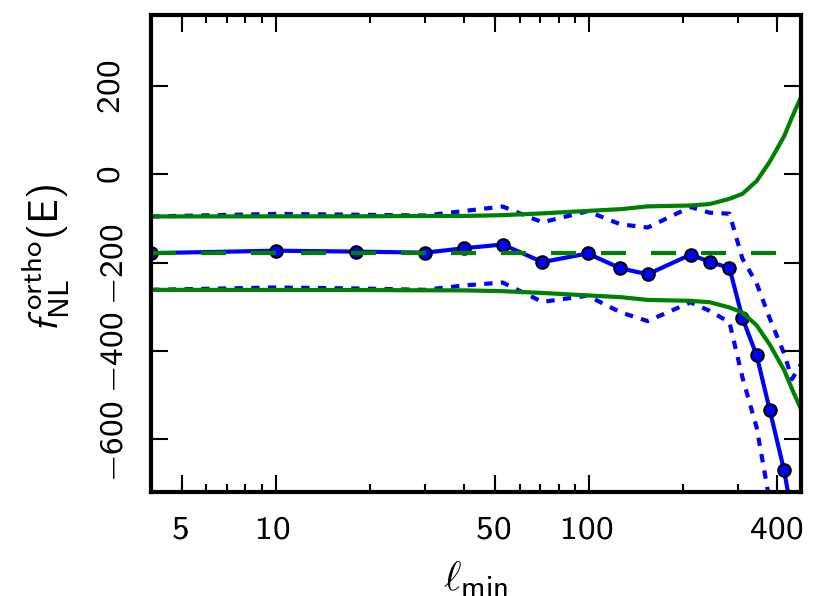}

\vspace{0.2cm}

\includegraphics[width=.32\linewidth]{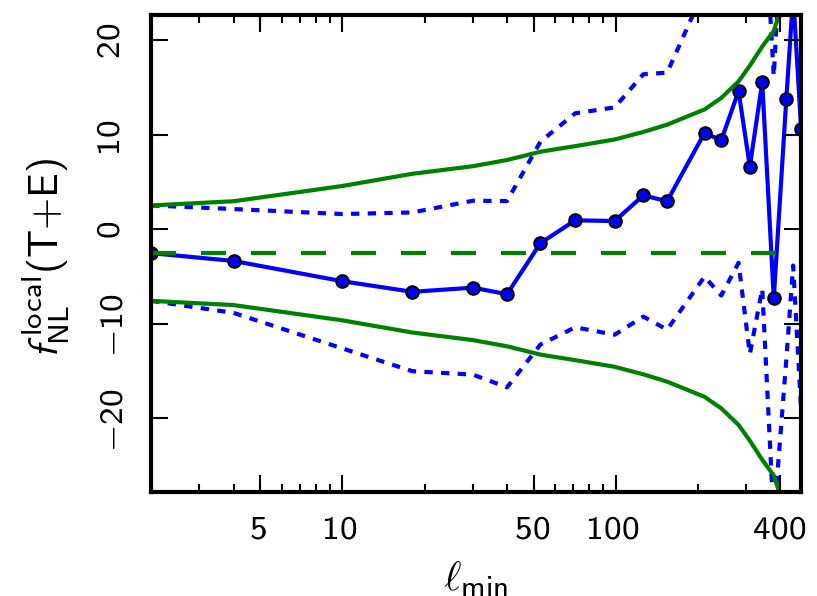}
\includegraphics[width=.32\linewidth]{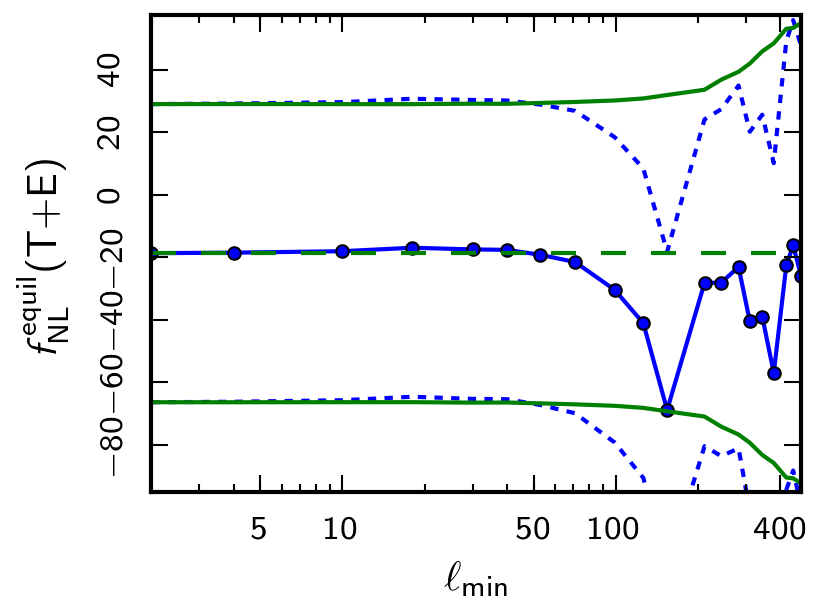}
\includegraphics[width=.32\linewidth]{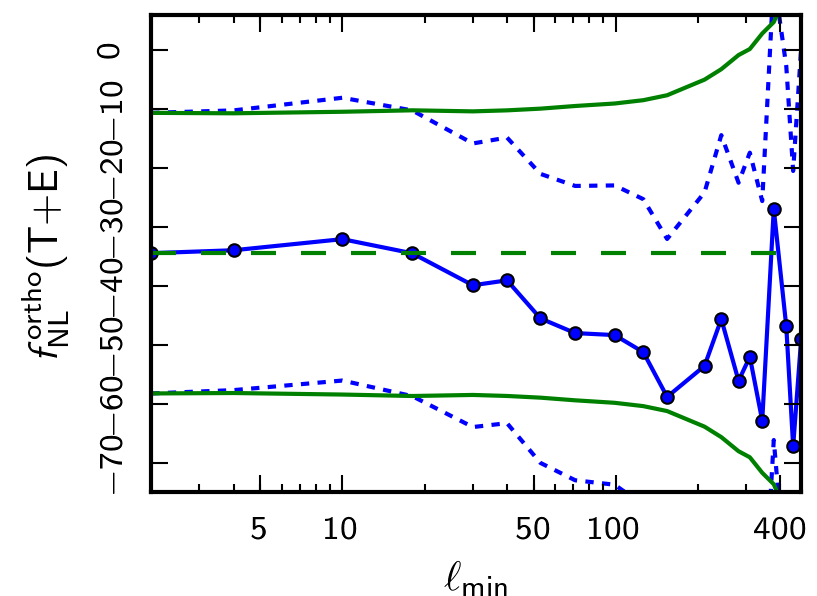}
\caption{Evolution of the $f_\mathrm{NL}$ parameters 
(solid blue line with data points) and their uncertainties (dotted blue lines) 
for the three primordial bispectrum templates as a function of the 
minimum multipole number $\ell_\mathrm{min}$ used in the analysis. From 
left to right the panels show local, 
equilateral, and orthogonal shape results, while the different rows from top to bottom
show results for \itT\ only, \itE\ only, and full \itTpE\ data.
To indicate more clearly the evolution of the uncertainties, they are also 
plotted around the final value of $f_\mathrm{NL}$ (solid green lines without 
data points, around the horizontal dashed green line).
The results here have been determined with the Binned bispectrum estimator for the \SMICA\ map, assume all shapes to be independent, and the
lensing bias has been subtracted.}
\label{Fig_lmindep}
\end{figure*}

\begin{figure*}[htbp!]
\centering
\includegraphics[width=.32\linewidth]{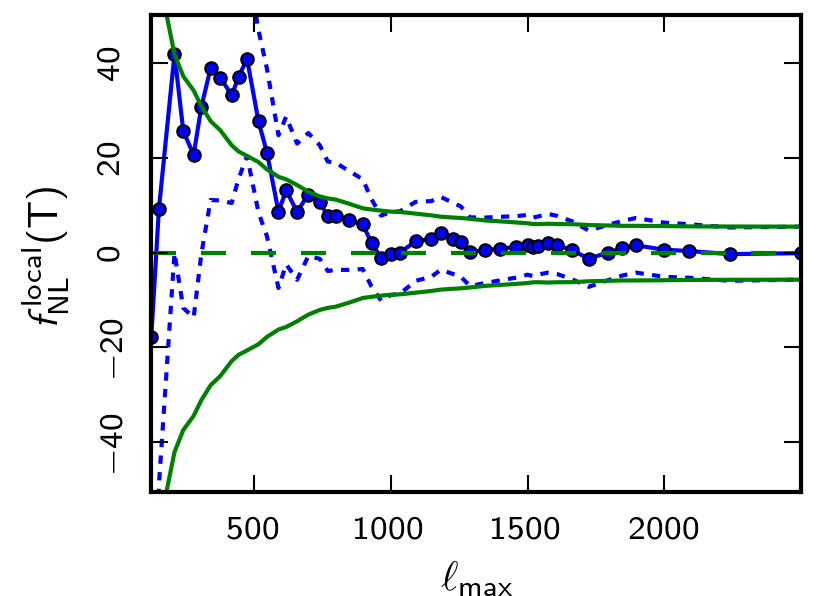}
\includegraphics[width=.32\linewidth]{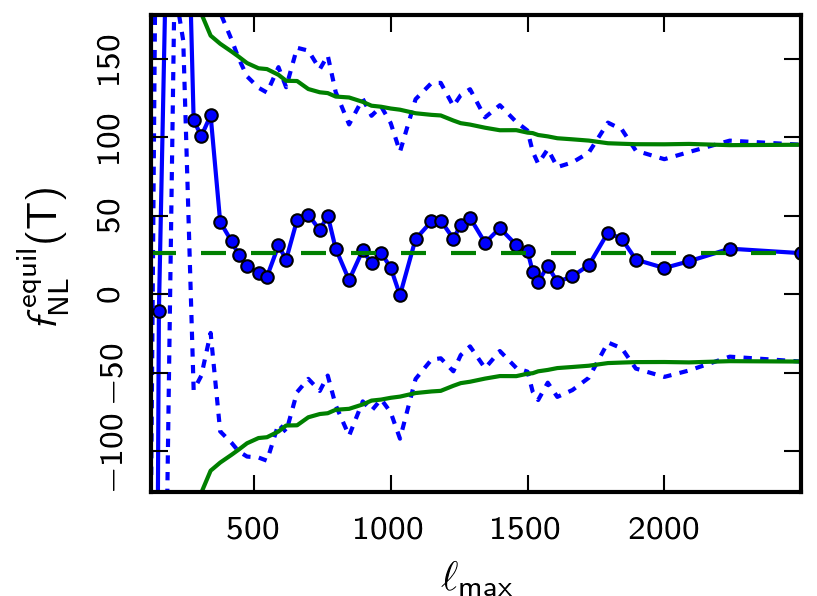}
\includegraphics[width=.32\linewidth]{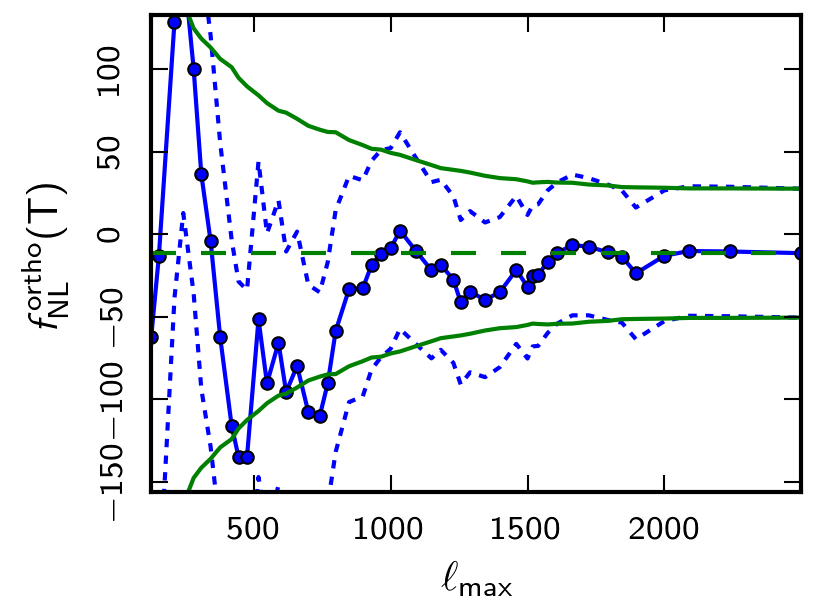}

\vspace{0.2cm}

\includegraphics[width=.32\linewidth]{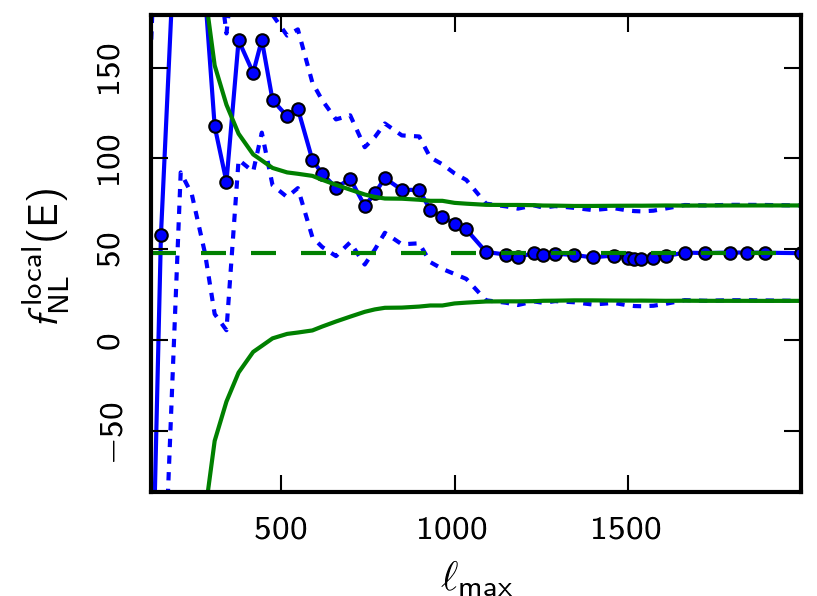}
\includegraphics[width=.32\linewidth]{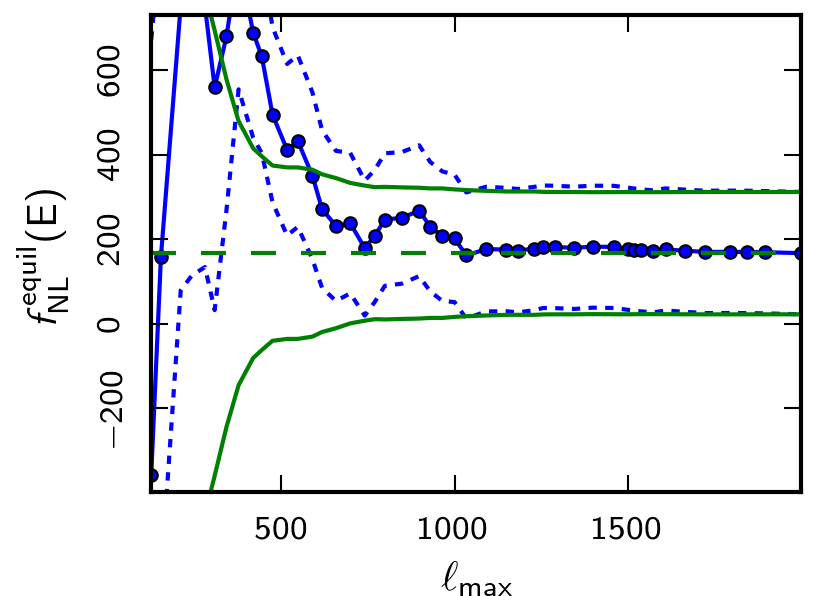}
\includegraphics[width=.32\linewidth]{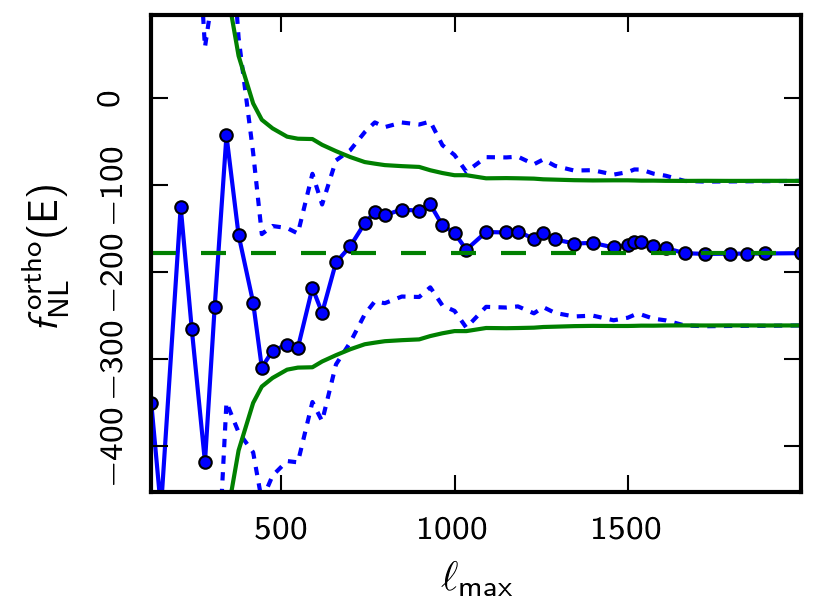}

\vspace{0.2cm}

\includegraphics[width=.32\linewidth]{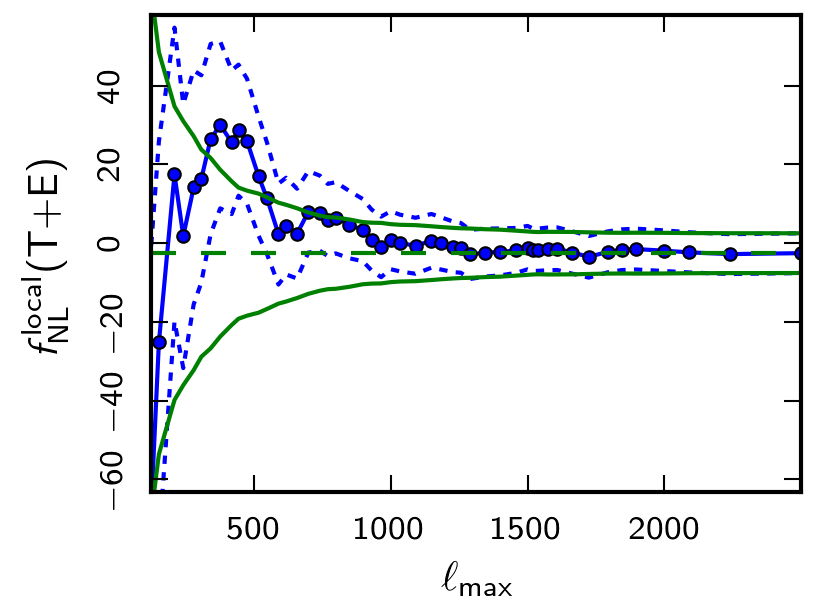}
\includegraphics[width=.32\linewidth]{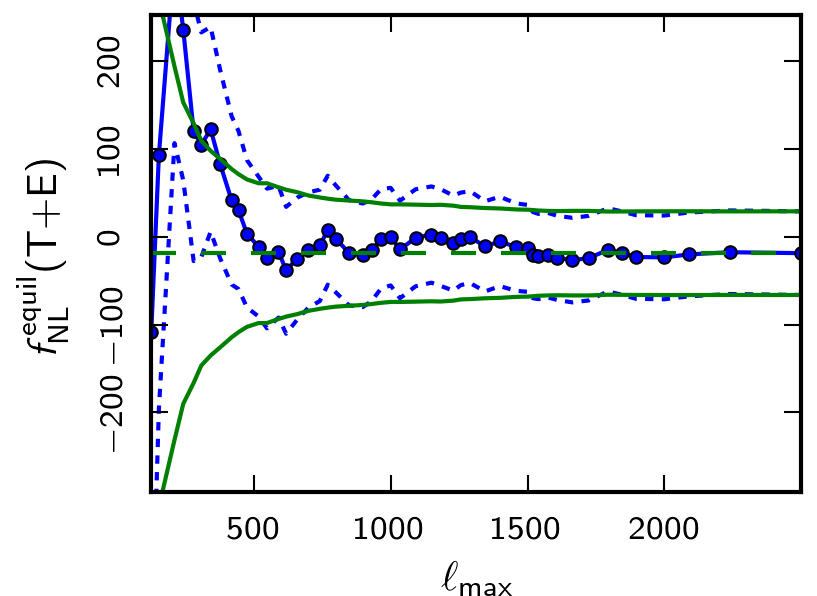}
\includegraphics[width=.32\linewidth]{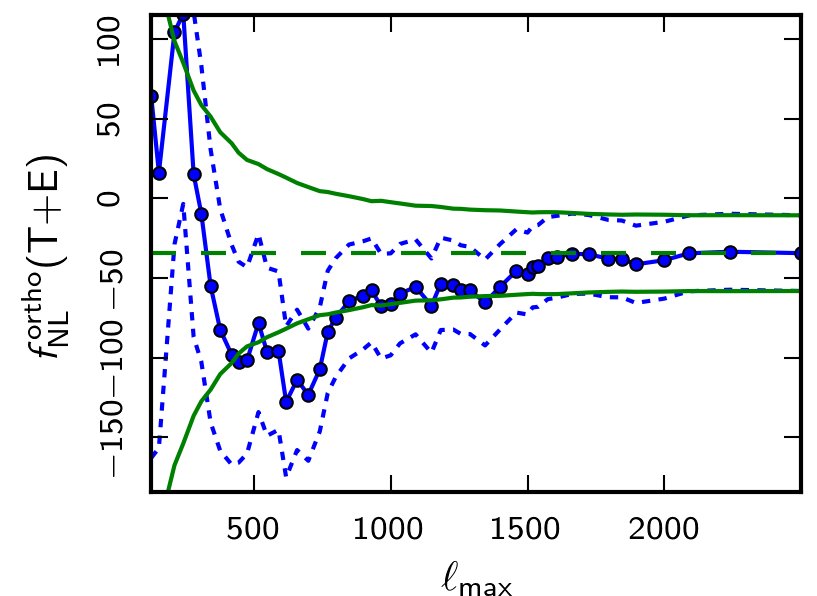}
\caption{Same as Fig.~\ref{Fig_lmindep}, but this time as a function of the 
maximum multipole number $\ell_\mathrm{max}$ used in the analysis.}
\label{Fig_lmaxdep}
\end{figure*}

As a second test,
to check the impact of the size of the polarization mask,
we return to the standard temperature mask, but this time
we change the polarization mask. It is altered as follows: each hole in the
common mask is grown by a region 20 pixels in width.
This reduces the sky fraction from 0.78 to 0.73. Results for this test can be found in
Table~\ref{tab:fNLpolextramask}, including the differences with the results
determined with the standard mask. We see that many results shift around somewhat, but nothing appears very significant (all less than
$1.5\,\sigma_{\Delta \fnl}$). The largest is again a $(2/3)\,\sigma_{\fnl}$ shift for
the \itE-only equilateral case, moving it closer to zero. This test was also
performed with the KSW and Modal~1 estimators, giving consistent (but not
identical) results. In particular, while some values shift a bit more, the shift
in the equilateral shape is smaller for both the KSW and Modal~1 estimators.
However, all estimators agree on the signs of the shifts.

Our third and final test of the effects of mask choice is very similar to the
previous one, except that this time the polarization mask is enlarged by an
additional 40 (instead of 20) pixels around every hole. This further reduces
$f_\mathrm{sky}$ to $0.66$. Results can also be found in
Table~\ref{tab:fNLpolextramask}, and we find similar conclusions as for
the previous test.

To summarize, while we see some effects on our $\fnl$ results when
considering different masks, none of these appear to be significant
(further reinforced
by the fact that the shifts vary somewhat between estimators), nor do they
change the conclusions of our paper.

\subsection{Dependence on multipole number}\label{sec:deplminlmax}

As in previous releases we also test the dependence of the results for $\fnl$
on the choice of $\ell_\mathrm{min}$ and $\ell_\mathrm{max}$ used in the
analysis. This test is most easily performed using the Binned estimator.
Results are shown in Fig.~\ref{Fig_lmindep} for the dependence on
$\ell_\mathrm{min}$ and in Fig.~\ref{Fig_lmaxdep} for the dependence on
$\ell_\mathrm{max}$.

Considering first Fig.~\ref{Fig_lmindep}, for $\ell_\mathrm{min}$, the plots
look very similar to
the ones in the paper investigating the 2015 \Planck\ data \citepalias{planck2014-a19}, with increased stability
for the \itE-only local results. As explained in Sect.~\ref{sec:settings}, it was
decided to use $\ell_\mathrm{min}=4$ for the polarization maps, since there
was an issue with bias and increased variance when the two lowest multipoles for
\itE\ were included. However, this still allows us to use many more low-$\ell$
polarization modes than for the 2015 data, when it was necessary to adopt
$\ell_\mathrm{min}=40$ for polarization.

Turning to Fig.~\ref{Fig_lmaxdep}, for $\ell_\mathrm{max}$, we also notice
good agreement with the analysis of the 2015 data, and slightly
more stable results (e.g., for $T$-only equilateral). 
We see that the \itT-only results have stabilized by
$\ell=2000$ and the \itE-only results by $\ell=1500$, so that there is no
problem with the KSW and Modal estimators using these lower values for
$\ell_\mathrm{max}$. As before we also confirm the ``\WMAP\ excess'' for the
local shape at $\ell\approx 500$ \citep{2012arXiv1212.5225B}, even more
clearly than in \citetalias{planck2014-a19}.
However, this does not appear to be a major outlier when considering all
values of $\ell_{\rm max}$ and all possible shapes.

\subsection{Summary of validation tests}

Throughout this section we have discussed a set of tests aimed at evaluating 
the robustness of our results. For convenience, we summarize here our main findings.
\begin{itemize}
\item[--] We find good consistency for $f_{\rm NL}$ local, equilateral, and orthogonal 
measurements, between all component-separation methods and with all bispectrum estimators, separately considering \itT-only, \itE-only, and \itTpE\ results.
The agreement for \itE-only results has significantly improved with respect to the previous release.
\item[--] The possible exception 
is provided by the orthogonal \itT-only estimates. In this case, a comparison with simulations 
shows significant differences in specific cases, namely \SMICA\ -- \Commander\ and, to a lesser extent, 
\SMICA\ -- \SEVEM. However, these differences are coming from fluctuations in $f_{\rm NL}$ for 
\Commander\ and \SEVEM\ with respect to the two previous \Planck\ releases, when all temperature maps were 
always in excellent agreement. 
Our main \SMICA\ results are, on the other hand, completely stable. Moreover, the discrepancy becomes much less significant when adding 
polarization, it is limited to one specific shape, and it is in any case at a level that does not alter the 
interpretation of the results. Therefore, in the end, we do not consider this issue to be problematic.
\item[--] Nevertheless, in light of this test, we are led to a slight preference for \SMICA\ and \NILC\ as methods of choice for primordial NG analysis.
Given that \SMICA\ also gave a slightly better performance for NG analysis in the two previous releases, and that results extracted from 
\SMICA\ maps have been quite stable over time, we maintain \SMICA\ as our final choice, as already justified in detail in the papers analysing the 2013 and 2015 \Planck\ data \citepalias{planck2013-p09a,planck2014-a19}.
\item[--] We find very good consistency between fully reconstructed bispectra, for different component-separation methods, using both
the Modal and the Binned approaches. Polarization-only bispectra again show a large improvement compared to the previous analysis of \Planck\ data.
\item[--] The observed noise mismatch between the data and the FFP10 simulations does 
not seem to impact our results. Our cubic statistics cannot be biased by this mismatch, and tests on simulations 
show that the effect on $f_{\rm NL}$ errors is negligible.
\item[--] Results are stable to changes in sky coverage and different cuts in the multipole domain. Restricting the analysis
to the multipole range probed by \WMAP\ shows good agreement between \WMAP\ and \Planck. 
\item[--] We find no sign of any residual Galactic thermal dust contamination
in the \Planck\ component-separated CMB maps.
\item[--] Contamination from the thermal SZ effect on the standard primordial and lensing \itT-only results is not at a significant level, compared 
to statistical errors.
\end{itemize}
Overall, the results display a high level of internal consistency, and are notably characterized by a large improvement in the quality of polarization-only bispectra 
with respect to the previous release. Whereas in 2015 we cautioned the reader to take polarization-based $f_{\rm NL}$ estimates as preliminary, 
we can now state that our \itTpE-based constraints are fully robust. This is one of the main conclusions of this paper.

\section{Limits on the primordial trispectrum}
\label{sec:tau_gnl}



We will now present constraints on three shapes for the primordial
four-point function or trispectrum, denoted $\gnlloc$, $\gnldotpi4$, 
and $\gnldpi4$, and described below.

The details of the analysis are mostly unchanged from the paper analysing the 2015 \Planck\ data (see section~9 of \citetalias{planck2014-a19}).
Nevertheless, we briefly review the four-point analysis here; for
more details, see \citetalias{planck2014-a19}, or \cite{long_trispectrum},
which contains technical details of the pipeline.

First, we describe the three different signals of interest.
The local-type trispectrum $\gnlloc$ arises if the initial adiabatic
curvature $\zeta$ is given by the following non-Gaussian model:
\be
\zeta(\vx) = \zeta_{\rm G}(\vx) + \frac{9}{25}\, \gnlloc \,\zeta_{\rm G}(\vx)^3
\ee
where $\zeta_{\rm G}$ is a Gaussian field.
The trispectrum in the local model is given by
\be
\langle \zeta_{\vk_1} \zeta_{\vk_2} \zeta_{\vk_3} \zeta_{\vk_4} \rangle'
  = \frac{54}{25}\, \gnlloc \left[ P_\zeta(k_1) P_\zeta(k_2) P_\zeta(k_3) + \mbox{3 perms.} \right].
\label{eq:gnlloc}
\ee
In this equation and throughout this section, a ``primed'' four-point
function denotes the four-point function without its momentum-conserving
delta function, i.e.,
\be
\langle \zeta_{\vk_1} \zeta_{\vk_2} \zeta_{\vk_3} \zeta_{\vk_4} \rangle
  = 
\langle \zeta_{\vk_1} \zeta_{\vk_2} \zeta_{\vk_3} \zeta_{\vk_4} \rangle'
  (2\pi)^3 \delta^{(3)}\Big( \sum\vk_i \Big) + \mbox{disc.},
\ee
where ``+\ disc.'' denotes disconnected contributions to the 4-point function.
It can be shown that the local-type trispectrum in Eq.~(\ref{eq:gnlloc})
is always negligibly small in single-field inflation \citep{Senatore:2012wy};
however, it can be large in 
multifield models of inflation in which a large bispectrum is forbidden by 
symmetry \citep{Senatore:2010wk}.

The next two shapes $\gnldotpi4$, $\gnldpi4$ are generated by the operators
$\dot\sigma^4$ and $(\partial_i\sigma)^2 (\partial_j\sigma)^2$ in the effective field
theory (EFT) of inflation \citep{2010JCAP...09..035B,Senatore:2010wk,long_trispectrum}.
For data analysis purposes, they can be defined by the following trispectra:
\begin{align}
\langle \zeta_{\vk_1} \zeta_{\vk_2} \zeta_{\vk_3} \zeta_{\vk_4} \rangle'
   =\ & \frac{9216}{25} \gnldotpi4 A_\zeta^3 \int_{-\infty}^0 d\tau_E^{\vphantom{4}} \, \tau_E^4 
              \left( \prod_{i=1}^4 \frac{e^{k_i\tau_E}}{k_i} \right)  \nn \\
   =\ & \frac{221184}{25} \gnldotpi4\, A_\zeta^3
          \frac{1}{k_1k_2k_3k_4 K^5}\,;
\label{eq:gnldotpi4}
\end{align}
\begin{align}
\langle \zeta_{\vk_1} \zeta_{\vk_2} \zeta_{\vk_3} \zeta_{\vk_4} \rangle'
   =\ & \frac{82944}{2575} \gnldpi4 A_\zeta^3 \int_{-\infty}^0 d\tau_E \nn
              \left[ \prod_{i=1}^4 \frac{(1-k_i\tau_E)e^{k_i\tau_E}}{k_i^3} \right] \nn \\
& \times\,\left[ (\vk_1\cdot\vk_2)(\vk_3\cdot\vk_4) + \mbox{2 perms.} \right]
  \nn \\
  =\ & 
    \frac{165888}{2575}  \gnldpi4 A_\zeta^3 \nn \\
& \times\,
        \left( \frac{2K^4 - 2K^2\sum k_i^2 + K \sum k_i^3 + 12 k_1 k_2 k_3 k_4}{k_1^3 k_2^3 k_3^3 k_4^3 K^5} \right) \nn \\
& \times\,\left[ (\vk_1\cdot\vk_2)(\vk_3\cdot\vk_4) + \mbox{2 perms.} \right].
\label{eq:gnldpi4}
\end {align}
Here $K=\sum_i k_i$, and numerical prefactors
have been chosen so that the trispectra have the same normalization as the
local shape in Eq.~(\ref{eq:gnlloc}) when restricted to tetrahedral
wavenumber configurations with $|\vk_i| = k$ and $(\vk_i\cdot\vk_j) = -k^2/3$.
We mention in advance that there is another shape $\gnlB$ that arises
at the same order in the EFT expansion, to be discussed at the end of
this section.

In Eqs.~(\ref{eq:gnldotpi4}) and (\ref{eq:gnldpi4}), we have written
each trispectrum in two algebraically equivalent ways, an integral
representation and an ``integrated'' form.
The integral representation arises naturally when evaluating the
Feynman diagram for the EFT operator.
It also turns out to be useful for data analysis, since the resulting
``factorizable'' representation for the trispectrum leads to an efficient
algorithm for evaluating the CMB trispectrum estimator
\citep{planck2014-a19,long_trispectrum}.

For simplicity in Eqs.~(\ref{eq:gnldotpi4}) and (\ref{eq:gnldpi4}), we
have assumed a scale-invariant power spectrum $P_\zeta(k) = A_\zeta/k^3$.
For the \Planck\ analysis, we slightly modify these trispectra to a
power-law spectrum $P_\zeta(k) \propto k^{n_{\rm s}-4}$, as described in
appendix~C of \cite{long_trispectrum}.

To estimate each $\gnl$ parameter from \Planck\ data,
we use the ``pure-MC'' trispectrum estimation pipeline
from section~IX.B of \cite{long_trispectrum}.
In this pipeline, the data are specified as a filtered
harmonic space map $\tilde d_{\ell m}$, and its covariance
is characterized via a set of 1000 filtered signal + noise
simulations.

The ``filter'' is an experiment-specific linear operation
whose input consists of one or more pixel-space maps,
and whose output is a single harmonic-space map, $d_{\ell m}$.
In the \Planck\ trispectrum analysis, we define the filter as follows.
First, we take the single-frequency pixel-space maps,
and combine them to obtain a single component-separated
pixel-space map, using one of the component-separation
algorithms, \SMICA, \SEVEM, \NILC, 
\Commander, or \SMICA\ no-SZ (a variant of
the \SMICA\ algorithm that guarantees zero response to
Compton-$y$ sources, at the expense of slightly higher
noise; see Sect.~\ref{sec:noSZ}).
Second, we subtract the best-fit monopole and dipole,
inpaint masked point sources, and apodize the Galactic
plane boundary.  The details of these steps are unchanged from the 2015
\Planck\ data analysis, and are described in section~9.1 of \citetalias{planck2014-a19}.
Third, we take the spherical transform of the pixel-space
map out to $l_{\rm max}=1600$, obtaining a harmonic-space map $d_{\ell m}$.
Finally, we define the filtered map $\tilde d_{\ell m}$ by
applying the multiplicative factor:
\be
\tilde d_{\ell m} = \frac{d_{\ell m}}{b_\ell C_\ell + b_\ell^{-1} N_\ell} .
\ee
This sequence of steps defines a linear operation, whose
input is a set of single-frequency pixel-space maps, and
whose output is a filtered harmonic-space map $\tilde d_{\ell m}$.
This filtering operation is used as a building block in
the trispectrum pipeline described in \citet{long_trispectrum},
and is the only part of the pipeline that is \Planck-specific.

The results of the analysis, for all three trispectrum shapes
and five different component-separation algorithms, are presented
in Table~\ref{tab:gnl}.
We do not find evidence for a nonzero primordial trispectrum.

\begin{table*}[htbp!]                 
\begingroup
\newdimen\tblskip \tblskip=5pt
\caption{\Planck\ 2018 constraints on the trispectrum parameters $\gnlloc$,
  $\gnldotpi4$, and $\gnldpi4$ from different component-separated maps.}
\label{tab:gnl}
\nointerlineskip
\vskip -3mm
\footnotesize
\setbox\tablebox=\vbox{
   \newdimen\digitwidth
   \setbox0=\hbox{\rm 0}
   \digitwidth=\wd0
   \catcode`*=\active
   \def*{\kern\digitwidth}
   \newdimen\signwidth
   \setbox0=\hbox{+}
   \signwidth=\wd0
   \catcode`!=\active
   \def!{\kern\signwidth}
\newdimen\dotwidth
\setbox0=\hbox{.}
\dotwidth=\wd0
\catcode`?=\active
\def?{\kern\dotwidth}
\halign{\hbox to 1.25in{#\leaderfil}\tabskip1em&
\hfil#\hfil\tabskip 2em&
\hfil#\hfil\tabskip 2em&
\hfil#\hfil\tabskip 0pt\cr
\noalign{\doubleline}
\noalign{\vskip -2pt}
\omit& $\gnlloc$& $\gnldotpi4$& $\gnldpi4$\cr
\noalign{\vskip 3pt\hrule\vskip 5pt}
   \SMICA&        $*(-5.8\pm6.5)\times10^4$& $*(-0.8\pm1.9)\times10^6$& $*(-3.9\pm3.9)\times10^5$\cr
   \SMICA\ no-SZ& $(-12.3\pm6.6)\times10^4$& $*(-0.6\pm1.9)\times10^6$& $*(-3.5\pm3.9)\times10^5$\cr
   \SEVEM&        $*(-5.5\pm6.5)\times10^4$& $*(-0.8\pm1.9)\times10^6$& $*(-3.2\pm3.9)\times10^5$\cr
   \NILC&         $*(-3.6\pm6.3)\times10^4$& $*(-0.8\pm1.9)\times10^6$& $*(-4.0\pm3.9)\times10^5$\cr
   \Commander&    $*(-8.1\pm6.5)\times10^4$& $*(-0.8\pm1.9)\times10^6$& $*(-3.5\pm3.9)\times10^5$\cr
\noalign{\vskip 3pt\hrule\vskip 4pt}}}
\endPlancktablewide
\endgroup
\end{table*}

Each entry in Table~\ref{tab:gnl} is a constraint on a single
$\gnl$-parameter with the others held fixed.
We next consider joint constraints involving multiple $\gnl$-parameters.
In this case, we need to know the covariance matrix between
$\gnl$-parameters.  We find that
\be
\mbox{Corr}(\gnldotpi4, \gnldpi4) = 0.61  \label{eq:gnl_correlation},
\ee
and the correlation between $\gnlloc$ and the other two $\gnl$-parameters is negligible.

Multifield models of inflation will generally predict a linear combination
of the quartic operators $\dot\sigma^4$ and $(\partial_i\sigma)^2 (\partial_j\sigma)^2$.
In addition, there is a third operator $\dot\sigma^2 (\partial_i\sigma)^2$ that arises
at the same order in the EFT expansion.
For completeness, its trispectrum is given by
\begin{align}
\langle \zeta_{\vk_1} \zeta_{\vk_2} \zeta_{\vk_3} \zeta_{\vk_4} \rangle'
  &= -\frac{13824}{325} \gnlB A_\zeta^3 \int_{-\infty}^0 d\tau_E^{\vphantom{2}} \, \tau_E^2 \qquad\qquad\qquad \nn \\
\noalign{$\qquad\qquad \times \,
        \bigg[ \frac{\displaystyle{(1-k_3\tau_E)(1-k_4\tau_E)}}{\displaystyle{k_1k_2k_3^3k_4^3}} (\vk_3\cdot\vk_4)\, e^{\sum k_i\tau_E}
+ \mbox{5 perms.} \bigg]$}
\noalign{\vskip 5pt}
  &= 
    -\frac{27648}{325} 
     \gnlB A_\zeta^3 \nn \\
\noalign{$\qquad\qquad \times \,
        \bigg[ \frac{\displaystyle{K^2 + 3(k_3+k_4)K + 12k_3k_4}}{\displaystyle{k_1 k_2 k_3^3 k_4^3 K^5}} (\vk_3\cdot\vk_4) + \mbox{5 perms.} \bigg].$}
   \label{eq:gnlB}
\end{align}
However, a Fisher matrix analysis shows that this trispectrum is nearly 100\,\% correlated
with the $\dot\sigma^4$ and $(\partial_i\sigma)^2 (\partial_j\sigma)^2$ trispectra.
If the parameter $\gnlB$ is non-zero, then we can absorb it into the ``effective'' values
of the parameters $\gnldotpi4$ and $\gnldpi4$ as
\begin{align}
(\gnldotpi4)_{\rm eff} &= 0.59\, \gnlB, \nn \\
(\gnldpi4)_{\rm eff} &= 0.091\, \gnlB.  \label{eq:gnlB_eff}
\end{align}
Therefore, to study joint constraints involving multiple $\gnl$ parameters,
it suffices to consider the two parameters, $\gnldotpi4$ and $\gnldpi4$,
with correlation coefficient given in Eq.~(\ref{eq:gnl_correlation}).

We define the two-component parameter vector
\be
g_i = \left( \begin{array}{c}
  \gnldotpi4  \\
  \gnldpi4
\end{array} \right).
\ee
and let $\hat g_i$ denote the two-component vector of single-$\gnl$
estimates from the \SMICA\ maps (Table~\ref{tab:gnl}):
\be
\hat g_i = \left( \begin{array}{c}
  -8.0 \\
  -3.9
\end{array} \right) \times 10^{-5}.
\ee
We also define a two-by-two Fisher matrix $F_{ij}$, whose diagonal is given by $F_{ii} = 1/\sigma_i^2$,
where $\sigma_i$ is the single-$\gnl$ statistical error in Table~\ref{tab:gnl}, and whose off-diagonal is
$F_{12} = r F_{11}^{1/2} F_{22}^{1/2}$, where $r$ is the correlation in Eq.~(\ref{eq:gnl_correlation}).
This gives:
\be
F_{ij} = \left( \begin{array}{cc}
  2.8  & \phantom{0}8.3  \\
  8.3 &  66.5
\end{array} \right) \times 10^{-13}.
\ee
Now, given a set of ``theory'' $\gnl$ values, represented by a two-vector $g_i$,
we compare to the \Planck\ data by computing the following quantity for the trispectrum:
\be
\chi^2(g_i) = \bigl[F_{ii} \hat g_i - (Fg)_i\ \bigr] \, F_{ij}^{-1} \, \bigl[F_{jj} \hat g_j - (Fg)_j\bigr]\label{eq:trispectrum_chi2}.
\ee

In a model-building context where the $g_{\rm NL}$ quantities $g_i$ depend on
model parameters, confidence regions on model parameters can be obtained by
appropriately thresholding $\chi^2$.
We give some examples in Sect.~\ref{sec:Implications}.

\section{Implications for early-Universe physics}
\label{sec:Implications}

We now want to convert constraints on primordial NG into constraints on parameters of various models of inflation. This allows us to highlight the constraining power of NG measurements, as an additional complementary observable beyond the CMB power spectra. In particular NG constraints can severely limit the parameter space of models that are alternatives to the standard single-field models of slow-roll inflation, since they typically feature a higher level of NG. 

Unless stated otherwise, we follow the same procedures adopted in \citetalias{planck2013-p09a} and \citetalias{planck2014-a19}. A posterior of the model parameters is built based on the following steps: we start from the assumption that the sampling distribution is Gaussian (which is supported by Gaussian simulations); the likelihood is approximated by the sampling distribution, but centred on the NG estimate (see \citealt{2009ApJS..184..264E}); we use uniform or Jeffreys' priors, over intervals of the model parameter space that are physically meaningful (or as otherwise stated); and in some cases where two or more parameters are involved, we marginalize the posterior to provide one-dimensional limits on the parameter under consideration.


\subsection{General single-field models of inflation}

\paragraph{DBI models}
Dirac-Born-Infeld (DBI)
models of inflation (\citealt{2004PhRvD..70j3505S,2004PhRvD..70l3505A}) arise from high-energy string-theory constructions and generate a nonlinearity parameter $f_{\rm NL}^{\rm DBI}=-(35/108) (c_\mathrm{s}^{-2}-1)$, where $c_\mathrm{s}$ is the sound speed of the inflaton perturbations (\citealt{2004PhRvD..70j3505S,2004PhRvD..70l3505A,2007JCAP...01..002C}). The enhancement of the NG amplitude due to a possible sound speed 
$c_\mathrm{s} < 1$ arises from a non-standard kinetic term of the inflaton field. Notice that we have constrained the exact theoretical (non-separable) shape (see equation~7 of \citetalias{planck2013-p09a}), even though it is very similar to the equilateral type. Our constraint $f_{\rm NL}^{\rm DBI}=46 \pm 58 $ from temperature data ($f_{\rm NL}^{\rm DBI}=14 \pm 38 $ from temperature and polarization) at 68\,\%~CL (with lensing and point sources subtracted, see Table~\ref{tab:equilmodels}) implies 
\begin{align}
\label{csDBIT}
c_\mathrm{s}^{\rm DBI} \geq\ & 0.079 \quad\quad \text{(95\,\%, \itT\ only)}\, ,\\
\noalign{\noindent and}
\label{csDBITE}
c_\mathrm{s}^{\rm DBI} \geq\ & 0.086 \quad\quad \text{(95\,\%, \textit{T+E})}\, .
\end{align}

\paragraph{Implications for the effective field theory of inflation}
 \label{Implications_EFT}
Now we can update CMB limits on the speed of sound $c_\mathrm{s}$ at which inflaton fluctuations propagated in the very early Universe. A very general constraint on this inflationary parameter can be obtained by employing the EFT approach to inflation \citep[][and see Sect.~\ref{sec:tau_gnl}]{2008JHEP...03..014C,2008PhRvD..77l3541W}. This approach allows us to obtain predictions for the parameter space of primordial NG through a general characterization of the inflaton field interactions. The Lagrangian of the system is expanded into the dominant operators that respect some underlying symmetries. The procedure thus determines a unifying scheme for classes of models featuring deviations from single-field slow-roll inflation. Typically the equilateral and orthogonal templates represent an accurate basis to describe the full parameter space of EFT single-field models of inflation, and therefore we will use the constraints on $f_{\rm NL}^{\rm equil}$ and $f_{\rm NL}^{\rm ortho}$.

As a concrete example, let us consider the Lagrangian of general single-field models of inflation (of the form $P(X,\varphi)$ models, where $X=g^{\mu \nu} \partial_\mu \phi \, \partial_\nu \phi$) written with the EFT approach: 
\begin{align}
 S&=\int d^4x \sqrt{-g} \left[\,-\,\frac{M^2_{\rm Pl} \dot{H}}{c_\mathrm{s}^2} \left( \dot{\pi}^2-c_\mathrm{s}^2 \frac{(\partial_i \pi)^2}{a^2} \right) \right. \nn \\
 &\left. - M_{\rm Pl}^2 \dot{H} (1-c_\mathrm{s}^{-2}) \dot{\pi} \frac{(\partial_i \pi)^2}{a^2}
  + \left( M_{\rm Pl}^2 \dot{H} (1-c_\mathrm{s}^{-2}) -\frac{4}{3} M_3^4 \right) \dot{\pi}^3 \right]\, .
\end{align}
The scalar perturbation $\pi$ generates the curvature perturbation $\zeta=- H \pi$. In this case there are two relevant inflaton interactions,
 $\dot{\pi} (\partial_i \pi)^2$ and $(\dot{\pi})^3$, producing two specific bispectra with amplitudes $f_{\rm NL}^{\rm EFT1}=-(85/324)(c_\mathrm{s}^{-2}-1)$ and $f_{\rm NL}^{\rm EFT2}=-(10/243)(c_\mathrm{s}^{-2}-1) \[\tilde{c}_3+(3/2) c_\mathrm{s}^2\]$, respectively.
 Here $M_3$ is the amplitude of the operator $\dot{\pi}^3$ \citep[see][]{2010JCAP...01..028S,2007JCAP...01..002C,2010AdAst2010E..72C}, with the dimensionless parameter ${\tilde c}_3 (c_\mathrm{s}^{-2}-1)=2 M_3^4 c_\mathrm{s}^2 /(\dot{H} M_{\rm Pl}^2)$ \citep{2010JCAP...01..028S}. The two EFT shapes can be projected onto the equilateral and orthogonal shapes, with the mean values of the estimators for $f^{\mathrm{equil}}_{\mathrm{NL}}$ and $f^{\mathrm{ortho}}_{\mathrm{NL}} $ expressed in terms of $c_\mathrm{s}$ and $\tilde{c}_3$ as
\begin{align}
\label{meanfNL}
f^{\mathrm{equil}}_{\mathrm{NL}} &=\frac{1-c_\mathrm{s}^2}{c_\mathrm{s}^2} \left [-0.275 - 0.0780 c_{\rm s}^2 - (2/3) \times 0.780 \tilde{c}_3 \right]\, , \nonumber \\
f^{\mathrm{ortho}}_{\mathrm{NL}} &=\frac{1-c_\mathrm{s}^2}{c_\mathrm{s}^2} \left[ 0.0159 - 0.0167 c_{\rm s}^2 - (2/3) \times 0.0167 \tilde{c}_3\right] \, ,
\end{align}
where the coefficients come from the Fisher matrix between the theoretical bispectra predicted by the two operators $\dot{\pi} (\nabla \pi)^2$ and $\dot{\pi}^3$ and the equilateral and orthogonal templates.
Notice that DBI models reduce to the condition $\tilde{c}_3= 3(1-c_\mathrm{s}^2)/2$, while the non-interacting (vanishing NG) case corresponds to 
$c_\mathrm{s}=1$ and $M_3=0$ (or ${\tilde c}_3 (c_\mathrm{s}^{-2}-1)=0)$. 

\begin{figure}[htbp!]
\includegraphics[width=0.9\hsize]{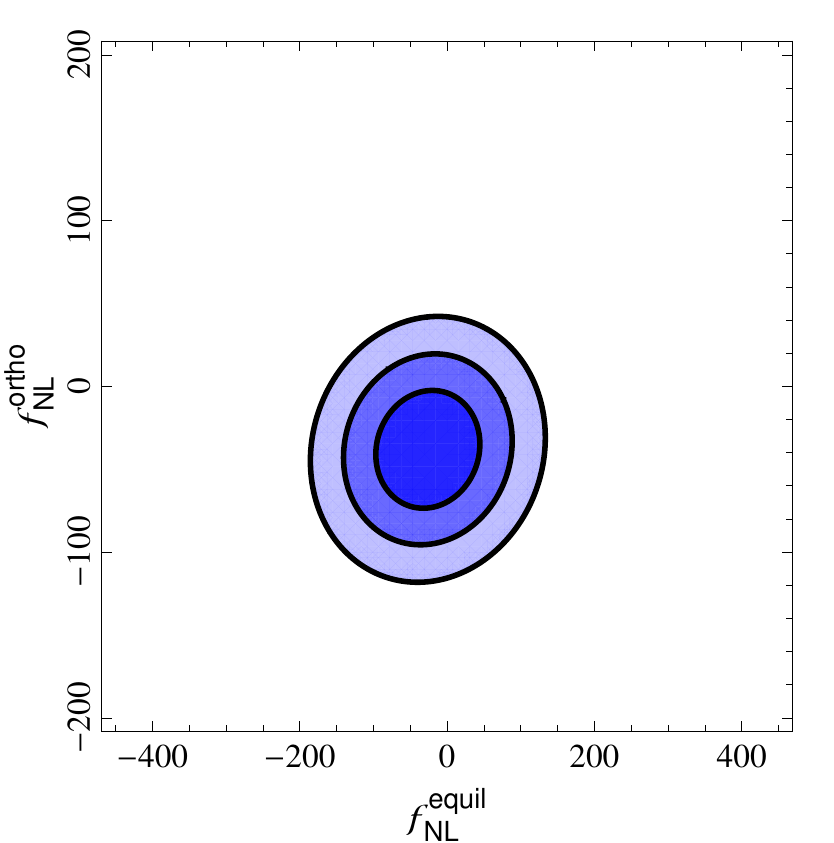}
\caption{68\,\%, 95\,\%, and 99.7\,\% confidence regions in the parameter space $(f_\mathrm{NL}^\mathrm{equil}, f_\mathrm{NL}^\mathrm{ortho})$, defined by thresholding $\chi^2$, as described in the text.}
\label{fig:eq_ort}
\end{figure}

\begin{figure}[htbp!]
\includegraphics[width=0.9\hsize]{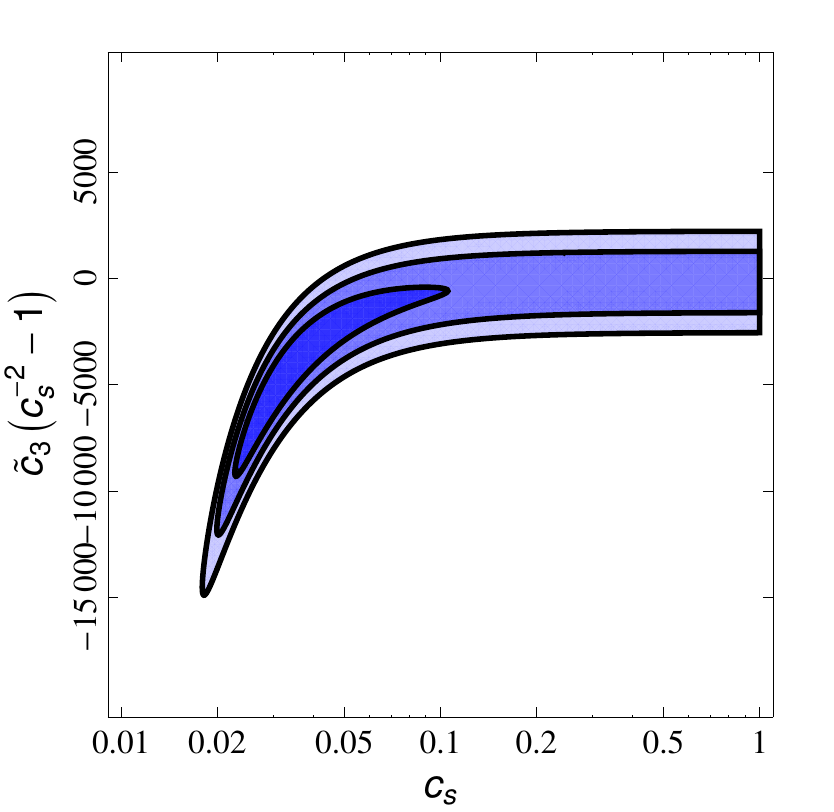}
\caption{68\,\%, 95\,\%, and 99.7\,\% confidence regions in the single-field inflation parameter space $(c_\mathrm{s}, \tilde{c}_3)$, obtained from Fig.~\ref{fig:eq_ort} via the change of variables in Eq.~(\ref{meanfNL}).}
\label{fig:cs_c3}
\end{figure}

We then proceed as in the two previous analyses \citepalias{planck2013-p09a,planck2014-a19}. We employ a 
$\chi^2$ statistic computed as $\chi^2(\tilde{c}_3,c_\mathrm{s})={\vec v}^{\sf T}(\tilde{c}_3,c_\mathrm{s}) \tens{C}^{-1} {\vec v}(\tilde{c}_3,c_\mathrm{s})$, with $v^i(\tilde{c}_3,c_\mathrm{s})=f^i(\tilde{c}_3,c_\mathrm{s})-f^i_{P}$ ($i$=\{equilateral, orthogonal\}), where $f^i_{P}$ are the joint estimates of equilateral and orthogonal $f_{\rm NL}$ values (see Table~\ref{Tab_KSW+SMICA}), while $f^i(\tilde{c}_3,c_\mathrm{s})$ are provided by Eq.~(\ref{meanfNL}) and $\tens{C}$ is the covariance matrix of the joint estimators. Figure~\ref{fig:eq_ort} shows the 68\,\%, 95\,\%, and 99.7\,\% confidence regions for $f^{\rm equil}_{\rm NL}$ and $f^{\rm ortho}_{\rm NL}$, as derived from
from the $T+E$ constraints, with the requirement $\chi^2 \leq 2.28$, 5.99, and 11.62, respectively (corresponding to a $\chi^2$ variable with two degrees of freedom). In Fig.~\ref{fig:cs_c3} we show the corresponding confidence regions in the $(\tilde{c}_3,c_\mathrm{s})$ parameter space.
Marginalizing over $\tilde{c}_3$ we find 
\begin{align}
c_\mathrm{s} \geq\ & 0.021 \quad\quad \text{(95\,\%, \itT\ only)}\, ,\\
\noalign{\noindent and}
c_\mathrm{s} \geq\ & 0.021 \quad\quad \text{(95\,\%, \textit{T+E})}\, .
\end{align}
There is a slight improvement in comparison with the constraints obtained in \citetalias{planck2014-a19} coming from the $T+E$ data. 


\subsection{Multi-field models}
\label{multifield}
Constraints on primordial NG of the local type lead to strong implications for models of inflation where scalar fields (different from the inflaton) are dynamically important for the generation of the primordial curvature perturbation. In the following we test two scenarios for curvaton models. 

\paragraph{Basic curvaton models}
The simplest adiabatic curvaton models predict primordial NG of the local shape with a nonlinearity parameter \citep{2004PhRvD..69d3503B,2004PhRvL..93w1301B}
\begin{equation}
\label{fNLcurv}
f_\mathrm{NL}^\mathrm{local} = \frac{5}{4r_{\rm D}} - \frac{5 r_{\rm D}}{6} - \frac{5}{3}\, ,
\end{equation}
in the case where the curvaton field has a quadratic potential (\citealt{2002PhLB..524....5L,2003PhRvD..67b3503L,2005PhRvL..95l1302L,2006JCAP...09..008M,2006PhRvD..74j3003S}). Here the parameter $r_{\rm D}=[3\rho_{\rm curvaton}/(3 \rho_{\rm curvaton}+4\rho_{\rm radiation})]_{\rm D}$ is the ``curvaton decay fraction'' at the time of the curvaton decay in the sudden decay approximation. We assume a uniform prior, $0<r_\mathrm{D}<1$. It is worth recalling that these models predict a lower bound for the level of NG, of the order of unity (corresponding precisely to $f_\mathrm{NL}^\mathrm{local} = -5/4$), which is considered as a typical threshold to distinguish between standard single-field and multi-field scenarios. 
Our constraint $f_{\rm NL}^{\rm local}=- 0.5 \pm 5.6$ at 68\,\%~CL (see Table~\ref{Tab_KSW+SMICA}) implies 
\begin{equation}
r_{\rm D} \geq 0.19 \quad\quad \text{(95\,\%, \itT\ only)}\, .
\end{equation}
The constraint $f_{\rm NL}^{\rm local}=-0.9 \pm 5.1$ at 68\,\%~CL obtained from temperature and polarization data yields the constraint
\begin{equation}
r_{\rm D} \geq 0.21 \quad\quad \text{(95\,\%, \textit{T +E})}\, . 
\end{equation}
These limits indicate that in such scenarios the curvaton field has a non-negligible energy density when it decays. 
Meaningful improvements are achieved with respect to previous bounds in \citetalias{planck2014-a19}, namely an almost 20\,\% improvement from $T$-only data and a 10\,\% improvement when including $E$-mode polarization. 

\paragraph{Decay into curvaton particles}
We reach similar improvements on the parameters of the second scenario of the curvaton models we consider. In this case one accounts for the possibility that the inflaton field can decay into curvaton particles \citep{2006JCAP...04..009L}, a possibility that is neglected in the above expression (Eq.~\ref{fNLcurv}) for $f_{\rm NL}^\mathrm{local}$. It might be the case that the classical curvaton field survives and begins to dominate. In this case also the curvaton particles produced during reheating are expected to survive and dominate over other species at the epoch of their decay (since thay have the same equation of state as the classical curvaton field). Primordial adiabatic perturbations are generated given that 
the classical curvaton field and the curvaton particles decay at the same time (see \citealt{2006JCAP...04..009L} for a detailed discussion). 
To interpret $\fnl$ in this scenario we employ the general formula for $f_\mathrm{NL}^\mathrm{local}$ derived in \cite{2006PhRvD..74j3003S}, which takes into account the possibility that the inflaton field decays into curvaton particles:
\label{fNLcurvextended}
\begin{equation}
f_\mathrm{NL}^\mathrm{local} = (1+\Delta_{\rm s}^2) \frac{5}{4r_{\rm D}} - \frac{5 r_{\rm D}}{6} - \frac{5}{3}\, .
\end{equation}
Here the parameter $\Delta_{\rm s}^2$ is the ratio of the energy density of curvaton particles to the energy density of the classical curvaton field \citep{2006JCAP...04..009L,2006PhRvD..74j3003S}, while now $\rho_{\rm curv}$ in the expression for $r_{\rm D}$ must be replaced by the sum of the densities of the curvaton particles and curvaton field.  As in \citetalias{planck2014-a19} we use uniform priors $0<r_\mathrm{D}<1$ and $0 < \Delta_{\rm s}^2 <10^2$. Our limits on $f_{\rm NL}^{\rm local}$ constrain 
\begin{align}
\Delta_{\rm s}^2 \leq\ & 6.9 \quad\quad \text{(95\,\%, \itT\ only)}\, ,\\
\noalign{\noindent and}
\Delta_{\rm s}^2 \leq\ & 6.2 \quad\quad \text{(95\,\%, \textit{T +E})}\, , 
\end{align}
which does not exclude a contribution of curvaton particles comparable to the one from the classical curvaton field. 


\subsection{Non-standard inflation models} \label{subsec:nonstand}

\paragraph{Directional-dependent NG}
Table~\ref{tab:dirdep} shows the constraints on directionally-dependent bispectra (Eq.~\ref{vectorBis}). This kind of NG 
is predicted by several different inflationary models. For example, it is a robust and (almost unavoidable) outcome of models of inflation where scale-invariant 
gauge fields are present during inflation. As summarized in Sect.~\ref{sec:models} they are also produced from partially massless higher-spin particles \citep{2018PhRvD..98d3533F} or from models of solid inflation \citep{2012arXiv1210.0569E,2014PhRvD..90f3506E,Shiraishi:2013vja}, as well as in models of inflation that break both rotational and parity invariance \citep{2015JCAP...07..039B}. To compare with the constraints obtained in the analysis of the 2015 \Planck\ data \citepalias{planck2014-a19}, we reconsider the specific model where the inflaton is coupled to the kinetic term $F^2$ of a gauge field via a term $\mathcal{L}\,{=}\,-I^2(\phi) F^2/4$, where $I(\phi)$ is a function that depends on the inflaton field, having an appropriate time evolution during inflation (see, e.g., \citealt{1992ApJ...391L...1R}). Specifically in these models the production of super-horizon vector field perturbations switches on the $L=0$ and $L=2$ modes in the bispectrum, with nonlinearity parameters $f_{\rm NL}^L=X_L (|g_*|/0.1)\, (N_{k_{3}}/60)$, 
with $X_{L=0}=(80/3)$ and $X_{L=2}=-(10/6)$, respectively \citep{Barnaby:2012tk,Bartolo:2012sd,Shiraishi:2013vja}. 
In these expressions $g_*$ is a parameter that measures the amplitude of a quadrupolar 
anisotropy in the power spectrum (see, e.g., \citealt{2007PhRvD..75h3502A}), while $N$ is the number of e-folds (from the 
the end of inflation) at which the relevant scales cross outside the Hubble scale. 
It is therefore interesting to set some limits on the parameter $g_*$ exploiting the constraints from primordial NG of this type. 
Using the \SMICA\ constraints from \itT\ (or \textit{T+E}) in Table~\ref{tab:dirdep}, marginalizing over a uniform prior $50 \leq N \leq 70$, and 
assuming uniform priors on $-1 \leq g_* \leq 1$, we obtain the 95\,\% bounds
$- 0.041 < g_* < 0.041 $ ($- 0.036 < g_* < 0.036$), and $- 0.35 < g_* < 0.35$ ($- 0.30 < g_* < 0.30$), from the $L=0$ and $L=2$ modes, respectively (considering $g_*$ to be scale independent).

\paragraph{Tensor NG and pseudoscalars}
Using the \SMICA\ {\it T+E} result $f_{\rm NL}^{\rm tens} = (8 \pm 11) \times 10^2$ (68\,\%~CL), we here place constraints on two specific inflation models, including either a U(1)-axion coupling or an SU(2)-axion one. The former U(1) model results in $f_{\rm NL}^{\rm tens} \approx 6.4 \times 10^{11} {\cal P}^3 \epsilon^3 e^{6\pi\xi} / \xi^9$, where ${\cal P}$ is the vacuum-mode curvature power spectrum, $\epsilon$ is a slow-roll parameter of the inflaton field, and $\xi$ expresses the strength of the U(1)-axion coupling \citep{2013JCAP...11..047C,2013JCAP...11..051S}. We then fix $\epsilon$ to be $0.01$ and marginalize ${\cal P}$ with the prior, $1.5 \times 10^{-9} < {\cal P} < 3.0 \times 10^{-9}$; assuming a prior, $0.1 < \xi < 7.0$, the upper bound on $\xi$ is derived as $\xi < 3.3$ (95\,\%~CL). In the latter SU(2) model, under one specific condition, the tensor nonlinear parameter is related to the tensor-to-scalar ratio $r$ and the energy density fraction of the gauge field $\Omega_{\rm A}$ as $f_{\rm NL}^{\rm tens} \approx 2.5 r^2 / \Omega_{\rm A}$ \citep{2018PhRvD..97j3526A}. The lower bound on $\Omega_{\rm A}$ is estimated under a prior, $0 < \Omega_{\rm A} < 1$. We find $\Omega_{\rm A} > 2.3 \times 10^{-7}$ and $2.7 \times 10^{-9}$ (95\,\%~CL) for $r = 10^{-2}$ and $10^{-3}$, respectively. 

\paragraph{Warm inflation}
As in previous analyses \citepalias{planck2013-p09a,planck2014-a19}, we adopt the expression $f_\mathrm{NL}^\mathrm{warm} = -15 \ln \left(1 + r_\mathrm{d}/14 \right) - 5/2$ \citep{2007JCAP...04..007M}. This is valid when dissipative effects are strong, i.e., for values $r_\mathrm{d} \gtrsim 2.5$ of the dissipation parameter $r_\mathrm{d} = \Gamma/(3H)$ (measuring the effectiveness of the energy transfer from the inflaton field to radiation).\footnote{The intermediate and weak dissipative regimes $(r_{\rm d} \leq 1)$ predict an NG amplitude with a strong dependence on the microscopic parameters ($T/H$ and $r_{\rm d}$), giving rise to a different additional bispectrum shape (see \citealt{2014JCAP...12..008B}).
} 
Assuming a constant prior $0 \leq \log_{10} r_\mathrm{d} \leq 4$, the \SMICA\ constraints $f_\mathrm{NL}^\mathrm{warmS} =
-39 \pm 44$
(at 68\,\%~CL) from \itT\
 and $f_\mathrm{NL}^\mathrm{warmS} =
 -48\pm 27$
 from \textit{T+E} (see Table~\ref{tab:fnlnonstandard}), yield $\log_{10} r_\mathrm{d} \leq 3.6$ and $\log_{10} r_\mathrm{d} \leq 3.5$, respectively, at 95 \,\%~CL. The results show that strong-dissipative effects in warm inflation models remain allowed. This is a regime where gravitino overproduction problems can be evaded \citep[for a discussion see][]{2008JCAP...01..027H}.

\subsection{Alternatives to inflation}
As an example we update the constraints on some ekpyrotic/cyclic models (e.g., see \citealt{2010AdAst2010E..67L} for a review). Typically local NG is produced through a conversion of ``intrinsic'' non-Gaussianity in the entropy fluctuations into the curvature perturbation. This conversion can proceed in different ways. The ``ekpyrotic conversion'' 
models, for which the conversion acts during the ekpyrotic phase, have already been ruled out \citep[][\citetalias{planck2013-p09a}]{Koyama:2007if}. 
On the other hand, in the ``kinetic conversion'' models the conversion takes place after the ekpyrotic phase and a local bispectrum is generated with an amplitude $f_{\rm NL}^\mathrm{local} = (3/2) \, \kappa_3 \sqrt{\epsilon} \pm 5$.\footnote{
There might also be the case where the intrinsic NG is vanishing and primordial NG is generated only by nonlinearities in the conversion process, reaching an amplitude $f_{\rm NL}^\mathrm{local} \approx \pm 5$ \citep{Qiu:2013eoa,Li:2013hga,Fertig:2013kwa}.
}
 The sign depends on the details of the conversion process \citep{Lehners:2007wc,2010AdAst2010E..67L,Lehners:2013cka}, and typical values of the parameter $\epsilon$ are $\epsilon \approx 50$ or greater. Assuming $\epsilon \approx 100$ and using a uniform prior on $-5 < \kappa_3 <5$ the constraints on $f_{\rm NL}^\mathrm{local}$ from \itT\ only (see Table~\ref{Tab_KSW+SMICA}), implies $-1.1 < \kappa_3 < 0.36$ and $-0.43 < \kappa_3 < 1.0$ at 95\,\%~CL, for the plus and minus sign in $f_{\rm NL}^\mathrm{local}$, respectively. The \textit{T+E} constraints on $f_{\rm NL}^\mathrm{local}$ (Table~\ref{Tab_KSW+SMICA}) yield $-1.05 < \kappa_3 < 0.27$ and $-0.38< \kappa_3 < 0.94$ at 95\,\%~CL, for the plus and minus sign, respectively.
If we take $\epsilon \approx 50$ as an example, we obtain the following limits: $-1.6 < \kappa_3 < 0.51$ and $-0.62 < \kappa_3 < 1.5$ at 95\,\%~CL from \itT\ only; and $-1.5 < \kappa_3 < 0.39$ and $-0.54 < \kappa_3 < 1.3$ at 95\,\%~CL from \textit{T+E} constraints.


\subsection{Inflationary interpretation of CMB trispectrum results}
\label{ssec:inflationary_trispectrum}

We briefly analyze inflationary implications of the \Planck\ trispectrum constraints,
using the \SMICA\ limits on $g_{NL}$ parameters from Table~\ref{tab:gnl}.

First, we consider single-field inflationary models, using the
effective action for the Goldstone boson $\pi$ \citep[see, e.g.,][]{long_trispectrum}:
\begin{align}
S_{\pi} =& \int d^4 x \sqrt{- g}\, \bigg\{ -M^2_{\rm Pl}\dot{H} \left(\partial_\mu \pi\right)^2 \nn \\
& \qquad
 + 2 M^4_2 \left[\dot\pi^2+\dot{\pi}^3-\dot\pi\frac{(\partial_i\pi)^2}{a^2}+(\partial_\mu\pi)^2(\partial_\nu\pi)^2 \nn \right] \nn \\
& \qquad
 - \frac{M_3^4}{3!} \left[ 8\,\dot{\pi}^3+12 \dot\pi^2(\partial_\mu\pi)^2+\cdots \right] \nn \\
& \qquad
 + \frac{M_4^4}{4!} \left[ 16\,\dot\pi^4+32\dot\pi^3(\partial_\mu\pi)^2 + \cdots \right]+\cdots \bigg\},
\end{align}
The mass scale $M_4$ is related to our previously-defined $\gnl$ parameters by:
\be
\gnldotpi4 = \frac{25}{288} \frac{M_4^4}{H^4}\, A_\zeta \, c_{\rm s}^3 \ .
\ee
Therefore, using the $\gnldotpi4$ limit from Table~\ref{tab:gnl}, we get the following constraint on single-field models:
\be
-12.8 \times 10^{14} < \frac{M_4^4}{H^4 c_{\rm s}^3} < 8.2 \times 10^{14}\hspace{1cm} \mbox{(95\,\%~CL).}
\ee
Next consider the case of multifield inflation.
Here, we consider an action of the more general form:
\begin{align}
S_\sigma =& \int d^4x\, \sqrt{-g} \,\, \bigg[ \frac{1}{2} (\partial_\mu\sigma)^2 
 + \frac{1}{\Lambda^4_1} \dot\sigma^4 \nn \\
& \hspace{1cm}
 + \frac{1}{\Lambda^4_2} \dot\sigma^2 (\partial_i\sigma)^2
 + \frac{1}{\Lambda^4_3} (\partial_i\sigma)^2 (\partial_j\sigma)^2
\bigg], \label{eq:S_multi}
\end{align}
where $\sigma$ is a light field that acquires quantum fluctuations
with power spectrum $P_\sigma(k) = H^2/(2k^3)$.
We assume that $\sigma$ converts to adiabatic curvature $\zeta$,
i.e.~$\zeta = (2 A_\zeta)^{1/2} H^{-1} \sigma$.
The model parameters $\Lambda_i$ are related to our previously-defined
$\gnl$ parameters by:
\begin{align}
\gnldotpi4 A_\zeta
 =\ & \frac{25}{768}\frac{H^4}{\Lambda_1^4}, \nn \\
\gnlB A_\zeta
 =\ & -\frac{325}{6912}\frac{H^4}{\Lambda_2^4},  \label{eq:gnl_lambda} \\
\gnldpi4 A_\zeta 
 =\ & \frac{2575}{20736} \frac{H^4}{\Lambda_3^4}, \nn
\end{align}
and can be constrained by thresholding the $\chi^2$-statistic
defined in Eq.~(\ref{eq:trispectrum_chi2}).
For example, to constrain the parameter $\Lambda$ in the Lorentz invariant model
\be
S = \int d^4x\, \sqrt{-g}\,
 \left[ \frac{1}{2} (\partial_\mu \sigma)^2 + \frac{1}{\Lambda^4} (\partial_\mu \sigma)^2 (\partial_\nu \sigma)^2 \right], \label{eq:S_lorentz}
\ee
we set $\Lambda_1^4 = -2 \Lambda_2^4 = \Lambda_3^4 = \Lambda^4$,
obtain $\gnl$-values using Eq.~(\ref{eq:gnl_lambda}),
and use Eq.~(\ref{eq:trispectrum_chi2}) to obtain $\chi^2$
as a function of $\Lambda$.
We then threshold at $\Delta\chi^2 = 4$ (as appropriate for one degree of freedom),
to obtain the following constraint:
\be
-0.33 < \frac{H^4}{\Lambda^4} < 0.11 \quad\quad \mbox{(95\,\%).}
\ee

\paragraph{DBI Trispectrum}
We use the trispectrum constraints on the shape $\dot\sigma^4$ in Table~\ref{tab:gnl} to determine a lower bound on the sound speed of the inflaton field in DBI models. In fact in these models 
the dominant contribution in the small-sound-speed limit \citep{2009JCAP...08..008C,2009PhRvD..80d3527A} to 
the contact interaction trispectrum \citep{2006PhRvD..74l1301H} produces such a 
shape, with an amplitude $g_\mathrm{NL}^{\dot\sigma^4}=-25/(768\, c_\mathrm{s}^4)$. 
We employ the same procedure described at the beginning of this section and, assuming a uniform prior in the range $0\leq c_\mathrm{s}\leq1/5$, we
derive the following constraint on $c_\mathrm{s}$:
\begin{equation} 
c_\mathrm{s}^{\rm DBI} \geq0.015 \quad\quad \text{(95\,\%)}\, . 
\end{equation} 
This constraint is independent from and consistent with the bounds of Eqs.~(\ref{csDBIT}) and (\ref{csDBITE}) obtained from the bispectrum measurements. 
Notice that in the trispectrum case we are ignoring the scalar exchange contribution, which turns out to be of the same order in $c_\mathrm{s}$. 

\paragraph{Curvaton trispectrum} 
A generic prediction of the simplest adiabatic curvaton scenario is also a local-type trispectrum with an amplitude $g_\mathrm{NL}^\mathrm{local}$ 
given by \citep{2006PhRvD..74j3003S}
\begin{equation} 
g_\mathrm{NL}^\mathrm{local}=\frac{25}{54}\left(-\frac{9}{r_{\rm D}}+\frac{1}{2}+10r_{\rm D}+3r_{\rm D}^2\right). 
\end{equation} 
Following the procedure described at the beginning of this section, we use the observational constraint obtained in Sect.~\ref{sec:tau_gnl} (see Table~\ref{tab:gnl}), and the 
same uniform prior ($0<r_\mathrm{D}<1$) as in Sect.~\ref{multifield}, to obtain a lower bound on the curvaton decay fraction 
\begin{equation} 
r_{\rm D} \geq0.05 \quad\quad \text{(95\,\%)}\,. 
\end{equation} 
This limit is consistent with the previous ones derived using the bispectrum 
measurements and it is about a factor of 4 weaker.

\section{Conclusions}
\label{sec:Conc}

In this paper, we have presented constraints on primordial NG, using
the \Planck\ full-mission CMB
temperature and \itE-mode polarization maps. 
Compared to the \Planck\ 2015 release we now include the low-$\ell$
($4 \leq \ell < 40$) polarization multipole range.

Our analysis produces the following final results
(68\,\%~CL, statistical): $f_{\rm NL}^{\rm local} = -0.9 \pm 5.1$; $f_{\rm NL}^{\rm equil} = -26 \pm 47$;
and $f_{\rm NL}^{\rm ortho} = - 38 \pm 24$.
These results are overall stable with respect to our
constraints from the 2015 \Planck\ data. They show no real improvement in errors, despite the
additional polarization modes. This is due to a combination of two
factors. Firstly, the local shape, which is most sensitive to
low-$\ell$ modes and where one would naively expect an improvement, is
actually less sensitive to polarization than the equilateral and orthogonal
shapes. This means that in the end none of the three shapes are very sensitive to
low-$\ell$ polarization modes. Secondly, the polarization simulations used to
determine the errors have a more realistic but slightly higher noise level than in
the previous release.

On the other hand, the quality of polarization data shows 
a clear improvement with respect to our previous analysis. This is confirmed by a large 
battery of tests on our data set, including comparisons between different estimator implementations (KSW, 
Binned, and two Modal estimators) and foreground-cleaning methods (\SMICA, \SEVEM, \NILC, and \Commander), 
studies of robustness under changes in sky coverage and multipole range, and an analysis of the 
impact of noise-related systematics.
While in our previous release we cautioned the reader to take polarization bispectra and related constraints 
as preliminary, in light of these tests we now consider our results based on the combined temperature and polarization data set to be fully reliable.
This also implies that polarization-only, \itE\itE\itE\ bispectra can now be used for independent tests, leading to primordial NG constraints at a sensitivity level
comparable to that of WMAP from temperature bispectra, and 
yielding statistical agreement.

As in the previous analyses, we go beyond the local, equilateral, and orthogonal 
$f_{\rm NL}$ constraints by considering a large number of additional
cases, such as scale-dependent feature and resonance bispectra, running
$f_{\rm NL}$ models, 
isocurvature primordial NG, and parity-breaking models. We set tight
constraints on all these scenarios, but do not detect any significant signals. 

On the other hand, the non-primordial lensing bispectrum
is now detected with an improved significance compared to 2015,
excluding the null hypothesis at $3.5\,\sigma$. The amplitude of the signal is
consistent with the expectation from the \Planck\ best-fit cosmological
parameters, further indicating the absence of significant foreground
contamination or spurious systematic effects.
We also explicitly checked for the presence of various non-primordial
contaminants, like unclustered extragalactic point sources, CIB, Galactic
thermal dust, and the thermal SZ effect, but apart from the first, none of these
were detected. The small amount of remaining point-source signal in the
cleaned maps has no impact on our other constraints because of its
negligible correlations.

We update our trispectrum constraints, now finding $g_{\rm NL}^{\rm local} = (-5.8 \pm 6.5) \times 10^4$ (68\,\%~CL, statistical),
while also constraining additional shapes, generated by different operators in an effective field-theory approach to inflation.

In addition to estimates of bispectrum and trispectrum amplitudes, we produce
model-independent reconstructions and analyses of the \Planck\ CMB
bispectrum. Finally, we use our measurements to obtain constraints on early-Universe scenarios 
that can generate primordial NG. We consider, for example, general single-field models of inflation, curvaton models,
models with axion fields producing parity-violating tensor bispectra, and inflationary scenarios generating directionally-dependent bispectra 
(such as those involving vector fields). 

In our data analysis efforts, which started with the 2013 release, we achieved a number of crucial scientific goals. In particular we reached an unprecedented level of sensitivity in the determination of the bispectrum and trispectrum amplitude parameters ($f_{\rm NL}$, $g_{\rm NL}$) and significantly extended the standard local, equilateral, and orthogonal analysis, encompassing a large number of additional shapes motivated by a variety of inflationary models. Moreover, we produced the first polarization-based CMB bispectrum constraints and the first detection of the (non-primordial) bispectrum induced by correlations between CMB lensing and secondary anisotropies.  
Our stringent tests of many types of non-Gaussianity are fully consistent with expectations from the standard 
single-field slow-roll paradigm and provide strong constraints on alternative scenarios. 
Nevertheless, the current level of sensitivity does not allow us to rule out or confirm most alternative scenarios. It is natural at this stage to ask ourselves what should be the $f_{\rm NL}$ sensitivity goal for future cosmological experiments. A number of studies has identified $f_{\rm NL} \sim 1$ as a target.  Achieving such sensitivity for local-type NG would enable us to either confirm or rule out a large class of multi-field models. 
 A similar target for equilateral, orthogonal, and scale-dependent shapes would allow us to distinguish standard slow-roll from more complex single-field scenarios, such as those characterized by higher-derivative kinetic terms or slow-roll-breaking features in the inflaton potential \citep[see e.g.,][and references therein]{2014arXiv1412.4671A,2018JCAP...04..016F}.
With this aim in mind,
the challenge for future cosmological observations will be therefore that of reducing the $f_{\rm NL}$ errors from this paper
by at least one order of magnitude.


\begin{acknowledgements}

The Planck Collaboration acknowledges the support of: ESA; CNES and CNRS/INSU-IN2P3-INP (France); ASI, CNR, and INAF (Italy); NASA and DoE (USA); STFC and UKSA (UK); CSIC, MINECO, JA, and RES (Spain); Tekes, AoF, and CSC (Finland); DLR and MPG (Germany); CSA (Canada); DTU Space (Denmark); SER/SSO (Switzerland); RCN (Norway); SFI (Ireland); FCT/MCTES (Portugal); and ERC and PRACE (EU). A description of the Planck Collaboration and a list of its members, indicating which technical or scientific activities they have been involved in, can be found at \url{http://www.cosmos.esa.int/web/planck}.
Some of the results in this paper have been derived using the {\tt{HEALPix}} package. We also acknowledge the use of CAMB.
Part of this work was undertaken on the STFC COSMOS@DiRAC HPC Facility at the University of Cambridge, funded by UK BIS NEI grants.
We gratefully acknowledge the IN2P3 Computer Center (\url{http://cc.in2p3.fr}) for
providing a significant amount of the computing resources and services needed 
for the analysis. This research used resources of the National Energy Research
Scientific Computing Center, a DOE Office of Science User Facility
supported by the Office of Science of the U.S. Department of Energy
under Contract No. DE-AC02-05CH11231.
Some computations were performed on the GPC and Niagara cluster at the SciNet HPC
Consortium; SciNet is funded by the Canada Foundation for Innovation under
the auspices of Compute Canada, the Government of Ontario, and the
University of Toronto.

 \end{acknowledgements}

\bibliographystyle{aat}
\bibliography{L09_bib,Planck_bib}


\raggedright
\end{document}